\pdfoutput=1
\documentclass[11pt]{article}
\usepackage[T1]{fontenc}
\usepackage[utf8]{inputenc}
\usepackage{lmodern}
\usepackage[margin=1in]{geometry}
\usepackage{amsmath,amssymb,amsthm}
\usepackage{mathtools}
\usepackage[round,authoryear]{natbib}
\usepackage{graphicx}
\usepackage{booktabs}
\usepackage{adjustbox}
\usepackage{xcolor}
\usepackage[ruled]{algorithm}
\usepackage[colorlinks=true,linkcolor=blue!50!black,citecolor=blue!50!black,urlcolor=blue!50!black]{hyperref}
\graphicspath{{figs/}}

\theoremstyle{plain}
\newtheorem{theorem}{Theorem}[section]
\newtheorem{lemma}[theorem]{Lemma}
\newtheorem{proposition}[theorem]{Proposition}
\newtheorem{corollary}[theorem]{Corollary}
\newtheorem{openproblem}[theorem]{Open problem}
\theoremstyle{definition}

\theoremstyle{remark}
\newtheorem{remark}{Remark}[section]

\newcommand{\R}{\mathbb{R}}
\newcommand{\E}{\mathbb{E}}
\newcommand{\Pcal}{\mathcal{P}}
\newcommand{\Ecal}{\mathcal{E}}
\newcommand{\Rcal}{\mathcal{R}}
\newcommand{\Wcal}{\mathcal{W}}
\newcommand{\Tcal}{\mathcal{T}}
\newcommand{\Ncal}{\mathcal{N}}
\newcommand{\Var}{\operatorname{Var}}
\newcommand{\KL}{\operatorname{KL}}
\newcommand{\ID}{\operatorname{ID}}
\newcommand{\osc}{\operatorname{osc}}
\newcommand{\gap}{\operatorname{gap}}
\newcommand{\Bin}{\operatorname{Bin}}
\newcommand{\VA}{\textup{VA}}
\newcommand{\rcut}{r_{\rm cut}}
\newcommand{\Lo}{\mathrm{Lo}}

\title{Variance-adaptive confidence regions for fixed-benchmark binomial mixtures, with application to LLM evaluation}
\author{Kuancheng Ye\\ Department of Statistics, University of California, Irvine\\ \texttt{kuanchey@uci.edu}}
\date{}

\input{xref-supplement}

\begin{document}
\sloppy
\maketitle

\begin{abstract}
We construct confidence intervals for linear functionals $\theta(F_m)=\int h\,dF_m$ of the empirical law $F_m$ of success probabilities $p_1,\dots,p_m$, from independent counts $c_i\sim\Bin(n,p_i)$, as for a language model run $n$ times per problem. Coverage is at least $1-\alpha$ for every $n$, $m$ and fixed $(p_i)$, simultaneously for all bounded $h$. The variance-adaptive (VA) region tests each candidate mixing law with the conditional variance of the tail counts it implies. This variance is linear in the law, so the region is a linear program, and cell envelopes keep it valid off-grid. Regions calibrated for iid sampling target a super-population law and are conservative for $\theta(F_m)$. With all $n$ tail counts, fixed $n$, continuous $h$ and a grid refined as $m\to\infty$, the identification part of VA's width is first-order worst-case optimal, as for any localization with the same index set. Under a rule fixed before the main comparator runs, VA was narrower than a two-stage plug-in band at every $n$ (median width ratio 0.60 at $n=128$), but not than the asymptotic, pointwise AMARI interval at $n=128$; on real rollout pools it was wider than both pointwise and Bonferroni-adjusted AMARI at $n\le128$. For $P(p>0)$ and pass@$k$ with $k\gg n$ the data support only lower bounds.
\end{abstract}

\noindent\textbf{Keywords:} binomial mixture; empirical Bayes; compound decision; conditional coverage; finite population; Bernstein inequality; partial identification; pass@$k$.

\section{Introduction}\label{sec:intro}

\subsection{Motivation}
Evaluations of large language models (LLMs) now routinely draw many rollouts per problem. On a fixed benchmark of $m$ problems, a model is run $n$ times on problem $i$ and $c_i$ of the $n$ answers are correct. \citet{brown2024large} report coverage curves from up to 10{,}000 samples per problem, \citet{schaeffer2025large} model how such curves scale, and \citet{yue2025does} compare base models with models trained from them by reinforcement learning with verifiable rewards (RLVR) through pass@$k$ curves up to $k=1024$, where pass@$k$ is the probability that at least one of $k$ samples is correct. Questions about a model's ``reasoning boundary'' concern the problem-level success probabilities $p_i$: the share of problems the model can ever solve, $P(p>0)$, the share it solves at least half the time, $P(p\ge0.5)$, or pass@$k_0$ for $k_0$ larger than the number of rollouts.

These targets differ in what $n$ rollouts can say about them. A share such as $P(p\ge0.5)$ can be bracketed closely once $n$ is moderate. For $P(p>0)$ and pass@$k$ with $k\gg n$ the counts give an informative lower bound, but the upper bound is close to $1$ whenever some problems are never solved, because such a problem may still have a small positive $p_i$ (Section~\ref{sec:ident}). We treat these targets as partially identified and report the bounds the data support.

Two features of the setting shape the statistics. First, the benchmark is fixed. For questions about the $m$ problems at hand, such as comparing two models on the same benchmark, the target is the empirical law $F_m=m^{-1}\sum_i\delta_{p_i}$, and coverage should hold conditionally on $(p_i)$ (Section~\ref{sec:estimand}). Second, success probabilities pile up near $0$ and $1$, where the counts are nearly deterministic, and a method calibrated as if the $p_i$ were drawn afresh does not use that information. \citet{zhou2026repeated} observe this for pass@1 in repeated-run code evaluation: the pooled binomial interval is about twice too wide for the fixed problem set and about twice too narrow for new problems.

\subsection{The F-localization template and the key observation}
\citet{ignatiadis2022confidence} build a confidence set for the marginal law of the observations (an F-localization) and optimise the target over all mixing laws whose marginal lies in it; on a grid this is a linear program (LP), and the interval covers every linear functional at once. For binomial counts the recipe goes back to \citet{lord1975empirical}. Ignatiadis and Wager state that ``how one should choose the F-localization'' is open \citep[Section~2.1]{ignatiadis2022confidence}, and \citet{brennan2020estimating} suggest that a variance-sensitive band could repair the lack of power of DKW localizations. This paper gives one answer for binomial counts with fixed success probabilities.

Let $P_k(p)=\Pr(\Bin(n,p)\ge k)$ and $T_k=m^{-1}\sum_i\mathbf 1\{c_i\ge k\}$. Under fixed $p_i$,
\[
\E T_k=\int P_k\,dF_m,\qquad m\Var(T_k)=V_k(F_m):=\int P_k(1-P_k)\,dF_m .
\]
If instead the $p_i$ are drawn from a law $G$, the indicator $\mathbf 1\{c_i\ge k\}$ has the iid variance $q_k(1-q_k)$ with $q_k=\int P_k\,dG$, the right calibration for the super-population target $\theta(G)$. The two differ by the between-problem term $\Var_G(P_k(p))\ge0$ of the law of total variance~\eqref{eq:vargap}, which is most of $q_k(1-q_k)$ when the $p_i$ cluster near $0$ and $1$: on the law fitted to the Monkey Business data of Section~\ref{sec:mb}, the median over $k$ of $V_k/(q_k(1-q_k))$ is 0.058 at $n=128$ (Table~\ref{S-tab:vargap}). Moreover $V_k(G)$ is linear in the candidate $G$, so a band calibrated to the candidate's own conditional variance can be kept inside an LP. Existing localizations adapt to the iid variance (KL and Berk--Jones bands through the curvature of the divergence, Pearson's $\chi^2$ through the multinomial covariance); the new element is the conditional, Poisson-binomial variance of the candidate law.

\subsection{Contributions}
The main property of the method is its guarantee: finite-sample coverage of the fixed-benchmark target, for every vector of success probabilities, simultaneously over all bounded functionals. Width is secondary, and the comparisons are mixed.
\begin{itemize}
\item[(C1)] \emph{A conditional variance-adaptive (VA) region and its coverage.} The region requires $|T_k-\mu_k(G)|\le r(V_k(G))$ for $k$ in a fixed index set, with a Bernstein or Bennett radius $r$. Cell envelopes, tangent cuts and two constraints that hold on every realisation make it an LP whose solution covers $\theta(F_m)$ with probability at least $1-\alpha$ for every $n$, $m$ and $(p_i)$, simultaneously for all bounded $h$, also when the partition, the tangent set and the targets depend on the data (Theorem~\ref{thm:coverage}, Algorithm~\ref{alg:va}).
\item[(C2)] \emph{Iid-calibrated and plug-in regions under fixed $p_i$.} The per-$k$ KL, iid-variance Bernstein, exact-binomial, Hoeffding and empirical-Bernstein bands remain valid for $\theta(F_m)$ but are conservative by the between-problem term (Proposition~\ref{prop:iid}). A two-stage band that bounds $V_k(F_m)$ from above by an unbiased statistic is also valid, and its width bound has the same order as VA's (Proposition~\ref{prop:twostage}).
\item[(C3)] \emph{Width theory.} A modulus bound for every bounded target (Theorem~\ref{thm:width}); worst-case optimality of the identification term for fixed $n$ and continuous $h$ when all $n$ tail counts are used and the grid is refined as $m\to\infty$ (Theorem~\ref{thm:id}), a property shared by every localization with the same index set; and, for laws with an atom at $0$ and the rest of the mass above $\delta$, lower bounds on the \emph{realised} widths of iid-calibrated regions (Proposition~\ref{prop:realised}). Our bounds give VA at most a constant-factor gain over these regions for fixed $n$, and a rate in a joint limit with $n$ growing (Corollaries~\ref{cor:sepA} and~\ref{cor:sepB}).
\item[(C4)] \emph{Evidence, judged by a rule fixed in advance.} A simulation study, a coverage check under resampling from the Monkey Business rollout pools and one pair of models (Sections~\ref{sec:sim}--\ref{sec:pair}). Coverage of $\theta(F_m)$ held in every replicate for every finite-sample region, while a plug-in bootstrap, the default in practice, undercovered in 41 of 135 simulation cells. Before the comparator runs we fixed a rule for claiming that VA is narrower than each of three comparators. The median width ratio of VA to the two-stage band was 0.746, 0.602 and 0.436 at $n=32$, 128 and 512, and to AMARI with a Bonferroni correction over three targets 0.808 and 0.513 at $n=128$ and 512. Against pointwise AMARI the rule failed at $n=128$ (0.977), VA was wider at $n=32$ (1.307), and narrower only at $n=512$ (0.626). On the real rollout pools VA was wider than both pointwise and Bonferroni-adjusted AMARI at $n\le128$.
\end{itemize}

\subsection{What we do not claim}
We do not claim that VA is shorter than AMARI, an asymptotic, pointwise interval for one functional at a time. Every width gain we report is for the conditional target $\theta(F_m)$: for the super-population target $\theta(G)$, VA followed by a concentration layer is wider than the best iid-calibrated localization and about twice as wide as AMARI (Table~\ref{tab:Glevel}). We do not claim minimax optimality of the estimation term; a matching lower bound in the conditional model is open (Open problem~\ref{op:estimation}). The illustration concerns one pair of models, RL after distillation, and supports no general statement about RL. The variance identity behind VA is classical; what is new is its use to calibrate each constraint at the candidate law.

\section{Related work}\label{sec:related}

\paragraph{Empirical Bayes intervals, compound decisions and finite populations.} Besides the F-localization and AMARI intervals of \citet{ignatiadis2022confidence}, \citet{armstrong2022robust} give robust empirical Bayes intervals in the normal-means problem with an average-coverage guarantee when the means are fixed; \citet{efron2016empirical}, \citet{chen2025empirical} and \citet{basu2026exact} give deconvolution standard errors, intervals for selectively reported data and exact pointwise intervals for the mixing CDF under iid sampling. For binomial data, \citet{kline2021reasonable} bound functionals of the mixing law and \citet{gu2025reasonable} compare F-localization, AMARI and the test inversion of \citet{fang2023inference}. These methods target the prior and are calibrated under iid sampling, mostly asymptotically. Coverage conditional on $(p_i)$ with target $F_m$ is the compound setting of \citet{robbins1951asymptotically,robbins1956empirical}; see \citet{zhang2003compound}, \citet{jiang2009general} and the compound guarantees of \citet{ignatiadis2024compound,ignatiadis2026compound}. For fixed binomial success probabilities, \citet{tian2017learning} combine per-moment concentration bounds with a union bound and an LP, with error bounds that hold conditionally on the $p_i$; \citet{vinayak2019maximum} study maximum likelihood for such populations, and \citet{jana2020extrapolating} study the related problem of estimating the profile of a finite population observed by subsampling. \citet{mao2007estimating} used a Kolmogorov--Smirnov ball and an LP for a lower confidence limit in a Poisson mixture. Our contribution within this line is a calibration: finite-sample intervals, conditional on the $p_i$ and simultaneous over all bounded functionals, whose localization uses the conditional variance of each candidate law.

\paragraph{Mixing laws and identification.} The nonparametric maximum likelihood estimator \citep{kiefer1956consistency,lindsay1983geometry,koenker2014convex,soloff2025multivariate} is one point of the identified set. Binomial counts carry information about the mixing law only through its first $n$ moments \citep{wood1999binomial}; \citet{singh2026fixed} work out the consequences for pass@$k$. Our identification results use the duality between moment matching and best polynomial approximation of \citet{wu2016minimax,wu2019chebyshev}. One-sided bounds for $P(p>0)$ parallel lower bounds on the proportion of non-null hypotheses \citep{genovese2004stochastic,meinshausen2006estimating}.

\paragraph{Concentration and score inversion.} The VA band inverts a test with the variance evaluated at the null, as the interval of \citet{wilson1927probable} does for one proportion. We use Bernstein and Bennett bounds \citep{bennett1962probability,boucheron2013concentration}. Sharper bounds \citep{bentkus2004hoeffding,pinelis2014bennett} and confidence sequences \citep{kuchibhotla2021near,waudbysmith2024estimating} could replace the Bernstein radius if concave and nondecreasing in the variance. \citet{mineiro2023time} give time-uniform bands for averaged conditional CDFs. A continuously updated GMM criterion \citep{hansen1996finite} also evaluates its weighting matrix at the candidate, but with the iid covariance.

\paragraph{Statistics of LLM evaluation.} \citet{chen2021evaluating} give the unbiased estimator of pass@$k$ for $k\le n$, and \citet{yang2026crossovers} paired bootstrap inference for it. \citet{miller2024adding} take the super-population view of evaluation, \citet{bowyer2025position} show that intervals based on the central limit theorem undercover on small benchmarks, \citet{dragoi2025beyond} propose the share of problems solved with frequency at least $\tau$, in our notation $P(p\ge\tau)$, and \citet{hariri2026dont} report posterior summaries of a model's success probability. Parametric extrapolations of pass@$k$ \citep{schaeffer2025large,kazdan2025efficient} owe their narrow width to a model for the law of the $p_i$. \citet{stroebl2024limits} show that imperfect verifiers cap the gains from resampling, and \citet{liu2025prorl} report that prolonged RL from a distilled model can expand the reasoning boundary.

\section{Setting, targets and estimand}\label{sec:setting}

\subsection{Model and targets}
Fix $p_1,\dots,p_m\in[0,1]$ and observe independent $c_i\sim\Bin(n,p_i)$. All probabilities are conditional on $(p_i)$. Write $F_m=m^{-1}\sum_i\delta_{p_i}$, abbreviated $F$, for the empirical mixing law, a law of success probabilities and not of the observations, and $\Pcal$ for the probability measures on $[0,1]$. The target is $\theta(F)=\int h\,dF$ for a bounded $h$. Our running examples are nondecreasing:
\[
h_\delta(p)=\min(p/\delta,1),\qquad h(p)=\mathbf 1\{p\ge\gamma\},\qquad h(p)=1-(1-p)^{k_0}\ \text{(pass@$k_0$)},
\]
with $k_0>n$ allowed. The functional $\theta_\delta=\int h_\delta\,dF$ is a Lipschitz surrogate for $P(p>0)$ \citep[compare][]{gu2025reasonable}, and it brackets pass@$k$: for $p\in[0,1]$ and integers $k\ge1$,
\begin{equation}\label{eq:passk-bracket}
(1-e^{-1})\,h_{1/k}(p)\ \le\ 1-(1-p)^k\ \le\ h_{1/k}(p),
\end{equation}
since $1-(1-p)^k\le\min(kp,1)$, $1-(1-p)^k\ge1-e^{-kp}\ge(1-e^{-1})kp$ for $kp\le1$, and $1-(1-p)^k\ge1-e^{-1}$ for $kp\ge1$. So $\theta_{0.01}$ is pass@100 up to a factor between $0.63$ and $1$, and $\delta$ can be set to the reciprocal of the largest sampling budget of interest, though not so small that guessing reaches it.

\subsection{Choice of estimand}\label{sec:estimand}
The target $\theta(F_m)$ describes the $m$ problems at hand. It is the scientific target when the benchmark itself is of interest: reporting performance on a fixed, published benchmark, comparing two models on the same problems, or auditing a fixed population such as an item bank or a corpus \citep{Ye2026prevalence}. A repeat of the evaluation would use the same problems, so the spread of the $p_i$ is not sampling noise. The super-population target $\theta(G)$, for $p_i$ drawn independently from $G$, is the target when the problems stand in for a population of tasks \citep{miller2024adding}; questions such as ``does RL narrow the reasoning boundary?'' are usually of this kind. The empirical Bayes intervals of \citet{ignatiadis2022confidence}, AMARI included, are built for $\theta(G)$.

Regions calibrated for iid sampling of the $p_i$ are calibrated for $\theta(G)$; under fixed $p_i$, those with a finite-sample proof (Proposition~\ref{prop:iid}) remain valid for $\theta(F_m)$ but are conservative. All width comparisons in Sections~\ref{sec:sim}--\ref{sec:pair} are for $\theta(F_m)$. To cover $\theta(G)$, a region for $\theta(F_m)$ must be enlarged by a layer that accounts for the draw of the $p_i$. Supplement~\ref{S-supp:glevel} gives three layers: Chernoff (KL) and empirical-Bernstein layers with a Bonferroni split over $J$ targets fixed before the data are seen, and a DKW layer \citep{massart1990tight} of width $\sqrt{\log(2/\alpha_2)/(2m)}$ that holds simultaneously for every monotone target, including targets chosen after seeing the data; simultaneity over all bounded $h$ is lost. A two-step interval pays the between-problem variance with a finite-sample constant (Proposition~\ref{S-prop:layer-floor}), whereas an interval built for $\theta(G)$ pays it once. Table~\ref{tab:Glevel} shows the result: for $\theta(G)$, VA with the KL layer was 1.08, 1.32 and 1.38 times as wide as the best non-AMARI localization at $n=32$, 128 and 512, narrower in only 2 of 45 cells, and 1.98, 2.14 and 2.33 times as wide as AMARI. What VA keeps at the level of $\theta(G)$ is finite-sample validity and, with the DKW layer, simultaneity over monotone targets.

\begin{table}[tbp]
\centering
\footnotesize
\caption{The super-population target $\theta_\delta(G)$, $m=1000$, medians over F1--F5 and three $\delta$ (15 cells per $n$). First column: VA for $\theta_\delta(F_m)$ at level $\alpha$, for reference. ``VA$+$KL'', ``VA$+$EB'', ``VA$+$DKW'': VA at $\alpha/2$ followed by a layer at $\alpha/2$ that covers $\theta(F_m)-\theta(G)$ (Propositions~\ref{S-prop:layer-kl}--\ref{S-prop:layer-dkw}; KL and EB with $J=3$ targets, DKW simultaneous over all monotone targets). ``Best FL'': the narrowest of DKW-Massart, Berk--Jones, KL-Bonf, Hoeff-$k$ and exact-binomial at level $\alpha$; AMARI: Empirikos-faithful, pointwise; oracle: $2z_{0.975}\,\mathrm{sd}_G(h)/\sqrt m$, the interval for an analyst who observes the $p_i$. Last two columns: median ratio of VA$+$KL to the best FL and to AMARI. All comparators were computed on counts drawn under the fixed $F_m$ of the simulation, as a proxy for redrawn $p_i$.}
\label{tab:Glevel}
\begin{adjustbox}{max width=\textwidth}
\begin{tabular}{rcccccccccc}
\toprule
 & $\theta(F_m)$ & \multicolumn{4}{c}{$\theta(G)$: VA-based} & \multicolumn{3}{c}{$\theta(G)$: other} & \multicolumn{2}{c}{VA$+$KL /}\\
\cmidrule(lr){2-2}\cmidrule(lr){3-6}\cmidrule(lr){7-9}\cmidrule(lr){10-11}
$n$ & VA & VA, $\alpha/2$ & VA$+$KL & VA$+$EB & VA$+$DKW & best FL & AMARI & oracle & best FL & AMARI\\
\midrule
32 & 0.149 & 0.153 & 0.253 & 0.293 & 0.246 & 0.241 & 0.123 & 0.057 & 1.08 & 1.98\\
128 & 0.060 & 0.063 & 0.151 & 0.189 & 0.156 & 0.111 & 0.065 & 0.057 & 1.32 & 2.14\\
512 & 0.032 & 0.034 & 0.122 & 0.156 & 0.128 & 0.089 & 0.056 & 0.057 & 1.38 & 2.33\\
\bottomrule
\end{tabular}
\end{adjustbox}
\end{table}

\subsection{Tail counts and the variance gap}
For $k=0,\dots,n+1$, $P_k$ is a nondecreasing polynomial of degree at most $n$, with $P_0\equiv1$ and $P_{n+1}\equiv0$. Put $u(x)=x(1-x)$ and, for $G\in\Pcal$, $\mu_k(G)=\int P_k\,dG$ and $V_k(G)=\int u(P_k)\,dG$. Then $\E T_k=\mu_k(F)$, $\Var(T_k)=V_k(F)/m$, we write $q_k=\mu_k(F)$, and for every $G$
\begin{equation}\label{eq:vargap}
\mu_k(G)\{1-\mu_k(G)\}-V_k(G)=\Var_G\big(P_k(p)\big)\ \ge 0 .
\end{equation}
We fix an index set $K\subseteq\{1,\dots,n\}$ before seeing the data, a level $\alpha\in(0,1)$, and put $L=\log(2|K|/\alpha)$, $a=\sqrt{2L/m}$ and $b=L/(3m)$. In the extreme case where half the problems have $p_i=0$ and half $p_i=1$, $T_k=1/2$ on every data set, so $V_k(F)=0$, while $q_k(1-q_k)=1/4$: a band calibrated with $q_k(1-q_k)$ has radius of order $m^{-1/2}$ for a target with no sampling noise at all. Benchmarks whose problems are nearly always or nearly never solved sit close to this case (Figure~\ref{fig:vargap}).

\subsection{Generalized F-localization}\label{sec:floc}
Under fixed, non-identical $p_i$ there is no single marginal law of the observations to localize. We call any data-dependent set $\Rcal\subseteq\Pcal$ with $\Pr(F_m\in\Rcal)\ge1-\alpha$ a \emph{generalized F-localization}, or a localization for short, and report $[\inf_{\Rcal}\theta,\sup_{\Rcal}\theta]$; since $\{F_m\in\Rcal\}$ does not involve $h$, the intervals are simultaneous over all $h$. Every region in Table~\ref{tab:regions} except VA and its variants is a pull-back of a set of tail-count or count-cell frequencies. The VA region is not: its constraint at $k$ depends on $G$ through $V_k(G)$, which is not a function of the means $\mu(G)$. DKW with Massart's constant, Berk--Jones \citep{berk1979goodness,owen1995nonparametric} and Pearson's $\chi^2$ are calibrated for iid data; the per-$k$ empirical-Bernstein band is finite-sample valid for independent, non-identical indicators by Theorem~11 of \citet{maurer2009empirical}.

\begin{table}[tbp]
\centering
\footnotesize
\caption{Confidence regions compared. All LP regions use the same cells and envelopes, and certified endpoints (except $\chi^2$, whose LP has auxiliary columns). $L=\log(2|K|/\alpha)$; $K$ is the pre-specified index set of Section~\ref{sec:K}; every LP region also has the sure constraints. ``Unproven'' means finite-sample valid for iid data, with no proof under fixed $p_i$. $^\dagger$Our re-implementation of AMARI (Supplement~\ref{S-supp:comparators}).}
\label{tab:regions}
\begin{adjustbox}{max width=\textwidth}
\begin{tabular}{lllll}
\toprule
Region & Constraint on candidate $G$ & Calibration & Valid, fixed $p_i$? & Reference \\
\midrule
VA & $|T_k-\mu_k(G)|\le r_B(V_k(G))$, $k\in K$ & Bernstein + Bonferroni & yes & Thm~\ref{thm:coverage} \\
VA-Bennett & as VA, Bennett radius $r_{Be}$ & Bennett + Bonferroni & yes & Thm~\ref{thm:coverage} \\
VA$\cap$KL & VA at $\alpha/2$ $\cap$ KL-Bonf at $\alpha/2$ & finite-sample & yes & Thm~\ref{thm:coverage}, Prop~\ref{prop:iid}(a) \\
VA-iid & $|T_k-\mu_k(G)|\le r_B(u(\mu_k(G)))$ (ablation) & Bernstein, iid variance & yes & Prop~\ref{prop:iid}(b) \\
KL-Bonf & $m\,\mathrm{KL}(T_k\|\mu_k(G))\le L$, $k\in K$ & Chernoff + Bonferroni & yes & Prop~\ref{prop:iid}(a) \\
Hoeff-$k$ & $|T_k-\mu_k(G)|\le\sqrt{L/(2m)}$, $k\in K$ & Hoeffding + Bonferroni & yes & \citet{hoeffding1963probability} \\
Hoeff-cell & Hoeffding on all $n+1$ count-cell frequencies & Hoeffding + Bonferroni & yes & \citet{hoeffding1963probability} \\
DKW-$4/\sqrt m$ & $\max_k|T_k-\mu_k(G)|\le 4/\sqrt m+\sqrt{\log(1/\alpha)/(2m)}$ & McDiarmid & yes & Prop~\ref{S-prop:dkw} \\
DKW-Massart & $\max_k|T_k-\mu_k(G)|\le\sqrt{\log(2/\alpha)/(2m)}$ & iid-exact (Massart) & unproven & Rem~\ref{rem:open-validity} \\
Berk--Jones & $\max_k \mathrm{KL}(T_k\|\mu_k(G))\le c_m$ & iid, Monte Carlo $c_m$ & unproven & Rem~\ref{rem:open-validity} \\
Two-stage & $|T_k-\mu_k(G)|\le r_{L_2}(\hat V^+_k)$, $\hat V^+_k$ a KL bound on $\E(T_k-T_{2k})$ & two stages at $\alpha/2$ & yes & Prop~\ref{prop:twostage} \\
Exact-binom & $c_{\rm lo}(\mu_k(G))<mT_k<c_{\rm hi}(\mu_k(G))$, KL where (c) fails & exact binomial + Bonferroni & yes & Prop~\ref{prop:iid}(a),(c) \\
EB-DKW & per-$k$ empirical Bernstein, plug-in $T_k(1-T_k)$, $k\in K$ & emp.\ Bernstein + Bonferroni & yes & \citet[Thm~11]{maurer2009empirical} \\
$\chi^2$ & Pearson $\chi^2$ on pooled count cells & asymptotic & asymptotic & Rem~\ref{rem:open-validity} \\
AMARI$^\dagger$ & affine minimax estimator, bias-aware CI & asymptotic, pointwise & asymptotic & \citet{ignatiadis2022confidence} \\
Plug-in bootstrap & not a region: $m^{-1}\sum_ih(c_i/n)$, percentile bootstrap & none & no & Section~\ref{sec:sim-design} \\
\bottomrule
\end{tabular}
\end{adjustbox}
\end{table}

\subsection{Identification}\label{sec:ident}
Even with exact tail-count means, $F$ is only partially identified. For $\Delta\in\R^K_{\ge0}$ let $\Ncal_F(\Delta)=\{G\in\Pcal:|\mu_k(G)-\mu_k(F)|\le\Delta_k,\ k\in K\}$ and define the one-sided moduli \citep[compare the modulus of continuity in][]{donoho1994statistical}
\[
\omega^+_F(\Delta)=\sup_{G\in\Ncal_F(\Delta)}\theta(G)-\theta(F),\qquad \omega^-_F(\Delta)=\theta(F)-\inf_{G\in\Ncal_F(\Delta)}\theta(G),
\]
and the identification width $\ID_K(F)=\omega^+_F(0)+\omega^-_F(0)$. With $K=\{1,\dots,n\}$ the tail-count means determine the first $n$ moments of $F$ and nothing more (Proposition~\ref{S-prop:moments}). $P(p>0)$ is lower semicontinuous but not continuous in $G$, so only lower confidence bounds are informative. Some bounds hold with probability one: $c_i>0$ implies $p_i>0$ and $c_i<n$ implies $p_i<1$, so $F(\{0\})\le\#\{c_i=0\}/m$ and $F(\{1\})\le\#\{c_i=n\}/m$. These \emph{sure constraints} cost no $\alpha$, and we add them to every LP region; they are the only linear constraints on the cell weights that hold for every parameter vector compatible with the counts (Proposition~\ref{S-prop:sure-only}). For box-monotone regions with a fixed cut set, which include VA and every band region, they change no endpoint for targets right-continuous at $0$ and left-continuous at $1$, such as $\theta_\delta$, $P(p\ge\gamma)$ with $0<\gamma<1$ and pass@$k_0$, and for $P(p>0)$ they raise the lower endpoint to at least $T_1$, the share of problems solved at least once (Proposition~\ref{S-prop:sure-noeffect}). That proposition does not cover the $\chi^2$ and Hoeff-cell regions, nor cut sets that differ between runs through adaptive refinement.

\section{The VA region and its coverage}\label{sec:va}

\subsection{One-dimensional bands}
Define the Bernstein radius $r_B(V)=\sqrt{2LV/m}+L/(3m)=a\sqrt V+b$ and the Bennett radius $r_{Be}(V)=V\,h_B^{-1}\{L/(mV)\}$ for $V>0$, $r_{Be}(0)=0$, where $h_B(x)=(1+x)\log(1+x)-x$ \citep{bennett1962probability,boucheron2013concentration}.

\begin{lemma}[Bands for one tail count]\label{lem:bands}
Fix $k$. For $r=r_B$ and for $r=r_{Be}$,
\[
\Pr\{T_k-\mu_k(F)> r(V_k(F))\}\le e^{-L},\qquad \Pr\{\mu_k(F)-T_k> r(V_k(F))\}\le e^{-L}.
\]
Moreover $r_{Be}\le r_B$, and both radii are concave and nondecreasing on $[0,1/4]$.
\end{lemma}

The inequalities are strict because $r_{Be}(0)=0$: if $V_k(F)=0$, every $p_i\in\{0,1\}$ and $T_k=\mu_k(F)$ almost surely. The proof (Supplement~\ref{S-supp:proofs-va}) needs only independence of the summands $\mathbf 1\{c_i\ge k\}-P_k(p_i)$. Let $\Ecal=\{|T_k-\mu_k(F)|\le r(V_k(F))\ \forall k\in K\}$ be the \emph{band event}. Its complement is the union of the $2|K|$ strict one-sided events, so $\Pr(\Ecal)\ge1-2|K|e^{-L}=1-\alpha$.

\subsection{The exact region and the two-stage band}
The exact VA region is
\[
\Rcal_{\VA}=\{G\in\Pcal:\ |T_k-\mu_k(G)|\le r(V_k(G))\ \ \forall k\in K\}.
\]
It is convex, because $\mu_k$ and $V_k$ are linear in $G$ and $r$ is concave and nondecreasing, and on $\Ecal$ it contains $F$. A candidate $G$ is tested with the variance that $T_k$ would have if $G$ were the truth, as the interval of \citet{wilson1927probable} evaluates the variance at the null; here the null is a mixing law and its variance is linear in the law, so the inversion can be carried out over all candidates at once.

The sample variance $T_k(1-T_k)$ has mean $q_k(1-q_k)-V_k(F)/m$, so empirical-Bernstein bands estimate the iid variance. In fact no statistic is unbiased for $V_k(F_m)$ for all $(p_i)$, since its mean has degree at most $n$ in each $p_i$ and $V_k$ has degree $2n$ (Proposition~\ref{S-prop:nounbiased}). A conservative plug-in is possible. By sub-multiplicativity of binomial tails (Lemma~\ref{S-lem:submult}), $V_k(F)\le\Psi_k(F):=\mu_k(F)-\mu_{2k}(F)$ when $2k\le n+1$, and by a mirror inequality $V_k(F)\le\mu_{2k-n-1}(F)-\mu_k(F)$ otherwise (Lemma~\ref{S-lem:mirror}); $\hat D_k=T_k-T_{2k}$ (or $T_{2k-n-1}-T_k$) is unbiased for $\Psi_k(F)$. The \emph{two-stage band} spends $\alpha/2$ on simultaneous KL upper confidence bounds $\hat V^+_k\ge\Psi_k(F)$, capped at $1/4$, and uses the fixed Bernstein radii $\rho_k=r_{L_2}(\hat V^+_k)$, $L_2=\log(4|K|/\alpha)$, for the remaining $\alpha/2$; with the sure constraints, its region is a polytope.

\begin{proposition}[Two-stage band]\label{prop:twostage}
For every $n$, $m$ and $(p_i)$, the two-stage band covers $\theta(F_m)$ with probability at least $1-\alpha$, simultaneously for all bounded $h$. Let $a_\ell=\sqrt{2L_\ell/m}$ and $b_2=L_2/(3m)$, with $L_1=L$. On an event of probability at least $1-3\alpha/2$, $\rho_k\le a_2\sqrt{\Psi_k(F)}+2.04\,a_1a_2+b_2$, and the conclusions (i)--(ii) of Theorem~\ref{thm:width} hold with $\bar D_k$ replaced by the tube half-width $D^{2S}_k=r_{L_2}(V_k(F))+\rho_k+\eta_k\le a_2\{\sqrt{V_k(F)}+\sqrt{\Psi_k(F)}\}+O(L/m)+\eta_k$.
\end{proposition}

The proof is in Supplement~\ref{S-supp:twostage}. The band pays twice, in the split of $\alpha$ and through $\Psi_k-V_k=\int(P_k^2-P_{2k})\,dF$, which is large when $F$ has mass between $k/n$ and $2k/n$. The width bounds do not separate the two methods: when the partner $2k$ lies in $K$, the leading term of VA's bound (Theorem~\ref{thm:width}(iii) before $\sqrt{x+y}\le\sqrt x+\sqrt y$ is applied) is within a factor $0.955$--$0.994$ of the two-stage band's at $|K|=31$ (Proposition~\ref{S-prop:twostage-compare}). With the index set $K_{\rm geo}$ below, which lacks most partners $2k$, the per-$k$ ratio of leading terms has median 0.976--1.022 on the simulation laws, so it can exceed $1$. Any advantage of VA must come from the realised candidate variances, and the comparison is empirical (Section~\ref{sec:sim-verdict}).

\subsection{The gridded LP}\label{sec:lp}
Take a finite partition $C_1,\dots,C_J$ of $[0,1]$ with closed hulls $[a_j,b_j]$ that contains the cells $\{0\}$ and $\{1\}$, and define the envelopes
\[
P^{lo}_{kj}=P_k(a_j),\quad P^{hi}_{kj}=P_k(b_j),\quad \bar V_{kj}=\max_{p\in[a_j,b_j]}u(P_k(p)),\quad h^{lo}_j=\inf_{C_j}h,\quad h^{hi}_j=\sup_{C_j}h .
\]
Near $p=0$, $dP_1/dp=n$, so evaluating $P_1$ at a grid node instead of the unknown $p_i$ can misstate its mean by $n$ times the cell width, exactly where $\theta_\delta$ is decided; the envelopes replace this error by a guaranteed bracket. Let $\Tcal\subset(0,1/4]$ be a finite set of tangent points and $\rcut(V)=\min_{V_0\in\Tcal}\{r(V_0)+r'(V_0)(V-V_0)\}\ge r(V)$. The variables are cell weights $w$ in the simplex $\Delta_J$, with $\bar V_k(w)=\sum_j\bar V_{kj}w_j$, and the feasible set $\Wcal$ is defined by
\begin{equation}\label{eq:lp}
T_k\le \sum_jP^{hi}_{kj}w_j+\rcut(\bar V_k(w)),\qquad T_k\ge\sum_jP^{lo}_{kj}w_j-\rcut(\bar V_k(w)),\qquad k\in K,
\end{equation}
one pair of linear inequalities per $(k,V_0)$, together with the sure constraints $w_{\{0\}}\le\#\{c_i=0\}/m$ and $w_{\{1\}}\le\#\{c_i=n\}/m$. The interval is $[\Lo,U]$ with $U=\max_{w\in\Wcal}\sum_jh^{hi}_jw_j$ and $\Lo=\min_{w\in\Wcal}\sum_jh^{lo}_jw_j$, and the empty set if $\Wcal=\emptyset$. Bennett and Bernstein cuts may be imposed together at no cost in $\alpha$, because the Bennett band event is contained in the Bernstein one. Endpoints are certified by weak LP duality, so solver tolerance cannot shrink the reported interval (Supplement~\ref{S-supp:impl}).

\paragraph{Choice of $K$.}\label{sec:K}
Validity holds for any $K$ fixed in advance. All intervals in this paper, except where stated, use
\[
K_{\rm geo}=\{1,\dots,\min(n,16)\}\cup\mathrm{round}(\mathrm{geomspace}(16,n,16)),\qquad |K_{\rm geo}|\in\{30,31\},
\]
the default index set of our software. The cost of a larger $K$ enters only through $\sqrt L$. $K_{\rm geo}$ and the coarse initial tangent set we use lie outside the hypotheses of the width theorems of Section~\ref{sec:width}, which use the indices $k\lesssim2n\delta$ densely together with partners $k'\ge2k$; a theorem-aligned rerun changed the median width by factors 1.012--1.098 (Supplement~\ref{S-supp:K}).

\begin{algorithm}[tbp]
\caption{VA confidence intervals}\label{alg:va}
\begin{enumerate}
\item[] \textbf{Input:} counts $c_1,\dots,c_m\in\{0,\dots,n\}$; level $\alpha$ (default $0.05$); index set $K$ fixed in advance (default $K_{\rm geo}$); envelope tolerance $\eta$ (default $10^{-3}$); bounded targets $h_1,\dots,h_J$, which may be chosen after seeing the data.
\item Compute $T_k=m^{-1}\#\{i:c_i\ge k\}$ for $k\in K$, and $L=\log(2|K|/\alpha)$.
\item Build a data-independent partition of $[0,1]$, including the cells $\{0\}$ and $\{1\}$, such that $P_k(b_j)-P_k(a_j)\le\eta$ on every cell for $k\in K$ and $h_j$ varies little on every cell. Compute the envelopes.
\item Form the LP~\eqref{eq:lp} with Bernstein (and optionally Bennett) tangent cuts at an initial tangent set, and the sure constraints.
\item For each target, maximise $\sum_jh^{hi}_jw_j$ and minimise $\sum_jh^{lo}_jw_j$; add tangent cuts at the current $\bar V_k(w)$ for every violated $k$ and re-solve, for a few rounds.
\item Certify each endpoint by weak duality from a dual-feasible vector computed over all cells.
\item[] \textbf{Output:} intervals $[\Lo_j,U_j]$ that cover $\theta_j(F_m)$, $j=1,\dots,J$, simultaneously with probability at least $1-\alpha$.
\item[] \textbf{Cost:} for one replicate with seven targets and both endpoints, the median over laws of the mean time per replicate was 12--34~s at $m=30$ and 21--130~s at $m=1000$, for $n=32$--$512$, in one single-threaded process on a shared host (Table~\ref{S-tab:s4-runtime}).
\end{enumerate}
\end{algorithm}

\subsection{Coverage}

\begin{theorem}[Finite-sample conditional coverage]\label{thm:coverage}
Let $K$, the levels and the radius family be fixed in advance. For every $n$ and $m$, every $(p_i)\in[0,1]^m$, every finite partition containing $\{0\}$ and $\{1\}$, every finite tangent set $\Tcal$ and every bounded $h$, monotone or not, where the partition, $\Tcal$ and $h$ may depend on the data,
\[
\Pr\big\{\Lo\le\theta(F_m)\le U\big\}\ \ge\ \Pr(\Ecal)\ \ge\ 1-\alpha .
\]
Since $\Ecal$ does not involve $h$, the intervals cover simultaneously for all bounded $h$.
\end{theorem}

\begin{proof}
Work on $\Ecal$. Let $w^*_j=F(C_j)$, the share of the $p_i$ in cell $C_j$, so $w^*\in\Delta_J$.

\emph{Step 1 (means).} $P_k$ is nondecreasing, so $P_k(a_j)\le P_k(p)\le P_k(b_j)$ for $p\in C_j$, and $\sum_jP^{lo}_{kj}w^*_j\le\mu_k(F)\le\sum_jP^{hi}_{kj}w^*_j$.

\emph{Step 2 (variances).} For $p\in C_j$, $u(P_k(p))\le\bar V_{kj}$, so $V_k(F)\le\bar V_k(w^*)$.

\emph{Step 3 (feasibility).} On $\Ecal$, since $r$ is nondecreasing and $r\le\rcut$,
\[
T_k\le\mu_k(F)+r(V_k(F))\le\textstyle\sum_jP^{hi}_{kj}w^*_j+\rcut(\bar V_k(w^*)),
\]
and the lower inequality follows in the same way. Moreover $w^*_{\{0\}}=F(\{0\})\le\#\{c_i=0\}/m$ and $w^*_{\{1\}}=F(\{1\})\le\#\{c_i=n\}/m$ on every realisation, because $p_i=0$ forces $c_i=0$ and $p_i=1$ forces $c_i=n$. Thus $w^*\in\Wcal$.

\emph{Step 4 (target).} Since $h^{lo}_j\le h(p)\le h^{hi}_j$ for $p\in C_j$, $\Lo\le\sum_jh^{lo}_jw^*_j\le\theta(F)\le\sum_jh^{hi}_jw^*_j\le U$.

Finally $\Pr(\Ecal)\ge1-\alpha$ by Lemma~\ref{lem:bands} and the union bound. Steps 1--4 are deterministic implications on $\Ecal$ that hold for every partition, tangent set and $h$ at once, so they hold at data-dependent choices; only $K$, the levels and the radius enter $\Ecal$.
\end{proof}

No regularity of $h$ or of the grid is needed; the grid affects only the width, through the envelope tolerance $\eta_k=\max_j\{P_k(b_j)-P_k(a_j)\}$. An LP built on point values $P_k(g_j)$ at grid nodes, without envelopes, has no coverage guarantee when $F_m$ is off the grid. Any finite family of valid per-$k$ constraints can be intersected, with $\alpha$ split by a union bound, as in the hybrid \VA$\cap$KL of Table~\ref{tab:regions}.

\section{Iid-calibrated localizations under fixed $p_i$}\label{sec:iid}

The familiar bands are calibrated for the super-population target. Are they valid for $\theta(F_m)$ when the $p_i$ are fixed, and how conservative are they?

\begin{proposition}[Iid calibration under fixed $p_i$]\label{prop:iid}
Under fixed $p_i$:
\begin{itemize}
\item[(a)] For each $k$ and $0<t\le1-q_k$, $\Pr(T_k\ge q_k+t)\le\exp\{-m\KL(q_k+t\|q_k)\}$, and symmetrically for the lower tail. The per-$k$ KL region with Bonferroni over $K$ has coverage at least $1-\alpha$.
\item[(b)] The iid-variance band $\{|T_k-q_k|\le r(u(q_k))\ \forall k\in K\}$ has probability at least $1-\alpha$. For every $G$, $V_k(G)\le u(\mu_k(G))$, so with the same $r$, $K$ and $L$ the exact regions satisfy $\Rcal_{\VA}\subseteq\Rcal_{\rm iid}:=\{G:|T_k-\mu_k(G)|\le r(u(\mu_k(G)))\ \forall k\}$, and at a candidate $G$ the square-root term of the radius shrinks by the factor
\begin{equation}\label{eq:factor}
\sqrt{V_k(G)/u(\mu_k(G))}=\sqrt{1-\Var_G(P_k(p))/[\mu_k(G)\{1-\mu_k(G)\}]}.
\end{equation}
\item[(c)] Let $c_{\rm lo}(q),c_{\rm hi}(q)$ be exact $\Bin(m,q)$ critical values at level $\alpha/(2|K|)$ in each tail. The per-$k$ test that keeps $G$ when $c_{\rm lo}(\mu_k(G))<mT_k<c_{\rm hi}(\mu_k(G))$ rejects $G=F$ with probability at most $\alpha/|K|$ whenever $c_{\rm lo}(q_k)\le mq_k-1$ and $c_{\rm hi}(q_k)-1\ge mq_k$.
\end{itemize}
\end{proposition}

Part (a) is Theorem~1 of \citet{hoeffding1963probability}, which allows non-identical summands; part (c) uses Theorem~4 of \citet{hoeffding1956distribution} (Supplement~\ref{S-supp:proofs-iid}). For the super-population target the between-problem term in~\eqref{eq:factor} is genuine sampling variability; for $\theta(F_m)$ it is not random, so iid-calibrated regions are valid but conservative. The same holds for the KL, Berk--Jones and empirical-Bernstein bands and for $\chi^2$, while the Hoeffding and DKW bands use the worst-case variance $1/4$. For DKW we can prove validity under fixed $p_i$ only with the constant $4/\sqrt m$ in place of Massart's (Proposition~\ref{S-prop:dkw}).

\begin{remark}[What remains open]\label{rem:open-validity}
We have no finite-sample proof under fixed $p_i$ for DKW with Massart's constant (exact enumeration for $m\le10$, $n\le3$ and Monte Carlo at $m=50,200$ found no violation), for the simultaneous Berk--Jones band calibrated by the iid null distribution of its supremum, or for the asymptotic Pearson $\chi^2$ (a sketch is in Supplement~\ref{S-supp:chi2}). These regions are labelled ``unproven'' or ``asymptotic'' in the experiments.
\end{remark}

\begin{figure}[tbp]
\centering
\includegraphics[width=0.92\textwidth]{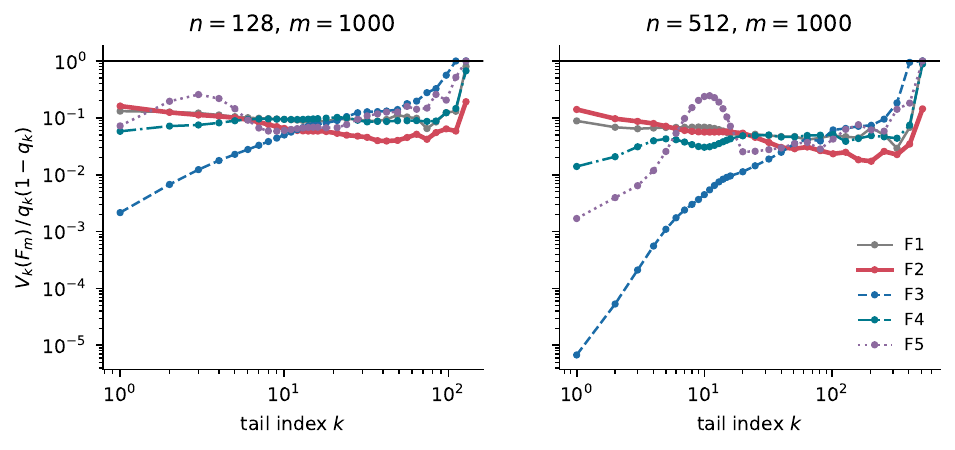}
\caption{The variance gap. Ratio of the conditional variance $V_k(F_m)$ to the iid variance $q_k(1-q_k)$ against $k\in K_{\rm geo}$, for the simulation draws of F1--F5 with $m=1000$. F2 (thick) is fitted to Monkey Business.}
\label{fig:vargap}
\end{figure}

Figure~\ref{fig:vargap} shows $V_k/(q_k(1-q_k))$ on the five simulation laws of Section~\ref{sec:sim}; its median over $k\in K$ ranges from 0.010 (F3, $n=512$) to 0.344 (F3, $n=32$). For laws with an atom at $0$ and a Beta component, the $k=1$ factor decays like $n^{-a_0/2}$, where $a_0$ is the exponent of the density at $0^+$ (Supplement~\ref{S-supp:spikebeta}), so mass close to, but not at, zero limits the gain.

\section{Width theory}\label{sec:width}

For $\lambda\in\R^K$ let $\psi_\lambda=h-\sum_{k\in K}\lambda_kP_k$, $\gap^+_\lambda(F)=\sup_{[0,1]}\psi_\lambda-\int\psi_\lambda\,dF$, $\gap^-_\lambda(F)=\int\psi_\lambda\,dF-\inf_{[0,1]}\psi_\lambda$ and $\osc(\psi)=\sup\psi-\inf\psi$. The tangent set for $r_B$ is $\{t^2:t\in\Tcal_0\}$, with $\Tcal_0$ geometric of ratio $\varrho$ from $t_{\min}$ to $1/2$, and $\kappa=(\varrho^{1/2}+\varrho^{-1/2})/2$. For $k\in K$ let $k'$ be the smallest element of $K\cup\{n+1\}$ with $k'\ge2k$, and $g_k(F)=\int(P_k^2-P_{k'})\,dF\ge0$ (Lemma~\ref{S-lem:submult}).

\subsection{A width bound}

\begin{theorem}[Width upper bound]\label{thm:width}
Let $h$ be bounded and $r=r_B$, with the tangent set above or any finite tangent set containing it. For $w\in\Wcal$ put $D_k(w)=r_B(V_k(F))+\rcut(\bar V_k(w))+\eta_k$, $\bar D_k=\sup_{w\in\Wcal}D_k(w)$ and $\bar D_{n+1}=0$. On $\Ecal$:
\begin{itemize}
\item[(i)] $U-\Lo\le\omega^+_F(\bar D)+\omega^-_F(\bar D)$;
\item[(ii)] for all $\lambda,\lambda'\in\R^K$, $U-\Lo\le\gap^+_\lambda(F)+\gap^-_{\lambda'}(F)+\sum_{k\in K}(|\lambda_k|+|\lambda'_k|)\bar D_k$, and in particular $U-\Lo\le\osc(\psi_\lambda)+2\sum_k|\lambda_k|\bar D_k$;
\item[(iii)] for every $k\in K$, $\bar D_k\le a(\sqrt{V_k(F)}+\tfrac12)+2b+\eta_k$ and
\[
\bar D_k\le\big(\sqrt{S_k}+a\kappa\big)^2,\qquad S_k=a(1+\kappa)\sqrt{V_k(F)}+a\kappa\sqrt{g_k(F)+\eta_k+\bar D_{k'}}+2b+a\,t_{\min}+\eta_k .
\]
\end{itemize}
Adding Bennett cuts, the sure constraints or further tangents shrinks $\Wcal$, and the conclusions remain true.
\end{theorem}

The proof (Supplement~\ref{S-supp:proofs-width}) places each cell's mass at a point of the cell where $h$ is nearly extremal; only monotonicity of $P_k$ is used, not of $h$. Part (iii) controls the candidate's variance $\bar V_k(w)$, which can exceed $V_k(F)$, by $V_k(G)\le\mu_k(G)-\mu_{k'}(G)$ (Lemma~\ref{S-lem:candvar}), and solves the resulting self-bounding inequality. For fixed $F$ the leading term is of order $\sqrt{L/m}$, and the remainder through $\bar D_{k'}$ is of order $a^{3/2}$. The bound adapts to $V_k(F)+g_k(F)$, not to $V_k(F)$ alone. For $k>n/2$ the upward partner gives $g_k\ge F(\{1\})$; with the mirror partner $k''=\max\{j\in K\cup\{0\}:j\le2k-n-1\}$, part (iii) holds with $g_k+\bar D_{k'}$ replaced by the smaller of the two partner terms (Proposition~\ref{S-prop:mirror-width}).

\subsection{Comparison with Bonferroni linear certificates}
For $\lambda\in\R^K$ let $\mathrm{bias}_\lambda=\min_c\|h-c-\sum_k\lambda_kP_k\|_\infty$. The linear estimator $c+\sum_k\lambda_kT_k$ gives, on $\Ecal$, a valid interval of length $\ell_\lambda=2\,\mathrm{bias}_\lambda+2\sum_k|\lambda_k|\,r_B(V_k(F))$, a \emph{Bonferroni linear certificate}.

\begin{corollary}[Comparison with Bonferroni linear certificates]\label{cor:oracle}
Define $D^{\det}_{n+1}=0$ and, downwards in $k\in K$, $D^{\det}_k=\min\{a(\sqrt{V_k(F)}+\frac12)+2b+\eta_k,\ (\sqrt{S^{\det}_k}+a\kappa)^2\}$, where $S^{\det}_k$ is $S_k$ with $\bar D_{k'}$ replaced by $D^{\det}_{k'}$. On $\Ecal$, $\bar D_k\le D^{\det}_k$ for all $k$ and
\[
U-\Lo\le\inf_\lambda\Big\{2\,\mathrm{bias}_\lambda+2\sum_k|\lambda_k|D^{\det}_k\Big\}\le\rho^{\det}\inf_\lambda\ell_\lambda,\qquad \rho^{\det}=\max_{k\in K}D^{\det}_k/r_B(V_k(F))\ge1 .
\]
For a data-independent partition, $\rho^{\det}$ is a deterministic constant. On the Bernstein band event $\Ecal$ without Bennett cuts, and on the Bennett band event if Bennett cuts are imposed, $\bar D_k\ge2r_B(V_k(F))$.
\end{corollary}

The factor $2$ is the price of test inversion: the region absorbs both the data's deviation (a radius at $F$) and the candidate's tolerance (a radius at $G$). The benchmark $\ell_\lambda$ is not the best linear-estimator interval, whose leading term is shorter by a factor of at least about $1.39$ at $|K|=31$ (Supplement~\ref{S-supp:widthdisc}). When $V_k(F)\to0$, $\rho^{\det}$ can grow like $a^{-1/2}$; the comparison is then with an estimator that knows $V_k(F)$ is nearly $0$, which no procedure can know, so $\rho^{\det}$ is a bound and not a measured inefficiency.

\subsection{Worst-case identification}
Let $E_n(h)=\min_{\deg Q\le n}\|h-Q\|_\infty$ be the error of best uniform approximation on $[0,1]$.

\begin{theorem}[Worst-case identification optimality, $K=\{1,\dots,n\}$]\label{thm:id}
Let $K=\{1,\dots,n\}$, let $h$ take values in $[0,1]$, and let the tangent set be as in Theorem~\ref{thm:width} (it contains $t=1/2$).
\begin{itemize}
\item[(i)] If $h$ is bounded, $\sup_F\ID_K(F)\le2E_n(h)$, and for fixed $n$ and every $(p_i)$, on $\Ecal$, $U-\Lo\le2E_n(h)+2\|\lambda^*\|_1(a+2b+\max_k\eta_k)$, where $c^*+\sum_k\lambda^*_kP_k$ is a best approximation. If the data-independent grid has tolerance $\eta_m\to0$, the right-hand side tends to $2E_n(h)$ as $m\to\infty$.
\item[(ii)] If $h$ is continuous, then for every $\zeta\in(0,1)$, every procedure $C$ with conditional coverage $1-\alpha$ on $[0,1]^m$ and every $\nu>0$, there is $(p_i)$ with $\Pr\{|C|\ge 2E_n(h)-\nu-1/\sqrt{\zeta m}\}\ge1-2\alpha-2\zeta$, where $|C|=\sup C-\inf C$.
\end{itemize}
\end{theorem}

Part (i) holds because $\{1,P_1,\dots,P_n\}$ spans the polynomials of degree at most $n$; part (ii) combines a fuzzy-hypothesis lemma with the moment-matching duality $\sup\{\theta(G)-\theta(G'):G,G'\text{ with equal moments of order }1,\dots,n\}=2E_n(h)$ \citep[Appendix~E]{wu2016minimax} (Supplement~\ref{S-supp:proofs-lower}). So, for fixed $n$, continuous $h$, all $n$ tail counts and a grid refined as $m$ grows, the identification part of VA's width is first-order worst-case optimal as $m\to\infty$, as it is for every localization with the same index set. The statement says nothing at the tolerance we use: at $\eta=10^{-3}$ and $\|\lambda^*\|_1\gtrsim10^3$ ($n=128$, $\delta=0.01$) the grid term is at least $2$. It is also vacuous for indicator targets, since a jump of size $J$ forces $E_n(h)\ge J/2$, and $E_n=0$ for pass@$k_0$ with $k_0\le n$. The implemented 30--31 element index sets are not covered, and $2E_K(h)\ge2E_n(h)$ (Supplement~\ref{S-supp:widthdisc}).

\begin{remark}[The constant for $h_\delta$]\label{rem:cn}
Write $2E_n(h_\delta)=c_n(\delta)\sqrt{1-\delta}/(n\sqrt\delta)$. Rigorous brackets for 40 pairs $(n,\delta)$ with $16\le n\le256$ (LP polynomial error plus a curvature remainder; de la Vall\'ee Poussin lower bound; Table~\ref{S-tab:cn}) give $c_n\in[0.2799,0.2891]$, that is $c_n/\beta\in[0.9993,1.0317]$ (rounded outward), in the 19 cases with $n\sqrt\delta\ge12$, numerically close to Bernstein's constant $\beta=0.2801694990\ldots$ \citep{bernstein1914meilleure,varga1985bernstein}. For small $n\sqrt\delta$ the constant is larger: $c_n=0.4459$ at $n=16$, $\delta=0.002$. We conjecture that $c_n\to\beta$ as $n\to\infty$ for fixed $\delta$; we know of no verified source, and nothing below depends on it.
\end{remark}

\subsection{Separated laws}\label{sec:separated}
LLM success probabilities often consist of an atom at $0$ (unsolved problems) and mass well above $0$. Let
\[
F=\pi_0\delta_0+(1-\pi_0)F_+,\qquad \pi_0>0,\quad\mathrm{supp}\,F_+\subseteq[p_{\min},1],\quad\delta<p_{\min},
\]
with $h=h_\delta$, $k_*=\lceil n\delta\rceil$, $K\ni1,2,k_*,k_*'$, and $\epsilon(j)=\Pr\{\Bin(n,p_{\min})<j\}$. Then $V_k(F)\le(1-\pi_0)\epsilon(k)$ is exponentially small in $n$ when $p_{\min}$ is well above $k/n$, whereas $q_k(1-q_k)\ge\pi_0(1-\pi_0)P_k(p_{\min})$ is not. Put
\[
A_k=(1+\kappa)\sqrt{V_k(F)}+\kappa\sqrt{g_k(F)}+t_{\min},\quad A_{\VA}=\frac{A_1}{P_1(\delta)}+A_{k_*},\quad C_{\rm ID}=(1-\pi_0)\Big[\frac{1-P_1(\delta)}{P_1(\delta)}+\epsilon(k_*)\Big],
\]
and let $R_{\VA}=O(a^{3/2})$ be the remainder of Lemma~\ref{S-lem:leading} (for fixed $\kappa$ and $\eta\le a^{3/2}$).

\begin{proposition}[Realised widths of iid-calibrated regions]\label{prop:realised}
Let $\mathcal R=\{G:\mu_k(G)\in A_k(T_k)\ \forall k\in K\}$ with intervals $A_k(T)$ (Hoeff-$k$, iid-Bernstein, KL-Bonf, exact-binomial), let $p_{\min}\ge\delta$ and $\rho_k\ge0$ be deterministic, and let $\epsilon_{\rm real}$ be the largest $\epsilon\le\pi_0/2$ such that $\mu_k(F)+\epsilon P_k(\delta)\in A_k(T)$ for every $T\in[\mu_k(F)-\rho_k,\mu_k(F)+\rho_k]$ and every $k\in K$, with $\epsilon_{\rm real}=0$ if there is no such $\epsilon$. On the event $\{|T_k-\mu_k(F)|\le\rho_k\ \forall k\}\cap\{F\in\mathcal R\}$, with or without the sure constraints and on any grid, the realised width is at least $\epsilon_{\rm real}$, and without the sure constraints at least $\ID_K(F)+\epsilon_{\rm real}\{1-\omega^+_F(0)/\pi_0\}$.
\end{proposition}

The proof moves mass $\epsilon$ from the atom at $0$ to $\delta$, which shifts each $\mu_k$ by $\epsilon P_k(\delta)$ and $\theta_\delta$ by $\epsilon$ (Supplement~\ref{S-supp:realised}). With $\rho_k=r_B(V_k(F))$ the first event is VA's band event $\Ecal$. For iid-Bernstein, $\Ecal$ implies $F\in\mathcal R$; for Hoeff-$k$ it does when $r_B(V_k(F))\le a/2$ for all $k$, which can fail when $V_k(F)$ is near $1/4$; for KL-Bonf, $\{F\in\mathcal R\}$ is a separate event of probability at least $1-\alpha$. Under conditions (J1)--(J3) below, $\epsilon_{\rm real}\ge(c_{\rm comp}-o(1))a$ with $c_{\rm comp}=\frac12$ for Hoeff-$k$ and, under a further condition (J4) that must be checked at each point, $c_{\rm comp}=\sqrt{\pi_0(1-\pi_0)/2}$ for iid-Bernstein; (J4) fails for large $n$ when $n\in K$, $\delta>p_{\max}^2$ and $F_+$ has no mass above $p_{\max}$ (Supplement~\ref{S-supp:realised}). For KL-Bonf we have only a sketch. Computed exactly at the separated laws of Supplement~\ref{S-supp:sepnum}, $\epsilon_{\rm real}/a$ lies in $[0.480,0.499]$ for Hoeff-$k$, $[0.486,0.500]$ for iid-Bernstein and $[0.478,0.499]$ for KL-Bonf: above the constants $0.354$ and $0.25$ of an oracle-centred argument (Proposition~\ref{S-prop:compexcess}) for the last two, but slightly below $\frac12$ for Hoeff-$k$.

\begin{corollary}[Constant-factor gain]\label{cor:sepA}
On $\Ecal$, $U-\Lo\le C_{\rm ID}+a\,A_{\VA}+R_{\VA}$. Let $\mathcal R$ be a region of Proposition~\ref{prop:realised} with $\epsilon_{\rm real}>0$ at $\rho_k=r_B(V_k(F))$, fix $\gamma>0$, and suppose $m\le m_{\rm ID}(\gamma):=2L\gamma^2/C_{\rm ID}^2$. Then on $\Ecal\cap\{F\in\mathcal R\}$ the ratio of the realised VA width to the realised width of $\mathcal R$ is at most $(A_{\VA}+\gamma+R_{\VA}/a)/(\epsilon_{\rm real}/a)$.
\end{corollary}

$A_{\VA}$ is exponentially small in $n$ when $p_{\min}$ is well above $2\delta$, apart from $t_{\min}$, which the user controls; $R_{\VA}/a=O((L/m)^{1/4})$ dominates below an explicit crossover. For fixed $(n,\delta,F)$ there is no rate separation as $m\to\infty$, and the window $m\le m_{\rm ID}$, which reflects the crude certificates behind $C_{\rm ID}$, can make the corollary vacuous: in the two fixed-$n$ configurations of Supplement~\ref{S-supp:sepnum} the ratio bound for Hoeff-$k$ is below $1$ only in the middle of one configuration's range and exceeds $0.95$ throughout the other.

\begin{corollary}[Rate separation in a joint limit]\label{cor:sepB}
Fix $F,\delta,\pi_0$ with $p_{\min}>2\delta$, and let $m\to\infty$ and $n=n_m\to\infty$ with $K\supseteq\{1,2,k_*,2k_*\}$, such that
(J1) $n\delta\ge\frac34\log\frac{m}{2L}+1$;
(J2) $n\KL\big(\frac{2k_*-1}{n}\big\|p_{\min}\big)\ge\frac34\log\frac{m}{2L}$;
(J3) $t_{\min}\le a^{1/2}$ and $\max_k\eta_k\le a^{3/2}$.
Then on $\Ecal$, $U-\Lo=O\{(L/m)^{3/4}\}$ with a constant depending only on $\kappa$, while the realised Hoeff-$k$ width is at least $(\frac12-o(1))\sqrt{2L/m}$, and, under (J4), the realised iid-Bernstein width is at least $(\sqrt{\pi_0(1-\pi_0)/2}-o(1))\sqrt{2L/m}$.
\end{corollary}

Conditions (J1)--(J2) are lower bounds on $n$, of order $\log m$; with $K=\{1,\dots,n\}$, $L$ grows like $\log n$, which is $O(\log\log m)$ when $n$ is polylogarithmic in $m$. So our bounds give at most a constant-factor gain over Hoeffding- and iid-Bernstein-calibrated regions for fixed $n$, inside the window $m\le m_{\rm ID}$, and a rate in a joint limit with $n$ growing (for iid-Bernstein under (J4)). We do not show that a rate \emph{requires} $n$ to grow: that would need a lower bound on VA's excess width, part of the open problem below. Supplement~\ref{S-supp:sepnum} reports realised widths on three separated laws: on an atomic grid that contains the atoms of $F$, VA's width falls with slope close to $-1$ in $m$ before the crossover, against $-1/2$ for the competitors.

\subsection{Lower bounds}
Supplement~\ref{S-supp:proofs-lower} gives local lower bounds for every procedure with conditional coverage (the parametric rate for a finite-population mean, and a bound from moving one problem off zero) and records which upper and lower bounds match (Table~\ref{S-tab:matched}). The estimation term is not matched.

\begin{openproblem}[Estimation-term lower bound]\label{op:estimation}
Find a lower bound for the estimation term in the conditional model that matches Theorem~\ref{thm:width} up to constants.
\end{openproblem}

\section{Simulation study}\label{sec:sim}

\subsection{Design}\label{sec:sim-design}
We follow the ADEMP structure of \citet{morris2019using}. \emph{Aims}: check coverage, compare widths, and apply the decision rule of Section~\ref{sec:sim-verdict}. \emph{Data-generating mechanisms}: F1 $=0.3\delta_0+0.7\,\mathrm{Beta}(0.3,1.5)$; F2 $=0.0323\delta_0+0.0205\delta_1+0.9472\,\mathrm{Beta}(0.282,0.401)$, fitted to the 1{,}022 problem--model units of Section~\ref{sec:mb}; F3 $=0.5\delta_0+0.5\,\mathrm{Beta}(2,5)$; F4 $=0.1\delta_0+0.9\,U(0,1)$; F5 $=0.2\delta_{0.02}+0.5\delta_0+0.3\,\mathrm{Beta}(1,3)$. The grid is $m\in\{250,1000,4000\}$, $n\in\{32,128,512\}$ and $\delta\in\{0.01,0.02,0.05\}$, 45 cells per $n$, plus F2 at $m\in\{30,50,100\}$. For each $(F,m)$ one $F_m$ is drawn and held fixed, and replicates redraw only the counts. \emph{Estimands}: $\theta_\delta(F_m)$, and $P(p\ge0.05)$, $P(p\ge0.5)$, pass@8 and pass@1024. \emph{Methods}: the regions of Table~\ref{tab:regions} with $K_{\rm geo}$, the $\eta=10^{-3}$ envelope grid and the sure constraints, and a plug-in bootstrap that estimates $\theta$ by $m^{-1}\sum_ih(c_i/n)$ (or the unbiased estimator of \citet{chen2021evaluating} for pass@$k$, $k\le n$) with a percentile bootstrap over problems. AMARI \citep[Sections~3--4]{ignatiadis2022confidence} is an affine estimator $m^{-1}\sum_iQ(c_i)$ with worst-case bias over a pilot localization and a bias-aware interval; we re-implemented it following Empirikos.jl, with the empirical standard deviation of $Q(c_i)$ and the interval intersected with the pilot interval (Supplement~\ref{S-supp:comparators}). The pilot is a DKW localization at the fixed level $0.01$, not adjusted for multiplicity, so the nominal error of ``AMARI at $\alpha$'' is up to $\alpha+0.01$; this favours AMARI. On the distilled model's counts of Section~\ref{sec:pair}, our endpoints agree with Empirikos.jl to within $2\times10^{-4}$. AMARI was run on the same count vectors as VA at $\alpha$ and at $\alpha/3$ (Bonferroni over the three $\delta$), at $n=512$ for $m=1000$ only. \emph{Performance measures}: coverage with Clopper--Pearson bounds, median width, and bootstrap Monte Carlo intervals that resample replicates jointly within each $(F,m,n)$ block across methods and targets.

The replicate counts are $R=30$ at $m=1000$ and $n\le128$, $20$ at $n=512$, $15$ at $m\in\{250,4000\}$ and $n\le128$, and $20$ at small $m$; AMARI has $R=30$ in every cell. They were reduced twice from the planned counts, both times because of run time: once before launch, after a timing pilot, and once during the run, after 800 tasks, when the run was 1.19 times slower than projected (among other changes, from $20$ to $15$ at $m\in\{250,4000\}$, $n\le128$, and from $30$ to $20$ at small $m$). Both reductions were made on run-time grounds, before any width ratio or verdict was computed; only tasks in the final plan are used (Supplement~\ref{S-supp:simextra}). One VA solve failed (F5, $m=1000$, $n=512$, $\theta_{0.05}$) and is reported as a failure.

\subsection{The decision rule and its outcome}\label{sec:sim-verdict}
Before the comparator runs we wrote down a rule for claiming that VA is narrower than each of three comparators: C1, the two-stage band; C2, AMARI at level $\alpha$; C3, AMARI at $\alpha/3$. For each comparator and $n$, take the median over cells of the ratio of VA's median width to the comparator's. VA may be called narrower only if this median is at most $0.90$ both at $n=128$ and at $n=512$; otherwise the abstract and introduction lead with validity and conditional coverage, and the comparator's numbers are reported in full. A comparator cell whose Clopper--Pearson upper bound on coverage is below $0.95$ does not count as narrower. The rule file is time-stamped 14 November 2026, 00:43 UTC, before the first comparator output, and its SHA-256 hash was recorded then; recomputing the hash shows only that the file is unchanged since, and the evidence for its timing is the time stamp. The rule is an internal protocol, not a public registration, and it covers the C1--C3 comparison only: the laws, grid, replicate counts and the targets of Section~\ref{sec:pair} were not pre-registered. The rule preceded the C1--C3 runs, not every comparison: when it was written, earlier comparisons were known, of a model-based AMARI with VA on a preliminary simulation (AMARI narrower in 59 of 90 cells at $n\le128$) and of the Empirikos-faithful AMARI with VA on the Monkey Business pools and the illustration data (Supplements~\ref{S-supp:verdict} and~\ref{S-supp:archival}). The rule names C2 ``AMARI at $n=512$'' but applies the threshold at $n=128$ and $512$; we apply the stricter, literal reading (Supplement~\ref{S-supp:verdict}).

No comparator cell was disqualified for coverage. The medians are (Table~\ref{tab:width}):
\begin{itemize}
\item C1, two-stage band: 0.746, 0.602 and 0.436 at $n=32$, 128 and 512, with Monte Carlo intervals 0.736--0.754, 0.581--0.609 and 0.427--0.440; VA was narrower in all 135 cells. The claim stands.
\item C3, AMARI at $\alpha/3$: 1.179, 0.808 (0.797--0.819) and 0.513 (0.496--0.524). The claim stands at $n=128$ and $512$, in the median; VA is wider at $n=32$, and at $n=128$ the cell ratios range over 0.214--1.175.
\item C2, pointwise AMARI: 1.307, 0.977 (0.963--0.986) and 0.626 (0.602--0.639). The rule fails at $n=128$, where the cell ratios range over 0.260--1.360 and VA is narrower in 25 of 45 cells. VA is wider at $n=32$ and narrower at $n=512$, in all 15 cells run.
\end{itemize}
Because the rule failed for C2, this paper does not present width as its main contribution. The width claims we make are those the rule allows: against the two-stage band at every $n$, against Bonferroni-adjusted AMARI at $n\ge128$ in the median, and against pointwise AMARI at $n=512$.

\subsection{Coverage}
Every finite-sample and iid-calibrated localization covered every target in every replicate of every cell (Table~\ref{tab:coverage}). Full coverage in $R=30$ (or $15$) replicates is consistent with true coverage as low as 0.884 (0.782), so these runs check the code; validity rests on Theorem~\ref{thm:coverage}. The plug-in bootstrap undercovered: in 41 of the 135 $\theta_\delta$ cells its Clopper--Pearson upper bound was below $0.95$, mostly at $n=32$ and for pass@1024 (Table~\ref{S-tab:s4-plugin}). A sharper check records whether the true $F_m$ satisfies each region's constraints, which implies coverage of every target at once, in $10^5$ draws per cell, including two adversarial laws. VA missed its band in at most 0.00207 of draws (median 0.00033) at nominal level $0.05$: it is valid but very conservative, through Bonferroni over 30--31 indices and the Bernstein tail. Calibrating the band constant on the true Poisson-binomial law, an oracle, shrank VA's width by factors 0.49--0.98 (Supplement~\ref{S-supp:simextra}). No region with finite-sample calibration exceeded $\alpha$; the asymptotic $\chi^2$ region did, in four cells at $n=512$.

\begin{table}[tbp]
\centering
\footnotesize
\caption{Coverage evidence at $\alpha=0.05$ (sure constraints in every LP region). Simulation: minimum empirical coverage of $\theta_\delta(F_m)$ over the 45 cells per $n$ (135 in all; AMARI 105, run at $n=512$ for $m=1000$ only), the number of cells whose two-sided 95\% Clopper--Pearson upper bound is below $0.95$, and the replicate range; ``other'' is the minimum over $P(p\ge0.05)$, $P(p\ge0.5)$, pass@8 and pass@1024. MB: minimum coverage of $\theta_\delta(\hat p)$ over the eight Monkey Business configurations and the pooled data, all $n$ (AMARI: $n\le128$). Band: non-coverage of the band event, $R=10^5$ per cell, maximum over cells and number of cells whose Clopper--Pearson lower bound exceeds $\alpha$. For the two-stage band the event is ``the true $F_m$ satisfies the region's constraints''. Full coverage in $R=30$ (or $R=15$) replicates is consistent with true coverage as low as 0.884 (0.782), the two-sided Clopper--Pearson lower bound; these columns check the code. All regions: Table~\ref{S-tab:s4-coverage-full}.}
\label{tab:coverage}
\begin{adjustbox}{max width=\textwidth}
\begin{tabular}{lccccccccc}
\toprule
 & \multicolumn{4}{c}{Simulation} & MB & \multicolumn{2}{c}{Band, F1--F5} & \multicolumn{2}{c}{Band, A1--A2}\\
\cmidrule(lr){2-5}\cmidrule(lr){6-6}\cmidrule(lr){7-8}\cmidrule(lr){9-10}
Region & min cov. & CP$_{\rm hi}<0.95$ & $R$ & other & min cov. & max & $>\alpha$ & max & $>\alpha$\\
\midrule
VA & 1.000 & 0/135 & 15--30 & 1.000 & 1.000 & 0.00207 & 0 & 0.00139 & 0\\
Two-stage & 1.000 & 0/135 & 15--30 & 1.000 & 1.000 & 0.00000 & 0 & 0.00000 & 0\\
Exact-binom & 1.000 & 0/135 & 15--30 & 1.000 & 1.000 & 0.00053 & 0 & 0.00669 & 0\\
KL-Bonf & 1.000 & 0/135 & 15--30 & 1.000 & 1.000 & 0.00025 & 0 & 0.00105 & 0\\
Berk--Jones & 1.000 & 0/135 & 15--30 & 1.000 & 1.000 & 0.00202 & 0 & 0.01010 & 0\\
$\chi^2$ & 1.000 & 0/135 & 15--30 & 1.000 & 1.000 & 0.05917 & 2 & 0.07930 & 2\\
Plug-in bootstrap & 0.000 & 41/135 & 15--30 & 0.000 & 0.000 & -- & -- & -- & --\\
AMARI & 1.000 & 0/105 & 30--30 & 0.967 & 0.950 & -- & -- & -- & --\\
AMARI ($\alpha/3$) & 1.000 & 0/105 & 30--30 & 0.967 & 0.950 & -- & -- & -- & --\\
\bottomrule
\end{tabular}
\end{adjustbox}
\end{table}

\subsection{Width}
VA was narrower than every non-VA localization in every cell except seven cells for $\chi^2$, all at $n=32$ (Table~\ref{tab:width}). The median ratio of VA to the best other localization in each cell was 0.824, 0.612 and 0.441 at $n=32$, 128 and 512 (Monte Carlo intervals 0.811--0.833, 0.604--0.621 and 0.436--0.445), and to the best non-VA region with a finite-sample proof 0.781, 0.607 and 0.441. The ablation \VA-iid, the same Bernstein LP with the iid variance $u(\mu_k(G))$, was wider than VA in all 135 cells (median ratio 0.416), so the gain comes from the conditional variance. It grows with $n$, as the variance gap does, and is largest for F3 and smallest for F1 and F5, whose mass sits close to zero (Supplement~\ref{S-supp:simextra}); this is an empirical finding on this grid. For $P(p\ge0.5)$ the median ratio was 0.820 and 0.695 at $n=32$ and 128 but 1.108 at $n=512$, where the target is nearly identified. On small benchmarks (F2, $m\in\{30,50,100,250\}$) VA was narrower than the best other localization in all 36 cells, but the intervals are wide: 0.32--0.39 for $\theta_{0.01}$ at $m=30$ (Table~\ref{S-tab:s4-smallm}).

\begin{table}[tbp]
\centering
\small
\caption{Median over the simulation cells (F1--F5, $m\in\{250,1000,4000\}$, three $\delta$; 45 cells per $n$, 15 at $n=512$ for AMARI) of the ratio of VA's median width to the comparator's median width for $\theta_\delta(F_m)$, $\alpha=0.05$. Below $1$: VA narrower. Brackets: bootstrap 95\% Monte Carlo intervals of the median (replicates resampled jointly within each $(F,m,n)$ simulation block, across methods and targets). AMARI is the Empirikos-faithful variant; C1--C3 are the comparators of the pre-specified rule (Section~\ref{sec:sim-verdict}). ``Best other localization'' is the narrowest non-VA localization in the cell: KL-Bonf, Hoeff-$k$, Hoeff-cell, both DKW bands, Berk--Jones, EB-DKW, $\chi^2$, exact-binomial, and the two-stage band with Bernstein or Bennett radius (\VA-iid, a VA ablation, and AMARI are excluded); ``best proven-valid'' is the narrowest of these with a finite-sample proof under fixed $p_i$, that is, all but DKW-Massart, Berk--Jones and $\chi^2$. The plug-in bootstrap is omitted because it undercovers (Table~\ref{tab:coverage}). All regions: Table~\ref{S-tab:s4-ratio-full}.}
\label{tab:width}
\begin{adjustbox}{max width=\textwidth}
\begin{tabular}{lcccc}
\toprule
Comparator & $n=32$ & $n=128$ & $n=512$ & VA narrower\\
\midrule
Two-stage & 0.746 & 0.602 & 0.436 & 135/135\\
Exact-binom & 0.721 & 0.517 & 0.337 & 135/135\\
KL-Bonf & 0.657 & 0.448 & 0.288 & 135/135\\
VA-iid & 0.636 & 0.421 & 0.264 & 135/135\\
Berk--Jones & 0.696 & 0.483 & 0.326 & 135/135\\
DKW-Massart & 0.745 & 0.543 & 0.352 & 135/135\\
$\chi^2$ & 0.803 & 0.334 & 0.145 & 128/135\\
\midrule
AMARI, pointwise (C2) & 1.307 [1.293, 1.321] & 0.977 [0.963, 0.986] & 0.626 [0.602, 0.639] & 47/105\\
AMARI, $\alpha/3$ (C3) & 1.179 [1.165, 1.211] & 0.808 [0.797, 0.819] & 0.513 [0.496, 0.524] & 65/105\\
Two-stage (C1), MC 95\% & [0.736, 0.754] & [0.581, 0.609] & [0.427, 0.440] & \\
\midrule
best other localization & 0.824 [0.811, 0.833] & 0.612 [0.604, 0.621] & 0.441 [0.436, 0.445] & \\
best proven-valid region & 0.781 [0.769, 0.785] & 0.607 [0.595, 0.617] & 0.441 [0.435, 0.443] & \\
\bottomrule
\end{tabular}
\end{adjustbox}
\end{table}

\paragraph{Where VA loses.} The asymptotic $\chi^2$ region was narrower in 7 cells, all at $n=32$ and six with $\delta=0.01$, by a ratio of at most 1.157; there $\theta_\delta$ depends on mass below $\delta$, where $P_k$ barely moves. AMARI was narrower than VA in 58 of the 105 cells where it was run: 38 of 45 at $n=32$, 20 of 45 at $n=128$ and none at $n=512$. AMARI gives an asymptotic interval for one functional at a time and targets $\theta(G)$; VA's intervals are finite-sample, conditional and simultaneous, so the eight targets of Section~\ref{sec:pair} need no multiplicity correction.

\paragraph{Computation.} A VA replicate for seven targets took 21, 105 and 130~s at $m=1000$ and $n=32$, 128 and 512 (median over laws of the mean per replicate), against about 1~s for the two-stage band and 6, 23 and 181~s for AMARI, in single-threaded processes on a shared host (Table~\ref{S-tab:s4-runtime}). $K=\{1,\dots,n\}$ was 6--14\% wider at $n=128$ and impractical at $n=512$ (Supplement~\ref{S-supp:simextra}).

\section{Coverage under resampling from real rollout pools}\label{sec:mb}

To check the implementation under a law nobody chose, we use the Monkey Business data \citep{brown2024large}, which store 10{,}000 rollouts per problem: GSM8K with Llama-3-8B-Instruct and Llama-3-70B-Instruct ($m=127$ each), and MATH with Llama-3-8B-Instruct, Llama-3-70B-Instruct, Llama-3-8B, Gemma-2B, Gemma-7B and Pythia-12B ($m=128$ each), and the pool of all eight ($m=1022$ problem--model units; the same problem appears under several models). The pseudo-truth $\hat p_i$ is the frequency of correct answers over the stored rollouts. Each replicate draws $n\in\{32,128,512\}$ rollouts per unit \emph{with replacement}, independently across units, so $c_i\sim\Bin(n,\hat p_i)$ holds exactly and $\theta(\hat p)$ is the conditional target of Theorem~\ref{thm:coverage}. The design tests the implementation under a realistic, heavily clustered $F_m$, not the binomial model itself: any dependence between rollouts of a problem is removed by the resampling. We used $R=20$ per configuration and $R=100$ ($n\le128$) or 60 ($n=512$) for the pool.

Every localization covered $\theta_\delta(\hat p)$ in every replicate, for every configuration and the pool. The plug-in bootstrap did not: its minimum coverage over configurations was 0.000, 0.020 and 0.700 at $n=32$, 128 and 512. On the pool (Table~\ref{tab:mb-wr}), VA was narrower than the two-stage band, KL-Bonf and Berk--Jones in all 81 (configuration, $n$, $\delta$) cells, and its widths fell from 0.137--0.067 at $n=128$ to 0.057--0.039 at $n=512$, against about 0.10 for Berk--Jones at $n=512$. \textbf{AMARI is shorter here.} On the pool, the median ratio VA/AMARI over $\delta$ is 1.412 at $n=32$ and 1.290 at $n=128$, and against AMARI at $\alpha/3$ it is 1.238 and 1.122; over the 54 cells with $n\le128$, VA is narrower than pointwise AMARI in 10 and than AMARI at $\alpha/3$ in 20. AMARI was not run on these data at $n=512$. In a check with $R=500$ replicates in 48 cells, AMARI's lowest coverage was 0.962 [0.941, 0.977], and no cell had a Clopper--Pearson upper bound below $0.95$ (Supplement~\ref{S-supp:mbextra}).

\begin{table}[tbp]
\centering
\footnotesize
\caption{Monkey Business, pooled data ($m=1022$ problem--model units), median width of the 95\% interval for $\theta_\delta(\hat p)$, rollouts resampled with replacement, $K_{\rm geo}$. AMARI: Empirikos-faithful (empirical standard error, intersected with the pilot localization), pointwise and at $\alpha/3$; run at $n\le128$ only. The plug-in bootstrap undercovers (minimum coverage over configurations 0.000, 0.020 and 0.700 at $n=32$, 128 and 512) and is shown for reference.}
\label{tab:mb-wr}
\begin{adjustbox}{max width=\textwidth}
\begin{tabular}{rrcccccccc}
\toprule
$n$ & $\delta$ & VA & Two-stage & KL-Bonf & Berk--Jones & $\chi^2$ & AMARI & AMARI $\alpha/3$ & Plug-in\\
\midrule
32 & 0.01 & 0.283 & 0.326 & 0.336 & 0.329 & 0.307 & 0.245 & 0.263 & 0.055\\
32 & 0.02 & 0.276 & 0.326 & 0.336 & 0.329 & 0.293 & 0.166 & 0.187 & 0.055\\
32 & 0.05 & 0.148 & 0.201 & 0.217 & 0.204 & 0.211 & 0.105 & 0.120 & 0.054\\
128 & 0.01 & 0.137 & 0.182 & 0.193 & 0.180 & 0.226 & 0.103 & 0.116 & 0.046\\
128 & 0.02 & 0.100 & 0.148 & 0.160 & 0.150 & 0.207 & 0.077 & 0.089 & 0.047\\
128 & 0.05 & 0.067 & 0.117 & 0.135 & 0.124 & 0.194 & 0.059 & 0.071 & 0.050\\
512 & 0.01 & 0.057 & 0.100 & 0.111 & 0.102 & 0.198 & -- & -- & 0.042\\
512 & 0.02 & 0.048 & 0.096 & 0.109 & 0.100 & 0.204 & -- & -- & 0.044\\
512 & 0.05 & 0.039 & 0.094 & 0.112 & 0.099 & 0.216 & -- & -- & 0.049\\
\bottomrule
\end{tabular}
\end{adjustbox}
\end{table}

\section{Illustration: a distilled model and an RL model trained from it}\label{sec:pair}

\citet{yue2025does} argue from pass@$k$ curves that RLVR from pretrained bases improves sampling efficiency but narrows the reasoning boundary; \citet{liu2025prorl} report that prolonged RL from a distilled model can expand it. Our pair is of the second kind: DeepSeek-R1-Distill-Qwen-1.5B (distilled) and DeepScaleR-1.5B-Preview, an RL model trained from it.

\paragraph{Data.} We use the per-problem counts released by \citet{yang2026crossovers} (repository \texttt{cafferychen777/passk-inference}, commit \texttt{ac7c3fc5}). The $m=1060$ problems come from OlympiadBench (633), Gaokao-2023-en (358), AMC23 (39) and AIME24 (30), as recorded per problem in both count files. Each model generated $n=128$ answers per problem (vLLM, temperature $0.6$, top-$p$ $0.95$, at most 32{,}768 tokens), scored by the \texttt{math-verify} package with a verification timeout scored as a failure; checkpoint revisions and generation seeds were not recorded upstream. Truncation, timeouts and guessing bias the measured $p_i$ most at the low end, so our coverage statements concern ``$p_i$ as measured by this verifier under this decoding configuration'' (Supplement~\ref{S-supp:pairextra}).

\paragraph{Targets.} We report $\theta_\delta$ for $\delta\in\{0.01,0.02,0.05\}$, $P(p\ge0.05)$, $P(p\ge0.5)$, pass@256, pass@1024 and $P(p>0)$. Table~\ref{tab:e3} shows four; all eight are in Table~\ref{S-tab:e3-all}. VA is the primary analysis and the other regions are comparators. Because the VA intervals hold simultaneously for all bounded $h$, they remain valid for targets chosen after seeing the data; which targets are emphasised here should be read as exploratory.

\begin{table}[tbp]
\centering
\scriptsize
\caption{Distilled model (DeepSeek-R1-Distill-Qwen-1.5B) and RL-from-distilled model (DeepScaleR-1.5B-Preview) on the counts of \citet{yang2026crossovers}, $m=1060$, $n=128$; sure constraints in every LP region. Marginal 95\% intervals (certified endpoints) and the Bonferroni difference (each marginal at $\alpha/2$). $^*$Excludes 0. AMARI: Empirikos-faithful, pointwise and Bonferroni-adjusted over the eight targets ($\alpha/8$). The plug-in bootstrap targets the observed-count functional $m^{-1}\sum_ih(c_i/n)$, has no coverage guarantee for $\theta(F_m)$ and undercovered in the simulations. AMARI is intersected with a pilot DKW localization at the fixed level 0.01, not adjusted for multiplicity; for $P(p\ge0.5)$ the pilot interval binds, so the pointwise and $\alpha/8$ intervals coincide. AMARI's pass@$k$ intervals use point evaluation on a finite node set. All regions and the targets $\theta_{0.02}$, $P(p\ge0.05)$, pass@256 and $P(p>0)$: Table~\ref{S-tab:e3-all}.}
\label{tab:e3}
\begin{adjustbox}{max width=\textwidth}
\begin{tabular}{llccc}
\toprule
Target & Region & Distilled & RL & RL$-$distilled (Bonferroni)\\
\midrule
$\theta_{0.01}$ & VA & [0.871, 0.947] & [0.854, 0.912] & [-0.096, 0.043]\\
 & Two-stage & [0.857, 0.968] & [0.841, 0.933] & [-0.132, 0.081]\\
 & Berk--Jones & [0.851, 0.973] & [0.829, 0.946] & [-0.150, 0.102]\\
 & AMARI & [0.868, 0.937] & [0.844, 0.905] & [-0.100, 0.043]\\
 & AMARI ($\alpha/8$) & [0.860, 0.946] & [0.835, 0.913] & [-0.115, 0.058]\\
 & Plug-in bootstrap & [0.866, 0.904] & [0.846, 0.885] & [-0.064, 0.024]\\
\addlinespace
$\theta_{0.05}$ & VA & [0.831, 0.867] & [0.831, 0.860] & [-0.038, 0.031]\\
 & Two-stage & [0.815, 0.886] & [0.815, 0.876] & [-0.075, 0.065]\\
 & Berk--Jones & [0.805, 0.892] & [0.803, 0.886] & [-0.094, 0.086]\\
 & AMARI & [0.825, 0.869] & [0.821, 0.865] & [-0.054, 0.046]\\
 & AMARI ($\alpha/8$) & [0.817, 0.877] & [0.812, 0.874] & [-0.070, 0.062]\\
 & Plug-in bootstrap & [0.823, 0.865] & [0.820, 0.861] & [-0.049, 0.043]\\
\addlinespace
$P(p\ge0.5)$ & VA & [0.569, 0.653] & [0.670, 0.724] & [0.015, 0.156]$^*$\\
 & Two-stage & [0.513, 0.713] & [0.625, 0.777] & [-0.095, 0.270]\\
 & Berk--Jones & [0.524, 0.695] & [0.625, 0.765] & [-0.077, 0.249]\\
 & AMARI & [0.527, 0.696] & [0.625, 0.770] & [-0.070, 0.243]\\
 & AMARI ($\alpha/8$) & [0.527, 0.696] & [0.625, 0.770] & [-0.070, 0.243]\\
 & Plug-in bootstrap & [0.586, 0.643] & [0.668, 0.724] & [0.016, 0.148]$^*$\\
\addlinespace
pass@1024 & VA & [0.885, 1.000] & [0.865, 0.996] & [-0.135, 0.112]\\
 & Two-stage & [0.861, 1.000] & [0.844, 0.999] & [-0.158, 0.139]\\
 & Berk--Jones & [0.855, 1.000] & [0.833, 0.999] & [-0.170, 0.147]\\
 & AMARI & [0.863, 0.976] & [0.857, 0.969] & [-0.127, 0.114]\\
 & AMARI ($\alpha/8$) & [0.853, 0.986] & [0.845, 0.981] & [-0.147, 0.134]\\
 & Plug-in bootstrap & [0.871, 0.909] & [0.851, 0.889] & [-0.065, 0.024]\\
\bottomrule
\end{tabular}
\end{adjustbox}
\end{table}

\paragraph{Marginal intervals.} Among the localizations, the VA family gave the narrowest interval for every target other than $P(p>0)$. With the sure constraints every LP region reports the lower endpoint 0.8906 (distilled) and 0.8708 (RL) for $P(p>0)$, the shares of problems solved at least once, and the upper endpoint $1$. For pass@1024 the VA interval is $[0.885,1.000]$ and $[0.865,0.996]$: only a lower bound is informative. Pointwise AMARI was shorter than VA for pass@256, pass@1024 and, for the distilled model, $\theta_{0.01}$ and $\theta_{0.02}$, and wider for the other targets.

\paragraph{The nearly identified target.} At $n=128$, $P(p\ge0.5)$ is nearly identified: only 5.9\% (distilled) and 3.7\% (RL) of the problems have counts between 54 and 74. VA gave $[0.569,0.653]$ and $[0.670,0.724]$. Two-index brackets show how little of the LP this target needs. For the three index pairs $K=\{64-d,64+d\}$, $d\in\{5,10,15\}$, all reported, VA gave widths 0.094, 0.077 and 0.099 (distilled) and 0.056, 0.051 and 0.068 (RL), against 0.084 and 0.054 with $K_{\rm geo}$: comparable, and narrower only with $\{54,74\}$, which gave $[0.572,0.649]$ and $[0.672,0.723]$. A closed-form bracket from the same two tail counts with KL bounds, $\theta\in[(\mu_{74}-P_{74}(\tfrac12))/(1-P_{74}(\tfrac12)),\,\mu_{54}/P_{54}(\tfrac12)]$, gave $[0.518,0.702]$ and $[0.623,0.777]$; over the three pairs the closed-form brackets were 1.8--6.0 times as wide as the VA brackets (Supplement~\ref{S-supp:pairextra}). For this target the conditional-variance calibration, not the number of indices, carries the width.

\paragraph{Paired difference.} For the difference we use each marginal at level $\alpha/2$ and take $[L_{\rm rl}-U_{\rm dist},\,U_{\rm rl}-L_{\rm dist}]$, valid and conservative because it ignores the pairing. Among the localizations, only the VA family excluded zero, and only for $P(p\ge0.5)$: $[0.015,0.156]$, of width $0.141$, against $0.365$ for the two-stage band. The Bonferroni split costs little (width $0.138$ with each marginal at $\alpha$). The two-index VA brackets gave $[0.022,0.152]$ with $K=\{54,74\}$, $[0.007,0.165]$ with $\{59,69\}$ and $[-0.001,0.167]$ with $\{49,79\}$, and the closed-form KL brackets contained zero for all three pairs. AMARI's difference, $[-0.070,0.243]$, and every other AMARI difference contained zero. By source, the VA difference for $P(p\ge0.5)$ was $[0.004,0.180]$ on OlympiadBench ($m=633$) and $[-0.022,0.138]$ on Gaokao-2023-en ($m=358$); on AMC23 and AIME24 ($m\le39$) the marginal intervals are about 0.3--0.4 wide (Table~\ref{S-tab:bysource}). These per-source intervals are not simultaneous over sources. No region certified a change in $\theta_\delta$, pass@256, pass@1024 or $P(p>0)$.

The findings should be attributed separately. VA certifies that, on this benchmark and with this verifier, the share of problems solved at least half the time is higher for the RL model. \citet{yang2026crossovers} certify a pass@$k$ crossover for this pair, a gain for $k\le10$ and a loss for $61\le k\le128$, a statement about a different functional. VA certifies nothing about $P(p>0)$ or about pass@$k$ for $k>n$, and one pair of models supports no general statement about RL.

\begin{figure}[tbp]
\centering
\includegraphics[width=0.92\textwidth]{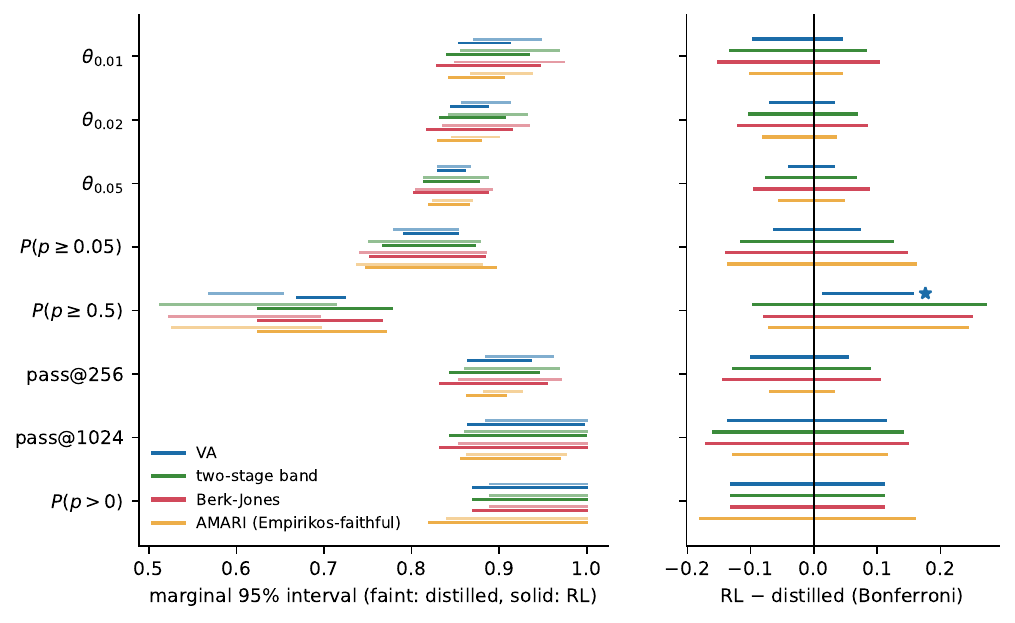}
\caption{Distilled and RL-from-distilled models on the counts of \citet{yang2026crossovers}. Left: marginal 95\% intervals (faint: distilled; solid: RL) for VA, the two-stage band, Berk--Jones and AMARI (pointwise). Right: Bonferroni intervals for the difference RL$-$distilled; the star marks the one interval that excludes zero (VA, $P(p\ge0.5)$).}
\label{fig:forest}
\end{figure}

\section{Discussion}\label{sec:disc}

The main contribution is a localization for binomial counts calibrated to the conditional variance of the candidate law, with finite-sample coverage conditional on the success probabilities and simultaneous over all bounded functionals. A fixed benchmark needs that guarantee, and the plug-in bootstrap most often reported in practice lacks it. Under a rule fixed before the main comparator runs, VA was narrower than the two-stage band at every $n$ and, in the median, than Bonferroni-adjusted AMARI at $n\ge128$ in simulations; it was not narrower than pointwise AMARI at $n=128$, it was wider at $n=32$, and on the Monkey Business pools it was wider than both pointwise and Bonferroni-adjusted AMARI at $n\le128$. For the super-population target its width advantage disappears.

VA is also very conservative: the true law violated one of its bands in at most 0.00207 of draws at nominal level $0.05$. Routes to recover the slack while keeping conditional validity include tail bounds that are sharper for indicators with small variance \citep{bentkus2004hoeffding,pinelis2014bennett}, a Chernoff band for the Poisson-binomial law of $mT_k$, whose log moment generating function is also linear in $G$, and a union bound weighted towards the indices that carry the width.

The evidence has limits. Each simulated law was drawn once per $m$; there is no atom-free law; the Berk--Jones threshold is a Monte Carlo quantile for continuous iid data (VA was at least 9\% narrower than Berk--Jones in every cell); AMARI is our re-implementation, with a fixed pilot level and node set whose sensitivity we did not study; the separated-law rates are shown on an atomic grid only; and the illustration concerns one pair of models measured by one verifier.

Open problems:
(1) a lower bound for the estimation term in the conditional model (Open problem~\ref{op:estimation}) and the limit of $c_n$ in Remark~\ref{rem:cn};
(2) finite-sample validity of DKW with Massart's constant and of the simultaneous Berk--Jones band under non-identical counts, and removing $\log|K|$ by chaining;
(3) a paired joint region with width theory, the most useful extension for comparing two models \citep[compare][]{kline2021reasonable,gu2025reasonable,vinayak2019optimal};
(4) a super-population version of VA that keeps simultaneity over a useful class of functionals, and a VA pilot inside AMARI;
(5) known verifier error rates, problem-dependent numbers of rollouts $n_i$, and other count models whose tail statistics have a conditional variance linear in the mixing law.

\paragraph{Practical guidance.} VA is a tool for questions about a fixed set of problems. It helps most when the $p_i$ pile up near $0$ and $1$ and when $n$ is large, and least at small $n$ and small $\delta$. Finite-sample simultaneous intervals are wide on small benchmarks: under F2, VA's interval for $\theta_\delta$ was below about $0.2$ from about $m=100$ with $n\ge128$ (one cell, at $m=250$ and $n=128$, gave 0.201, because each $m$ has its own draw of $F_m$) and below $0.1$ in every $\theta_\delta$ cell at $m=250$ with $n=512$ and at $m\ge1000$ with $n\ge128$, but in no cell with $m\le100$. Do not use $K=\{1,\dots,n\}$ at $n=512$. To plan $n$, the identification rule of Supplement~\ref{S-supp:design} uses LP values of $E_n$; to plan $m$, an oracle-centred VA LP at a pilot law can be solved. For the pair of Section~\ref{sec:pair}, certifying a difference of $0.025$ in $\theta_{0.01}$ would need about four times the observed number of rollouts, with $(m,n)=(1060,512)$ or $(2120,256)$, or sixteen times as many problems at $n=128$; the oracle-centred width is a planning proxy, not a guarantee. Table~\ref{tab:which} summarises when to use which method.

\begin{table}[tbp]
\centering\small
\caption{When to use which method.}
\label{tab:which}
\begin{tabular}{p{5.0cm}p{9.5cm}}
\toprule
Situation & Suggested method\\
\midrule
Super-population target, one pre-specified functional & AMARI with an empirical standard error (asymptotic, pointwise)\\
Fixed benchmark, several functionals, or functionals chosen after seeing the data & VA (finite-sample, conditional, simultaneous)\\
Fixed benchmark, one pre-specified functional & VA if conditional finite-sample coverage is required; otherwise pointwise AMARI, which was as short or shorter at $n\le128$ but is asymptotic and built for $\theta(G)$\\
pass@$k$ with $k\le n$ & The unbiased estimator of \citet{chen2021evaluating}, with paired inference as in \citet{yang2026crossovers}\\
pass@$k$ with $k\gg n$, or $P(p>0)$ & VA lower bound only; a parametric extrapolation only with its assumptions stated\\
Threshold shares $P(p\ge\gamma)$, $\gamma$ inside $(0,1)$ & VA, or VA with two indices around $n\gamma$ fixed in advance; at $n=512$ other localizations were narrower in our simulations\\
Small benchmark ($m$ below about 100) & VA, expecting widths of 0.1--0.4 for $\theta_\delta$; plug-in bootstrap intervals are shorter but can undercover\\
Two models on the same problems & Bonferroni over the two VA marginals for now; a paired joint region is open\\
\bottomrule
\end{tabular}
\end{table}

\section*{Declarations}\label{sec:decl}
\paragraph{Data and code availability.} The Monkey Business data \citep{brown2024large} are available from the Hugging Face dataset \texttt{ScalingIntelligence/monkey\_business} under the MIT licence; we read only the per-sample correctness indicators, and the terms for the underlying problems and model outputs may differ. The counts for the two models are from \texttt{github.com/cafferychen777/passk-inference} (MIT), commit \texttt{ac7c3fc58d5e325447607e778bdcbc897182cf7e}, files \texttt{examples/deepscaler32k/\{base,rl\}.jsonl}. The Python package \texttt{valoc}, the experiment scripts, the numerical checks, pinned requirements and the scripts that generate every table and figure are available at \url{https://github.com/YYKKCC-kevin/valoc} under the MIT licence. All computations used CPUs only.

\paragraph{Use of AI assistance.} AI assistants (Claude, Anthropic; Codex, OpenAI) were used in drafting and checking proofs, code and text; the author checked the final results and takes full responsibility for the content.

\phantomsection\label{end:body}
\bibliographystyle{plainnat-nonote}
\bibliography{refs}

\end{document}


\sloppy
\maketitle
\tableofcontents

\noindent Numbers such as Theorem~\ref{M-thm:coverage} or equation~\eqref{M-eq:vargap} refer to the main text. Results labelled S are stated and proved here. Notation is as in Sections~\ref{M-sec:setting}--\ref{M-sec:width} of the main text: $L=\log(2|K|/\alpha)$, $a=\sqrt{2L/m}$, $b=L/(3m)$, $u(x)=x(1-x)$, and $F=F_m$ is the empirical mixing law. Experiment labels: E1 is the main simulation study, E2 the Monkey Business resampling check, E3 the illustration with the distilled and RL models, E4 the summary of E1 widths, E5 the band-event Monte Carlo study, E6 the index-set study, E7 the high-replicate AMARI coverage study and E8 the separated-law computations. Numerical checks are labelled CHK-1 to CHK-9; each is a script in the released code (\texttt{checks/}) with a fixed seed or a deterministic computation, and its output is stored with the results. Unless a table or figure says otherwise, it comes from the main runs, in which every LP region has the sure constraints. Section~\ref{supp:archival} collects, for reference only, tables and figures from a preliminary run without the sure constraints and with a model-based AMARI variant; the main text does not use them.

\section{Proofs for Section~\ref{M-sec:va}}\label{supp:proofs-va}

\begin{proof}[Proof of Lemma~\ref{M-lem:bands}]
\emph{Step 1 (setup).} Let $X_i=\mathbf 1\{c_i\ge k\}-P_k(p_i)$. These are independent, with $\E X_i=0$, $X_i\le1$ and $\Var X_i=u(P_k(p_i))$. Let $S=\sum_iX_i=m\{T_k-\mu_k(F)\}$ and $v=\sum_i\Var X_i=mV_k(F)$. If $v=0$, every $X_i=0$ almost surely, so $S=0$ and the two events $\{S>s\}$ and $\{-S>s\}$ with $s=m\,r(0)\ge0$ have probability $0$; this is why the lemma uses strict inequalities, since $r_{Be}(0)=0$. Assume $v>0$. For $\lambda\ge0$ the function $x\mapsto(e^{\lambda x}-1-\lambda x)/x^2$ equals $\lambda^2\int_0^1(1-s)e^{\lambda xs}\,ds$ and is nondecreasing on $\R$, so for $x\le1$, $e^{\lambda x}\le1+\lambda x+x^2(e^\lambda-1-\lambda)$. Taking expectations and using $1+y\le e^y$,
\[
\log\E e^{\lambda S}\le v\,(e^\lambda-1-\lambda).
\]
\emph{Step 2 (Bennett).} By Chernoff's bound with $\lambda=\log(1+s/v)$, $\Pr(S\ge s)\le\exp\{-v\,h_B(s/v)\}$. Setting $v\,h_B(s/v)=L$ gives $s=v\,h_B^{-1}(L/v)=m\,r_{Be}(V_k(F))$.

\emph{Step 3 (Bernstein).} For $0\le\lambda<3$, since $j!\ge2\cdot3^{j-2}$, $e^\lambda-1-\lambda\le\lambda^2/\{2(1-\lambda/3)\}$. Let $c=1/3$, $s=\sqrt{2vL}+cL$, $\rho=\sqrt{2L/v}$ and $\lambda=\rho/(1+c\rho)<1/c$. Then $1-c\lambda=1/(1+c\rho)$ and
\[
\lambda s-\frac{v\lambda^2}{2(1-c\lambda)}=\frac{\rho(v\rho+cL)}{1+c\rho}-\frac{v\rho^2}{2(1+c\rho)}=\frac{2L+c\rho L-L}{1+c\rho}=L,
\]
using $v\rho^2=2L$ and $v\rho=\sqrt{2vL}$. Chernoff gives $\Pr(S\ge s)\le e^{-L}$, hence $\Pr(S>s)\le e^{-L}$, and $s/m=r_B(V_k(F))$; the same holds in Step 2.

\emph{Step 4 (other tail).} Apply Steps 1--3 to $-X_i\le1$.

\emph{Step 5 ($r_{Be}\le r_B$).} The Bennett bound is the optimised Chernoff bound for the moment generating function bound $v(e^\lambda-1-\lambda)$, which is smaller than the sub-gamma bound used in Step 3. So at $s=mr_B$, $v\,h_B(s/v)\ge L$, and since $h_B$ is increasing, $mr_{Be}\le mr_B$.

\emph{Step 6 (shape).} $r_B$ is concave and nondecreasing. $r_{Be}(V)=m^{-1}x\,h_B^{-1}(L/x)$ with $x=mV$ is the perspective of the concave increasing function $h_B^{-1}$, hence concave on $(0,\infty)$. It is positive on all of $(0,\infty)$, since $h_B^{-1}>0$ there. A concave function that is bounded below on $[0,\infty)$ is nondecreasing: if $f(x)>f(y)$ for some $x<y$, concavity gives $f(z)\le f(y)-\{f(x)-f(y)\}(z-y)/(y-x)\to-\infty$ as $z\to\infty$. Hence $r_{Be}$ is nondecreasing on $[0,\infty)$, in particular on $[0,1/4]$. Positivity on $[0,1/4]$ alone would not suffice: $V(0.3-V)$ is concave, vanishes at $0$ and is positive on $(0,1/4]$, but decreases on $(0.15,1/4]$.
\end{proof}

In a Monte Carlo check (CHK-3, \texttt{checks/chk3\_bernstein\_band\_mc.py}) with $n=64$, $m=400$, $|K|=64$, 150 zeros and 250 draws from $\mathrm{Beta}(0.3,1.5)$, and $2\times10^4$ replications, the simultaneous non-coverage of the Bernstein band was $0.0001$ (2 of $2\times10^4$; Clopper--Pearson interval $[0.00001,0.00036]$) at nominal $0.05$. The upper-side summands satisfy $X_i\le1-P_k(p_i)$, so the range constant $1$ in Step 1 is loose; a sharper range would tighten the band (Section~\ref{M-sec:disc}).

\begin{proposition}[No unbiased estimator of the conditional variance]\label{prop:nounbiased}
For $n\ge1$ and $1\le k\le n$, no statistic $\phi(c_1,\dots,c_m)$ satisfies $\E\phi=V_k(F_m)$ for all $(p_i)\in[0,1]^m$.
\end{proposition}
\begin{proof}
$\E\phi$ is a polynomial of degree at most $n$ in each $p_i$ separately, whereas $V_k(F_m)$, a sum of the terms $m^{-1}\{P_k(p_i)-P_k(p_i)^2\}$, has degree $2n$ in $p_i$, since the leading coefficient of $P_k$ is $\pm\binom{n-1}{k-1}\ne0$.
\end{proof}

\begin{lemma}[Monotone bracketing]\label{lem:bracket}
Let $k_-<k<k_+$ with $k_\pm$ consecutive in $K$, and let $r_\pm$ be the radii at $k_\pm$. For every $G$ in the exact VA region, $\mu_k(G)\in[\mu_{k_+}(G),\mu_{k_-}(G)]\subseteq[T_{k_+}-r_+,T_{k_-}+r_-]$, while $T_k\in[T_{k_+},T_{k_-}]$. Hence $|T_k-\mu_k(G)|\le(T_{k_-}-T_{k_+})+\max(r_-,r_+)$.
\end{lemma}
\begin{proof}
$P_k$ and $T_k$ are nonincreasing in $k$.
\end{proof}

\subsection{The sure constraints}\label{supp:sure}
Let $N_0=\#\{i:c_i=0\}$, $N_n=\#\{i:c_i=n\}$, $n\ge1$, and let the partition contain $\{0\}$ (index $j_0$) and $\{1\}$ (index $j_1$) as cells. The sure constraints are $w_{j_0}\le N_0/m$ and $w_{j_1}\le N_n/m$. Their validity is the extra sentence in Step 3 of the proof of Theorem~\ref{M-thm:coverage}; the same sentence makes them valid for every LP region whose coverage proof is ``$w^*$ is feasible on the region's event'', which covers all regions of Table~\ref{M-tab:regions} and the two-stage band.

\begin{proposition}[No other sure linear constraints]\label{prop:sure-only}
Fix $m$, a partition as above and an observed count vector $c$. Let $\mathcal S(c)$ be the set of cell-weight vectors $w^*(p)$ of parameter vectors $p\in[0,1]^m$ with $\Pr_p(c)>0$, and $\Pi(c)=\{w\in\Delta_J:w_{j_0}\le N_0/m,\ w_{j_1}\le N_n/m\}$. Then $\mathrm{conv}\,\mathcal S(c)=\Pi(c)$. Consequently every linear inequality that holds on $\mathcal S(c)$ is implied by $w\in\Pi(c)$, and no sure linear constraint bounds the mass of a cell inside $(0,1)$.
\end{proposition}
\begin{proof}
$\Pr_p(c)>0$ if and only if $p_i\in[0,1)$ when $c_i=0$, $p_i\in(0,1]$ when $c_i=n$, and $p_i\in(0,1)$ otherwise. So $\mathcal S(c)\subseteq\Pi(c)\cap m^{-1}\mathbb Z^J$, by the argument of Step 3. Conversely, given an integer vector $x\ge0$ with $\sum_jx_j=m$, $x_{j_0}\le N_0$ and $x_{j_1}\le N_n$, put $x_{j_0}$ of the problems with $c_i=0$ at $p=0$, $x_{j_1}$ of those with $c_i=n$ at $p=1$, and the remaining problems at interior points of the other cells according to $x$; each count then has positive probability. Hence $\mathcal S(c)=\Pi(c)\cap m^{-1}\mathbb Z^J$. The constraint matrix of $\{x\ge0,\ \mathbf 1^\top x=m,\ x_{j_0}\le N_0,\ x_{j_1}\le N_n\}$ has in each column at most one entry in the all-ones row and at most one in a unit row, so it is totally unimodular (split its rows into these two classes), and with integer right-hand sides every vertex of $m\Pi(c)$ is integral. Hence $\mathrm{conv}\,\mathcal S(c)=\Pi(c)$, and the last statements follow from Farkas' lemma.
\end{proof}

Integrality, $mw^*\in\mathbb Z^J$, is a further sure constraint, but it is not convex and its convex hull is $\Pi(c)$. Statements such as ``most problems with $c_i=0$ have small $p_i$'' are probabilistic and already enter through the bands.

\begin{proposition}[When the sure constraints change nothing]\label{prop:sure-noeffect}
Call an LP region box-monotone if each constraint depends on $w$ only through sums $\sum_j\phi^{lo}_{\ell j}w_j$, $\sum_j\phi^{hi}_{\ell j}w_j$ and $\sum_j\bar\phi_{\ell j}w_j$ of lower, upper and maximal cell envelopes of functions $\phi_\ell$, and stays satisfied when the lower sums decrease and the other sums increase. VA, VA-Bennett, \VA$\cap$KL and every band region $\{\sum_jP^{hi}_{kj}w_j\ge\ell_k,\ \sum_jP^{lo}_{kj}w_j\le u_k\}$ (Hoeff-$k$, KL-Bonf, iid-Bernstein, exact-binomial, DKW variants, Berk--Jones, EB-DKW and the two-stage band) are box-monotone. Let $C_1=(0,g_1)$ and $C_{J-1}=[g_{R-1},1)$ be the cells next to $\{0\}$ and $\{1\}$. If $\inf_{C_1}h\le h(0)\le\sup_{C_1}h$ and $\inf_{C_{J-1}}h\le h(1)\le\sup_{C_{J-1}}h$, which holds when $h$ is right-continuous at $0$ and left-continuous at $1$, then $U$ and $\Lo$ of a box-monotone region are the same with and without the sure constraints. For $h=\mathbf 1\{p>0\}$, $U=1$ and $\Lo_{\rm new}=\max(\Lo_{\rm old},T_1)$; for $h=\mathbf 1\{p=1\}$, $U_{\rm new}=\min(U_{\rm old},N_n/m)$.
\end{proposition}
\begin{proof}
Given a feasible $w$, move the mass $w_{j_0}$ to $C_1$ and $w_{j_1}$ to $C_{J-1}$. The hull $[0,g_1]$ of $C_1$ contains $0$, so its lower envelopes are at most, and its upper and maximal envelopes at least, the values at $0$; the same holds at $1$. By box-monotonicity the new vector is feasible, it satisfies the sure constraints trivially, and by the condition on $h$ its objective is at least as extreme. Adding constraints cannot make the optimum more extreme, so the endpoints coincide. For $\mathbf 1\{p>0\}$, every cell other than $\{0\}$ has $h^{lo}=1$, so $\Lo=1-\max_ww_{j_0}$; the constraint caps $w_{j_0}$ at $N_0/m=1-T_1$. The case $\mathbf 1\{p=1\}$ is symmetric.
\end{proof}

An independent implementation with $n=16$, $m=150$ and 40 data sets found endpoint changes of at most $3\times10^{-7}$ for $\theta_{0.05}$, $\mathbf 1\{p\ge0.5\}$ and pass@64, and confirmed the closed forms for $\mathbf 1\{p>0\}$ and $\mathbf 1\{p=1\}$ to $10^{-13}$. In the illustration, all targets except $P(p>0)$ changed by at most $10^{-6}$ in every region when the sure constraints were added. The proposition assumes a fixed cut set, so it does not cover runs whose cutting-plane refinement adds different cuts with and without the constraints, and the $\chi^2$ and Hoeff-cell regions constrain cell probabilities, which are not monotone in $p$, so it does not cover them either; in a regression check on one simulated data set (F2, $m=1000$, $n=128$, 13 regions) no width other than that for $P(p>0)$ changed by more than $10^{-6}$.

\section{The two-stage band}\label{supp:twostage}

\begin{lemma}[Mirror inequality]\label{lem:mirror}
If $2k\ge n+1$ and $0\le j\le2k-n-1$, then $u(P_k)\le P_j-P_k$ on $[0,1]$.
\end{lemma}
\begin{proof}
Let $Q_i(p)=\Pr\{\Bin(n,1-p)\ge i\}=1-P_{n-i+1}(p)$. Lemma~\ref{lem:submult} at $1-p$ gives $Q_{2i}\le Q_i^2$. With $i=n-k+1$, $Q_i=1-P_k$ and $Q_{2i}=1-P_{2k-n-1}$, valid since $2i\le n+1$. Hence $u(P_k)=u(Q_i)=Q_i-Q_i^2\le Q_i-Q_{2i}=P_{2k-n-1}-P_k\le P_j-P_k$.
\end{proof}
A numerical check over 13 values of $n$ up to 512, all admissible $k$ and 4001 values of $p$ gave $\max\{u(P_k)-(P_{2k-n-1}-P_k)\}=0$.

\emph{Construction.} For $k\in K$ let $j_k=2k$ and $\hat D_k=T_k-T_{2k}$ if $2k\le n+1$, and $j_k=2k-n-1$ and $\hat D_k=T_{j_k}-T_k$ otherwise; $\Psi_k(F)=\E\hat D_k$, so $V_k(F)\le\Psi_k(F)$ by Lemma~\ref{lem:submult} or~\ref{lem:mirror}. With $L_1=\log(2|K|/\alpha)$, let $\Psi^+_k=\max\{q\in[\hat D_k,1]:m\KL(\hat D_k\|q)\le L_1\}$ and $\hat V^+_k=\min(\Psi^+_k,1/4)$. With $L_2=\log(4|K|/\alpha)$, $a_2=\sqrt{2L_2/m}$ and $b_2=L_2/(3m)$, the radii are $\rho_k=a_2\sqrt{\hat V^+_k}+b_2$, and the region is $\{w\in\Delta_J:\sum_jP^{hi}_{kj}w_j\ge T_k-\rho_k,\ \sum_jP^{lo}_{kj}w_j\le T_k+\rho_k,\ k\in K\}$ with the sure constraints. A Bennett variant uses $r_{Be}$ at level $L_2$.

\begin{proof}[Proof of Proposition~\ref{M-prop:twostage}]
\emph{Coverage.} Let $\Ecal_1=\{V_k(F)\le\hat V^+_k\ \forall k\}$ and $\Ecal_2=\{|T_k-\mu_k(F)|\le r_{L_2}(V_k(F))\ \forall k\}$. If $\Psi^+_k<\Psi_k$, then $\hat D_k<\Psi_k$ and $m\KL(\hat D_k\|\Psi_k)>L_1$, because $q\mapsto\KL(\hat D_k\|q)$ is increasing on $[\hat D_k,1]$. Since $m\hat D_k$ is a sum of independent Bernoulli variables with mean $m\Psi_k$, Theorem~1 of \citet{hoeffding1963probability}, applied to the complements, bounds the probability of this event by $e^{-L_1}=\alpha/(2|K|)$. A union bound gives $\Pr(\Ecal_1^c)\le\alpha/2$, and Lemma~\ref{M-lem:bands} at level $L_2$ gives $\Pr(\Ecal_2^c)\le\alpha/2$. On $\Ecal_1\cap\Ecal_2$, monotonicity of $r_{L_2}$ gives $|T_k-\mu_k(F)|\le\rho_k$, so $w^*$ satisfies every row, and the sure constraints by Step 3 of Theorem~\ref{M-thm:coverage}; Step 4 gives coverage. The comparison of the random radius with the truth's variance is deterministic on $\Ecal_1\cap\Ecal_2$, so no independence between the stages is needed.

\emph{Size of the radii.} Let $\Ecal_{\rm up}=\{\hat D_k\le\Psi_k+a_1\sqrt{\Psi_k}+L_1/(3m)\ \forall k\}$ with $a_1=\sqrt{2L_1/m}$. The summands of $m\hat D_k$ have variance at most their mean, so the Bernstein bound of Lemma~\ref{M-lem:bands} and a union bound give $\Pr(\Ecal_{\rm up})\ge1-\alpha/2$. From $\KL(x\|q)\ge(q-x)^2/(2q)$ for $x\le q$, the constraint defining $\Psi^+_k$ implies $\Psi^+_k\le\hat D_k+a_1\sqrt{\hat D_k}+2L_1/m$. On $\Ecal_{\rm up}$, with $b_1=L_1/(3m)$ and $a_1^2=6b_1$, this gives $\Psi^+_k\le(\sqrt{\Psi_k}+a_1)^2+6.45\,b_1$, so $\sqrt{\hat V^+_k}\le\sqrt{\Psi_k}+2.04\,a_1$, and $\rho_k\le a_2\sqrt{\Psi_k}+2.04\,a_1a_2+b_2$. On $\Ecal_1\cap\Ecal_2\cap\Ecal_{\rm up}$, which has probability at least $1-3\alpha/2$, the tube argument of Theorem~\ref{M-thm:width} with $D^{2S}_k=r_{L_2}(V_k(F))+\rho_k+\eta_k$ in place of $D_k(w)$ gives the conclusions of Theorem~\ref{M-thm:width}(i)--(ii), with $D^{2S}_k\le a_2\{\sqrt{V_k(F)}+\sqrt{\Psi_k(F)}\}+2b_2+2.04\,a_1a_2+\eta_k$. The remainder is $O(L/m)$.
\end{proof}

\begin{proposition}[The two width bounds]\label{prop:twostage-compare}
When the partner $2k$ lies in $K$, the leading term of VA's bound in Theorem~\ref{M-thm:width}(iii), before $\sqrt{x+y}\le\sqrt x+\sqrt y$ is applied, is $a\{\sqrt{V_k}+\kappa\sqrt{V_k+g_k}\}$, and that of the two-stage band is $a_2\{\sqrt{V_k}+\sqrt{\Psi_k}\}$ with $\Psi_k=V_k+g_k$. Their ratio lies in $[\sqrt{L/L_2},\,\kappa\sqrt{L/L_2}]$, which is $[0.955,0.994]$ at $|K|=31$, $\alpha=0.05$ and $\kappa=1.041$. VA's remainder is $O(a^{3/2})$, the two-stage band's $O(L/m)$.
\end{proposition}
The proof is algebra, since $\Psi_k\ge V_k$. So the width theory does not separate VA from the two-stage band. On the simulation laws with $K_{\rm geo}$, which lacks most partners $2k$, the per-$k$ ratio of leading terms has median 0.976--1.022, while the inflation $\sqrt{\Psi_k/V_k}$ of the plug-in variance has median 2.1--3.2 over $k$. This inflation is the room VA has to gain, since VA's radii are evaluated at candidate laws whose conditional variance can be far below $\Psi_k(F)$.

A certified comparison is possible law by law. Work with exact regions and let $\Ecal_2'=\{\hat D_k\ge\Psi_k-a_1\sqrt{\Psi_k}-b_1\ \forall k\}$. On $\Ecal\cap\Ecal_2'$, which has probability at least $1-3\alpha/2$, VA's width is at most $\sup\{\theta(G)-\theta(G'):G,G'\in\mathcal O_{\VA}\}$ with $\mathcal O_{\VA}=\{G:|\mu_k(G)-\mu_k(F)|\le r_L(V_k(F))+r_L(V_k(G))\ \forall k\}$, and the two-stage width is at least $\sup\{\theta(G)-\theta(G'):G,G'\in\mathcal I_{2S}\}$ with $\mathcal I_{2S}=\{G:|\mu_k(G)-\mu_k(F)|\le r_{L_2}(\Psi^-_k)-r_L(V_k(F))\ \forall k\}$ and $\Psi^-_k=\min\{(\Psi_k-a_1\sqrt{\Psi_k}-b_1)_+,1/4\}$ (on $\Ecal_2'$, $\hat V^+_k\ge\Psi^-_k$, and Lemma~\ref{lem:reduction} transfers the bound to the gridded region). Both quantities are deterministic and computable by one LP each. A first-order expansion of both realised widths as $m\to\infty$, which would show VA at least as narrow up to the factor $\sqrt{L/L_2}$, needs an envelope theorem for LPs whose right-hand sides depend on $w$ and uniqueness of the dual solutions; we have only a sketch.

\section{Proofs for Section~\ref{M-sec:iid}}\label{supp:proofs-iid}

\begin{proof}[Proof of Proposition~\ref{M-prop:iid}]
(a) This is Theorem~1 of \citet{hoeffding1963probability}, stated for independent, not necessarily identically distributed $X_i\in[0,1]$ with $q$ the average mean. Apply it to $X_i=\mathbf 1\{c_i\ge k\}$, and to $1-X_i$ for the lower tail. At $t=1-q_k$ the bound reads $\prod_iP_k(p_i)\le q_k^m$, which is the AM--GM inequality. The per-$k$ KL constraint is convex in $G$ because $y\mapsto\KL(x\|y)$ is convex, and for fixed $T_k$ it reduces to an interval for $\mu_k(G)$.

(b) By~\eqref{M-eq:vargap}, $V_k(F)\le u(q_k)$. Since $r$ is nondecreasing, the band event $\Ecal$ of Lemma~\ref{M-lem:bands} is contained in $\{|T_k-q_k|\le r(u(q_k))\ \forall k\}$, which therefore has probability at least $1-\alpha$. For every $G$, \eqref{M-eq:vargap} gives $V_k(G)\le u(\mu_k(G))$, so every constraint of $\Rcal_{\VA}$ implies the corresponding constraint of $\Rcal_{\rm iid}$. The factor~\eqref{M-eq:factor} is algebra.

(c) Write $S=mT_k$, $\bar s=q_k(F)$ and $B\sim\Bin(m,\bar s)$. Theorem~4 of \citet{hoeffding1956distribution} gives $\Pr(S\le c)\le\Pr(B\le c)$ for integers $0\le c\le m\bar s-1$ and $\Pr(S\le c)\ge\Pr(B\le c)$ for integers $m\bar s\le c\le m$. Lower side: if $c_{\rm lo}\le mq_k-1$, then $\Pr(S\le c_{\rm lo})\le\Pr(B\le c_{\rm lo})\le\alpha/(2|K|)$. Upper side: if $c_{\rm hi}-1\ge mq_k$, then $\Pr(S\ge c_{\rm hi})=1-\Pr(S\le c_{\rm hi}-1)\le1-\Pr(B\le c_{\rm hi}-1)=\Pr(B\ge c_{\rm hi})\le\alpha/(2|K|)$.
\end{proof}

Hoeffding's Theorem~5 gives the two-sided form $\Pr(b\le S\le c)\ge\Pr(b\le B\le c)$ for $0\le b\le m\bar s\le c\le m$, and his Theorem~3 gives $\E g(S)\le\E g(B)$ for strictly convex $g$. A numerical check over 2000 random instances with $m\le24$ and 26{,}341 inequalities (CHK-4, \texttt{checks/chk4\_hoeffding1956.py}) found no violation of the direction used in (c), while the reversed direction failed in 24{,}339 of them. The condition in (c) concerns the true $q_k$. A valid implementation checks it at each candidate $\mu_k(G)$ and falls back to the KL band of (a) when it fails; at $G=F$ the rule then selects a valid test.

For the DKW band, let $Z=\max_{1\le k\le n}|T_k-q_k|$, the Kolmogorov distance between the empirical count CDF and the averaged CDF $m^{-1}\sum_i\Pr(c_i\le\cdot)$.

\begin{proposition}[DKW for non-identical counts]\label{prop:dkw}
For every $\varepsilon>0$, $\Pr(Z\ge 4/\sqrt m+\varepsilon)\le e^{-2m\varepsilon^2}$. Hence $Z\le 4/\sqrt m+\sqrt{\log(1/\alpha)/(2m)}$ with probability at least $1-\alpha$.
\end{proposition}

At $\alpha=0.05$ the radius is $5.22/\sqrt m$, against $1.36/\sqrt m$ for Massart's constant. Per-$k$ Hoeffding with Bonferroni, of radius $\sqrt{\log(2|K|/\alpha)/(2m)}$, is already tighter, and the region of Proposition~\ref{prop:dkw} serves only as a proven-valid DKW-type comparator.

\begin{proof}[Proof of Proposition~\ref{prop:dkw}]
\emph{Step 1 (bounded differences).} Changing one $c_i$ changes every $T_k$, and therefore $Z$, by at most $1/m$. McDiarmid's inequality gives $\Pr(Z\ge\E Z+\varepsilon)\le e^{-2m\varepsilon^2}$.

\emph{Step 2 (symmetrisation).} Let $(c_i')$ be an independent copy and $\sigma_i$ independent Rademacher signs. By Jensen's inequality conditional on $(c_i)$, and because the pairs $(c_i,c_i')$ are independent and each pair is exchangeable,
\begin{align*}
\E Z&\le\E\max_k\Big|\frac1m\sum_i(\mathbf 1\{c_i\ge k\}-\mathbf 1\{c'_i\ge k\})\Big|=\E\max_k\Big|\frac1m\sum_i\sigma_i(\mathbf 1\{c_i\ge k\}-\mathbf 1\{c'_i\ge k\})\Big|\\
&\le2\,\E\max_k\Big|\frac1m\sum_i\sigma_i\mathbf 1\{c_i\ge k\}\Big|.
\end{align*}
\emph{Step 3 (maximal inequality).} Conditional on $(c_i)$, order the indices by decreasing $c_i$. Each set $\{i:c_i\ge k\}$ is a prefix of this order (tied blocks enter together), and the signs remain iid Rademacher in this order. So the maximum is at most $\max_{0\le j\le m}|S_j|$ for the partial sums $S_j$ of the reordered signs. Doob's $L^2$ inequality gives $\E\max_jS_j^2\le4\,\E S_m^2=4m$, so $\E Z\le2\cdot2\sqrt m/m=4/\sqrt m$.
\end{proof}

Replacing Doob's inequality by L\'evy's reflection inequality gives no improvement, and integrating a sub-Gaussian maximal tail gives $\E\max_j|S_j|\lesssim2.15\sqrt m$, which is worse. A Monte Carlo check at $m=20,100,500$ for a spike-plus-uniform law (CHK-1, \texttt{checks/chk1\_dkw\_EZ.py}) gives $\sqrt m\,\E Z=0.366$, $0.362$ and $0.358$, so the constant $4$ is loose by about an order of magnitude.

\subsection{Pearson $\chi^2$ (sketch)}\label{supp:chi2}
Let $\hat f$ be the empirical pmf of the counts and $\bar f=\int b\,dF$, with $b(p)$ the vector of binomial probabilities. Under fixed $p_i$, $\mathrm{Cov}(\hat f)=m^{-1}\Sigma_c$ with $\Sigma_c=\mathrm{diag}(\bar f)-\int bb^\top dF$, whereas the multinomial covariance is $\Sigma_{\rm iid}=\mathrm{diag}(\bar f)-\bar f\bar f^\top$, and $\Sigma_{\rm iid}-\Sigma_c=\int bb^\top dF-\bar f\bar f^\top\succeq0$, the matrix version of~\eqref{M-eq:vargap}. Along a triangular array with $\bar f$ bounded away from $0$ on the cells used, and under Lindeberg conditions (the step not written out), the Pearson statistic converges to $\|Y_c\|^2$ with $Y_c\sim N(0,D^{-1/2}\Sigma_cD^{-1/2})$, $D=\mathrm{diag}(\bar f)$. Writing $Y_{\rm iid}\overset d=Y_c+W$ with independent Gaussian $W$, Anderson's inequality \citep{anderson1955integral} for the symmetric convex set $\{\|y\|^2\le c\}$ gives $\Pr(\|Y_c+W\|^2\le c)\le\Pr(\|Y_c\|^2\le c)$, so the iid $\chi^2$ calibration would be asymptotically conservative. This is a sketch, not a proof. The band-event runs (Table~\ref{tab:s4-e5}) show that at $n=512$ the pooled-cell $\chi^2$ region can miss its nominal level in finite samples.

\section{Proofs for Section~\ref{M-sec:width}: the width bound}\label{supp:proofs-width}

\begin{lemma}[Sub-multiplicativity of binomial tails]\label{lem:submult}
For all $p\in[0,1]$ and integers $k,l\ge0$, $P_{k+l}(p)\le P_k(p)P_l(p)$. In particular $u(P_k)\le P_k-P_{k'}$ for every $k'\ge2k$.
\end{lemma}
\begin{proof}
The cases $p\in\{0,1\}$ are immediate. Fix $p\in(0,1)$, let $b_j=\binom nj p^j(1-p)^{n-j}$ and $S(j)=\sum_{i\ge j}b_i=P_j(p)$. The ratio $\varrho_i=b_{i+1}/b_i=(n-i)p/\{(i+1)(1-p)\}$ is nonincreasing in $i$. Then $S(j+1)/S(j)=\E[\varrho_I\mid I\ge j]$ for $I\sim b$. The law of $I$ given $I\ge j+1$ stochastically dominates the law given $I\ge j$, and $\varrho$ is nonincreasing, so $S(j+2)/S(j+1)\le S(j+1)/S(j)$. Hence $j\mapsto\log S(j)$ is concave on $\{0,\dots,n+1\}$ with $\log S(0)=0$, and therefore subadditive. Finally $u(P_k)=P_k-P_k^2\le P_k-P_{2k}\le P_k-P_{k'}$.
\end{proof}

A numerical check over all $k+l\le n+1$, $n\le256$ and 451 values of $p$ (CHK-5, \texttt{checks/chk5\_submultiplicativity.py}) found $\max(P_{k+l}-P_kP_l)=0$.

\begin{lemma}[Candidate variance]\label{lem:candvar}
For every $G\in\Pcal$ and $k\in K$, with $k'$ the partner index of Section~\ref{M-sec:width},
\[
V_k(G)\le\mu_k(G)-\mu_{k'}(G)=V_k(F)+g_k(F)+\{\mu_k(G)-\mu_k(F)\}-\{\mu_{k'}(G)-\mu_{k'}(F)\}.
\]
\end{lemma}
\begin{proof}
Integrate Lemma~\ref{lem:submult} against $G$. The identity $\mu_k(F)-\mu_{k'}(F)=V_k(F)+g_k(F)$ is the definition of $g_k$.
\end{proof}

For $F$ with tiny $V_k(F)$ the candidate variance can be much larger than $V_k(F)$ at the estimation scale: in an LP check at $n=128$, $m=1000$, the largest $V_k(G)$ over $G\in\Ncal_F(2r_B(V_k(F)))$ exceeded $V_k(F)$ by a factor of up to 41 at $k=1$ for F3. Lemma~\ref{lem:candvar} is how the proof of Theorem~\ref{M-thm:width} controls this inflation.

\begin{proof}[Proof of Theorem~\ref{M-thm:width}]
\emph{Step 1 (tube).} For $w\in\Wcal$ let $G^{lo}_w=\sum_jw_j\delta_{a_j}$ and $G^{hi}_w=\sum_jw_j\delta_{b_j}$. Then $\mu_k(G^{lo}_w)\le\mu_k(G^{hi}_w)\le\mu_k(G^{lo}_w)+\eta_k$. By~\eqref{M-eq:lp} and $\Ecal$,
\[
\mu_k(G^{hi}_w)\ge T_k-\rcut(\bar V_k(w))\ge\mu_k(F)-r_B(V_k(F))-\rcut(\bar V_k(w)),
\]
\[
\mu_k(G^{hi}_w)\le\mu_k(G^{lo}_w)+\eta_k\le T_k+\rcut(\bar V_k(w))+\eta_k\le\mu_k(F)+D_k(w).
\]
The same holds for $G^{lo}_w$ with $\eta_k$ on the other side, so $G^{lo}_w,G^{hi}_w\in\Ncal_F(D(w))\subseteq\Ncal_F(\bar D)$.

\emph{Step 2 (part (i)).} On $\Ecal$, $\Wcal$ is a nonempty polytope (it contains $w^*$), so $U$ is attained at some $w_U$. Fix $\nu>0$. Since $h$ is bounded and $C_j\ne\emptyset$, choose $x_j\in C_j$ with $h(x_j)\ge h^{hi}_j-\nu$ and put $G=\sum_j(w_U)_j\delta_{x_j}$. Because $P_k$ is nondecreasing and $x_j\in[a_j,b_j]$, $\mu_k(G^{lo}_{w_U})\le\mu_k(G)\le\mu_k(G^{hi}_{w_U})$, so by Step 1, $|\mu_k(G)-\mu_k(F)|\le D_k(w_U)\le\bar D_k$ and $G\in\Ncal_F(\bar D)$. Moreover $\theta(G)\ge U-\nu$, so $U-\theta(F)\le\omega^+_F(\bar D)+\nu$. Symmetrically, choosing $x'_j\in C_j$ with $h(x'_j)\le h^{lo}_j+\nu$ at the minimiser gives $\theta(F)-\Lo\le\omega^-_F(\bar D)+\nu$. Let $\nu\downarrow0$. Monotonicity of $h$ is not used; only monotonicity of $P_k$ is.

\emph{Step 3 (part (ii)).} For $G\in\Ncal_F(\Delta)$,
\[
\theta(G)-\theta(F)=\int\psi_\lambda\,d(G-F)+\sum_k\lambda_k\{\mu_k(G)-\mu_k(F)\}\le\gap^+_\lambda(F)+\sum_k|\lambda_k|\Delta_k,
\]
and the lower side is symmetric. For the single-$\lambda$ form, apply the identity to two measures $G_1,G_2\in\Ncal_F(\Delta)$ and use $\int\psi_\lambda\,d(G_1-G_2)\le\osc(\psi_\lambda)$.

\emph{Step 4 (crude radius).} $\bar V_k(w)\le1/4$, and the cut at $t=1/2$ gives $\rcut(1/4)\le a/2+b$.

\emph{Step 5 (cut error).} We show $\rcut(V)\le a\kappa\sqrt V+a\,t_{\min}+b$ for $V\le1/4$. If $\sqrt V<t_{\min}$, the cut at $t_{\min}$ gives $a(V/t_{\min}+t_{\min})/2+b\le a\,t_{\min}+b$. Otherwise $t\le\sqrt V\le\varrho t$ for consecutive tangent points. Writing $\sqrt V=tx$ with $x\in[1,\varrho]$, the two cuts give $\sqrt V(x+x^{-1})/2$ and $\sqrt V(x/\varrho+\varrho/x)/2$ (times $a$, plus $b$), and the smaller is at most $\kappa\sqrt V$, with the worst case at $x=\sqrt\varrho$.

\emph{Step 6 (variance of a feasible $w$).} $\bar V_{kj}\le u(P_k(b_j))+\eta_k$, since $u$ is 1-Lipschitz on $[0,1]$ and $P_k$ varies by at most $\eta_k$ on $[a_j,b_j]$. So $\bar V_k(w)\le V_k(G^{hi}_w)+\eta_k$, and by Lemma~\ref{lem:candvar} and Step 1, $V_k(G^{hi}_w)\le V_k(F)+g_k(F)+D_k(w)+D_{k'}(w)$. If $k'=n+1$, $\mu_{n+1}\equiv0$ and $D_{n+1}=0$.

\emph{Step 7 (self-bounding).} Combining Steps 5 and 6 with $\sqrt{x+y}\le\sqrt x+\sqrt y$ and $D_{k'}(w)\le\bar D_{k'}$,
\[
D_k(w)\le a\sqrt{V_k(F)}+b+a\kappa\sqrt{V_k(F)+g_k+\eta_k+D_k(w)+\bar D_{k'}}+a\,t_{\min}+b+\eta_k\le S_k+a\kappa\sqrt{D_k(w)} .
\]
\emph{Step 8.} From $x\le S+c\sqrt x$, $\sqrt x\le(c+\sqrt{c^2+4S})/2\le c+\sqrt S$. Take the supremum over $w$.
\end{proof}

\begin{proposition}[Mirror partner in Theorem~\ref{M-thm:width}(iii)]\label{prop:mirror-width}
For $k\in K$ with $2k\ge n+1$ let $k''=\max\{j\in K\cup\{0\}:j\le2k-n-1\}$, with $\mu_0\equiv1$ and $\bar D_0=0$, and $g^\downarrow_k(F)=\mu_{k''}(F)-\mu_k(F)-V_k(F)\ge0$. Then Theorem~\ref{M-thm:width}(iii) holds with $\sqrt{g_k+\eta_k+\bar D_{k'}}$ in $S_k$ replaced by $\sqrt{\min\{g_k+\bar D_{k'},\,g^\downarrow_k+\bar D_{k''}\}+\eta_k}$.
\end{proposition}
\begin{proof}
In Step 6, Lemma~\ref{lem:mirror} integrated against $G^{hi}_w$ gives $V_k(G^{hi}_w)\le\mu_{k''}(G^{hi}_w)-\mu_k(G^{hi}_w)\le V_k(F)+g^\downarrow_k+D_{k''}(w)+D_k(w)$ by Step 1. Take the smaller of the two bounds.
\end{proof}
With the upward partner alone, $g_k\ge F(\{1\})$ for every $k>n/2$, since $P_k(1)^2-P_{n+1}(1)=1$; the mirror removes this term for laws with an atom at $1$. If $\bar D_{k''}$ is itself bounded through its own partner, the recursion of Corollary~\ref{M-cor:oracle} may become circular; use the crude bound for $\bar D_{k''}$ there.

\begin{remark}[Doubling chain; sketch]
With the crude bound $\bar D_{k'}\le a+2b+\eta_{k'}$, (iii) gives $\bar D_k\lesssim a\sqrt{V_k(F)+g_k(F)}+a^{3/2}+a^2$. Iterating along the chain $k\to k'\to k''\to\cdots$ $j$ times replaces $a^{3/2}$ by $a^{2-2^{-j}}$, with constants growing in $j$ that we have not written out. We conjecture $\bar D_k\asymp a\sqrt{V_k(F)+g_k(F)}+O(a^2)$ for small $V_k,g_k$.
\end{remark}

\begin{proof}[Proof of Corollary~\ref{M-cor:oracle}]
$\bar D_k\le D^{\det}_k$ by downward induction in $k$: the bounds of Theorem~\ref{M-thm:width}(iii) are nondecreasing in $\bar D_{k'}$, and $k'>k$, so the induction hypothesis $\bar D_{k'}\le D^{\det}_{k'}$ applies. $\osc(\psi_\lambda)=2\min_c\|\psi_\lambda-c\|_\infty=2\,\mathrm{bias}_\lambda$, so the first inequality is Theorem~\ref{M-thm:width}(ii). $\rho^{\det}\ge1$ because each branch of $D^{\det}_k$ is at least $r_B(V_k(F))=a\sqrt{V_k(F)}+b$: the first since $a(\sqrt{V_k(F)}+\frac12)+2b+\eta_k\ge a\sqrt{V_k(F)}+b$, and the second since $(\sqrt{S}+a\kappa)^2\ge S\ge a(1+\kappa)\sqrt{V_k(F)}+2b$. Since $D^{\det}_k\le\rho^{\det}r_B(V_k(F))$ and $\rho^{\det}\ge1$, $2\,\mathrm{bias}_\lambda+2\sum_k|\lambda_k|D^{\det}_k\le\rho^{\det}\ell_\lambda$. The constant $\rho^{\det}$ involves $\eta_k$, so it is deterministic only when the partition does not depend on the data. If $w^*\in\Wcal$, then $\bar D_k\ge D_k(w^*)\ge r_B(V_k(F))+\rcut(\bar V_k(w^*))\ge2r_B(V_k(F))$ because $\rcut\ge r_B$, $\bar V_k(w^*)\ge V_k(F)$ and $r_B$ is nondecreasing. Without Bennett cuts, $w^*\in\Wcal$ on $\Ecal$ (proof of Theorem~\ref{M-thm:coverage}); with Bennett cuts it needs the Bennett band event, which is contained in $\Ecal$. Restricting the maximum in $\rho^{\det}$ to the support of $\lambda$ is also valid. The plug-in case follows from the Abel summation $\sum_j\beta_jb_j=\beta_0+\sum_{k\ge1}(\beta_k-\beta_{k-1})P_k$ for the Bernstein basis $b_j$.
\end{proof}

\begin{remark}[Concavity of the modulus]\label{rem:concave}
$\omega^\pm_F$ are concave and nondecreasing on $\R^K_{\ge0}$: if $G_i\in\Ncal_F(\Delta_i)$, then $sG_1+(1-s)G_2\in\Ncal_F(s\Delta_1+(1-s)\Delta_2)$, and $\theta$ is linear. Hence $\omega^\pm_F(s\Delta)-\omega^\pm_F(0)\ge s\{\omega^\pm_F(\Delta)-\omega^\pm_F(0)\}$, so shrinking all radii by a factor $s$ shrinks the excess of the bound in Theorem~\ref{M-thm:width}(i) over $\ID_K$ by at most $s$. This limits the \emph{bound} when $\bar D$ scales as a vector; it does not limit the true width, which can decay faster in a window (Corollary~\ref{M-cor:sepA}).
\end{remark}

\subsection{Further discussion of the width results}\label{supp:widthdisc}

\emph{Theorem~\ref{M-thm:width}.} The leading term of $S_k$ is of order $\sqrt{L/m}$; the term $a\,t_{\min}$ is of the same order, so the tangent floor must be small (we use $t_{\min}=10^{-7}$). For laws with mass where $P_k$ is near $1/2$, $g_k(F)$ can be of order $1/4$, and the bound is then no better than one for an iid-variance band, even when VA's realised width is much smaller. Since Theorem~\ref{M-thm:width} holds for every bounded $h$, so do Corollary~\ref{M-cor:oracle} and the certificate bounds, in particular for $\mathbf 1\{p\ge\gamma\}$, pass@$k_0$ and $\mathbf 1\{p>0\}$.

\emph{Corollary~\ref{M-cor:oracle}.} The benchmark $\ell_\lambda$ uses per-index Bernstein radii at the union-bound level $L$. The best linear-estimator interval has half-width about $\sqrt{2v_\lambda\log(2/\alpha)/m}$ with $\sqrt{v_\lambda}\le\sum_k|\lambda_k|\sqrt{V_k(F)}$, so the leading term of $\ell_\lambda$ is at least $\sqrt{L/\log(2/\alpha)}\approx1.39$ times as long at $|K|=31$, $\alpha=0.05$, and longer when the $\lambda_k$ alternate in sign. Besides the factor $2$, $\rho^{\det}$ contains the candidate-variance inflation and the cut error of Theorem~\ref{M-thm:width}(iii). The plug-in estimator $m^{-1}\sum_ih(c_i/n)$ is the special case $K=\{1,\dots,n\}$, $\lambda_k=h(k/n)-h((k-1)/n)$, with $\|\lambda\|_1=h(1)-h(0)$ and bias equal to the Bernstein-polynomial error.

\emph{Theorem~\ref{M-thm:id}.} Randomising the $p_i$ in part (ii) does not weaken the conclusion for conditional coverage: $C$ must cover at every fixed $(p_i)$ in the support of the prior, so the bound obtained on average over the prior is attained at some fixed $(p_i)$. The grid term $2\|\lambda^*\|_1\eta$ is at most $\epsilon$ only if $\eta\le\epsilon/(2\|\lambda^*\|_1)$; with $\eta_m=\sqrt{2L/m}$ the excess over $2E_n(h)$ is $O(\|\lambda^*\|_1\sqrt{L/m})$, and an adaptive grid needs at most $|K|/\eta_m+3$ cells. The constant $\|\lambda^*\|_1$ is very large: at $n=128$ and $\delta=0.01$, restricting $\|\lambda\|_1\le1000$ leaves an approximation term of 0.067 against $2E_n=0.022$ (Table~\ref{tab:tradeoff}). For the targets with $E_n(h)\ge1/2$ the content of the theory is in Theorem~\ref{M-thm:width}(i)--(ii). The theorem concerns the population problem with all $n$ tail counts; the 30--31 element sets used in the experiments are not covered, $2E_K(h)\ge2E_n(h)$ for a smaller $K$, and $K=\{1,\dots,n\}$ widened the intervals at $n=128$ by 6--14\% in our runs and was not practical at $n=512$ (Section~\ref{supp:simextra}).

\section{Proofs for Section~\ref{M-sec:separated}: separated laws}\label{supp:proofs-sep}

Throughout, $F=\pi_0\delta_0+(1-\pi_0)F_+$ with $\mathrm{supp}F_+\subseteq[p_{\min},1]$, $\delta<p_{\min}$, $h=h_\delta$, $k_*=\lceil n\delta\rceil$ and $\epsilon(j)=\Pr\{\Bin(n,p_{\min})<j\}\le\exp\{-n\KL((j-1)/n\|p_{\min})\}$ for $j-1\le np_{\min}$.

\begin{lemma}[Certificate bound]\label{lem:certificate}
On $\Ecal$, $U-\Lo\le C_{\rm ID}+\bar D_1/P_1(\delta)+\bar D_{k_*}$, and $V_k(F)\le(1-\pi_0)\epsilon(k)$, $g_k(F)\le(1-\pi_0)\epsilon(k')$.
\end{lemma}
\begin{proof}
\emph{Upper certificate.} $P_1$ is concave with $P_1(0)=0$, so $P_1(p)\ge(p/\delta)P_1(\delta)$ for $p\le\delta$ and $P_1(p)\ge P_1(\delta)$ for $p\ge\delta$; hence $h_\delta\le P_1/P_1(\delta)$. With $\lambda=e_1/P_1(\delta)$, $\sup\psi_\lambda=0$ (at $p=0$) and $\gap^+_\lambda(F)=(1-\pi_0)\int\{P_1/P_1(\delta)-1\}\,dF_+\le(1-\pi_0)\{1-P_1(\delta)\}/P_1(\delta)$.
\emph{Lower certificate.} Markov's inequality gives $P_{k_*}(p)\le np/k_*\le p/\delta$, so $h_\delta\ge P_{k_*}$. With $\lambda'=e_{k_*}$, $\inf\psi_{\lambda'}=0$ and $\gap^-_{\lambda'}(F)=(1-\pi_0)\int(1-P_{k_*})\,dF_+\le(1-\pi_0)\epsilon(k_*)$.
\emph{Variances.} On $[p_{\min},1]$, $u(P_k)\le1-P_k\le\epsilon(k)$ and $P_k^2-P_{k'}\le1-P_{k'}\le\epsilon(k')$; both vanish at $0$.
Apply Theorem~\ref{M-thm:width}(ii).
\end{proof}

\begin{lemma}[Leading term and remainder]\label{lem:leading}
For $k\in K$ let $A_k=(1+\kappa)\sqrt{V_k(F)}+\kappa\sqrt{g_k(F)}+t_{\min}$, $Q_k=a\kappa\sqrt{a+2b+\eta_{k'}}+a\kappa\sqrt{\eta_k}+2b+\eta_k$ and $R_k=Q_k+2a\kappa\sqrt{aA_k+Q_k}+a^2\kappa^2$. On $\Ecal$, $\bar D_k\le aA_k+R_k$, and $R_k=\kappa(1+2\sqrt{A_k})a^{3/2}+O(a^{7/4}+a\sqrt\eta+\eta)$ with absolute constants given $\kappa$. Set $R_{\VA}=R_1/P_1(\delta)+R_{k_*}$.
\end{lemma}
\begin{proof}
In $S_k$ of Theorem~\ref{M-thm:width}(iii), use $\sqrt{x+y+z}\le\sqrt x+\sqrt y+\sqrt z$ on $g_k+\eta_k+\bar D_{k'}$ and the crude bound $\bar D_{k'}\le a+2b+\eta_{k'}$ (since $V\le1/4$). This gives $S_k\le aA_k+Q_k$, and $(\sqrt{S_k}+a\kappa)^2=S_k+2a\kappa\sqrt{S_k}+a^2\kappa^2$. For the expansion use $\sqrt{aA+Q}\le\sqrt{aA}+\sqrt Q$ and $Q_k=\kappa a^{3/2}+O(a^2+a\sqrt\eta+\eta)$.
\end{proof}

\begin{proof}[Proof of Corollary~\ref{M-cor:sepA}]
Lemmas~\ref{lem:certificate} and~\ref{lem:leading} give $U-\Lo\le C_{\rm ID}+aA_{\VA}+R_{\VA}$ on $\Ecal$. For a competitor $\mathcal R$, Proposition~\ref{M-prop:realised} with $\rho_k=r_B(V_k(F))$ gives a realised width of at least $\epsilon_{\rm real}$ on $\Ecal\cap\{F\in\mathcal R\}$. If $m\le m_{\rm ID}(\gamma)$ then $C_{\rm ID}\le\gamma a$, and the ratio is at most $(A_{\VA}+\gamma+R_{\VA}/a)/(\epsilon_{\rm real}/a)$. The bound on $A_{\VA}$ in terms of $\epsilon(\cdot)$ uses $V_k\le\epsilon(k)$, $g_1\le\epsilon(2)$ and $g_{k_*}\le\epsilon(k_*')$. The crossover $m^*_1=2L\kappa'^4/A_{\VA}^4$, with $\kappa'=\kappa[(1+2\sqrt{A_1})/P_1(\delta)+1+2\sqrt{A_{k_*}}]$ the leading coefficient of $R_{\VA}$, is where $aA_{\VA}=\kappa'a^{3/2}$.
\end{proof}

\begin{proof}[Proof of Corollary~\ref{M-cor:sepB}]
(1) By (J2) and Chernoff, $\epsilon(2k_*)\le(2L/m)^{3/4}=a^{3/2}$; this uses $(2k_*-1)/n\le2\delta+1/n<p_{\min}$. Since $\epsilon$ is nondecreasing, $\epsilon(1),\epsilon(2),\epsilon(k_*)\le a^{3/2}$.
(2) By (J1), $1-P_1(\delta)=(1-\delta)^n\le e^{-n\delta}\le e^{-1}a^{3/2}$, so $P_1(\delta)\ge1-e^{-1}$, $\{1-P_1(\delta)\}/P_1(\delta)\le0.59\,a^{3/2}$ and $C_{\rm ID}\le1.59\,a^{3/2}$.
(3) The partner of $1$ is $2$ and that of $k_*$ is $2k_*$, so $A_1,A_{k_*}\le(1+2\kappa)a^{3/4}+a^{1/2}$ using (J3), and $aA_{\VA}=O(a^{3/2})$.
(4) $R_{\VA}=O(a^{3/2})$ by Lemma~\ref{lem:leading} and (J3).
(5) The competitor bounds are proved in Section~\ref{supp:realised}.
\end{proof}

The following oracle-centred bound came first; Proposition~\ref{M-prop:realised} is its realised version and supersedes it in the main text. We keep it because its construction is used in that proof and because it gives the constants $c_{\rm comp}$.

\begin{proposition}[Excess of iid-calibrated regions, oracle-centred]\label{prop:compexcess}
Consider the per-$k$ KL, iid-Bernstein or Hoeffding region with $c_{\rm comp}=\frac12\sqrt{\pi_0(1-\pi_0)}$, $\sqrt{\pi_0(1-\pi_0)/2}$ and $\frac12$, centred at $T_k=\mu_k(F)$, and let $\epsilon_0=c_{\rm comp}a\le\pi_0/2$. If $p_{\min}\ge\delta$, its width is at least $\ID_K(F)+\epsilon_0\{1-\omega^+_F(0)/\pi_0\}$.
\end{proposition}

So the competitors' oracle-centred excess over the identification width is of order $\sqrt{L/m}$ whenever $\omega^+_F(0)<\pi_0$. The competitor regions are exact and un-gridded; the envelope grid only enlarges them (Lemma~\ref{lem:reduction}). In a numerical check over three separated laws and $m=10^3,\dots,10^7$ (CHK-2), the ratio of the observed excess to this lower bound ranged over 4.00--8.22 (KL), 2.89--5.81 (iid-Bernstein) and 2.01--4.04 (Hoeffding), with no violation in 45 cases (Table~\ref{tab:propcheck}).

\begin{proof}[Proof of Proposition~\ref{prop:compexcess}]
\emph{Step 1.} Each competitor's constraint at $k$ depends on $G$ only through $\mu_k(G)$: the radius is a constant (Hoeffding), $r(u(\mu_k(G)))$ (iid-Bernstein), or a KL ball. With $T_k=\mu_k(F)$, every $G\in\Ncal_F(0)$ is feasible, so $\inf\theta\le\theta(F)-\omega^-_F(0)$.

\emph{Step 2.} Take $G^+\in\Ncal_F(0)$ with $\theta(G^+)\ge\theta(F)+\omega^+_F(0)-\nu$. Let $s=\epsilon_0/\pi_0\le1/2$ and $\tilde G=F+\pi_0(\delta_\delta-\delta_0)$, a probability measure because $F$ has an atom $\pi_0$ at $0$. Set $G=(1-s)G^++s\tilde G$. Since $P_k(0)=0$, $h_\delta(0)=0$ and $h_\delta(\delta)=1$,
\[
\mu_k(G)=\mu_k(F)+\epsilon_0P_k(\delta),\qquad\theta(G)\ge\theta(F)+(1-s)\{\omega^+_F(0)-\nu\}+\epsilon_0 .
\]
\emph{Step 3 (feasibility).} We have $\mu_k(G)\ge\mu_k(F)\ge(1-\pi_0)P_k(p_{\min})\ge(1-\pi_0)P_k(\delta)$, using $p_{\min}\ge\delta$, and $1-q_k(G)\ge\pi_0-\epsilon_0\ge\pi_0/2$. The deviation is $|T_k-\mu_k(G)|=\epsilon_0P_k(\delta)$.
Hoeffding ($c_{\rm comp}=1/2$): the radius is $\sqrt{L/(2m)}=a/2\ge\epsilon_0P_k(\delta)$.
Iid-Bernstein ($c_{\rm comp}=\sqrt{\pi_0(1-\pi_0)/2}$): the radius is at least $a\sqrt{u(q_k(G))}\ge a\sqrt{\pi_0(1-\pi_0)/2}\sqrt{P_k(\delta)}\ge\epsilon_0P_k(\delta)$, since $P_k(\delta)\le\sqrt{P_k(\delta)}$.
KL ($c_{\rm comp}=\frac12\sqrt{\pi_0(1-\pi_0)}$): using $\KL(x\|y)\le(x-y)^2/u(y)$,
\[
m\KL(T_k\|\mu_k(G))\le\frac{m\epsilon_0^2P_k(\delta)^2}{(1-\pi_0)P_k(\delta)\pi_0/2}\le\frac{2m\epsilon_0^2}{\pi_0(1-\pi_0)}=\frac{ma^2}{2}=L .
\]
\emph{Step 4.} Subtracting, the width is at least $\omega^-_F(0)+(1-s)\{\omega^+_F(0)-\nu\}+\epsilon_0$. Let $\nu\to0$ and use $s=\epsilon_0/\pi_0$.
\end{proof}

\subsection{Realised widths of the competitors}\label{supp:realised}

\begin{lemma}[Reduction to exact regions]\label{lem:reduction}
Let $\Rcal=\{G:\mu_k(G)\in A_k(T_k)\ \forall k\in K\}$ with intervals $A_k(T)$, and let its gridded version be $\{w:\sum_jP^{hi}_{kj}w_j\ge\inf A_k,\ \sum_jP^{lo}_{kj}w_j\le\sup A_k\}$. If $G\in\Rcal$, then $w_j=G(C_j)$ is feasible for the gridded region, $\sum_jh^{hi}_jw_j\ge\theta(G)$ and $\sum_jh^{lo}_jw_j\le\theta(G)$. Hence the gridded width is at least $\sup\{\theta(G)-\theta(G'):G,G'\in\Rcal\}$. With the sure constraints the same holds for $G$ with $G(\{0\})\le N_0/m$ and $G(\{1\})\le N_n/m$.
\end{lemma}
\begin{proof}
The cell envelopes give $\sum_jP^{lo}_{kj}w_j\le\mu_k(G)\le\sum_jP^{hi}_{kj}w_j$ and $\sum_jh^{lo}_jw_j\le\theta(G)\le\sum_jh^{hi}_jw_j$.
\end{proof}
This covers Hoeff-$k$, KL-Bonf, iid-Bernstein and the exact-binomial band; for iid-Bernstein and KL, $A_k(T)=\{q:|T-q|\le r(u(q))\}$ and $\{q:m\KL(T\|q)\le L\}$ are intervals. It also shows that the competitor regions of Proposition~\ref{prop:compexcess}, which are exact and un-gridded, are contained in the implemented ones.

\begin{proof}[Proof of Proposition~\ref{M-prop:realised}]
For $\epsilon\in[0,\pi_0]$ let $G_\epsilon=F+\epsilon(\delta_\delta-\delta_0)$, a probability measure, with $\mu_k(G_\epsilon)=\mu_k(F)+\epsilon P_k(\delta)$ and $\theta(G_\epsilon)=\theta(F)+\epsilon$, since $h_\delta(0)=0$, $h_\delta(\delta)=1$ and $P_k(0)=0$. On $\{|T_k-\mu_k(F)|\le\rho_k\ \forall k\}$, every $\epsilon$ in the set that defines $\epsilon_{\rm real}$ puts $G_\epsilon$ in $\Rcal$, and $F\in\Rcal$ by assumption. Also $G_\epsilon(\{0\})=F(\{0\})-\epsilon\le N_0/m$ and $G_\epsilon(\{1\})=F(\{1\})\le N_n/m$ by Step 3 of Theorem~\ref{M-thm:coverage}, so the sure constraints do not exclude $G_\epsilon$. By Lemma~\ref{lem:reduction}, $U\ge\theta(F)+\epsilon$ and $\Lo\le\theta(F)$; the region is closed, so the supremum over feasible $\epsilon$ is attained. For the second bound use the construction $G=(1-s)G^++s\tilde G$ of the proof of Proposition~\ref{prop:compexcess} at the realised data; its mean shift is $\epsilon P_k(\delta)$ as for $G_\epsilon$, so the same $\epsilon_{\rm real}$ works, and Step 4 of that proof gives the bound.
\end{proof}

\emph{Explicit bounds.} Take the minimum over $k$ of $\min(\pi_0/2,e_k)$, where:
\begin{itemize}
\item Hoeff-$k$, $A_k(T)=[T-a/2,T+a/2]$: $e_k=(a/2-\rho_k)/P_k(\delta)$, exactly. If $\rho_k>a/2$ for some $k$, which can happen when $V_k(F)$ is near $1/4$, the defining set is empty and $\epsilon_{\rm real}=0$.
\item iid-Bernstein, $A_k(T)=\{q:|T-q|\le a\sqrt{u(q)}+b\}$, $\rho_k=r_L(V_k(F))$: $e_k\ge\max\{e^{(i)}_k,e^{(ii)}_k\}$ with $e^{(i)}_k=a\{c_B\sqrt{P_k(\delta)}-\sqrt{V_k(F)}\}/P_k(\delta)$, $c_B=\sqrt{\pi_0(1-\pi_0)/2}$, which uses $u(\mu_k(G_\epsilon))\ge\pi_0(1-\pi_0)P_k(\delta)/2$ for $\epsilon\le\pi_0/2$; and $e^{(ii)}_k=aW_k/\{(1+a)P_k(\delta)\}$ with $W_k=\Var_F(P_k)\ge\pi_0(1-\pi_0)\mu^+_k{}^2$, $\mu^+_k=\int P_k\,dF_+$. For (ii), write $x=\epsilon P_k(\delta)$; since $u$ is 1-Lipschitz, $u(\mu_k+x)\ge V_k+W_k-x$, and $\sqrt{V+W-x}-\sqrt V\ge W-x$ because each root is at most $1/2$, so $a\sqrt V+x\le a\sqrt{V+W-x}$ when $x\le aW/(1+a)$.
\item KL-Bonf: $\KL(x\|q)\le(x-q)^2/(2\min_{[x,q]}u)$ gives $A_k(T)\supseteq\{q:|T-q|\le a\sqrt{\min(u(T),u(q))}\}$, and the iid-Bernstein argument applies at the worse endpoint. When $\mu_k\ll L/m$ the Bernstein event does not imply that $T_k$ is KL-acceptable at $\mu_k(F)$; for those $k$ one uses the KL deviation event at level $L-s$, at the cost of probability $1-\alpha e^s$. We have not written this switch out uniformly in $m$.
\end{itemize}

\begin{proof}[Proof of the competitor bounds in Corollary~\ref{M-cor:sepB}]
\emph{Hoeff-$k$.} For $k\le2k_*$, $V_k\le(1-\pi_0)\epsilon(k)\le\epsilon(2k_*)\le a^{3/2}$ by step (1) of the proof of Corollary~\ref{M-cor:sepB}, so $e_k\ge a/2-a^{7/4}-b$. For $k>2k_*$, $V_k\le(1-\pi_0)/4$ because the atom at $0$ has $u(P_k(0))=0$, so $a/2-\rho_k\ge\frac a2(1-\sqrt{1-\pi_0})-b>0$ for large $m$; and $P_k(\delta)\le\Pr\{\Bin(n,\delta)\ge2n\delta\}\le e^{-0.386\,n\delta}\to0$ by (J1), using $\KL(x\|q)\ge q\varphi(x/q)$ with $\varphi(t)=t\log t-t+1$ and $\varphi(2)=2\log2-1$. Hence $e_k\ge a/2$ eventually, and $\epsilon_{\rm real}\ge a/2-o(a)$.

\emph{iid-Bernstein, under} (J4): $\liminf_m\min_{k\in K,k>2k_*}\pi_0(1-\pi_0)\mu^+_k(F)^2/P_k(\delta)\ge c_B$. For $k\le2k_*$, $\mu^+_k\ge1-a^{3/2}$, so $W_k\ge\pi_0(1-\pi_0)(1-o(1))$; if $P_k(\delta)\le\sqrt{2\pi_0(1-\pi_0)}(1-o(1))$ then $e^{(ii)}_k\ge c_Ba(1-o(1))$, and otherwise $P_k(\delta)$ is bounded below and $e^{(i)}_k\ge a(c_B-o(1))$. For $k>2k_*$, (J4) and $e^{(ii)}_k$ give the bound.
\end{proof}

Condition (J4) is a deterministic function of $(F_+,\delta,n,K)$ and must be checked at each point; along the joint path of Figure~\ref{fig:separated} its logarithm is at least 6.8 against the threshold $\log c_B=-1.04$. It is a real restriction: if $n\in K$ and $F_+$ has no mass above $p_{\max}$, then at $k=n$ the ratio in (J4) is at most $\pi_0(1-\pi_0)(p_{\max}^2/\delta)^n\to0$, so (J4) fails for large $n$ whenever $\delta>p_{\max}^2$. A sufficient condition in terms of $\KL(x\|\delta)-2\KL(x\|p_{\max})$ is plausible but not proved. For KL-Bonf the corresponding statement is only sketched; Table~\ref{tab:realised} reports the exact values of $\epsilon_{\rm real}/a$ instead.

\begin{table}[tbp]
\centering\small
\caption{Realised lower bounds at the points of Figure~\ref{fig:separated}: range of $\epsilon_{\rm real}/a$ over configs A and B ($m=10^3,\dots,10^8$, $n=64$) and the joint path, computed exactly on a grid of 20{,}001 values of $\epsilon$, with $\rho_k=r_L(V_k(F))$. Probabilities: at least $1-\alpha$ (Hoeffding, iid-Bernstein) and $1-1.051\alpha$ (KL, with $s=0.05$). In no case does the lower bound exceed the observed median width.}
\label{tab:realised}
\begin{tabular}{lcc}
\toprule
Competitor & $\epsilon_{\rm real}/a$ & constant of the oracle-centred argument\\
\midrule
Hoeff-$k$ & 0.480--0.499 & 0.5\\
iid-Bernstein & 0.486--0.500 & 0.354\\
KL-Bonf & 0.478--0.499 & 0.25\\
\bottomrule
\end{tabular}
\end{table}

\section{Lower bounds}\label{supp:proofs-lower}

A procedure $C$ has conditional coverage $1-\alpha$ on $\Theta\subseteq[0,1]^m$ if $\Pr_p\{\theta(F_m(p))\in C\}\ge1-\alpha$ for all $p\in\Theta$; $|C|=\sup C-\inf C$, so the bounds apply to set-valued procedures.

\begin{proposition}[Local lower bounds]\label{prop:lower}
Let $C$ have conditional coverage $1-\alpha$ on $[0,1]^m$ and $h=h_\delta$.
\begin{itemize}
\item[(a)] Fix $p$ and $\tau\in(0,1-2\alpha)$. Let $A=\{i:p_i\le\delta/2\}$ and $\sigma_A^2=m^{-1}\sum_{i\in A}p_i(1-p_i)$, and assume $nm\sigma_A^2\ge2\tau^2$. Then $\Pr_p\{|C|\ge(\tau/\delta)\sqrt{2\sigma_A^2/(nm)}\}\ge1-2\alpha-\tau$.
\item[(b)] If $p$ has a coordinate equal to $0$ and $\tau\le\min(1,n\delta)$, then $\Pr_p\{|C|\ge\tau/(n\delta m)\}\ge1-2\alpha-\tau$.
\end{itemize}
\end{proposition}

Part (a) is the parametric rate for a finite-population mean; part (b) moves one problem off zero. Table~\ref{tab:matched} records what the upper and lower bounds match.

\begin{table}[tbp]
\centering\small
\caption{What the upper and lower bounds match. $\ID_{[n]}$ is the identification width with $K=\{1,\dots,n\}$.}
\label{tab:matched}
\begin{tabular}{p{3.0cm}p{4.2cm}p{4.2cm}p{2.6cm}}
\toprule
Term & Upper bound (VA) & Lower bound (any valid method) & Matched?\\
\midrule
Worst-case identification, $n$ fixed, $m\to\infty$ & $2E_n(h)$ with $K=\{1,\dots,n\}$ (Thm~\ref{M-thm:id}(i)) & $2E_n(h)-O(m^{-1/2})$, $h$ continuous (Thm~\ref{M-thm:id}(ii)) & Yes, with $K=\{1,\dots,n\}$\\
$F$-specific identification & $\ID_K(F)$ via $\omega_F(0)$ (Thm~\ref{M-thm:width}) & $(s/m)\ID_{[n]}(F_S)/2-\sqrt{s/\zeta}/m$ at a resampled $p'$, for any $s$-subset $S$ (Lemma~\ref{lem:fuzzy}; see the text) & Up to a factor $2$ and the scale $s/m$\\
Estimation, $F$ with mass in $(0,\delta)$ & $2\sum_k|\lambda_k|\bar D_k$ (Cor~\ref{M-cor:oracle}) & $\delta^{-1}\sqrt{\sigma_A^2/(nm)}$ (Prop~\ref{prop:lower}(a)) & Rate $m^{-1/2}$ only\\
Estimation, separated laws & $C_{\rm ID}+aA_{\VA}+O(a^{3/2})$; joint limit $O((L/m)^{3/4})$ (Cors~\ref{M-cor:sepA}--\ref{M-cor:sepB}) & $\tau/(n\delta m)$ (Prop~\ref{prop:lower}(b)); iid-calibrated regions: $\epsilon_{\rm real}\approx a/2$ (Prop~\ref{M-prop:realised}) & No\\
\bottomrule
\end{tabular}
\end{table}

Matching the estimation term (Open problem~\ref{M-op:estimation}) meets two obstacles. Lower bounds that use mixture structure randomise the $p_i$, and under the randomisation the count law has the iid covariance; by Cauchy--Schwarz such constructions see only alternatives in the iid-scaled box $|\Delta\mu_k|\lesssim\sqrt{q_k(1-q_k)\chi^2}$, while the VA upper bound lives in the $V_k$-scaled box. And the fixed-$p$ model carries information that VA does not use: with $n=1$ and $F=\frac12(\delta_0+\delta_1)$, $T_1=1/2$ exactly, while under $G=\delta_{1/2}$ that event has probability of order $m^{-1/2}$. A construction that keeps $\theta(F_m)$ deterministic, for example random permutations of a fixed multiset, would need total-variation bounds for Poisson-multinomial laws with matched moments. An approximate moment-matching bound of the Wu--Yang type is given in Theorem~\ref{thm:wuyang}; it is not active at $\delta\ge0.01$.

For distributions $P,Q$ on a finite set, $\TV$ is total variation and $H^2(P,Q)=\sum(\sqrt P-\sqrt Q)^2\in[0,2]$.

\begin{lemma}[Two-point bound for intervals]\label{lem:lecam}
If $C$ has coverage $1-\alpha$ at $p$ and at $p'$, then $\Pr_p\{|C|\ge|\theta(p)-\theta(p')|\}\ge1-2\alpha-\TV(\Pr_p,\Pr_{p'})$.
\end{lemma}
\begin{proof}
$\Pr_p(\theta,\theta'\in C)\ge\Pr_p(\theta\in C)-\Pr_p(\theta'\notin C)\ge1-\alpha-\{\Pr_{p'}(\theta'\notin C)+\TV\}\ge1-2\alpha-\TV$. On this event $\operatorname{diam}C\ge|\theta(p)-\theta(p')|$, whether or not $C$ is an interval; throughout, $|C|$ denotes $\operatorname{diam}C=\sup C-\inf C$.
\end{proof}

\begin{lemma}[Fuzzy hypotheses with a randomised subset]\label{lem:fuzzy}
Fix $S\subseteq[m]$ with $|S|=s$ and values $p_i$ for $i\notin S$. Under the prior $\Pi_G$ the $p_i$, $i\in S$, are iid from $G$. Let $Q_G=\int\Pr_p\,d\Pi_G(p)$, $\bar\theta_G=m^{-1}\{\sum_{i\notin S}h(p_i)+s\,\theta(G)\}$, and let $h\in[0,1]$. If $C$ has coverage $1-\alpha$ on $\mathrm{supp}\,\Pi_G\cup\mathrm{supp}\,\Pi_{G'}$, then for $\zeta\in(0,1)$ and $t=\sqrt{s/(4\zeta)}/m$,
\[
Q_G\Big\{|C|\ge\tfrac sm|\theta(G)-\theta(G')|-2t\Big\}\ge1-2\alpha-2\zeta-\TV(Q_G,Q_{G'}),
\]
and $\TV(Q_G,Q_{G'})\le\min\{s\TV(f_G,f_{G'}),\sqrt s\,H(f_G,f_{G'})\}$, where $f_G=\int b\,dG$. The same bound holds under $\Pr_p$ for some $p\in\mathrm{supp}\,\Pi_G$.
\end{lemma}
\begin{proof}
Under $\Pi_G$, $\theta(F_m)-\bar\theta_G=m^{-1}\sum_{i\in S}\{h(p_i)-\theta(G)\}$ has variance at most $s/(4m^2)$, so by Chebyshev $\Pr\{|\theta(F_m)-\bar\theta_G|>t\}\le\zeta$. Since coverage holds for every $p$ in the support, $Q_G\{C\cap[\bar\theta_G\pm t]\ne\emptyset\}\ge1-\alpha-\zeta$, and likewise for $G'$. Move the $G'$ event to $Q_G$ at cost $\TV(Q_G,Q_{G'})$ and argue as in Lemma~\ref{lem:lecam}; on the event that $C$ meets both intervals, $\operatorname{diam}C=\sup C-\inf C\ge|\bar\theta_G-\bar\theta_{G'}|-2t$, and no convexity of $C$ is used. Under $Q_G$ the coordinates are independent, those outside $S$ have the same law under both priors, and those in $S$ are iid $f_G$; subadditivity of $\TV$ over products and $H^2(P^{\otimes s},Q^{\otimes s})=2\{1-(1-H^2/2)^s\}\le sH^2$, $\TV\le H$ give the bounds. $Q_G$ is an average of $\Pr_p$ over the prior, which gives the last claim.
\end{proof}

The prior $\Pi_G$ is a device of the proof, not a change of target. The procedure is assumed to have conditional coverage at every fixed $p$ in $\mathrm{supp}\,\Pi_G\cup\mathrm{supp}\,\Pi_{G'}$. The events used, $\{C\cap[\bar\theta_G\pm t]\ne\emptyset\}$, depend on the data only, since $\bar\theta_G$ is deterministic, so they can be transported between $Q_G$ and $Q_{G'}$ at cost $\TV(Q_G,Q_{G'})$. The conclusion $Q_G(A)\ge x$ for a data event $A$ is an average over $p\sim\Pi_G$ of $\Pr_p(A)$, so $\Pr_p(A)\ge x$ for some fixed $p$, which is the conditional setting. \emph{The localized entry of Table~\ref{tab:matched}.} Let $|S|=s<m$, $G=F_S$ the empirical law of $\{p_i\}_{i\in S}$, and $G'$ any law with the same moments of orders $1,\dots,n$, so that $f_G=f_{G'}$ and $\TV(Q_G,Q_{G'})=0$. Lemma~\ref{lem:fuzzy} gives, at a parameter vector whose $S$-coordinates are a resample from $F_S$ (a point of $\mathrm{supp}\,\Pi_{F_S}$, which depends on $G'$), $\Pr\{|C|\ge(s/m)|\theta(F_S)-\theta(G')|-\sqrt{s/\zeta}/m\}\ge1-2\alpha-2\zeta$. A single comparison law $G'$ gives only the one-sided gap $|\theta(F_S)-\theta(G')|$, not the full width $\ID_{[n]}(F_S)$. Taking the supremum over $G'$, the best constant is $\max\{\omega^+_{F_S}(0),\omega^-_{F_S}(0)\}\ge\ID_{[n]}(F_S)/2$, with the moduli of Section~\ref{M-sec:ident} at $K=\{1,\dots,n\}$; for every $\nu>0$ some $G'$ attains it to within $\nu$. Hence, for each $\nu>0$, there is such a parameter vector with $\Pr\{|C|\ge(s/m)\ID_{[n]}(F_S)/2-\nu-\sqrt{s/\zeta}/m\}\ge1-2\alpha-2\zeta$. Using the two extremal laws of the fiber as $G$ and $G'$ instead gives $(s/m)\ID_{[n]}(F_S)$, but at a parameter vector drawn from an extremal law rather than from $F_S$, so it is not a bound localized at $F_S$.

\begin{proof}[Proof of Theorem~\ref{M-thm:id}]
(i) $P_k$ has lowest-order term $\binom nkp^k$, so $\{1,P_1,\dots,P_n\}$ is a basis of the polynomials of degree at most $n$. For bounded $h$ the minimum defining $E_n(h)$ is attained, since the polynomials of degree at most $n$ form a finite-dimensional subspace of the bounded functions. For the best approximant $c^*+\sum_k\lambda^*_kP_k$, $\osc(\psi_{\lambda^*})\le2E_n(h)$, so $\ID_K(F)=\omega^+_F(0)+\omega^-_F(0)\le2E_n(h)$ by Theorem~\ref{M-thm:width}(ii) with $\Delta=0$. On $\Ecal$, the crude bound $\bar D_k\le a(\sqrt{V_k}+\frac12)+2b+\eta_k\le a+2b+\eta_k$ in Theorem~\ref{M-thm:width}(ii), which holds for bounded $h$, gives the stated width bound. If $h$ has a jump of size $J$, every polynomial differs from $h$ by at least $J/2$ near the jump, so $E_n(h)\ge J/2$.
(ii) Let $G,G'$ have equal moments of orders $1,\dots,n$. Each binomial probability is a polynomial of degree at most $n$, so $f_G=f_{G'}$ and $\TV(Q_G,Q_{G'})=0$. Lemma~\ref{lem:fuzzy} with $s=m$ ($2t=1/\sqrt{\zeta m}$) gives $p$ with $\Pr_p\{|C|\ge|\theta(G)-\theta(G')|-1/\sqrt{\zeta m}\}\ge1-2\alpha-2\zeta$. The supremum of $\theta(G)-\theta(G')$ over such pairs is $2E_n(h)$ for continuous $h$, and this duality step (Hahn--Banach in the dual of $C[0,1]$) is where continuity is used: the direction ``$\le$'' is weak duality, and ``$\ge$'' is the Hahn--Banach dual of best uniform approximation, since a signed measure annihilating the polynomials of degree at most $n$ is of the form $G-G'$ \citep[Appendix~E]{wu2016minimax}. Choose a pair within $\nu$ of the supremum.
\end{proof}

At $n=128$, $\delta=0.01$, explicit pairs from a two-measure LP certify $\theta(G)-\theta(G')\ge0.0221$, independently of the duality argument.

\begin{proof}[Proof of Proposition~\ref{prop:lower}]
(a) Let $x_i=c\,p_i(1-p_i)/n$ for $i\in A$ and $0$ otherwise, $p'=p+x$, and $c=\tau\sqrt{2n/(m\sigma_A^2)}$. The assumption gives $c\le n$, so $x_i\le p_i\le\delta/2$ and $p'_i\le\delta$, and $\theta(p')-\theta(p)=\sum_ix_i/(m\delta)=(\tau/\delta)\sqrt{2\sigma_A^2/(nm)}$. Using $\KL(\mathrm{Bern}(q')\|\mathrm{Bern}(q))\le(q'-q)^2/u(q)$, $\KL(\Pr_{p'}\|\Pr_p)\le\sum_inx_i^2/u(p_i)=c^2m\sigma_A^2/n=2\tau^2$, and Pinsker's inequality gives $\TV\le\tau$. Apply Lemma~\ref{lem:lecam}.
(b) Move one zero coordinate to $x=\tau/n\le\delta$. Then $\TV\{\Bin(n,0),\Bin(n,x)\}=1-(1-x)^n\le nx=\tau$, and $\theta$ changes by $x/(\delta m)$. Apply Lemma~\ref{lem:lecam}.
\end{proof}

For general smooth monotone $h$, the argument of (a) gives the rate $\{\sum_ih'(p_i)^2p_i(1-p_i)\}^{1/2}/(m\sqrt n)$, the usual parametric rate for a finite-population mean of $h(p_i)$.

\begin{theorem}[Approximate moment matching]\label{thm:wuyang}
Let $G,G'$ be probability measures on $[0,\varsigma]$ with equal moments of orders $1,\dots,d_0$. Then
\[
\TV(f_G,f_{G'})\le\frac12\sum_{d=d_0+1}^{n}\binom nd(2\varsigma)^d\le\frac12\sum_{d>d_0}\frac{(2n\varsigma)^d}{d!}.
\]
\end{theorem}
\begin{proof}
Expanding $(1-p)^{n-j}$, $f_G(j)-f_{G'}(j)=\binom nj\sum_l\binom{n-j}{l}(-1)^l\{m_{j+l}(G)-m_{j+l}(G')\}$, where $m_d$ is the $d$-th moment; the terms with $j+l\le d_0$ vanish. Use $|m_d(G)-m_d(G')|\le\varsigma^d$ and $\sum_j\binom nj\binom{n-j}{d-j}=\binom nd2^d$.
\end{proof}

\begin{remark}
Combined with Lemma~\ref{lem:fuzzy} ($s=m$, $\varsigma=2\delta$), Theorem~\ref{thm:wuyang} gives a lower bound of order $1/\max\{4e^2n\delta,\log(m/\tau)\}$ minus $1/\sqrt{\zeta m}$, in the spirit of \citet{wu2019chebyshev}. Its constant is the best approximation error $2E_{d_0}(\min(2x,1);[0,1])$. With these crude constants the bound exceeds the identification width only for $\delta$ of order $10^{-3}$ or smaller, so it is not active at the values of $\delta$ used in the paper. A Hellinger version would be sharper; we have not evaluated it.
\end{remark}

\begin{remark}[Numerical certificates]
The dual variables of the VA LP's mean rows define a $\lambda$ that decomposes the realised width into $\gap_\lambda$ plus weighted radii; this is exact as a split of the optimal value and approximate as a reading of Theorem~\ref{M-thm:width}(ii), because the cut rows also carry $\bar V$ terms. For the lower side, any feasible pair $(G,G')$ of an LP that bounds $H(f_G,f_{G'})$ from above, via $\chi^2(f\|f_{\rm ref})\le(\sum_j|f_j-f_{{\rm ref},j}|/\sqrt{f_{{\rm ref},j}})^2$, yields a valid lower bound through Lemma~\ref{lem:fuzzy}. We have not implemented it.
\end{remark}

\section{The super-population target}\label{supp:glevel}
Here $p_i\overset{iid}{\sim}G$, the counts are as before given $p$, and $\theta(G)=\int h\,dG$ with $h\in[0,1]$. Split $\alpha=\alpha_1+\alpha_2$, let $[\Lo,U]$ be the VA interval at level $\alpha_1$, and let $J$ be the number of targets for which joint coverage of $\theta(G)$ is claimed. In Propositions~\ref{prop:layer-kl} and~\ref{prop:layer-eb} the $J$ targets must be fixed before the data are seen, because the Hoeffding and Maurer--Pontil events concern fixed functions $h$; only the DKW layer of Proposition~\ref{prop:layer-dkw} allows monotone targets chosen after seeing the data.

\begin{proposition}[KL layer]\label{prop:layer-kl}
Put $c_2=\log(2J/\alpha_2)/m$, $\ell=\min\{t\in[0,\Lo]:\KL(\Lo\|t)\le c_2\}$ and $\upsilon=\max\{t\in[U,1]:\KL(U\|t)\le c_2\}$. Then $\Pr\{\theta(G)\in[\ell,\upsilon]$ for all $J$ targets$\}\ge1-\alpha_1-\alpha_2$, unconditionally over $p\sim G^{\otimes m}$.
\end{proposition}
\begin{proof}
$\Pr(\Ecal^c\mid p)\le\alpha_1$ for every $p$ (Theorem~\ref{M-thm:coverage}). If $\theta(G)>\upsilon$, then on $\Ecal$, $\theta(G)>U\ge\theta(F_m)$ and $\KL(U\|\theta(G))>c_2$; as $\KL(\cdot\|\theta)$ decreases on $[0,\theta]$, $\KL(\theta(F_m)\|\theta(G))>c_2$. Since $\theta(F_m)=m^{-1}\sum_ih(p_i)$ is a mean of iid $[0,1]$ variables with mean $\theta(G)$, Theorem~1 of \citet{hoeffding1963probability} bounds the probability of this event by $e^{-mc_2}=\alpha_2/(2J)$. The lower side is symmetric; take a union bound over sides, targets and $\Ecal^c$.
\end{proof}

\begin{proposition}[Empirical-Bernstein layer]\label{prop:layer-eb}
For $c\in[0,1]$ let $s_j(c)=\max\{(h^{hi}_j-c)^2,(h^{lo}_j-c)^2\}$ and $\bar V=\frac m{m-1}\inf_{c\in[0,1]}\max_{w\in\Wcal}\sum_js_j(c)w_j$. With $e=\sqrt{2\bar V\log(4J/\alpha_2)/m}+7\log(4J/\alpha_2)/\{3(m-1)\}$, the intervals $[\Lo-e,U+e]$ cover $\theta(G)$ for all $J$ targets with probability at least $1-\alpha_1-\alpha_2$.
\end{proposition}
\begin{proof}
On $\Ecal$, for every $c$, $\Var_{F_m}(h)\le\int(h-c)^2\,dF_m\le\sum_jw^*_js_j(c)\le\max_{w\in\Wcal}\sum_js_j(c)w_j$, because a convex function on $[h^{lo}_j,h^{hi}_j]$ is maximised at an endpoint; this holds for all $c$ at once, so also at a data-chosen $c$. So the unbiased sample variance of the $h(p_i)$ is at most $\bar V$. Theorem~4 of \citet{maurer2009empirical} for iid $h(p_i)\in[0,1]$, applied to $h$ and $1-h$ at level $\alpha_2/(2J)$ each, gives the claim, since the bound is increasing in the sample variance.
\end{proof}

\begin{proposition}[Simultaneous layer for monotone targets]\label{prop:layer-dkw}
Let $e_D=\sqrt{\log(2/\alpha_2)/(2m)}$. With probability at least $1-\alpha_1-\alpha_2$, simultaneously for every monotone $h:[0,1]\to[0,1]$, $\theta_h(G)\in[\Lo_h-e_D,\,U_h+e_D]$.
\end{proposition}
\begin{proof}
By \citet{massart1990tight}, $\Pr\{\sup_x|F_m(x)-G(x)|>e_D\}\le\alpha_2$ for iid $p_i$ from any $G$; atoms of $G$ only make the statistic stochastically smaller. A nondecreasing $h$ with values in $[0,1]$ can be written $h(x)=h(0)+\int_{[0,1)}\mathbf 1\{x>t\}\,d\nu_1(t)+\int_{(0,1]}\mathbf 1\{x\ge t\}\,d\nu_2(t)$ with finite measures of total mass $h(1)-h(0)\le1$, where $\nu_1$ carries the continuous part and the right jumps, including a jump at $0$, and $\nu_2$ the left jumps. By Fubini, $\int h\,d(F_m-G)=\int\{G(t)-F_m(t)\}\,d\nu_1+\int\{G(t^-)-F_m(t^-)\}\,d\nu_2$, and both integrands are bounded by $\sup_x|F_m-G|$. For nonincreasing $h$ use $1-h$. On $\Ecal$ the VA intervals hold for all $h$ at once.
\end{proof}

The Hoeffding layer $\sqrt{\log(2J/\alpha_2)/(2m)}$ is also valid. Proposition~\ref{prop:layer-dkw} keeps simultaneity over the whole monotone class, $\theta_\delta$ for all $\delta$, $P(p\ge\gamma)$ for all $\gamma$, pass@$k$ for all $k$ and $P(p>0)$, without a $\log J$ term; simultaneity is lost only for non-monotone or unrestricted $h$.

\begin{proposition}[Floor of a two-step interval]\label{prop:layer-floor}
Any interval $[\Lo-e^-,U+e^+]$ for $\theta(G)$ whose layer is computed from $(m,\alpha_2)$ and a bound $\bar V\ge\Var_{F_m}(h)$ has width at least $W_{\VA}(\alpha_1)+e^-+e^+$. For the KL layer, as $m\to\infty$ with $U$ and $\Lo$ bounded away from $0$ and $1$, $\upsilon-\ell=W_{\VA}+\{\sqrt{2U(1-U)c_2}+\sqrt{2\Lo(1-\Lo)c_2}\}(1+o(1))$. If moreover $W_{\VA}\to0$, the ratio to the oracle interval $2z_{1-\alpha/2}\sqrt{\Var_G(h)/m}$ for an analyst who observes the $p_i$ is at least $\sqrt{2\log(2J/\alpha_2)}\,z_{1-\alpha/2}^{-1}\sqrt{\theta(1-\theta)/\Var_G(h)}\ge1.51$ at $J=1$, $\alpha_1=\alpha_2=0.025$.
\end{proposition}
\begin{proof}
$\upsilon-U$ solves $\KL(U\|U+x)=c_2$, and $\KL(U\|U+x)=x^2/\{2U(1-U)\}+O(x^3)$ uniformly for $U$ in a compact subset of $(0,1)$. The lower side is the same. Under $W_{\VA}\to0$, $U,\Lo\to\theta(G)$; use $\Var_G(h)\le\theta(1-\theta)$.
\end{proof}

For fixed $n$, $W_{\VA}\to\ID_K(F)$, which is positive in general, so no general comparison with the oracle holds. Between-problem variance is genuine sampling noise for $\theta(G)$, and a two-step construction pays it with a finite-sample constant, added to VA's width rather than in quadrature; AMARI and the iid-calibrated localizations target $\theta(G)$ directly and pay it once. Proposition~\ref{M-prop:iid} and the identity $\Var_p(m^{-1}\sum_i\phi(c_i))=m^{-1}\int\Var_{\Bin(n,p)}\phi\,dF_m$, against $m^{-1}[\int\Var_{\Bin(n,p)}\phi\,dF_m+\Var_{F_m}(\E_{\Bin(n,p)}\phi)]$ when the $p_i$ are drawn from $F_m$, make the same point for the localization step (Proposition~\ref{prop:compound}).

\emph{Computation.} Table~\ref{M-tab:Glevel} uses $m=1000$, VA rerun at $\alpha/2$ on the simulation replicates, and the layers at $\alpha/2$: KL with $J=3$ (primary) and $J=7$, Hoeffding with $J=3$, DKW, and the empirical-Bernstein layer computed cheaply, by the VA LP with the default tangent set and no refinement, which is an outer approximation of $\Wcal$ and so conservative, and a golden-section search over $c$. The comparators are at level $\alpha$ on the same counts, which are drawn under the fixed $F_m$; a coverage study with the $p_i$ redrawn was not run. The KL layer with $J=7$ was wider than with $J=3$, and the empirical-Bernstein layer was wider than the KL layer in the median (0.293, 0.189 and 0.156 at $n=32$, 128 and 512, against 0.253, 0.151 and 0.122). Table~\ref{tab:s4-glevel-cells} gives every cell.

\begin{table}[tbp]
\centering
\scriptsize
\caption{G-level comparison per cell ($m=1000$): width of VA$+$KL, of the best non-AMARI localization and of AMARI, and the two ratios.}
\label{tab:s4-glevel-cells}
\begin{adjustbox}{max width=\textwidth}
\begin{tabular}{lrrccccc}
\toprule
$F$ & $n$ & $\delta$ & VA$+$KL & best FL & AMARI & /best FL & /AMARI\\
\midrule
F1 & 32 & 0.01 & 0.594 & 0.588 & 0.330 & 1.01 & 1.80\\
F1 & 32 & 0.02 & 0.383 & 0.341 & 0.193 & 1.12 & 1.98\\
F1 & 32 & 0.05 & 0.248 & 0.186 & 0.108 & 1.33 & 2.28\\
F1 & 128 & 0.01 & 0.231 & 0.179 & 0.106 & 1.29 & 2.17\\
F1 & 128 & 0.02 & 0.196 & 0.142 & 0.083 & 1.38 & 2.37\\
F1 & 128 & 0.05 & 0.167 & 0.111 & 0.065 & 1.51 & 2.58\\
F1 & 512 & 0.01 & 0.159 & 0.110 & 0.065 & 1.45 & 2.44\\
F1 & 512 & 0.02 & 0.151 & 0.100 & 0.061 & 1.51 & 2.48\\
F1 & 512 & 0.05 & 0.142 & 0.090 & 0.056 & 1.58 & 2.51\\
F2 & 32 & 0.01 & 0.299 & 0.289 & 0.211 & 1.03 & 1.41\\
F2 & 32 & 0.02 & 0.286 & 0.288 & 0.142 & 0.99 & 2.01\\
F2 & 32 & 0.05 & 0.203 & 0.165 & 0.092 & 1.23 & 2.20\\
F2 & 128 & 0.01 & 0.173 & 0.150 & 0.085 & 1.16 & 2.04\\
F2 & 128 & 0.02 & 0.151 & 0.120 & 0.067 & 1.26 & 2.26\\
F2 & 128 & 0.05 & 0.144 & 0.107 & 0.056 & 1.34 & 2.57\\
F2 & 512 & 0.01 & 0.122 & 0.093 & 0.050 & 1.32 & 2.45\\
F2 & 512 & 0.02 & 0.119 & 0.089 & 0.048 & 1.34 & 2.47\\
F2 & 512 & 0.05 & 0.124 & 0.090 & 0.049 & 1.38 & 2.52\\
F3 & 32 & 0.01 & 0.352 & 0.415 & 0.158 & 0.85 & 2.23\\
F3 & 32 & 0.02 & 0.245 & 0.241 & 0.123 & 1.02 & 2.00\\
F3 & 32 & 0.05 & 0.175 & 0.137 & 0.085 & 1.28 & 2.06\\
F3 & 128 & 0.01 & 0.138 & 0.124 & 0.066 & 1.11 & 2.09\\
F3 & 128 & 0.02 & 0.127 & 0.096 & 0.062 & 1.32 & 2.04\\
F3 & 128 & 0.05 & 0.128 & 0.092 & 0.061 & 1.39 & 2.08\\
F3 & 512 & 0.01 & 0.117 & 0.086 & 0.059 & 1.35 & 1.97\\
F3 & 512 & 0.02 & 0.116 & 0.086 & 0.059 & 1.35 & 1.96\\
F3 & 512 & 0.05 & 0.120 & 0.087 & 0.059 & 1.38 & 2.03\\
F4 & 32 & 0.01 & 0.144 & 0.136 & 0.121 & 1.06 & 1.19\\
F4 & 32 & 0.02 & 0.147 & 0.136 & 0.085 & 1.08 & 1.73\\
F4 & 32 & 0.05 & 0.137 & 0.112 & 0.072 & 1.22 & 1.92\\
F4 & 128 & 0.01 & 0.101 & 0.090 & 0.054 & 1.12 & 1.88\\
F4 & 128 & 0.02 & 0.089 & 0.070 & 0.045 & 1.28 & 2.00\\
F4 & 128 & 0.05 & 0.095 & 0.071 & 0.040 & 1.35 & 2.35\\
F4 & 512 & 0.01 & 0.075 & 0.058 & 0.035 & 1.28 & 2.12\\
F4 & 512 & 0.02 & 0.077 & 0.060 & 0.035 & 1.29 & 2.23\\
F4 & 512 & 0.05 & 0.083 & 0.063 & 0.035 & 1.32 & 2.33\\
F5 & 32 & 0.01 & 0.642 & 0.634 & 0.350 & 1.01 & 1.83\\
F5 & 32 & 0.02 & 0.389 & 0.355 & 0.203 & 1.09 & 1.92\\
F5 & 32 & 0.05 & 0.253 & 0.192 & 0.105 & 1.32 & 2.41\\
F5 & 128 & 0.01 & 0.245 & 0.189 & 0.121 & 1.29 & 2.03\\
F5 & 128 & 0.02 & 0.202 & 0.146 & 0.095 & 1.39 & 2.14\\
F5 & 128 & 0.05 & 0.164 & 0.106 & 0.059 & 1.54 & 2.76\\
F5 & 512 & 0.01 & 0.123 & 0.089 & 0.062 & 1.39 & 2.00\\
F5 & 512 & 0.02 & 0.149 & 0.103 & 0.069 & 1.44 & 2.16\\
F5 & 512 & 0.05 & 0.136 & 0.085 & 0.051 & 1.60 & 2.64\\
\bottomrule
\end{tabular}
\end{adjustbox}
\end{table}

\section{Two classical identities}\label{supp:classical}

\begin{proposition}[Moment identifiability of binomial mixtures]\label{prop:moments}
For $G,G'\in\Pcal$ the following are equivalent: (a) $f_G=f_{G'}$, where $f_G(j)=\int\binom nj p^j(1-p)^{n-j}\,dG(p)$, $j=0,\dots,n$; (b) $\int p^d\,dG=\int p^d\,dG'$ for $d=1,\dots,n$; (c) $\mu_k(G)=\mu_k(G')$ for $k=1,\dots,n$. Consequently $\ID_{\{1,\dots,n\}}(F)=\sup\{\theta(G)-\theta(G'):G,G'$ with the moments of $F$ up to order $n\}$.
\end{proposition}
\begin{proof}
The three vectors are images of one another under invertible affine maps (affine because of the normalisations $\mu_0=1$ and $\int p^0\,dG=1$). (a)$\Leftrightarrow$(c): $\mu_k=\sum_{j\ge k}f(j)$ and $f(j)=\mu_j-\mu_{j+1}$. (b)$\Leftrightarrow$(a): expanding $(1-p)^{n-j}$ writes $f(j)$ as a combination of moments of orders $j,\dots,n$ with leading coefficient $\binom nj\ne0$, a triangular system; conversely $p^d=\sum_{j\ge d}\{\binom jd/\binom nd\}\binom nj p^j(1-p)^{n-j}$.
\end{proof}

The proposition is why Theorem~\ref{M-thm:id} is a moment-matching statement, $\mathrm{span}\{1,P_1,\dots,P_n\}=\mathrm{span}\{1,p,\dots,p^n\}$, and why the NPMLE \citep{lindsay1983geometry} does not resolve partial identification: its fitted law is one point of the identified set. \citet{wood1999binomial} studies the geometry of the moment curve and the consequences of non-identifiability for maximum likelihood.

\begin{proposition}[Compound identity]\label{prop:compound}
For fixed $p_i$ and any $\phi:\{0,\dots,n\}\to\R$: (a) $\E_p[m^{-1}\sum_i\phi(c_i)]=\int\E_{\Bin(n,p)}\phi\,dF_m(p)$; (b) $\Var_p[m^{-1}\sum_i\phi(c_i)]=m^{-1}\int\Var_{\Bin(n,p)}\phi\,dF_m(p)$; (c) if instead $p_i\overset{iid}{\sim}F_m$, the variance is $m^{-1}[\int\Var_{\Bin(n,p)}\phi\,dF_m+\Var_{F_m}(\E_{\Bin(n,p)}\phi)]$.
\end{proposition}
\begin{proof}
(a) and (b) follow from linearity and independence; (c) is the law of total variance.
\end{proof}
Part (a) is the observation behind compound decision theory \citep{robbins1951asymptotically,zhang2003compound}: under fixed parameters, the average behaviour of a separable statistic is its Bayes behaviour with the empirical law of the parameters as prior. With $\phi=\mathbf 1\{c\ge k\}$ it reads $\E T_k=\mu_k(F_m)$. Parts (b) and (c) give~\eqref{M-eq:vargap}: compound first moments coincide with the Bayes ones under the prior $F_m$, and compound second moments are smaller by $\Var_{F_m}(\E_p\phi)$. VA is calibrated with the compound variance (b), the iid-calibrated regions with (c), and (c) is the right variance for $\theta(G)$.

\section{Implementation}\label{supp:impl}

\paragraph{Grid and envelopes.} All LP regions share one data-independent partition $\{0\},(0,g_1),[g_1,g_2),\dots,[g_{R-1},1),\{1\}$, refined adaptively until the envelope tolerance is at most $\eta=10^{-3}$ on every function a region constrains. Regions constrained on $k\in K$ use a grid refined on $\{P_k:k\in K\}$ and $h$ (12.7k cells at $n=128$, 16.0k at $n=512$). Regions using all $k=1,\dots,n$ (DKW variants, Berk--Jones, Hoeff-cell, $\chi^2$) use a grid refined on $\sup_k|P_k(b)-P_k(a)|$ (15.6k and 29.8k cells). Means use $P_k$ at the cell endpoints, variances the cell maximum $\bar V$, and the objective $\inf_Ch$ and $\sup_Ch$. Table~\ref{tab:grid} compares grids on one data set per case: with the tolerance $\eta=10^{-3}$ used throughout, the width exceeds the oracle width obtained by inserting the true $p_i$ as nodes by at most 0.0008 in five $(F,m,n)$ cases (F1 at $m=1000$ and $n\in\{32,128,512\}$, F2 at $m=1000$, $n=128$, and F3 at $m=4000$, $n=512$), each with VA, KL-Bonf and DKW-Massart and three targets. An LP built on point values at grid nodes, without envelopes, has no coverage guarantee when $F_m$ is off the grid.

\paragraph{Tangent cuts.} Initial tangent points $t\in\{10^{-7},10^{-6},10^{-5},10^{-4}\}\cup\mathrm{geomspace}(10^{-3},0.5,6)$ ($V_0=t^2$), followed by up to five cutting-plane rounds that add tangents at the current $\bar V_k(w)$ for every violated $k$. Every tangent is a valid outer approximation, so coverage is unchanged. In unit tests the tangent-cut LP contains the exact second-order cone program and is within 2\% of its width.

\paragraph{Exact intervals for iid-calibrated bands.} For fixed $T_k$, the admissible set of $q_k$ under the iid-Bernstein and per-$k$ KL constraints is an interval, found by bisection and rounded outward. The envelope rows are then linear. The exponential-cone form of Berk--Jones agrees with the LP form in tests.

\paragraph{LP solution and certificates.} HiGHS solves the LP with column generation over cells, row scaling and a fallback chain (default, interior point, dual simplex). Entries below $10^{-10}$ after scaling are zeroed and the right-hand side is relaxed by $10^{-10}$, which is valid on the simplex. Reported endpoints are certified by weak duality from a dual-feasible vector computed over all cells, so they are valid even if pricing stopped early. Two solver issues were fixed before the runs: HiGHS's time limit is cumulative on a model and is re-armed before each solve, and warm starts, which could stall on degenerate optima (435~s warm against 8~s cold on one case), are capped at three times the cold time. VA results computed before these fixes were discarded.

\subsection{Index sets}\label{supp:K}
The width bounds of Section~\ref{M-sec:width} use the indices $k\lesssim2n\delta$ densely, together with partners $k'\ge2k$. This motivates the rule-based set
\[
K_{\rm rule}=\{1,\dots,k_0'\}\cup\mathrm{round}(\mathrm{geomspace}(k_0',n,16))\cup\{n\},
\]
\[
k_0'=\max(16,\lceil2n\delta_{\max}\rceil+2),\quad \delta_{\max}=0.05,
\]
a dense block that contains the doubling partner of $\lceil n\delta\rceil$, followed by 16 geometrically spaced indices. It has $|K_{\rm rule}|=31$ at $n=128$ and $69$ at $n=512$. $K_{\rm geo}$ is the same formula with $k_0'=16$. The two coincide for $n\le140$ ($k_0'=54$ at $n=512$), so they differ only at $n=512$ in our experiments. Indices spaced like $\sqrt k$ after the dense block are a natural alternative; the index-set study (E6, Table~\ref{tab:e6}) includes two such variants, a dyadic set and $K=\{1,\dots,n\}$. Going from $|K|=32$ to $|K|=512$ at $\alpha=0.05$ multiplies the square-root term of the radius by $1.18$, and an index omitted from $K$ is controlled by its neighbours through Lemma~\ref{lem:bracket}.

\emph{Theorems and experiments.} The width theory uses the indices $k\lesssim2n\delta$ densely together with partners $k'\ge2k$, and $K_{\rm geo}$ does not always contain them; the experiments also use a coarse initial tangent set with refinement. So the experiments lie outside the hypotheses of the width theorems: the theorems describe the method, and the experiments measure it in the configuration a user would run. To measure the gap, we reran VA at $m=1000$ with $K_{\rm rule}\cup\{2k\}$ and the theorem's fixed tangent set $\mathrm{geomspace}(10^{-7},0.5,28)$, without refinement: its median width was 1.012, 1.016 and 1.098 times that of VA as run, at $n=32$, 128 and 512 (range 0.970--1.210; Table~\ref{tab:s4-thm}).

\emph{Time line.} Our package returned $K_{\rm geo}$ as its default index set until 12 November 2026, 09:02, $K_{\rm rule}$ from then until 14 November 2026, and $K_{\rm geo}$ again since. The preliminary simulation (E1, Section~\ref{supp:archival}) ran from 05:10 to 08:36 that day with $K_{\rm geo}$, so $K_{\rm geo}$ was fixed before any simulation result existed and was not changed in response to any result. A with-replacement Monkey Business run that used the then default $K_{\rm rule}$ is kept only as a sensitivity analysis (Table~\ref{tab:mb-wr-K}). Every run of the main simulation, the band-event study, the Monkey Business check and the illustration passes its index set explicitly. Table~\ref{tab:Kper} records which set each experiment used; the assignments of the preliminary runs were checked by re-running stored replicates.

\begin{table}[tbp]
\centering\small
\caption{Index set $K$ used in each experiment. $|K|$ is given for $n=32/128/512$ unless stated.}
\label{tab:Kper}
\begin{adjustbox}{max width=\textwidth}
\begin{tabular}{lll}
\toprule
Experiment & $K$ & $|K|$\\
\midrule
Main simulation, AMARI, two-stage, exact-binomial, plug-in & $K_{\rm geo}$ (explicit) & 30 / 31 / 31\\
Main simulation, theorem-aligned VA & $K_{\rm rule}\cup\{2k\}$ & explicit\\
Band events (primary rows), Monkey Business, illustration & $K_{\rm geo}$ (explicit) & 30 / 31 / 31\\
Band events, rows VA ($K_{\rm rule}$) and VA ($K=[n]$) & as labelled & --\\
Illustration, two-index bracket & $\{64-d,64+d\}$, $d\in\{5,10,15\}$ & 2\\
Illustration by source & $K_{\rm geo}$ (explicit) & 31\\
E7 AMARI & not used & --\\
E8 separated laws (Figure~\ref{fig:separated}) & $K_{\rm rule}\cup\{1,2,k_*,2k_*\}$ & 31 at $n=64$; 43--121 on the joint path\\
\midrule
\multicolumn{3}{l}{\emph{Preliminary runs (Section~\ref{supp:archival})}}\\
Grid slack (GR) & $K_{\rm geo}$ & 30 / 31 / 31\\
E1 simulation (widths, coverage, identification) & $K_{\rm geo}$ & 30 / 31 / 31\\
E2 Monkey Business, without replacement & $K_{\rm geo}$ & 30 / 31 / 31\\
E2 Monkey Business, with replacement & $K_{\rm geo}$ & 30 / 31 / 31\\
E2 with replacement, sensitivity row & $K_{\rm rule}$ & 30 / 31 / 69\\
E5 band-event Monte Carlo, unsuffixed rows & $K_{\rm rule}$ & 30 / 31 / 69\\
E5, row VA ($K_{\rm geo}$) & $K_{\rm geo}$ & 30 / 31 / 31\\
E5, row VA ($K=[n]$) & $\{1,\dots,n\}$ & $n$\\
E6 $K$ sensitivity & each variant explicit; baseline $K_{\rm rule}$ & Table~\ref{tab:e6}\\
\bottomrule
\end{tabular}
\end{adjustbox}
\end{table}

\subsection{Tests and software}
\paragraph{Unit tests.} Eight tests pass: brute-force exact coverage ($m=20$, $n=2$, 231 histograms, $\alpha=0.5$); envelope intervals contain atom-inserted ones and the true cell weights are feasible off-grid; certified endpoints lie outside the primal ones; LP versus SOCP; Berk--Jones LP versus exponential cone; and reproduction of small hand-computed examples.

\paragraph{Software.} Python with numpy 2.4.6, scipy 1.17.1, highspy 1.16.0, cvxpy 1.9.3 with Clarabel 0.11.1 and pandas 3.1.1. Threads were pinned to one per process. Wall-clock times of the main runs: simulation sweep 2~h~56~min on 22 workers (9{,}240 tasks, one restart from the cache after the second reduction of replicates), band events 1{,}691~s on 2 workers, illustration 1{,}126~s, G-level empirical-Bernstein layer 1{,}826~s, design rule 191~s, oracle calibration 2{,}595~s, Monkey Business 1{,}515~s on 22 workers for the last segment, and the two-index bracket and per-source analyses 130~s on 8 workers. Preliminary runs: E1 2.44~h on 18 workers; E2 1.74~h on 6 workers. At most 24 worker processes ran at once.

\section{Comparators}\label{supp:comparators}

\paragraph{Localizations.} All comparators share the grid, envelopes and certificates of Section~\ref{supp:impl}. KL-Bonf: $m\KL(T_k\|\mu_k(G))\le L$ on the relevant side, $k\in K$. Hoeff-$k$: $|T_k-\mu_k(G)|\le\sqrt{L/(2m)}$, $k\in K$, by \citet{hoeffding1963probability}. Hoeff-cell: Hoeffding bounds on all $n+1$ count-cell frequencies with Bonferroni. Both are in the spirit of the union-bound-plus-LP construction of \citet{tian2017learning}, who use Bernstein-type per-moment bounds and an LP estimator; they are not that method. DKW-$4/\sqrt m$: Proposition~\ref{prop:dkw}. DKW-Massart: radius $\sqrt{\log(2/\alpha)/(2m)}$ \citep{massart1990tight}, valid for iid data. Berk--Jones \citep{berk1979goodness,owen1995nonparametric}: $\KL(T_k\|\mu_k(G))\le c_m$ for all $k$, with $c_m$ the Monte Carlo 0.95-quantile of the two-sided Berk--Jones statistic for continuous iid data of size $m$ (20{,}000 seeded draws); exact thresholds \citep{moscovich2016exact,moscovich2023fast} were not implemented. EB-DKW: per-$k$ empirical-Bernstein band with variance $\hat V=m(m-1)^{-1}T_k(1-T_k)$ and $L=\log(4|K|/\alpha)$; it is finite-sample valid under fixed $p_i$ by Theorem~11 of \citet{maurer2009empirical}, which requires only independence and values in $[0,1]$, applied to $X_i$ and $1-X_i$ with a union bound over $K$. Theorem~11 of \citet{maurer2009empirical} states, for a vector of independent, not necessarily identically distributed variables in $[0,1]$ and $n\ge2$, that with probability at least $1-\delta$ the average mean exceeds the sample mean by at most $\sqrt{2V_n\ln(2/\delta)/n}+7\ln(2/\delta)/\{3(n-1)\}$, with $V_n$ the unbiased sample variance; for indicators $V_n=mT_k(1-T_k)/(m-1)$. It is a Bonferroni band of per-$k$ empirical-Bernstein bounds, not a uniform variance-adaptive band such as those of \citet{howard2022sequential}, to which \citet{brennan2020estimating} point. $\chi^2$: Pearson statistic on count cells pooled from left to right until each observed count is at least $5$, critical value $\chi^2_{G-1,0.95}$, with a tangent-cut LP outer approximation.

Two-stage band: Section~\ref{supp:twostage}; the Bennett variant uses $r_{Be}$ at level $L_2$. Exact-binom: the per-$k$ test of Proposition~\ref{M-prop:iid}(c) at level $\alpha/|K|$, with the KL band of part (a) at candidates where Hoeffding's condition fails. Plug-in bootstrap: $m^{-1}\sum_ih(c_i/n)$, or the unbiased estimator $1-\binom{n-c_i}{k}/\binom nk$ of \citet{chen2021evaluating} averaged over problems for pass@$k$ with $k\le n$, with a percentile interval from $1000$ bootstrap resamples of the problems; it is not a localization and has no coverage guarantee for $\theta(F_m)$.

\paragraph{AMARI.} Our implementation follows Sections~3--4 of \citet{ignatiadis2022confidence} for a linear functional: the estimator $m^{-1}\sum_iQ(c_i)$; the worst-case bias $\max_{G\in C}|\int(A^\top Q-h)\,dG|$ over a pilot DKW localization at level $0.01$, computed by LP duality on 450 fixed nodes; $Q$ minimising squared bias plus $\rho$ times variance over $\rho\in\{0.25,1,4\}$, keeping the shortest bias-aware interval; and the interval $\hat\theta\pm\mathrm{cva}_\alpha(B/\widehat{\rm se})\,\widehat{\rm se}$. The Empirikos-faithful variant (following \texttt{src/amari.jl} of Empirikos.jl at commit \texttt{4c96a39}, 2026-08-30) uses the empirical standard deviation of $Q(c_i)$; an additional option intersects the interval with the pilot localization's interval. A model-based variant estimates the variance from a minimum-Kolmogorov-distance pilot fit; we report it only as an ablation. Differences from Empirikos.jl that remain: point evaluation on a fixed node set, the $\rho$ grid in place of the modulus path, a fixed pilot level, and the selection of $\rho$ by the model-based standard error before the empirical one is computed. The pilot localization has the fixed level $0.01$ and is not adjusted under Bonferroni, so ``AMARI at $\alpha$'' has nominal error up to $\alpha+0.01$ per target and ``AMARI at $\alpha/3$'' up to $\alpha/3+0.01$; this favours AMARI. Where the pilot interval binds, as for $P(p\ge0.5)$ in the illustration, the pointwise and Bonferroni-adjusted intervals coincide. The main simulation, Monkey Business and illustration results use the Empirikos-faithful variant with the outer intersection, at level $\alpha$ and with Bonferroni corrections ($\alpha/3$ over the three $\delta$, $\alpha/8$ over the eight targets of the illustration); the per-cell tables of the preliminary run (Section~\ref{supp:archival}) use the model-based variant. E7 compares the variants on Monkey Business (Table~\ref{tab:e7}).

\emph{Check against Empirikos.jl (CHK-9).} We ran Empirikos.jl (commit \texttt{4c96a39}) and our implementation on the distilled model's counts of Section~\ref{M-sec:pair} ($m=1060$, $n=128$, $\alpha=0.05$) for $\theta_{0.05}$ and $P(p\ge0.5)$ (scripts \texttt{checks/chk9\_empirikos.jl} and \texttt{checks/chk9\_empirikos\_compare.py}). For $\theta_{0.05}$, Empirikos.jl's default interval is $[0.82527,0.86901]$ and ours $[0.82512,0.86885]$; for $P(p\ge0.5)$ both are $[0.52663,0.69560]$, where the pilot localization's interval binds. The model-based variant gives $[0.82505,0.86891]$ and $[0.52416,0.71145]$. A finer $\rho$ grid of 25 values changed the $P(p\ge0.5)$ interval to $[0.52729,0.69560]$ and left $\theta_{0.05}$ unchanged. Sensitivity to the node set was not studied.

\paragraph{Omitted: FSST.} The test of \citet{fang2023inference} requires, for each target, data set and grid point of a test inversion, a bootstrap of the moment vector and a regularised cone projection. With $n+1=129$ moments and about $10^4$ grid columns this is several orders of magnitude more LPs than one F-localization solve, which was not feasible for the replicate sweeps.

\section{The decision rule}\label{supp:verdict}
The decision rule of Section~\ref{M-sec:sim-verdict} is an internal protocol: a decision rule fixed and hashed before the C1--C3 comparator runs, not an entry in a public registry. The file is time-stamped 14 November 2026, 00:43:11 UTC; the earliest comparator output (the AMARI timing pilot) is time-stamped six minutes later, and the simulation sweep finished at 04:47 UTC. Its SHA-256 hash, \texttt{31f1cb2b\ldots42492865}, was stored next to it and recomputed after the runs. The recomputed hash shows only that the file has not changed since the hash was stored; the evidence that the rule preceded the comparator runs is the file time stamp. The rule file and its hash are in the directory \texttt{prereg/} of the code release. Its operative part reads:
\begin{quote}\small
For each new comparator, compute the median width ratio VA / comparator over the E1 cells where both ran:
C1: the two-stage plug-in band built on $T_k-T_{k'}$ [\ldots];
C2: Empirikos-faithful AMARI at $n=512$;
C3: Bonferroni-adjusted (simultaneous) AMARI.
The median is taken per $n$, for $n=32$, 128 and 512.
If the median ratio is at most 0.90 at $n=128$ and at $n=512$ (VA at least 10\% narrower), the headline ``VA is narrower'' claim stands against that comparator, with the size reported.
If the median ratio is above 0.90 at $n=128$ or at $n=512$, the paper must not claim a width advantage over that comparator at that $n$. The abstract and introduction are rewritten to make finite-sample validity plus conditional coverage the main selling point, not width, and the comparator's numbers are reported in full.
[\ldots] A narrower interval that undercovers does not count as narrower. The rule is: a coverage upper Clopper--Pearson bound below 0.95 disqualifies the comparator in that cell.
\end{quote}
The omissions ``[\ldots]'' remove an attribution of C1 and a sentence on how AMARI's coverage is reported. The rule covers the C1--C3 comparison only. The laws, grid, replicate counts, other comparators and the targets of the illustration were not part of it. The rule was frozen before the C1--C3 runs, not before every comparison of VA with AMARI: when it was written, two earlier comparisons were known. On the preliminary simulation (Section~\ref{supp:archival}), a model-based AMARI was narrower than VA in 59 of the 90 $\theta_\delta$ cells at $n\le128$ (37 of 45 at $n=32$, 22 of 45 at $n=128$), and the Empirikos-faithful AMARI had been run on the Monkey Business pools (E7 and the with-replacement run) and on the illustration data. The rule, which names C2 ``AMARI at $n=512$'', was therefore written with this knowledge; the literal reading we apply below works against VA.

The rule was applied by a script (\texttt{experiments/verdict\_s4.py}) that recomputes the hash, takes VA and the two-stage band from the simulation sweep and AMARI from a separate run on the same count vectors (replicate $r$ of one is replicate $r$ of the other), and computes the medians and their bootstrap Monte Carlo intervals: $B=2000$, replicates resampled jointly within each $(F,m,n)$ block across methods and targets; AMARI's additional replicates, which share no counts with a VA replicate, form a separate stratum. Restricting AMARI to the replicates it shares with VA changes the medians by at most 0.005 and does not change any verdict.

\emph{Reading note.} The rule defines C2 as ``Empirikos-faithful AMARI at $n=512$'' but says that the threshold applies ``at $n=128$ and at $n=512$''. Read literally, it applies at every $n$ where C2 was run, and then it fails at $n=128$ (median 0.977). Read as a statement about $n=512$ only, C2 passes (0.626). We report the stricter, literal reading as the verdict, and it is the reason the abstract and introduction lead with validity and conditional coverage. Under either reading VA is not narrower than pointwise AMARI at $n\le128$.

\begin{table}[tbp]
\centering
\footnotesize
\caption{The pre-specified decision rule (Section~\ref{M-sec:sim-verdict}). Median over the $\theta_\delta$ simulation cells of VA's median width divided by the comparator's, with a bootstrap 95\% Monte Carlo interval, the range over cells and the share of cells with ratio at most $0.90$. ``$\le0.90$'': whether the median meets the threshold. No comparator cell was disqualified for coverage. The threshold applies at $n=128$ and $n=512$.}
\label{tab:s4-verdict}
\begin{adjustbox}{max width=\textwidth}
\begin{tabular}{lccccccc}
\toprule
Comparator & $n$ & cells & median & MC 95\% & range & share $\le0.90$ & $\le0.90$\\
\midrule
Two-stage band & 32 & 45 & 0.746 & 0.736--0.754 & 0.559--0.923 & 0.956 & yes\\
 & 128 & 45 & 0.602 & 0.581--0.609 & 0.405--0.802 & 1.000 & yes\\
 & 512 & 45 & 0.436 & 0.427--0.440 & 0.309--0.639 & 1.000 & yes\\
AMARI, pointwise & 32 & 45 & 1.307 & 1.293--1.321 & 0.795--1.646 & 0.111 & no\\
 & 128 & 45 & 0.977 & 0.963--0.986 & 0.260--1.360 & 0.400 & no\\
 & 512 & 15 & 0.626 & 0.602--0.639 & 0.187--0.875 & 1.000 & yes\\
AMARI, $\alpha/3$ & 32 & 45 & 1.179 & 1.165--1.211 & 0.699--1.477 & 0.156 & no\\
 & 128 & 45 & 0.808 & 0.797--0.819 & 0.214--1.175 & 0.667 & yes\\
 & 512 & 15 & 0.513 & 0.496--0.524 & 0.153--0.722 & 1.000 & yes\\
\bottomrule
\end{tabular}
\end{adjustbox}
\end{table}

\section{Further empirical results}\label{supp:empirical}

\subsection{Simulation study}\label{supp:simextra}

\paragraph{Replicate plan.} The plan written before a timing pilot (for example $R=60$ at $m=1000$ and $n\le128$, and $30$ at $m\in\{250,4000\}$) projected to 95 CPU-hours. Before launch it was cut, on run time alone, to $R=30$ at $m=1000$ and $n\le128$, $20$ at $n=512$ and at $m\in\{250,4000\}$, $30$ at small $m$, $20$ for VA at $\alpha/2$, $10$ for the theorem-aligned VA, and $10$ for VA$\cap$KL and VA-Bennett. After 800 tasks the run was 1.19 times slower than projected, because the $m=4000$ tasks were slower than assumed; it was stopped after a cache save and resumed with $R=15$ at $m\in\{250,4000\}$ and $n\le128$, $20$ at small $m$, and $6$ for VA$\cap$KL and VA-Bennett. Both reductions were made on run-time grounds, before any width ratio or verdict was computed. Tasks finished under the larger plan but outside the final one are not used.

\paragraph{Exact check and identification LPs.} For $m=20$ fixed off-grid $p_i$ and $n=2$, the exact law of the 231 possible count histograms gives coverage at least $1-\alpha$ for the eight regions tested (the four VA variants, KL-Bonf, both Hoeffding regions and DKW-$4/\sqrt m$); we used $\alpha=0.5$ so that a violation would be large enough to detect. The unit test asserts the bound and does not record the minimum exact coverage. In the preliminary run, 11 of the 30 oracle LPs that compute $\ID_K(F_m)$ at $n=512$ failed and remained missing after a repair pass with a longer time limit; these are not confidence regions, and no interval depends on them.

\paragraph{Coverage of AMARI and the plug-in bootstrap.} AMARI covered $\theta_\delta$ in every replicate; for the other targets it missed once, in 29 of 30 replicates for pass@1024 under F3 at $m=1000$, $n=128$. The plug-in bootstrap's undercoverage concentrates at $n=32$ (11 of the 18 cells for $\theta_{0.01}$, small $m$ included) and for pass@1024, where it extrapolates beyond $n$ (Table~\ref{tab:s4-plugin}); its coverage reached $0$ in some cells. All regions: Table~\ref{tab:s4-coverage-full}.

\paragraph{Band events.} For every cell we drew $R=10^5$ count vectors and recorded whether the true $F_m$ satisfies each region's constraints. With the envelope, this event implies coverage of every target at once. We added two adversarial laws, A1 (half the $p_i$ at $0$, half at $0.5$) and A2 (half at $0$, half uniform on $(0.45,0.55)$), for 63 cells in total. VA missed the band in at most 0.00207 of draws over F1--F5 (median 0.00033) and 0.00139 over A1--A2 (Table~\ref{tab:s4-e5}). Two sources contribute: Bonferroni over 30--31 indices, and the Bernstein tail bound, which is loose for indicators with small variance. The two-stage band's constraints were never violated; its joint event of Proposition~\ref{M-prop:twostage} missed in at most 0.00082. Berk--Jones reached 0.01010 on A1--A2. No region with finite-sample calibration exceeded $\alpha$. The asymptotic $\chi^2$ region exceeded $\alpha$ in four cells, all at $n=512$, with miss rates up to 0.079 and Clopper--Pearson lower bounds above $0.05$.

\paragraph{Calibration slack.} On F2 and F3 at $m=1000$ we calibrated the Bonferroni--Bernstein constant on the true Poisson-binomial law by Monte Carlo, an oracle that no user has. The nominal constant $L_0=7.09$--$7.12$ gives a band miss rate of 0.00005--0.00075; the constant with miss rate $0.05$ is $L^*=3.21$--$3.97$, so the radii shrink to 0.67--0.75 of the nominal ones. VA rerun at $L^*$ was 0.49--0.98 times as wide as VA, with the largest gain for F3 at $n=512$ (0.49--0.61) and the smallest for F2 at $n=32$ (0.83--0.98) (Table~\ref{tab:s4-oracle}).

\paragraph{Where the gain comes from.} The median width ratio VA/\VA-iid was 0.416 (range 0.060--0.891), and VA was narrower in all 135 cells. The Bennett radius changed little (median ratio VA/\VA-Bennett 1.013 over 30 cells), and the \VA$\cap$KL hybrid, which splits $\alpha$, was slightly wider than VA (median VA/\VA$\cap$KL 0.972 over 45 cells). The gain is largest for F3, half of whose mass sits at $0$ and the rest well inside $(0,1)$: at $n=512$ the $k=1$ variance ratio is below $10^{-5}$, the median ratio to the best other localization over the 27 F3 cells is 0.518, and at $m=4000$, $n=512$ the median VA width for $\theta_{0.01}$ is 0.0037 against 0.0082 for the two-stage band, 0.0435 for DKW-Massart and 0.0552 for Berk--Jones. It is smallest for F1 and F5 (medians 0.665 and 0.660): F5 has an atom at $0.02$, inside the range where $h_\delta$ bends, and the Beta component of F1 has density exponent $0.3$ at $0^+$ (Section~\ref{supp:spikebeta}). For F2 at $m=4000$, $n=512$ the median VA widths are 0.027, 0.021 and 0.016 for $\delta=0.01,0.02,0.05$. All regions: Table~\ref{tab:s4-ratio-full}.

\paragraph{Other targets.} The median ratio of VA to the best other localization was 0.801, 0.649 and 0.507 for $P(p\ge0.05)$, 0.560, 0.413 and 0.331 for pass@8, and 0.944, 0.876 and 0.774 for pass@1024, at $n=32$, 128 and 512. For $P(p\ge0.5)$ it was 0.820 and 0.695 at $n=32$ and 128, but 1.108 at $n=512$: there the target is nearly identified, few problems sit near $p=0.5$, and the band at the few relevant indices is cheaper without VA's candidate-variance cuts.

\paragraph{Small benchmarks.} For F2 with $m\in\{30,50,100,250\}$ (Table~\ref{tab:s4-smallm}), VA was narrower than the best other localization in all 36 cells, with median ratios 0.791, 0.759, 0.665 and 0.647. At $m=30$ the VA width for $\theta_{0.01}$ is 0.386, 0.359 and 0.320 at $n=32$, 128 and 512, and at $m=100$ it is 0.203, 0.183 and 0.119; among these $m$, widths below $0.1$ for $\theta_\delta$ appear only at $m=250$ with $n=512$ (at $m\ge1000$ they are below $0.1$ for every $\delta$ already at $n=128$; Section~\ref{M-sec:disc}). Each $m$ has its own draw of $F_m$, so the widths are not monotone in $m$. The plug-in bootstrap is narrower at small $m$ but undercovered at $m=250$, $n=32$.

\paragraph{Computation and the index set.} A VA replicate for seven targets took 21, 105 and 130~s at $m=1000$ and $n=32$, 128 and 512 (median over laws of the mean per replicate), against about 1~s for the two-stage band and KL-Bonf and 6, 23 and 181~s for AMARI (Table~\ref{tab:s4-runtime}). Each task ran in one single-threaded process on a shared 64-core host with 22--24 concurrent workers, so these are upper bounds for an idle machine. The index-set study of the preliminary run (Table~\ref{tab:e6}) is unaffected by the sure constraints, which do not change $\theta_\delta$ (Proposition~\ref{prop:sure-noeffect}). Relative to $K_{\rm rule}$ ($|K|=31$ at $n=128$ and 69 at $n=512$) on F2, F3 and F5 with $m=1000$, $K_{\rm geo}$ gave width ratios between 0.909 and 1.000, the $\sqrt k$-spaced, geometric and weighted variants between 0.969 and 1.024, and a dyadic set with only 8--10 indices between 0.811 and 1.148. Which indices carry the width depends on the law, and a smaller $K$ saves through $L$. $K=\{1,\dots,n\}$ was 6--14\% wider at $n=128$ and took 368.7~s per replicate; at $n=512$ it averaged 5931.5~s per replicate with 11 failed solves.

\begin{table}[tbp]
\centering
\footnotesize
\caption{Coverage evidence for all regions at $\alpha=0.05$ (sure constraints in every LP region). Simulation: minimum empirical coverage of $\theta_\delta(F_m)$ over the 45 cells per $n$ (135 in all; AMARI 105, run at $n=512$ for $m=1000$ only), the number of cells whose two-sided 95\% Clopper--Pearson upper bound is below $0.95$, and the replicate range; ``other'' is the minimum over $P(p\ge0.05)$, $P(p\ge0.5)$, pass@8 and pass@1024. MB: minimum coverage of $\theta_\delta(\hat p)$ over the eight Monkey Business configurations and the pooled data, all $n$ (AMARI: $n\le128$). Band: non-coverage of the band event, $R=10^5$ per cell, maximum over cells and number of cells whose Clopper--Pearson lower bound exceeds $\alpha$. For the two-stage band the event is ``the true $F_m$ satisfies the region's constraints''. Full coverage in $R=30$ (or $R=15$) replicates is consistent with true coverage as low as 0.884 (0.782), the two-sided Clopper--Pearson lower bound; these columns check the code.}
\label{tab:s4-coverage-full}
\begin{adjustbox}{max width=\textwidth}
\begin{tabular}{lccccccccc}
\toprule
 & \multicolumn{4}{c}{Simulation} & MB & \multicolumn{2}{c}{Band, F1--F5} & \multicolumn{2}{c}{Band, A1--A2}\\
\cmidrule(lr){2-5}\cmidrule(lr){6-6}\cmidrule(lr){7-8}\cmidrule(lr){9-10}
Region & min cov. & CP$_{\rm hi}<0.95$ & $R$ & other & min cov. & max & $>\alpha$ & max & $>\alpha$\\
\midrule
VA & 1.000 & 0/135 & 15--30 & 1.000 & 1.000 & 0.00207 & 0 & 0.00139 & 0\\
VA$\cap$KL & 1.000 & 0/45 & 6--6 & 1.000 & -- & 0.00082 & 0 & 0.00088 & 0\\
VA-Bennett & 1.000 & 0/30 & 6--6 & 1.000 & -- & 0.00216 & 0 & 0.00200 & 0\\
VA-iid & 1.000 & 0/135 & 15--30 & 1.000 & 1.000 & 0.00003 & 0 & 0.00053 & 0\\
Two-stage & 1.000 & 0/135 & 15--30 & 1.000 & 1.000 & 0.00000 & 0 & 0.00000 & 0\\
Exact-binom & 1.000 & 0/135 & 15--30 & 1.000 & 1.000 & 0.00053 & 0 & 0.00669 & 0\\
KL-Bonf & 1.000 & 0/135 & 15--30 & 1.000 & 1.000 & 0.00025 & 0 & 0.00105 & 0\\
Hoeff-$k$ & 1.000 & 0/135 & 15--30 & 1.000 & 1.000 & 0.00000 & 0 & 0.00000 & 0\\
Hoeff-cell & 1.000 & 0/135 & 15--30 & 1.000 & 1.000 & 0.00000 & 0 & 0.00000 & 0\\
DKW-$4/\sqrt m$ & 1.000 & 0/135 & 15--30 & 1.000 & 1.000 & 0.00000 & 0 & 0.00000 & 0\\
DKW-Massart & 1.000 & 0/135 & 15--30 & 1.000 & 1.000 & 0.00000 & 0 & 0.00059 & 0\\
Berk--Jones & 1.000 & 0/135 & 15--30 & 1.000 & 1.000 & 0.00202 & 0 & 0.01010 & 0\\
EB-DKW & 1.000 & 0/135 & 15--30 & 1.000 & 1.000 & 0.00000 & 0 & 0.00000 & 0\\
$\chi^2$ & 1.000 & 0/135 & 15--30 & 1.000 & 1.000 & 0.05917 & 2 & 0.07930 & 2\\
Plug-in bootstrap & 0.000 & 41/135 & 15--30 & 0.000 & 0.000 & -- & -- & -- & --\\
AMARI & 1.000 & 0/105 & 30--30 & 0.967 & 0.950 & -- & -- & -- & --\\
AMARI ($\alpha/3$) & 1.000 & 0/105 & 30--30 & 0.967 & 0.950 & -- & -- & -- & --\\
\bottomrule
\end{tabular}
\end{adjustbox}
\end{table}

\begin{table}[tbp]
\centering
\small
\caption{Variance gap on the E1 draws $F_m$ ($m=1000$): $V_k(F_m)/(q_k(1-q_k))$ over $k\in K$.}
\label{tab:vargap}
\begin{adjustbox}{max width=\textwidth}
\begin{tabular}{lrccccc}
\toprule
$F$ & $n$ & median & min & max & median $\mathrm{Var}_F(P_k)$ & $k=1$ \\
\midrule
F1 & 32 & 0.188 & 0.158 & 0.545 & 0.055 & 0.212 \\
F1 & 128 & 0.095 & 0.065 & 0.799 & 0.175 & 0.131 \\
F1 & 512 & 0.062 & 0.029 & 0.995 & 0.226 & 0.088 \\
F2 & 32 & 0.101 & 0.081 & 0.242 & 0.212 & 0.205 \\
F2 & 128 & 0.058 & 0.039 & 0.192 & 0.216 & 0.161 \\
F2 & 512 & 0.056 & 0.017 & 0.144 & 0.182 & 0.140 \\
F3 & 32 & 0.344 & 0.034 & 1.000 & 0.052 & 0.034 \\
F3 & 128 & 0.078 & 0.002 & 1.000 & 0.222 & 0.002 \\
F3 & 512 & 0.010 & 0.000 & 1.000 & 0.247 & 0.000 \\
F4 & 32 & 0.178 & 0.154 & 0.560 & 0.170 & 0.154 \\
F4 & 128 & 0.093 & 0.058 & 0.670 & 0.131 & 0.058 \\
F4 & 512 & 0.043 & 0.014 & 0.883 & 0.088 & 0.014 \\
F5 & 32 & 0.283 & 0.133 & 0.991 & 0.034 & 0.284 \\
F5 & 128 & 0.094 & 0.058 & 1.000 & 0.157 & 0.072 \\
F5 & 512 & 0.061 & 0.002 & 1.000 & 0.187 & 0.002 \\
\bottomrule
\end{tabular}
\end{adjustbox}
\end{table}

\begin{table}[tbp]
\centering
\footnotesize
\caption{Median ratio VA/region of median widths over the $\theta_\delta$ simulation cells (main simulation, $K_{\rm geo}$, sure constraints), by $n$ and overall, with range and the number of cells where VA is narrower. VA$\cap$KL and VA-Bennett: $m=1000$ only, 6 replicates; VA at $\alpha/2$ and VA with the theorem's index and tangent sets: $m=1000$ only. AMARI: Empirikos-faithful, run at $n=512$ for $m=1000$ only.}
\label{tab:s4-ratio-full}
\begin{adjustbox}{max width=\textwidth}
\begin{tabular}{lcccccccc}
\toprule
Region & cells & $n=32$ & $n=128$ & $n=512$ & all & min & max & VA narrower\\
\midrule
VA-iid & 135 & 0.636 & 0.421 & 0.264 & 0.416 & 0.060 & 0.891 & 135/135\\
VA$\cap$KL & 45 & 0.980 & 0.975 & 0.938 & 0.972 & 0.897 & 1.015 & 42/45\\
VA-Bennett & 30 & 1.006 & 1.015 & -- & 1.013 & 0.956 & 1.051 & 6/30\\
VA at $\alpha/2$ & 45 & 0.969 & 0.963 & 0.942 & 0.954 & 0.917 & 0.999 & 45/45\\
VA, theorem $K$ and tangents & 45 & 0.988 & 0.984 & 0.911 & 0.958 & 0.827 & 1.031 & 41/45\\
Two-stage & 135 & 0.746 & 0.602 & 0.436 & 0.588 & 0.309 & 0.923 & 135/135\\
Two-stage (Bennett) & 135 & 0.747 & 0.605 & 0.439 & 0.590 & 0.314 & 0.923 & 135/135\\
Exact-binom & 135 & 0.721 & 0.517 & 0.337 & 0.496 & 0.072 & 0.912 & 135/135\\
KL-Bonf & 135 & 0.657 & 0.448 & 0.288 & 0.429 & 0.061 & 0.895 & 135/135\\
Hoeff-$k$ & 135 & 0.634 & 0.412 & 0.252 & 0.385 & 0.061 & 0.878 & 135/135\\
Hoeff-cell & 135 & 0.649 & 0.224 & 0.069 & 0.224 & 0.015 & 0.945 & 135/135\\
DKW-$4/\sqrt m$ & 135 & 0.351 & 0.171 & 0.097 & 0.163 & 0.022 & 0.751 & 135/135\\
EB-DKW & 135 & 0.548 & 0.320 & 0.183 & 0.319 & 0.051 & 0.860 & 135/135\\
DKW-Massart & 135 & 0.745 & 0.543 & 0.352 & 0.519 & 0.084 & 0.917 & 135/135\\
Berk--Jones & 135 & 0.696 & 0.483 & 0.326 & 0.471 & 0.066 & 0.908 & 135/135\\
$\chi^2$ & 135 & 0.803 & 0.334 & 0.145 & 0.334 & 0.026 & 1.157 & 128/135\\
AMARI & 105 & 1.307 & 0.977 & 0.626 & 1.066 & 0.187 & 1.646 & 47/105\\
AMARI ($\alpha/3$) & 105 & 1.179 & 0.808 & 0.513 & 0.925 & 0.153 & 1.477 & 65/105\\
Plug-in bootstrap & 135 & 3.096 & 1.160 & 0.597 & 1.128 & 0.119 & 12.177 & 58/135\\
\bottomrule
\end{tabular}
\end{adjustbox}
\end{table}

\begin{table}[tbp]
\centering
\footnotesize
\caption{Minimum coverage over the 45 simulation cells ($m\in\{250,1000,4000\}$) for the additional targets of the main simulation.}
\label{tab:s4-cov-other}
\begin{adjustbox}{max width=\textwidth}
\begin{tabular}{lcccc}
\toprule
Region & $P(p\ge0.05)$ & $P(p\ge0.5)$ & pass@8 & pass@1024\\
\midrule
VA & 1.000 & 1.000 & 1.000 & 1.000\\
Two-stage & 1.000 & 1.000 & 1.000 & 1.000\\
Exact-binom & 1.000 & 1.000 & 1.000 & 1.000\\
KL-Bonf & 1.000 & 1.000 & 1.000 & 1.000\\
DKW-Massart & 1.000 & 1.000 & 1.000 & 1.000\\
Berk--Jones & 1.000 & 1.000 & 1.000 & 1.000\\
Plug-in bootstrap & 0.000 & 0.000 & 1.000 & 0.000\\
\bottomrule
\end{tabular}
\end{adjustbox}
\end{table}

\begin{table}[tbp]
\centering
\footnotesize
\caption{Plug-in bootstrap: plug-in estimate $m^{-1}\sum_ih(c_i/n)$, or the unbiased estimator of \citet{chen2021evaluating} for pass@$k$ with $k\le n$, with a percentile bootstrap interval over problems. Minimum coverage over the 18 $(F,m)$ cells (F1--F5 at $m\in\{250,1000,4000\}$ and F2 at $m\in\{30,50,100\}$), and the number of cells whose Clopper--Pearson upper bound is below $0.95$.}
\label{tab:s4-plugin}
\begin{adjustbox}{max width=\textwidth}
\begin{tabular}{lcccccc}
\toprule
 & \multicolumn{2}{c}{$n=32$} & \multicolumn{2}{c}{$n=128$} & \multicolumn{2}{c}{$n=512$}\\
Target & min cov. & cells & min cov. & cells & min cov. & cells\\
\midrule
$\theta_{0.01}$ & 0.000 & 11/18 & 0.000 & 6/18 & 1.000 & 0/18\\
$\theta_{0.02}$ & 0.000 & 10/18 & 0.000 & 5/18 & 0.000 & 2/18\\
$\theta_{0.05}$ & 0.000 & 7/18 & 1.000 & 0/18 & 1.000 & 0/18\\
$P(p\ge0.05)$ & 0.000 & 3/18 & 1.000 & 0/18 & 1.000 & 0/18\\
$P(p\ge0.5)$ & 0.000 & 6/18 & 1.000 & 0/18 & 1.000 & 0/18\\
pass@8 & 1.000 & 0/18 & 1.000 & 0/18 & 1.000 & 0/18\\
pass@1024 & 0.000 & 14/18 & 0.000 & 7/18 & 0.000 & 3/18\\
\bottomrule
\end{tabular}
\end{adjustbox}
\end{table}

\begin{table}[tbp]
\centering
\scriptsize
\caption{Small benchmarks: law F2, one draw of $F_m$ per $m$, median width of the 95\% interval for $\theta_\delta(F_m)$. ``Best other'' is the narrowest non-VA localization, the two-stage band included. The plug-in bootstrap is narrower but has no coverage guarantee; its empirical coverage is in the last column.}
\label{tab:s4-smallm}
\begin{adjustbox}{max width=\textwidth}
\begin{tabular}{rrlrccclcc}
\toprule
$m$ & $n$ & Target & $R$ & VA & Two-stage & Best other & (region) & Plug-in & Plug-in cov.\\
\midrule
30 & 32 & $\theta_{0.01}$ & 20 & 0.386 & 0.680 & 0.481 & DKW-Massart & 0.300 & 1.000\\
30 & 32 & $\theta_{0.02}$ & 20 & 0.407 & 0.680 & 0.481 & DKW-Massart & 0.300 & 1.000\\
30 & 32 & $\theta_{0.05}$ & 20 & 0.426 & 0.699 & 0.509 & DKW-Massart & 0.300 & 1.000\\
30 & 128 & $\theta_{0.01}$ & 20 & 0.359 & 0.655 & 0.454 & DKW-Massart & 0.278 & 1.000\\
30 & 128 & $\theta_{0.02}$ & 20 & 0.374 & 0.668 & 0.469 & DKW-Massart & 0.281 & 1.000\\
30 & 128 & $\theta_{0.05}$ & 20 & 0.333 & 0.689 & 0.480 & Berk--Jones & 0.294 & 1.000\\
30 & 512 & $\theta_{0.01}$ & 20 & 0.320 & 0.616 & 0.417 & DKW-Massart & 0.253 & 1.000\\
30 & 512 & $\theta_{0.02}$ & 20 & 0.297 & 0.650 & 0.443 & Berk--Jones & 0.272 & 1.000\\
30 & 512 & $\theta_{0.05}$ & 20 & 0.273 & 0.681 & 0.458 & Berk--Jones & 0.290 & 1.000\\
50 & 32 & $\theta_{0.01}$ & 20 & 0.315 & 0.531 & 0.412 & DKW-Massart & 0.220 & 1.000\\
50 & 32 & $\theta_{0.02}$ & 20 & 0.326 & 0.531 & 0.412 & DKW-Massart & 0.220 & 1.000\\
50 & 32 & $\theta_{0.05}$ & 20 & 0.348 & 0.550 & 0.423 & DKW-Massart & 0.226 & 1.000\\
50 & 128 & $\theta_{0.01}$ & 20 & 0.273 & 0.471 & 0.358 & DKW-Massart & 0.199 & 1.000\\
50 & 128 & $\theta_{0.02}$ & 20 & 0.282 & 0.484 & 0.372 & DKW-Massart & 0.204 & 1.000\\
50 & 128 & $\theta_{0.05}$ & 20 & 0.239 & 0.518 & 0.371 & Berk--Jones & 0.214 & 1.000\\
50 & 512 & $\theta_{0.01}$ & 20 & 0.211 & 0.445 & 0.326 & Berk--Jones & 0.192 & 1.000\\
50 & 512 & $\theta_{0.02}$ & 20 & 0.199 & 0.463 & 0.334 & Berk--Jones & 0.201 & 1.000\\
50 & 512 & $\theta_{0.05}$ & 20 & 0.187 & 0.516 & 0.346 & Berk--Jones & 0.212 & 1.000\\
100 & 32 & $\theta_{0.01}$ & 20 & 0.203 & 0.343 & 0.286 & DKW-Massart & 0.140 & 1.000\\
100 & 32 & $\theta_{0.02}$ & 20 & 0.216 & 0.343 & 0.286 & DKW-Massart & 0.140 & 1.000\\
100 & 32 & $\theta_{0.05}$ & 20 & 0.245 & 0.364 & 0.306 & DKW-Massart & 0.139 & 1.000\\
100 & 128 & $\theta_{0.01}$ & 20 & 0.183 & 0.286 & 0.257 & Berk--Jones & 0.125 & 1.000\\
100 & 128 & $\theta_{0.02}$ & 20 & 0.168 & 0.297 & 0.252 & Berk--Jones & 0.130 & 1.000\\
100 & 128 & $\theta_{0.05}$ & 20 & 0.144 & 0.340 & 0.248 & Berk--Jones & 0.137 & 1.000\\
100 & 512 & $\theta_{0.01}$ & 20 & 0.119 & 0.264 & 0.215 & Berk--Jones & 0.120 & 1.000\\
100 & 512 & $\theta_{0.02}$ & 20 & 0.111 & 0.283 & 0.215 & Berk--Jones & 0.124 & 1.000\\
100 & 512 & $\theta_{0.05}$ & 20 & 0.112 & 0.332 & 0.231 & Berk--Jones & 0.132 & 1.000\\
250 & 32 & $\theta_{0.01}$ & 15 & 0.267 & 0.348 & 0.330 & DKW-Massart & 0.104 & 0.000\\
250 & 32 & $\theta_{0.02}$ & 15 & 0.274 & 0.348 & 0.330 & DKW-Massart & 0.104 & 0.200\\
250 & 32 & $\theta_{0.05}$ & 15 & 0.194 & 0.329 & 0.276 & DKW-Massart & 0.104 & 1.000\\
250 & 128 & $\theta_{0.01}$ & 15 & 0.201 & 0.267 & 0.253 & Exact-binom & 0.093 & 1.000\\
250 & 128 & $\theta_{0.02}$ & 15 & 0.137 & 0.258 & 0.212 & Exact-binom & 0.094 & 1.000\\
250 & 128 & $\theta_{0.05}$ & 15 & 0.102 & 0.219 & 0.193 & DKW-Massart & 0.099 & 1.000\\
250 & 512 & $\theta_{0.01}$ & 20 & 0.090 & 0.195 & 0.165 & Exact-binom & 0.086 & 1.000\\
250 & 512 & $\theta_{0.02}$ & 20 & 0.081 & 0.198 & 0.167 & Exact-binom & 0.090 & 1.000\\
250 & 512 & $\theta_{0.05}$ & 20 & 0.072 & 0.194 & 0.172 & DKW-Massart & 0.097 & 1.000\\
\bottomrule
\end{tabular}
\end{adjustbox}
\end{table}

\begin{table}[tbp]
\centering
\footnotesize
\caption{Theorem-aligned configuration, $m=1000$, F1--F5 and three $\delta$: VA with $K_{\rm rule}\cup\{2k\}$ and the fixed tangent set $\mathrm{geomspace}(10^{-7},0.5,28)$ without refinement, against VA as run in the experiments ($K_{\rm geo}$, ten initial tangents and refinement). Ratio of median widths.}
\label{tab:s4-thm}
\begin{adjustbox}{max width=\textwidth}
\begin{tabular}{rcccccc}
\toprule
$n$ & cells & median ratio & range & min coverage & $R$\\
\midrule
32 & 15 & 1.012 & 0.994--1.050 & 1.000 & 10\\
128 & 15 & 1.016 & 0.970--1.088 & 1.000 & 10\\
512 & 15 & 1.098 & 1.054--1.210 & 1.000 & 10\\
\bottomrule
\end{tabular}
\end{adjustbox}
\end{table}

\begin{table}[tbp]
\centering
\footnotesize
\caption{Calibration slack, $m=1000$. $L_0=\log(2|K|/\alpha)$ is the nominal Bonferroni--Bernstein constant and $L^*$ the constant at which the band event of the true Poisson-binomial law has miss rate $0.05$ ($2\times10^4$ Monte Carlo count vectors per cell). ``radius'' is the ratio of the square-root terms, $\sqrt{L^*/L_0}$. Last three columns: median over 10 paired replicates of the width of VA run at $L^*$ divided by VA's width. The oracle-calibrated region uses the unknown law and is a diagnostic, not a method.}
\label{tab:s4-oracle}
\begin{adjustbox}{max width=\textwidth}
\begin{tabular}{lrrrrrrrccc}
\toprule
$F$ & $n$ & $|K|$ & $L_0$ & $L^*$ & miss at $L_0$ & miss at $L^*$ & radius & $\theta_{0.01}$ & $\theta_{0.02}$ & $\theta_{0.05}$\\
\midrule
F2 & 32 & 30 & 7.09 & 3.97 & 0.00075 & 0.0496 & 0.748 & 0.977 & 0.831 & 0.850\\
F2 & 128 & 31 & 7.12 & 3.81 & 0.00020 & 0.0493 & 0.732 & 0.789 & 0.828 & 0.766\\
F2 & 512 & 31 & 7.12 & 3.65 & 0.00020 & 0.0496 & 0.715 & 0.741 & 0.702 & 0.700\\
F3 & 32 & 30 & 7.09 & 3.76 & 0.00075 & 0.0494 & 0.729 & 0.710 & 0.750 & 0.771\\
F3 & 128 & 31 & 7.12 & 3.70 & 0.00020 & 0.0494 & 0.721 & 0.642 & 0.642 & 0.716\\
F3 & 512 & 31 & 7.12 & 3.21 & 0.00005 & 0.0483 & 0.671 & 0.488 & 0.506 & 0.607\\
\bottomrule
\end{tabular}
\end{adjustbox}
\end{table}

\begin{table}[tbp]
\centering
\footnotesize
\caption{Band-event study: non-coverage $1-\hat P(\text{true }F_m\text{ satisfies the region's constraints})$, $R=10^5$ per cell; maximum and median over the 45 cells of F1--F5 and the 18 cells of A1--A2, and the number of cells whose Clopper--Pearson lower bound exceeds $\alpha=0.05$. VA rows use $K_{\rm geo}$ unless stated.}
\label{tab:s4-e5}
\begin{adjustbox}{max width=\textwidth}
\begin{tabular}{lcccccc}
\toprule
 & \multicolumn{3}{c}{F1--F5} & \multicolumn{3}{c}{A1--A2}\\
\cmidrule(lr){2-4}\cmidrule(lr){5-7}
Region & max & median & $>\alpha$ & max & median & $>\alpha$\\
\midrule
VA & 0.00207 & 0.00033 & 0 & 0.00139 & 0.00028 & 0\\
VA, $K_{\rm rule}$ & 0.00207 & 0.00031 & 0 & 0.00139 & 0.00028 & 0\\
VA, $K=\{1,\dots,n\}$ & 0.00199 & 0.00009 & 0 & 0.00124 & 0.00035 & 0\\
VA-Bennett & 0.00216 & 0.00043 & 0 & 0.00200 & 0.00035 & 0\\
VA$\cap$KL & 0.00082 & 0.00014 & 0 & 0.00088 & 0.00020 & 0\\
VA-iid & 0.00003 & 0.00000 & 0 & 0.00053 & 0.00004 & 0\\
Two-stage & 0.00000 & 0.00000 & 0 & 0.00000 & 0.00000 & 0\\
Two-stage, stage-1 event & 0.00000 & 0.00000 & 0 & 0.00000 & 0.00000 & 0\\
Two-stage, joint event $\mathcal E_1\cap\mathcal E_2$ & 0.00082 & 0.00012 & 0 & 0.00063 & 0.00010 & 0\\
Two-stage (Bennett) & 0.00000 & 0.00000 & 0 & 0.00000 & 0.00000 & 0\\
Exact-binom & 0.00053 & 0.00000 & 0 & 0.00669 & 0.00122 & 0\\
KL-Bonf & 0.00025 & 0.00000 & 0 & 0.00105 & 0.00014 & 0\\
Hoeff-$k$ & 0.00000 & 0.00000 & 0 & 0.00000 & 0.00000 & 0\\
Hoeff-cell & 0.00000 & 0.00000 & 0 & 0.00000 & 0.00000 & 0\\
DKW-$4/\sqrt m$ & 0.00000 & 0.00000 & 0 & 0.00000 & 0.00000 & 0\\
DKW-Massart & 0.00000 & 0.00000 & 0 & 0.00059 & 0.00015 & 0\\
Berk--Jones & 0.00202 & 0.00008 & 0 & 0.01010 & 0.00602 & 0\\
EB-DKW & 0.00000 & 0.00000 & 0 & 0.00000 & 0.00000 & 0\\
$\chi^2$ & 0.05917 & 0.01300 & 2 & 0.07930 & 0.03950 & 2\\
\bottomrule
\end{tabular}
\end{adjustbox}
\end{table}

\begin{table}[tbp]
\centering
\footnotesize
\caption{Worked design rule. Oracle-centred VA at level $\alpha/2$ per model, at the plug-in pilot laws $c_i/128$ of the two models of Section~\ref{M-sec:pair}, target $\theta_{0.01}$. $e_{\rm lo}$: distance from the pilot value to the lower endpoint for the distilled model; $e_{\rm hi}$: to the upper endpoint for the RL model. A difference $D$ is certified when $e_{\rm lo}+e_{\rm hi}<D$. Problems are replicated to reach $m>1060$.}
\label{tab:s4-design}
\begin{adjustbox}{max width=\textwidth}
\begin{tabular}{rrrcccc}
\toprule
$m$ & $n$ & $mn$ & $e_{\rm lo}$ & $e_{\rm hi}$ & $e_{\rm lo}+e_{\rm hi}$ & certifies $0.025$\\
\midrule
16960 & 1024 & 17{,}367{,}040 & 0.0021 & 0.0034 & 0.0054 & yes\\
16960 & 512 & 8{,}683{,}520 & 0.0031 & 0.0048 & 0.0079 & yes\\
8480 & 1024 & 8{,}683{,}520 & 0.0030 & 0.0044 & 0.0073 & yes\\
4240 & 1024 & 4{,}341{,}760 & 0.0044 & 0.0056 & 0.0100 & yes\\
8480 & 512 & 4{,}341{,}760 & 0.0042 & 0.0053 & 0.0095 & yes\\
16960 & 256 & 4{,}341{,}760 & 0.0054 & 0.0057 & 0.0111 & yes\\
4240 & 512 & 2{,}170{,}880 & 0.0059 & 0.0062 & 0.0121 & yes\\
2120 & 1024 & 2{,}170{,}880 & 0.0066 & 0.0070 & 0.0136 & yes\\
8480 & 256 & 2{,}170{,}880 & 0.0067 & 0.0067 & 0.0134 & yes\\
16960 & 128 & 2{,}170{,}880 & 0.0123 & 0.0103 & 0.0225 & yes\\
2120 & 512 & 1{,}085{,}440 & 0.0084 & 0.0080 & 0.0164 & yes\\
1060 & 1024 & 1{,}085{,}440 & 0.0099 & 0.0096 & 0.0195 & yes\\
4240 & 256 & 1{,}085{,}440 & 0.0086 & 0.0083 & 0.0169 & yes\\
8480 & 128 & 1{,}085{,}440 & 0.0143 & 0.0132 & 0.0275 & no\\
1060 & 512 & 542{,}720 & 0.0122 & 0.0114 & 0.0237 & yes\\
2120 & 256 & 542{,}720 & 0.0112 & 0.0110 & 0.0223 & yes\\
4240 & 128 & 542{,}720 & 0.0166 & 0.0176 & 0.0342 & no\\
1060 & 256 & 271{,}360 & 0.0156 & 0.0161 & 0.0316 & no\\
2120 & 128 & 271{,}360 & 0.0196 & 0.0240 & 0.0437 & no\\
1060 & 128 & 135{,}680 & 0.0236 & 0.0352 & 0.0587 & no\\
\bottomrule
\end{tabular}
\end{adjustbox}
\end{table}

\begin{table}[tbp]
\centering
\footnotesize
\caption{Run time in seconds per replicate (all seven targets, each with both endpoints), main simulation; median over F1--F5 of the mean per cell. Each task ran in one single-threaded worker process on a shared 64-core host with 22--24 concurrent workers, so the times are upper bounds for an idle machine.}
\label{tab:s4-runtime}
\begin{adjustbox}{max width=\textwidth}
\begin{tabular}{lccccccccc}
\toprule
 & \multicolumn{3}{c}{$m=250$} & \multicolumn{3}{c}{$m=1000$} & \multicolumn{3}{c}{$m=4000$}\\
\cmidrule(lr){2-4}\cmidrule(lr){5-7}\cmidrule(lr){8-10}
Region & 32 & 128 & 512 & 32 & 128 & 512 & 32 & 128 & 512\\
\midrule
VA & 45.2 & 162.3 & 176.8 & 21.2 & 104.7 & 130.4 & 15.9 & 83.1 & 87.7\\
VA-iid & 1.0 & 0.7 & 0.6 & 1.2 & 0.9 & 0.6 & 1.1 & 1.2 & 0.6\\
Two-stage & 0.8 & 0.6 & 0.6 & 1.0 & 0.9 & 0.7 & 1.1 & 1.1 & 0.8\\
Exact-binom & 1.4 & 1.1 & 0.9 & 1.5 & 1.4 & 1.1 & 1.7 & 1.8 & 1.3\\
KL-Bonf & 1.1 & 0.7 & 0.5 & 1.2 & 0.9 & 0.6 & 1.1 & 1.1 & 0.6\\
Hoeff-$k$ & 0.8 & 0.6 & 0.5 & 0.9 & 0.9 & 0.6 & 1.0 & 0.9 & 0.6\\
DKW-Massart & 0.9 & 2.5 & 13.2 & 1.0 & 3.3 & 18.0 & 1.1 & 5.0 & 26.7\\
Berk--Jones & 1.1 & 2.7 & 15.3 & 1.2 & 4.2 & 26.8 & 1.2 & 5.2 & 39.3\\
$\chi^2$ & 3.1 & 10.9 & 19.5 & 5.0 & 23.2 & 54.9 & 5.0 & 35.6 & 143.8\\
Plug-in bootstrap & 0.0 & 0.0 & 0.0 & 0.0 & 0.0 & 0.0 & 0.1 & 0.1 & 0.1\\
\bottomrule
\end{tabular}
\end{adjustbox}
\end{table}

\clearpage

\subsection{Monkey Business}\label{supp:mbextra}
At $n=128$ the pooled VA widths were 0.137, 0.100 and 0.067 for $\delta=0.01,0.02,0.05$, against 0.182, 0.148 and 0.117 for the two-stage band and 0.180, 0.150 and 0.124 for Berk--Jones; at $n=512$ they were 0.057, 0.048 and 0.039, against 0.102, 0.100 and 0.099 for Berk--Jones. Over the 81 (configuration, $n$, $\delta$) cells VA was narrower than the two-stage band, the exact-binomial band, KL-Bonf and Berk--Jones in all 81, than DKW-Massart in 79 and than $\chi^2$ in 77 (Table~\ref{tab:s4-mb-cov} gives coverage). A preliminary run that subsampled without replacement, so that the counts are hypergeometric, gave the same picture and full coverage (Table~\ref{tab:mb-wr-wor}).

AMARI's coverage was at least 0.95 in every cell at $R=20$. For a sharper check we ran $R=500$ replicates of AMARI in 48 cells: five configurations, $n\in\{32,128\}$ and three values of $\delta$ with replacement, and three configurations also without replacement (Table~\ref{tab:e7}; AMARI does not use the sure constraints). With the Empirikos-faithful standard error no cell had a Clopper--Pearson upper bound below $0.95$; the lowest coverage was 0.962 [0.941, 0.977]. The model-based variant undercovered in ten GSM8K cells at $n=128$, which shows that the variance estimate matters for AMARI at these sample sizes; it is an ablation of our implementation, not a property of AMARI.

The truth $\theta(\hat p)$ differs from $\theta(p)$ by Monte Carlo error. A parametric bootstrap gives a standard deviation of at most 0.0028 per configuration and 0.0007 for the pooled data (Table~\ref{tab:mb-truth}). Because the counts are drawn from the stored pool, this error does not affect coverage of $\theta(\hat p)$. Replicates built from disjoint blocks of stored rollouts, which would keep any dependence between rollouts, were not run.

\begin{table}[tbp]
\centering
\footnotesize
\caption{Monkey Business with replacement: minimum coverage of $\theta_\delta(\hat p)$ over the eight configurations and the pooled data, by $n$ ($R=20$ per configuration, 100 or 60 pooled; AMARI $R=20$).}
\label{tab:s4-mb-cov}
\begin{adjustbox}{max width=\textwidth}
\begin{tabular}{lccc}
\toprule
Region & $n=32$ & $n=128$ & $n=512$\\
\midrule
VA & 1.000 & 1.000 & 1.000\\
Two-stage & 1.000 & 1.000 & 1.000\\
Exact-binom & 1.000 & 1.000 & 1.000\\
KL-Bonf & 1.000 & 1.000 & 1.000\\
Hoeff-$k$ & 1.000 & 1.000 & 1.000\\
Hoeff-cell & 1.000 & 1.000 & 1.000\\
DKW-$4/\sqrt m$ & 1.000 & 1.000 & 1.000\\
DKW-Massart & 1.000 & 1.000 & 1.000\\
Berk--Jones & 1.000 & 1.000 & 1.000\\
EB-DKW & 1.000 & 1.000 & 1.000\\
$\chi^2$ & 1.000 & 1.000 & 1.000\\
VA-iid & 1.000 & 1.000 & 1.000\\
Plug-in bootstrap & 0.000 & 0.020 & 0.700\\
AMARI, model-based SE & 0.900 & 0.700 & --\\
AMARI, model-based SE, $\alpha/3$ & 1.000 & 0.800 & --\\
AMARI & 0.950 & 1.000 & --\\
AMARI ($\alpha/3$) & 0.950 & 1.000 & --\\
\bottomrule
\end{tabular}
\end{adjustbox}
\end{table}

\begin{table}[tbp]
\centering
\scriptsize
\caption{Monkey Business: Monte Carlo error of the 10{,}000-sample truth $\theta_\delta(\hat p)$ (parametric bootstrap). With replacement, $\theta(\hat p)$ is the exact conditional target; these numbers matter only for reading $\theta(\hat p)$ as $\theta(p)$.}
\label{tab:mb-truth}
\begin{adjustbox}{max width=\textwidth}
\begin{tabular}{lrrrrrr}
\toprule
Configuration & $m$ & $\delta$ & $\theta_\delta(\hat p)$ & boot.\ SD & boot.\ bias & frac.\ $\hat p\in(0,2\delta)$ \\
\midrule
GSM8K\_Llama-3-70B-Instruct & 127 & 0.01 & 0.9931 & 0.0003 & 0.0000 & 0.0079 \\
GSM8K\_Llama-3-70B-Instruct & 127 & 0.02 & 0.9926 & 0.0001 & -0.0000 & 0.0079 \\
GSM8K\_Llama-3-70B-Instruct & 127 & 0.05 & 0.9923 & 0.0001 & -0.0000 & 0.0079 \\
GSM8K\_Llama-3-8B-Instruct & 127 & 0.01 & 0.9972 & 0.0006 & -0.0000 & 0.0079 \\
GSM8K\_Llama-3-8B-Instruct & 127 & 0.02 & 0.9947 & 0.0003 & -0.0000 & 0.0157 \\
GSM8K\_Llama-3-8B-Instruct & 127 & 0.05 & 0.9902 & 0.0005 & 0.0000 & 0.0236 \\
MATH\_Gemma-2B & 128 & 0.01 & 0.7184 & 0.0028 & -0.0006 & 0.4375 \\
MATH\_Gemma-2B & 128 & 0.02 & 0.6399 & 0.0022 & -0.0001 & 0.5703 \\
MATH\_Gemma-2B & 128 & 0.05 & 0.5070 & 0.0014 & -0.0002 & 0.7266 \\
MATH\_Gemma-7B & 128 & 0.01 & 0.8706 & 0.0023 & -0.0003 & 0.2656 \\
MATH\_Gemma-7B & 128 & 0.02 & 0.8193 & 0.0019 & -0.0003 & 0.3438 \\
MATH\_Gemma-7B & 128 & 0.05 & 0.7278 & 0.0015 & -0.0003 & 0.5391 \\
MATH\_Llama-3-70B-Instruct & 128 & 0.01 & 0.9402 & 0.0012 & 0.0000 & 0.1016 \\
MATH\_Llama-3-70B-Instruct & 128 & 0.02 & 0.9230 & 0.0012 & -0.0002 & 0.1797 \\
MATH\_Llama-3-70B-Instruct & 128 & 0.05 & 0.8664 & 0.0012 & -0.0001 & 0.2734 \\
MATH\_Llama-3-8B & 128 & 0.01 & 0.8176 & 0.0024 & -0.0006 & 0.3203 \\
MATH\_Llama-3-8B & 128 & 0.02 & 0.7530 & 0.0020 & -0.0002 & 0.4609 \\
MATH\_Llama-3-8B & 128 & 0.05 & 0.6247 & 0.0015 & -0.0001 & 0.6250 \\
MATH\_Llama-3-8B-Instruct & 128 & 0.01 & 0.8649 & 0.0021 & -0.0012 & 0.2422 \\
MATH\_Llama-3-8B-Instruct & 128 & 0.02 & 0.8128 & 0.0017 & -0.0001 & 0.2969 \\
MATH\_Llama-3-8B-Instruct & 128 & 0.05 & 0.7461 & 0.0012 & -0.0001 & 0.4297 \\
MATH\_Pythia-12B & 128 & 0.01 & 0.5616 & 0.0028 & -0.0009 & 0.6094 \\
MATH\_Pythia-12B & 128 & 0.02 & 0.4551 & 0.0027 & -0.0007 & 0.7266 \\
MATH\_Pythia-12B & 128 & 0.05 & 0.2878 & 0.0015 & -0.0002 & 0.8125 \\
POOLED & 1022 & 0.01 & 0.8452 & 0.0007 & -0.0004 & 0.2495 \\
POOLED & 1022 & 0.02 & 0.7984 & 0.0006 & -0.0002 & 0.3258 \\
POOLED & 1022 & 0.05 & 0.7173 & 0.0004 & -0.0001 & 0.4305 \\
\bottomrule
\end{tabular}
\end{adjustbox}
\end{table}

{\tiny
\setlength{\tabcolsep}{3pt}
\begin{longtable}{llrlccc}
\caption{E7: AMARI coverage of $\theta_\delta(\hat p)$ on Monkey Business, $R=500$, with Clopper--Pearson 95\% intervals. Empirikos-faithful: empirical standard error of $Q(c_i)$; ``+ outer'': also intersected with the pilot F-localization interval; model-based SE: the model-based variant of our implementation.}\label{tab:e7}\\
\toprule
Sampling & Configuration & $n$ & Target & Empirikos-faithful & + outer & model-based SE (ablation) \\
\midrule
\endfirsthead
\toprule
Sampling & Configuration & $n$ & Target & Empirikos-faithful & + outer & model-based SE (ablation) \\
\midrule
\endhead
\bottomrule
\endlastfoot
with & GSM8K\_Llama-3-70B-Instruct & 32 & $\theta_{0.01}$ & 1.000 [0.993, 1.000] & 1.000 [0.993, 1.000] & 0.944 [0.920, 0.962] \\
with & GSM8K\_Llama-3-70B-Instruct & 32 & $\theta_{0.02}$ & 1.000 [0.993, 1.000] & 1.000 [0.993, 1.000] & 0.948 [0.925, 0.966] \\
with & GSM8K\_Llama-3-70B-Instruct & 32 & $\theta_{0.05}$ & 0.994 [0.983, 0.999] & 0.994 [0.983, 0.999] & 0.964 [0.944, 0.979] \\
with & GSM8K\_Llama-3-70B-Instruct & 128 & $\theta_{0.01}$ & 0.996 [0.986, 1.000] & 0.996 [0.986, 1.000] & 0.914 [0.886, 0.937] \\
with & GSM8K\_Llama-3-70B-Instruct & 128 & $\theta_{0.02}$ & 0.984 [0.969, 0.993] & 0.984 [0.969, 0.993] & 0.902 [0.873, 0.927] \\
with & GSM8K\_Llama-3-70B-Instruct & 128 & $\theta_{0.05}$ & 0.998 [0.989, 1.000] & 0.998 [0.989, 1.000] & 0.904 [0.875, 0.928] \\
with & GSM8K\_Llama-3-8B-Instruct & 32 & $\theta_{0.01}$ & 0.972 [0.953, 0.985] & 0.972 [0.953, 0.985] & 0.974 [0.956, 0.986] \\
with & GSM8K\_Llama-3-8B-Instruct & 32 & $\theta_{0.02}$ & 0.994 [0.983, 0.999] & 0.994 [0.983, 0.999] & 0.988 [0.974, 0.996] \\
with & GSM8K\_Llama-3-8B-Instruct & 32 & $\theta_{0.05}$ & 0.962 [0.941, 0.977] & 0.962 [0.941, 0.977] & 0.972 [0.953, 0.985] \\
with & GSM8K\_Llama-3-8B-Instruct & 128 & $\theta_{0.01}$ & 0.992 [0.980, 0.998] & 0.992 [0.980, 0.998] & 0.860 [0.826, 0.889] \\
with & GSM8K\_Llama-3-8B-Instruct & 128 & $\theta_{0.02}$ & 0.984 [0.969, 0.993] & 0.984 [0.969, 0.993] & 0.904 [0.875, 0.928] \\
with & GSM8K\_Llama-3-8B-Instruct & 128 & $\theta_{0.05}$ & 1.000 [0.993, 1.000] & 1.000 [0.993, 1.000] & 0.954 [0.932, 0.971] \\
with & MATH\_Gemma-7B & 32 & $\theta_{0.01}$ & 1.000 [0.993, 1.000] & 1.000 [0.993, 1.000] & 1.000 [0.993, 1.000] \\
with & MATH\_Gemma-7B & 32 & $\theta_{0.02}$ & 1.000 [0.993, 1.000] & 1.000 [0.993, 1.000] & 1.000 [0.993, 1.000] \\
with & MATH\_Gemma-7B & 32 & $\theta_{0.05}$ & 1.000 [0.993, 1.000] & 1.000 [0.993, 1.000] & 1.000 [0.993, 1.000] \\
with & MATH\_Gemma-7B & 128 & $\theta_{0.01}$ & 0.998 [0.989, 1.000] & 0.998 [0.989, 1.000] & 1.000 [0.993, 1.000] \\
with & MATH\_Gemma-7B & 128 & $\theta_{0.02}$ & 1.000 [0.993, 1.000] & 1.000 [0.993, 1.000] & 1.000 [0.993, 1.000] \\
with & MATH\_Gemma-7B & 128 & $\theta_{0.05}$ & 1.000 [0.993, 1.000] & 1.000 [0.993, 1.000] & 1.000 [0.993, 1.000] \\
with & MATH\_Llama-3-8B-Instruct & 32 & $\theta_{0.01}$ & 1.000 [0.993, 1.000] & 1.000 [0.993, 1.000] & 1.000 [0.993, 1.000] \\
with & MATH\_Llama-3-8B-Instruct & 32 & $\theta_{0.02}$ & 1.000 [0.993, 1.000] & 1.000 [0.993, 1.000] & 1.000 [0.993, 1.000] \\
with & MATH\_Llama-3-8B-Instruct & 32 & $\theta_{0.05}$ & 0.996 [0.986, 1.000] & 0.996 [0.986, 1.000] & 1.000 [0.993, 1.000] \\
with & MATH\_Llama-3-8B-Instruct & 128 & $\theta_{0.01}$ & 1.000 [0.993, 1.000] & 1.000 [0.993, 1.000] & 1.000 [0.993, 1.000] \\
with & MATH\_Llama-3-8B-Instruct & 128 & $\theta_{0.02}$ & 1.000 [0.993, 1.000] & 1.000 [0.993, 1.000] & 1.000 [0.993, 1.000] \\
with & MATH\_Llama-3-8B-Instruct & 128 & $\theta_{0.05}$ & 1.000 [0.993, 1.000] & 1.000 [0.993, 1.000] & 1.000 [0.993, 1.000] \\
with & MATH\_Pythia-12B & 32 & $\theta_{0.01}$ & 1.000 [0.993, 1.000] & 1.000 [0.993, 1.000] & 1.000 [0.993, 1.000] \\
with & MATH\_Pythia-12B & 32 & $\theta_{0.02}$ & 0.998 [0.989, 1.000] & 0.998 [0.989, 1.000] & 0.998 [0.989, 1.000] \\
with & MATH\_Pythia-12B & 32 & $\theta_{0.05}$ & 1.000 [0.993, 1.000] & 1.000 [0.993, 1.000] & 0.996 [0.986, 1.000] \\
with & MATH\_Pythia-12B & 128 & $\theta_{0.01}$ & 1.000 [0.993, 1.000] & 1.000 [0.993, 1.000] & 0.998 [0.989, 1.000] \\
with & MATH\_Pythia-12B & 128 & $\theta_{0.02}$ & 1.000 [0.993, 1.000] & 1.000 [0.993, 1.000] & 0.998 [0.989, 1.000] \\
with & MATH\_Pythia-12B & 128 & $\theta_{0.05}$ & 1.000 [0.993, 1.000] & 1.000 [0.993, 1.000] & 0.996 [0.986, 1.000] \\
without & GSM8K\_Llama-3-70B-Instruct & 32 & $\theta_{0.01}$ & 1.000 [0.993, 1.000] & 1.000 [0.993, 1.000] & 0.964 [0.944, 0.979] \\
without & GSM8K\_Llama-3-70B-Instruct & 32 & $\theta_{0.02}$ & 1.000 [0.993, 1.000] & 1.000 [0.993, 1.000] & 0.956 [0.934, 0.972] \\
without & GSM8K\_Llama-3-70B-Instruct & 32 & $\theta_{0.05}$ & 0.992 [0.980, 0.998] & 0.992 [0.980, 0.998] & 0.954 [0.932, 0.971] \\
without & GSM8K\_Llama-3-70B-Instruct & 128 & $\theta_{0.01}$ & 0.994 [0.983, 0.999] & 0.994 [0.983, 0.999] & 0.928 [0.902, 0.949] \\
without & GSM8K\_Llama-3-70B-Instruct & 128 & $\theta_{0.02}$ & 0.988 [0.974, 0.996] & 0.988 [0.974, 0.996] & 0.912 [0.884, 0.935] \\
without & GSM8K\_Llama-3-70B-Instruct & 128 & $\theta_{0.05}$ & 1.000 [0.993, 1.000] & 1.000 [0.993, 1.000] & 0.914 [0.886, 0.937] \\
without & GSM8K\_Llama-3-8B-Instruct & 32 & $\theta_{0.01}$ & 0.974 [0.956, 0.986] & 0.974 [0.956, 0.986] & 0.958 [0.937, 0.974] \\
without & GSM8K\_Llama-3-8B-Instruct & 32 & $\theta_{0.02}$ & 0.980 [0.964, 0.990] & 0.980 [0.964, 0.990] & 0.978 [0.961, 0.989] \\
without & GSM8K\_Llama-3-8B-Instruct & 32 & $\theta_{0.05}$ & 0.966 [0.946, 0.980] & 0.966 [0.946, 0.980] & 0.990 [0.977, 0.997] \\
without & GSM8K\_Llama-3-8B-Instruct & 128 & $\theta_{0.01}$ & 0.996 [0.986, 1.000] & 0.996 [0.986, 1.000] & 0.840 [0.805, 0.871] \\
without & GSM8K\_Llama-3-8B-Instruct & 128 & $\theta_{0.02}$ & 0.982 [0.966, 0.992] & 0.982 [0.966, 0.992] & 0.900 [0.870, 0.925] \\
without & GSM8K\_Llama-3-8B-Instruct & 128 & $\theta_{0.05}$ & 0.998 [0.989, 1.000] & 0.998 [0.989, 1.000] & 0.934 [0.909, 0.954] \\
without & MATH\_Gemma-7B & 32 & $\theta_{0.01}$ & 1.000 [0.993, 1.000] & 1.000 [0.993, 1.000] & 1.000 [0.993, 1.000] \\
without & MATH\_Gemma-7B & 32 & $\theta_{0.02}$ & 1.000 [0.993, 1.000] & 1.000 [0.993, 1.000] & 1.000 [0.993, 1.000] \\
without & MATH\_Gemma-7B & 32 & $\theta_{0.05}$ & 0.998 [0.989, 1.000] & 0.998 [0.989, 1.000] & 0.998 [0.989, 1.000] \\
without & MATH\_Gemma-7B & 128 & $\theta_{0.01}$ & 1.000 [0.993, 1.000] & 1.000 [0.993, 1.000] & 1.000 [0.993, 1.000] \\
without & MATH\_Gemma-7B & 128 & $\theta_{0.02}$ & 1.000 [0.993, 1.000] & 1.000 [0.993, 1.000] & 1.000 [0.993, 1.000] \\
without & MATH\_Gemma-7B & 128 & $\theta_{0.05}$ & 1.000 [0.993, 1.000] & 1.000 [0.993, 1.000] & 1.000 [0.993, 1.000] \\
\end{longtable}
}

\clearpage

\subsection{Illustration}\label{supp:pairextra}

\paragraph{Scoring.} Truncation at the token cap and verifier timeouts are false negatives, and they fall mostly on problems with few successes. Correct final answers reached by guessing, which are easy on short numerical answers, count as successes and inflate small $p_i$, and an imperfect verifier adds false positives that resampling cannot remove \citep{stroebl2024limits}. Both change the $p_i$ that the method takes as given, and they affect the low-$p$ targets most: $P(p>0)$, $\theta_{0.01}$ and pass@1024. The upstream metadata record that the exact checkpoint revisions, generation seeds and runtime versions were not kept, and that a truncation diagnostic was removed from the export. The mean count is 77.33 for the distilled model and 87.05 for the RL model; the share of problems never solved is 0.1094 and 0.1292, and the share always solved is 0.1717 and 0.3302. We have not checked whether any problem overlaps with the RL model's training data; the per-source results below let a reader set aside any source.

\paragraph{Lower bounds only.} No region certifies anything about $P(p>0)$ beyond ``solved at least once'': envelope regions cannot tell $p=0$ from $p\in(0,g_1)$, and the identified set of this lower semicontinuous functional is one-sided. For pass@1024 the upper endpoint stays near $1$ whenever some problems are never solved. Parametric extrapolations of pass@$k$ \citep{kazdan2025efficient} are much narrower, but their width comes from the assumed form of the mixing law. The informative relative of pass@$k$ is $\theta_{1/k}$, which brackets it within a factor $1-e^{-1}$ by~\eqref{M-eq:passk-bracket}.

\paragraph{Other comparators for $P(p\ge0.5)$.} The two-stage band gave $[0.513,0.713]$ and $[0.625,0.777]$. For the paired difference the widths were $0.317$ for the exact-binomial band, $0.326$ for Berk--Jones, $0.354$ for KL-Bonf, $0.365$ for the two-stage band and $0.484$ for $\chi^2$, against $0.141$ for VA (Table~\ref{tab:e3-all}); AMARI at $\alpha/8$ gave the same difference interval as pointwise AMARI, because the pilot interval binds. The plug-in bootstrap of the observed-count share $\#\{c_i\ge64\}/m$ gives $[0.586,0.643]$ and $[0.668,0.724]$ and the difference $[0.016,0.148]$, but it targets an observed-count quantity, not a functional of $F_m$, and it undercovered in the simulations. Three sources make the VA difference interval wider than a paired analysis of one pre-specified target would be: the Bonferroni split between the two marginals, simultaneity over all $h$, and the slack of the Bernstein band. The first costs $0.003$ in width. The two-index brackets below bound the cost of simultaneity over the 31 indices of $K_{\rm geo}$ for this target; on the simulation laws, calibrating the band on the true law shrank VA's width by a factor 0.49--0.98, an upper bound on what the Bernstein slack costs there.

\paragraph{Two-index brackets.} Fix $\gamma=\frac12$ and indices $k_{\rm lo}<n\gamma<k_{\rm hi}$. Because $P_k$ is nondecreasing, every $F$ satisfies $\mu_k(F)\le P_k(\gamma)\{1-\theta\}+\theta$ and $\mu_k(F)\ge P_k(\gamma)\,\theta$, with $\theta=F([\gamma,1])$; so $\theta\in[(\mu_{k_{\rm hi}}-P_{k_{\rm hi}}(\gamma))/(1-P_{k_{\rm hi}}(\gamma)),\ \mu_{k_{\rm lo}}/P_{k_{\rm lo}}(\gamma)]$. Table~\ref{tab:bracket} reports two brackets for the three index pairs $(64-d,64+d)$, $d\in\{5,10,15\}$, fixed by that rule and all reported: the closed form with one-sided KL bounds on the two tail means (valid under fixed $p_i$ by Proposition~\ref{M-prop:iid}(a)), and the VA LP with the two-element index set, which computes the exact identified set of the two bands and is valid by Theorem~\ref{M-thm:coverage}. The VA bracket with $\{54,74\}$ is slightly narrower than VA with $K_{\rm geo}$ for both models (widths 0.077 and 0.051 against 0.084 and 0.054); with $\{59,69\}$ (0.094, 0.056) and $\{49,79\}$ (0.099, 0.068) it is wider. The VA difference interval excludes zero with $\{54,74\}$ and with $\{59,69\}$, but not with $\{49,79\}$. The closed-form KL brackets are 1.8--6.0 times as wide as the VA brackets and never exclude zero. For this nearly identified target, most of VA's advantage over the other regions comes from the conditional-variance calibration and little from the many indices; with $K_{\rm geo}$ the extra indices cost some width through $L$ and buy simultaneity over all bounded targets.

\paragraph{By source.} Table~\ref{tab:bysource} gives $P(p\ge0.5)$, $\theta_{0.05}$ and $\theta_{0.01}$ by benchmark source. The VA difference for $P(p\ge0.5)$ excludes zero on OlympiadBench, $[0.004,0.180]$ ($m=633$), but not on Gaokao-2023-en, $[-0.022,0.138]$ ($m=358$), and the marginal VA intervals on AMC23 ($m=39$) and AIME24 ($m=30$) are 0.31--0.40 wide. The direction of the pooled finding is the same in the two large sources. No source gives a VA difference that excludes zero for $\theta_{0.05}$ or $\theta_{0.01}$. The per-source intervals are not simultaneous over sources, and the stratification was chosen after the pooled analysis.

\begin{table}[tbp]
\centering
\footnotesize
\caption{Two-index brackets for $P(p\ge0.5)$ on the counts of Section~\ref{M-sec:pair} ($m=1060$, $n=128$). Marginal 95\% intervals and the Bonferroni difference RL$-$distilled (each marginal at $\alpha/2$); $^*$excludes $0$. VA with $K=\{k_{\rm lo},k_{\rm hi}\}$ is the VA LP of Theorem~\ref{M-thm:coverage} with a two-element index set. The closed form is $[(l_{\rm hi}-P_{k_{\rm hi}}(\tfrac12))/(1-P_{k_{\rm hi}}(\tfrac12)),\ u_{\rm lo}/P_{k_{\rm lo}}(\tfrac12)]$ with one-sided KL bounds $l_{\rm hi}\le\mu_{k_{\rm hi}}(F)$ and $u_{\rm lo}\ge\mu_{k_{\rm lo}}(F)$ at level $\alpha/2$ each. The three pairs are fixed by the rule $(64-d,64+d)$ and all are reported.}
\label{tab:bracket}
\begin{adjustbox}{max width=\textwidth}
\begin{tabular}{llccc}
\toprule
Method & Indices & Distilled & RL & RL$-$distilled\\
\midrule
VA, $K=\{k_{\rm lo},k_{\rm hi}\}$ & $\{59,69\}$ & [0.563, 0.658] & [0.669, 0.724] & [0.007, 0.165]$^*$\\
VA, $K=\{k_{\rm lo},k_{\rm hi}\}$ & $\{54,74\}$ & [0.572, 0.649] & [0.672, 0.723] & [0.022, 0.152]$^*$\\
VA, $K=\{k_{\rm lo},k_{\rm hi}\}$ & $\{49,79\}$ & [0.565, 0.664] & [0.663, 0.731] & [-0.001, 0.167]\\
Closed form, KL & $\{59,69\}$ & [0.434, 0.803] & [0.557, 0.890] & [-0.255, 0.464]\\
Closed form, KL & $\{54,74\}$ & [0.518, 0.702] & [0.623, 0.777] & [-0.086, 0.266]\\
Closed form, KL & $\{49,79\}$ & [0.526, 0.700] & [0.626, 0.766] & [-0.081, 0.247]\\
VA, $K_{\rm geo}$ (main analysis) & 31 indices & [0.569, 0.653] & [0.670, 0.724] & [0.015, 0.156]$^*$\\
\bottomrule
\end{tabular}
\end{adjustbox}
\end{table}

{\scriptsize
\setlength{\tabcolsep}{4pt}
\begin{longtable}{lllccc}
\caption{Intervals by benchmark source for the two models of Section~\ref{M-sec:pair} ($n=128$, $\alpha=0.05$; VA and the two-stage band with $K_{\rm geo}$ and the sure constraints). Marginal 95\% intervals and the Bonferroni difference RL$-$distilled (each marginal at $\alpha/2$); $^*$excludes $0$. The intervals are not simultaneous over sources. The plug-in bootstrap targets the observed-count functional $m^{-1}\sum_ih(c_i/n)$ and has no coverage guarantee for $\theta(F_m)$. ``All'' reproduces the main analysis.}\label{tab:bysource}\\
\toprule
Source & Target & Region & Distilled & RL & RL$-$distilled\\
\midrule
\endfirsthead
\toprule
Source & Target & Region & Distilled & RL & RL$-$distilled\\
\midrule
\endhead
\bottomrule
\endlastfoot
OlympiadBench ($m=633$) & $P(p\ge0.5)$ & VA & [0.476, 0.580] & [0.587, 0.653] & [0.004, 0.180]$^*$\\
 & $P(p\ge0.5)$ & Two-stage & [0.400, 0.657] & [0.523, 0.725] & [-0.143, 0.332]\\
 & $P(p\ge0.5)$ & Plug-in bootstrap & [0.491, 0.566] & [0.588, 0.662] & [0.011, 0.181]$^*$\\
 & $\theta_{0.05}$ & VA & [0.783, 0.837] & [0.787, 0.831] & [-0.053, 0.050]\\
 & $\theta_{0.05}$ & Two-stage & [0.758, 0.868] & [0.761, 0.853] & [-0.113, 0.100]\\
 & $\theta_{0.05}$ & Plug-in bootstrap & [0.774, 0.829] & [0.773, 0.833] & [-0.066, 0.069]\\
 & $\theta_{0.01}$ & VA & [0.840, 0.950] & [0.815, 0.898] & [-0.140, 0.063]\\
 & $\theta_{0.01}$ & Two-stage & [0.816, 0.978] & [0.794, 0.931] & [-0.192, 0.123]\\
 & $\theta_{0.01}$ & Plug-in bootstrap & [0.832, 0.884] & [0.804, 0.860] & [-0.088, 0.037]\\
\addlinespace
Gaokao-2023-en ($m=358$) & $P(p\ge0.5)$ & VA & [0.739, 0.823] & [0.805, 0.874] & [-0.022, 0.138]\\
 & $P(p\ge0.5)$ & Two-stage & [0.659, 0.909] & [0.738, 0.934] & [-0.182, 0.283]\\
 & $P(p\ge0.5)$ & Plug-in bootstrap & [0.740, 0.830] & [0.796, 0.874] & [-0.042, 0.142]\\
 & $\theta_{0.05}$ & VA & [0.896, 0.937] & [0.888, 0.929] & [-0.051, 0.036]\\
 & $\theta_{0.05}$ & Two-stage & [0.861, 0.968] & [0.856, 0.956] & [-0.119, 0.102]\\
 & $\theta_{0.05}$ & Plug-in bootstrap & [0.881, 0.941] & [0.875, 0.935] & [-0.075, 0.061]\\
 & $\theta_{0.01}$ & VA & [0.908, 0.987] & [0.899, 0.985] & [-0.094, 0.082]\\
 & $\theta_{0.01}$ & Two-stage & [0.876, 1.000] & [0.866, 1.000] & [-0.138, 0.128]\\
 & $\theta_{0.01}$ & Plug-in bootstrap & [0.896, 0.951] & [0.891, 0.947] & [-0.067, 0.057]\\
\addlinespace
AMC23 ($m=39$) & $P(p\ge0.5)$ & VA & [0.500, 0.889] & [0.690, 1.000] & [-0.220, 0.509]\\
 & $P(p\ge0.5)$ & Two-stage & [0.232, 1.000] & [0.437, 1.000] & [-0.584, 0.788]\\
 & $P(p\ge0.5)$ & Plug-in bootstrap & [0.564, 0.846] & [0.769, 0.974] & [-0.128, 0.410]\\
 & $\theta_{0.05}$ & VA & [0.862, 1.000] & [0.881, 1.000] & [-0.127, 0.146]\\
 & $\theta_{0.05}$ & Two-stage & [0.623, 1.000] & [0.632, 1.000] & [-0.397, 0.405]\\
 & $\theta_{0.05}$ & Plug-in bootstrap & [0.892, 1.000] & [0.928, 1.000] & [-0.080, 0.130]\\
 & $\theta_{0.01}$ & VA & [0.874, 1.000] & [0.901, 1.000] & [-0.107, 0.134]\\
 & $\theta_{0.01}$ & Two-stage & [0.623, 1.000] & [0.666, 1.000] & [-0.363, 0.404]\\
 & $\theta_{0.01}$ & Plug-in bootstrap & [0.972, 1.000] & [1.000, 1.000] & [0.000, 0.034]\\
\addlinespace
AIME24 ($m=30$) & $P(p\ge0.5)$ & VA & [0.065, 0.453] & [0.191, 0.589] & [-0.291, 0.546]\\
 & $P(p\ge0.5)$ & Two-stage & [0.000, 0.729] & [0.000, 0.884] & [-0.755, 0.907]\\
 & $P(p\ge0.5)$ & Plug-in bootstrap & [0.067, 0.333] & [0.200, 0.567] & [-0.200, 0.533]\\
 & $\theta_{0.05}$ & VA & [0.568, 0.917] & [0.601, 0.920] & [-0.338, 0.374]\\
 & $\theta_{0.05}$ & Two-stage & [0.288, 1.000] & [0.303, 1.000] & [-0.722, 0.741]\\
 & $\theta_{0.05}$ & Plug-in bootstrap & [0.562, 0.867] & [0.580, 0.900] & [-0.323, 0.371]\\
 & $\theta_{0.01}$ & VA & [0.621, 1.000] & [0.635, 1.000] & [-0.376, 0.392]\\
 & $\theta_{0.01}$ & Two-stage & [0.369, 1.000] & [0.341, 1.000] & [-0.693, 0.655]\\
 & $\theta_{0.01}$ & Plug-in bootstrap & [0.659, 0.930] & [0.600, 0.933] & [-0.345, 0.296]\\
\addlinespace
All ($m=1060$) & $P(p\ge0.5)$ & VA & [0.569, 0.653] & [0.670, 0.724] & [0.015, 0.156]$^*$\\
 & $P(p\ge0.5)$ & Two-stage & [0.513, 0.713] & [0.625, 0.777] & [-0.095, 0.270]\\
 & $P(p\ge0.5)$ & Plug-in bootstrap & [0.586, 0.643] & [0.668, 0.724] & [0.016, 0.148]$^*$\\
 & $\theta_{0.05}$ & VA & [0.831, 0.867] & [0.831, 0.860] & [-0.038, 0.031]\\
 & $\theta_{0.05}$ & Two-stage & [0.815, 0.886] & [0.815, 0.876] & [-0.075, 0.065]\\
 & $\theta_{0.05}$ & Plug-in bootstrap & [0.823, 0.865] & [0.820, 0.861] & [-0.049, 0.043]\\
 & $\theta_{0.01}$ & VA & [0.871, 0.947] & [0.854, 0.912] & [-0.096, 0.043]\\
 & $\theta_{0.01}$ & Two-stage & [0.857, 0.968] & [0.841, 0.933] & [-0.132, 0.081]\\
 & $\theta_{0.01}$ & Plug-in bootstrap & [0.866, 0.904] & [0.846, 0.885] & [-0.064, 0.024]\\
\end{longtable}
}

\clearpage
{\scriptsize
\setlength{\tabcolsep}{4pt}
\begin{longtable}{llccc}
\caption{All eight targets for the two models; layout as Table~\ref{M-tab:e3}.}\label{tab:e3-all}\\
\toprule
Target & Region & Distilled & RL & RL$-$distilled\\
\midrule
\endfirsthead
\toprule
Target & Region & Distilled & RL & RL$-$distilled\\
\midrule
\endhead
\bottomrule
\endlastfoot
$\theta_{0.01}$ & VA & [0.871, 0.947] & [0.854, 0.912] & [-0.096, 0.043]\\
 & VA-Bennett & [0.872, 0.947] & [0.854, 0.911] & [-0.095, 0.042]\\
 & VA$\cap$KL & [0.871, 0.949] & [0.854, 0.914] & [-0.098, 0.046]\\
 & VA-iid & [0.842, 0.981] & [0.820, 0.954] & [-0.166, 0.118]\\
 & Two-stage & [0.857, 0.968] & [0.841, 0.933] & [-0.132, 0.081]\\
 & Exact-binom & [0.853, 0.969] & [0.831, 0.942] & [-0.144, 0.095]\\
 & KL-Bonf & [0.846, 0.979] & [0.825, 0.952] & [-0.159, 0.111]\\
 & Hoeff-$k$ & [0.828, 1.000] & [0.809, 0.989] & [-0.194, 0.168]\\
 & Hoeff-cell & [0.821, 1.000] & [0.801, 0.999] & [-0.202, 0.182]\\
 & DKW-$4/\sqrt m$ & [0.725, 1.000] & [0.706, 1.000] & [-0.298, 0.279]\\
 & DKW-Massart & [0.845, 0.990] & [0.825, 0.959] & [-0.175, 0.125]\\
 & Berk--Jones & [0.851, 0.973] & [0.829, 0.946] & [-0.150, 0.102]\\
 & EB-DKW & [0.831, 1.000] & [0.809, 0.990] & [-0.195, 0.169]\\
 & $\chi^2$ & [0.807, 0.975] & [0.794, 0.943] & [-0.188, 0.143]\\
 & AMARI & [0.868, 0.937] & [0.844, 0.905] & [-0.100, 0.043]\\
 & AMARI ($\alpha/8$) & [0.860, 0.946] & [0.835, 0.913] & [-0.115, 0.058]\\
 & Plug-in bootstrap & [0.866, 0.904] & [0.846, 0.885] & [-0.064, 0.024]\\
\addlinespace
$\theta_{0.02}$ & VA & [0.858, 0.911] & [0.845, 0.887] & [-0.068, 0.031]\\
 & VA-Bennett & [0.858, 0.911] & [0.846, 0.887] & [-0.067, 0.031]\\
 & VA$\cap$KL & [0.857, 0.912] & [0.845, 0.888] & [-0.069, 0.033]\\
 & VA-iid & [0.828, 0.937] & [0.810, 0.918] & [-0.132, 0.095]\\
 & Two-stage & [0.843, 0.931] & [0.833, 0.906] & [-0.102, 0.067]\\
 & Exact-binom & [0.839, 0.931] & [0.821, 0.911] & [-0.114, 0.077]\\
 & KL-Bonf & [0.832, 0.937] & [0.814, 0.917] & [-0.126, 0.089]\\
 & Hoeff-$k$ & [0.816, 0.965] & [0.800, 0.943] & [-0.171, 0.133]\\
 & Hoeff-cell & [0.768, 0.979] & [0.752, 0.956] & [-0.235, 0.195]\\
 & DKW-$4/\sqrt m$ & [0.713, 1.000] & [0.697, 1.000] & [-0.307, 0.291]\\
 & DKW-Massart & [0.832, 0.946] & [0.816, 0.924] & [-0.138, 0.100]\\
 & Berk--Jones & [0.836, 0.933] & [0.818, 0.914] & [-0.119, 0.082]\\
 & EB-DKW & [0.816, 0.962] & [0.798, 0.943] & [-0.172, 0.134]\\
 & $\chi^2$ & [0.786, 0.947] & [0.782, 0.922] & [-0.172, 0.143]\\
 & AMARI & [0.847, 0.899] & [0.830, 0.878] & [-0.080, 0.033]\\
 & AMARI ($\alpha/8$) & [0.843, 0.912] & [0.822, 0.887] & [-0.091, 0.052]\\
 & Plug-in bootstrap & [0.851, 0.890] & [0.836, 0.876] & [-0.058, 0.030]\\
\addlinespace
$\theta_{0.05}$ & VA & [0.831, 0.867] & [0.831, 0.860] & [-0.038, 0.031]\\
 & VA-Bennett & [0.831, 0.866] & [0.831, 0.860] & [-0.037, 0.030]\\
 & VA$\cap$KL & [0.830, 0.868] & [0.830, 0.861] & [-0.039, 0.032]\\
 & VA-iid & [0.796, 0.896] & [0.794, 0.891] & [-0.106, 0.099]\\
 & Two-stage & [0.815, 0.886] & [0.815, 0.876] & [-0.075, 0.065]\\
 & Exact-binom & [0.808, 0.889] & [0.805, 0.884] & [-0.088, 0.081]\\
 & KL-Bonf & [0.800, 0.895] & [0.798, 0.890] & [-0.101, 0.094]\\
 & Hoeff-$k$ & [0.787, 0.916] & [0.785, 0.910] & [-0.137, 0.129]\\
 & Hoeff-cell & [0.623, 0.955] & [0.624, 0.935] & [-0.342, 0.323]\\
 & DKW-$4/\sqrt m$ & [0.685, 1.000] & [0.683, 1.000] & [-0.321, 0.319]\\
 & DKW-Massart & [0.803, 0.899] & [0.802, 0.893] & [-0.105, 0.097]\\
 & Berk--Jones & [0.805, 0.892] & [0.803, 0.886] & [-0.094, 0.086]\\
 & EB-DKW & [0.784, 0.919] & [0.782, 0.914] & [-0.144, 0.137]\\
 & $\chi^2$ & [0.755, 0.914] & [0.765, 0.906] & [-0.156, 0.157]\\
 & AMARI & [0.825, 0.869] & [0.821, 0.865] & [-0.054, 0.046]\\
 & AMARI ($\alpha/8$) & [0.817, 0.877] & [0.812, 0.874] & [-0.070, 0.062]\\
 & Plug-in bootstrap & [0.823, 0.865] & [0.820, 0.861] & [-0.049, 0.043]\\
\addlinespace
$P(p\ge0.05)$ & VA & [0.781, 0.852] & [0.792, 0.852] & [-0.062, 0.073]\\
 & VA-Bennett & [0.781, 0.852] & [0.792, 0.852] & [-0.062, 0.072]\\
 & VA$\cap$KL & [0.780, 0.853] & [0.791, 0.853] & [-0.064, 0.074]\\
 & VA-iid & [0.732, 0.890] & [0.743, 0.887] & [-0.152, 0.160]\\
 & Two-stage & [0.752, 0.877] & [0.767, 0.872] & [-0.114, 0.123]\\
 & Exact-binom & [0.745, 0.882] & [0.756, 0.880] & [-0.132, 0.140]\\
 & KL-Bonf & [0.736, 0.890] & [0.747, 0.887] & [-0.147, 0.155]\\
 & Hoeff-$k$ & [0.728, 0.910] & [0.737, 0.907] & [-0.179, 0.185]\\
 & Hoeff-cell & [0.167, 0.955] & [0.267, 0.935] & [-0.693, 0.795]\\
 & DKW-$4/\sqrt m$ & [0.623, 1.000] & [0.632, 1.000] & [-0.372, 0.381]\\
 & DKW-Massart & [0.744, 0.890] & [0.754, 0.889] & [-0.145, 0.153]\\
 & Berk--Jones & [0.742, 0.885] & [0.753, 0.883] & [-0.138, 0.147]\\
 & EB-DKW & [0.718, 0.914] & [0.729, 0.911] & [-0.193, 0.200]\\
 & $\chi^2$ & [0.693, 0.906] & [0.717, 0.902] & [-0.196, 0.217]\\
 & AMARI & [0.738, 0.880] & [0.748, 0.896] & [-0.135, 0.160]\\
 & AMARI ($\alpha/8$) & [0.738, 0.889] & [0.748, 0.898] & [-0.143, 0.160]\\
 & Plug-in bootstrap & [0.784, 0.832] & [0.795, 0.841] & [-0.043, 0.063]\\
\addlinespace
$P(p\ge0.5)$ & VA & [0.569, 0.653] & [0.670, 0.724] & [0.015, 0.156]$^*$\\
 & VA-Bennett & [0.569, 0.653] & [0.670, 0.724] & [0.016, 0.156]$^*$\\
 & VA$\cap$KL & [0.568, 0.654] & [0.669, 0.725] & [0.014, 0.158]$^*$\\
 & VA-iid & [0.513, 0.705] & [0.614, 0.773] & [-0.097, 0.266]\\
 & Two-stage & [0.513, 0.713] & [0.625, 0.777] & [-0.095, 0.270]\\
 & Exact-binom & [0.526, 0.694] & [0.627, 0.763] & [-0.073, 0.244]\\
 & KL-Bonf & [0.516, 0.704] & [0.618, 0.772] & [-0.092, 0.262]\\
 & Hoeff-$k$ & [0.515, 0.708] & [0.615, 0.780] & [-0.099, 0.271]\\
 & Hoeff-cell & [0.111, 0.954] & [0.267, 0.934] & [-0.692, 0.831]\\
 & DKW-$4/\sqrt m$ & [0.408, 0.810] & [0.508, 0.887] & [-0.311, 0.487]\\
 & DKW-Massart & [0.535, 0.686] & [0.634, 0.761] & [-0.061, 0.234]\\
 & Berk--Jones & [0.524, 0.695] & [0.625, 0.765] & [-0.077, 0.249]\\
 & EB-DKW & [0.494, 0.728] & [0.597, 0.797] & [-0.140, 0.312]\\
 & $\chi^2$ & [0.466, 0.735] & [0.589, 0.787] & [-0.155, 0.329]\\
 & AMARI & [0.527, 0.696] & [0.625, 0.770] & [-0.070, 0.243]\\
 & AMARI ($\alpha/8$) & [0.527, 0.696] & [0.625, 0.770] & [-0.070, 0.243]\\
 & Plug-in bootstrap & [0.586, 0.643] & [0.668, 0.724] & [0.016, 0.148]$^*$\\
\addlinespace
pass@256 & VA & [0.885, 0.960] & [0.865, 0.936] & [-0.098, 0.053]\\
 & VA-Bennett & [0.885, 0.960] & [0.865, 0.935] & [-0.097, 0.053]\\
 & VA$\cap$KL & [0.884, 0.962] & [0.865, 0.938] & [-0.100, 0.056]\\
 & VA-iid & [0.847, 0.974] & [0.824, 0.959] & [-0.153, 0.117]\\
 & Two-stage & [0.861, 0.968] & [0.844, 0.945] & [-0.127, 0.087]\\
 & Exact-binom & [0.857, 0.968] & [0.835, 0.952] & [-0.137, 0.099]\\
 & KL-Bonf & [0.851, 0.973] & [0.829, 0.958] & [-0.148, 0.111]\\
 & Hoeff-$k$ & [0.833, 0.989] & [0.813, 0.978] & [-0.180, 0.150]\\
 & Hoeff-cell & [0.827, 0.993] & [0.807, 0.981] & [-0.189, 0.159]\\
 & DKW-$4/\sqrt m$ & [0.730, 1.000] & [0.710, 1.000] & [-0.294, 0.274]\\
 & DKW-Massart & [0.849, 0.979] & [0.829, 0.963] & [-0.156, 0.122]\\
 & Berk--Jones & [0.855, 0.970] & [0.833, 0.954] & [-0.142, 0.104]\\
 & EB-DKW & [0.835, 0.988] & [0.813, 0.978] & [-0.180, 0.148]\\
 & $\chi^2$ & [0.813, 0.974] & [0.804, 0.951] & [-0.177, 0.144]\\
 & AMARI & [0.884, 0.926] & [0.864, 0.908] & [-0.068, 0.030]\\
 & AMARI ($\alpha/8$) & [0.875, 0.934] & [0.855, 0.916] & [-0.083, 0.046]\\
 & Plug-in bootstrap & [0.868, 0.905] & [0.848, 0.886] & [-0.064, 0.024]\\
\addlinespace
pass@1024 & VA & [0.885, 1.000] & [0.865, 0.996] & [-0.135, 0.112]\\
 & VA-Bennett & [0.885, 1.000] & [0.866, 0.996] & [-0.135, 0.112]\\
 & VA$\cap$KL & [0.885, 1.000] & [0.865, 0.997] & [-0.135, 0.113]\\
 & VA-iid & [0.847, 1.000] & [0.824, 1.000] & [-0.178, 0.156]\\
 & Two-stage & [0.861, 1.000] & [0.844, 0.999] & [-0.158, 0.139]\\
 & Exact-binom & [0.857, 1.000] & [0.835, 0.999] & [-0.167, 0.144]\\
 & KL-Bonf & [0.851, 1.000] & [0.829, 1.000] & [-0.174, 0.151]\\
 & Hoeff-$k$ & [0.833, 1.000] & [0.813, 1.000] & [-0.190, 0.170]\\
 & Hoeff-cell & [0.827, 1.000] & [0.807, 1.000] & [-0.195, 0.175]\\
 & DKW-$4/\sqrt m$ & [0.730, 1.000] & [0.710, 1.000] & [-0.294, 0.274]\\
 & DKW-Massart & [0.849, 1.000] & [0.829, 1.000] & [-0.175, 0.155]\\
 & Berk--Jones & [0.855, 1.000] & [0.833, 0.999] & [-0.170, 0.147]\\
 & EB-DKW & [0.835, 1.000] & [0.813, 1.000] & [-0.191, 0.168]\\
 & $\chi^2$ & [0.815, 1.000] & [0.806, 0.998] & [-0.198, 0.188]\\
 & AMARI & [0.863, 0.976] & [0.857, 0.969] & [-0.127, 0.114]\\
 & AMARI ($\alpha/8$) & [0.853, 0.986] & [0.845, 0.981] & [-0.147, 0.134]\\
 & Plug-in bootstrap & [0.871, 0.909] & [0.851, 0.889] & [-0.065, 0.024]\\
\addlinespace
$P(p>0)$ & VA & [0.891, 1.000] & [0.871, 1.000] & [-0.129, 0.109]\\
 & VA-Bennett & [0.891, 1.000] & [0.871, 1.000] & [-0.129, 0.109]\\
 & VA$\cap$KL & [0.891, 1.000] & [0.871, 1.000] & [-0.129, 0.109]\\
 & VA-iid & [0.891, 1.000] & [0.871, 1.000] & [-0.129, 0.109]\\
 & Two-stage & [0.891, 1.000] & [0.871, 1.000] & [-0.129, 0.109]\\
 & Exact-binom & [0.891, 1.000] & [0.871, 1.000] & [-0.129, 0.109]\\
 & KL-Bonf & [0.891, 1.000] & [0.871, 1.000] & [-0.129, 0.109]\\
 & Hoeff-$k$ & [0.891, 1.000] & [0.871, 1.000] & [-0.129, 0.109]\\
 & Hoeff-cell & [0.891, 1.000] & [0.871, 1.000] & [-0.129, 0.109]\\
 & DKW-$4/\sqrt m$ & [0.891, 1.000] & [0.871, 1.000] & [-0.129, 0.109]\\
 & DKW-Massart & [0.891, 1.000] & [0.871, 1.000] & [-0.129, 0.109]\\
 & Berk--Jones & [0.891, 1.000] & [0.871, 1.000] & [-0.129, 0.109]\\
 & EB-DKW & [0.891, 1.000] & [0.871, 1.000] & [-0.129, 0.109]\\
 & $\chi^2$ & [0.891, 1.000] & [0.871, 1.000] & [-0.129, 0.109]\\
 & AMARI & [0.841, 1.000] & [0.821, 1.000] & [-0.179, 0.159]\\
 & AMARI ($\alpha/8$) & [0.841, 1.000] & [0.821, 1.000] & [-0.179, 0.159]\\
 & Plug-in bootstrap & [0.871, 0.909] & [0.851, 0.889] & [-0.065, 0.024]\\
\end{longtable}
}

\clearpage

\section{Archival: preliminary runs}\label{supp:archival}
The tables and figures in this section come from preliminary runs and are kept for reference; the main text does not use them except where it says so (the index-set study, Table~\ref{tab:e6}, and the without-replacement Monkey Business run). They differ from the main runs in three ways: the LP regions have no sure constraints, which by Proposition~\ref{prop:sure-noeffect} do not change the $\theta_\delta$ widths of box-monotone regions; the replicate counts are larger ($R=30$--$200$); and AMARI is the model-based variant, not the Empirikos-faithful one. ``Best other localization'' in these tables excludes the two-stage and exact-binomial bands, which were added later. AMARI numbers here are therefore not comparable with those of the main text.

\subsection{Simulation (E1, E4, E5, E6) and grid check}
\begin{table}[tbp]
\centering
\scriptsize
\caption{E1 replicates per cell (minimum over $F$ and $\delta$).}
\label{tab:e1-R}
\begin{adjustbox}{max width=\textwidth}
\begin{tabular}{rrccccccccccccc}
\toprule
$m$ & $n$ & VA & VA$\cap$KL & VA-Bennett & VA-iid & KL-Bonf & Hoeff-$k$ & Hoeff-cell & DKW-$4/\sqrt m$ & DKW-Massart & Berk--Jones & EB-DKW & $\chi^2$ & AMARI$^\dagger$ \\
\midrule
250 & 32 & 40 & -- & -- & 40 & 40 & 40 & 40 & 40 & 40 & 40 & 40 & 40 & 30 \\
250 & 128 & 40 & -- & -- & 40 & 40 & 40 & 40 & 40 & 40 & 40 & 40 & 40 & 30 \\
250 & 512 & 30 & -- & -- & 30 & 30 & 30 & 30 & 30 & 30 & 30 & 30 & 30 & -- \\
1000 & 32 & 100 & 30 & 30 & 100 & 100 & 100 & 100 & 100 & 100 & 100 & 100 & 100 & 30 \\
1000 & 128 & 100 & 30 & 30 & 100 & 100 & 100 & 100 & 100 & 100 & 100 & 100 & 100 & 30 \\
1000 & 512 & 60 & 30 & -- & 60 & 60 & 60 & 60 & 60 & 60 & 60 & 60 & 60 & -- \\
4000 & 32 & 40 & -- & -- & 40 & 40 & 40 & 40 & 40 & 40 & 40 & 40 & 40 & 30 \\
4000 & 128 & 40 & -- & -- & 40 & 40 & 40 & 40 & 40 & 40 & 40 & 40 & 40 & 30 \\
4000 & 512 & 30 & -- & -- & 30 & 30 & 30 & 30 & 30 & 30 & 30 & 30 & 30 & -- \\
\bottomrule
\end{tabular}
\end{adjustbox}
\end{table}

\begin{table}[tbp]
\centering
\scriptsize
\caption{E1 median width, $m=250$, $n=32$, $\alpha=0.05$.}
\label{tab:e1-m250-n32}
\begin{adjustbox}{max width=\textwidth}
\begin{tabular}{lrccccccccccc}
\toprule
$F$ & $\delta$ & VA & VA-iid & KL-Bonf & Hoeff-$k$ & Hoeff-cell & DKW-$4/\sqrt m$ & DKW-Massart & Berk--Jones & EB-DKW & $\chi^2$ & AMARI$^\dagger$ \\
\midrule
F1 & 0.01 & 0.587 & 0.684 & 0.678 & 0.683 & 0.684 & 0.894 & 0.650 & 0.666 & 0.761 & 0.652 & 0.431 \\
F1 & 0.02 & 0.404 & 0.645 & 0.616 & 0.639 & 0.525 & 0.894 & 0.495 & 0.562 & 0.761 & 0.439 & 0.275 \\
F1 & 0.05 & 0.195 & 0.366 & 0.348 & 0.361 & 0.410 & 0.869 & 0.277 & 0.317 & 0.551 & 0.323 & 0.169 \\
F2 & 0.01 & 0.267 & 0.368 & 0.355 & 0.363 & 0.364 & 0.574 & 0.330 & 0.341 & 0.424 & 0.364 & 0.306 \\
F2 & 0.02 & 0.274 & 0.368 & 0.355 & 0.363 & 0.364 & 0.574 & 0.330 & 0.341 & 0.424 & 0.371 & 0.217 \\
F2 & 0.05 & 0.192 & 0.337 & 0.318 & 0.358 & 0.397 & 0.599 & 0.275 & 0.289 & 0.451 & 0.322 & 0.156 \\
F3 & 0.01 & 0.516 & 0.621 & 0.614 & 0.615 & 0.616 & 0.826 & 0.582 & 0.600 & 0.694 & 0.514 & 0.328 \\
F3 & 0.02 & 0.289 & 0.588 & 0.557 & 0.567 & 0.454 & 0.826 & 0.423 & 0.501 & 0.694 & 0.354 & 0.214 \\
F3 & 0.05 & 0.133 & 0.339 & 0.321 & 0.327 & 0.371 & 0.837 & 0.243 & 0.288 & 0.524 & 0.292 & 0.148 \\
F4 & 0.01 & 0.155 & 0.244 & 0.226 & 0.251 & 0.252 & 0.462 & 0.218 & 0.214 & 0.290 & 0.272 & 0.208 \\
F4 & 0.02 & 0.160 & 0.244 & 0.226 & 0.251 & 0.252 & 0.462 & 0.218 & 0.214 & 0.290 & 0.273 & 0.150 \\
F4 & 0.05 & 0.149 & 0.257 & 0.237 & 0.263 & 0.289 & 0.474 & 0.229 & 0.220 & 0.304 & 0.264 & 0.124 \\
F5 & 0.01 & 0.691 & 0.777 & 0.772 & 0.787 & 0.788 & 0.998 & 0.754 & 0.761 & 0.859 & 0.689 & 0.471 \\
F5 & 0.02 & 0.416 & 0.622 & 0.591 & 0.651 & 0.546 & 0.998 & 0.507 & 0.541 & 0.859 & 0.433 & 0.288 \\
F5 & 0.05 & 0.199 & 0.351 & 0.333 & 0.364 & 0.411 & 0.821 & 0.281 & 0.304 & 0.533 & 0.293 & 0.160 \\
\bottomrule
\end{tabular}
\end{adjustbox}
\end{table}

\begin{table}[tbp]
\centering
\scriptsize
\caption{E1 median width, $m=250$, $n=128$, $\alpha=0.05$.}
\label{tab:e1-m250-n128}
\begin{adjustbox}{max width=\textwidth}
\begin{tabular}{lrccccccccccc}
\toprule
$F$ & $\delta$ & VA & VA-iid & KL-Bonf & Hoeff-$k$ & Hoeff-cell & DKW-$4/\sqrt m$ & DKW-Massart & Berk--Jones & EB-DKW & $\chi^2$ & AMARI$^\dagger$ \\
\midrule
F1 & 0.01 & 0.203 & 0.398 & 0.378 & 0.386 & 0.427 & 0.829 & 0.293 & 0.342 & 0.606 & 0.375 & 0.171 \\
F1 & 0.02 & 0.137 & 0.315 & 0.299 & 0.306 & 0.432 & 0.763 & 0.233 & 0.271 & 0.476 & 0.337 & 0.136 \\
F1 & 0.05 & 0.099 & 0.270 & 0.255 & 0.262 & 0.537 & 0.696 & 0.192 & 0.227 & 0.424 & 0.310 & 0.120 \\
F2 & 0.01 & 0.201 & 0.289 & 0.272 & 0.291 & 0.305 & 0.503 & 0.257 & 0.259 & 0.339 & 0.337 & 0.148 \\
F2 & 0.02 & 0.135 & 0.258 & 0.239 & 0.302 & 0.413 & 0.526 & 0.229 & 0.217 & 0.368 & 0.311 & 0.120 \\
F2 & 0.05 & 0.107 & 0.246 & 0.228 & 0.266 & 0.656 & 0.578 & 0.196 & 0.204 & 0.390 & 0.306 & 0.104 \\
F3 & 0.01 & 0.104 & 0.353 & 0.333 & 0.338 & 0.386 & 0.816 & 0.245 & 0.296 & 0.557 & 0.312 & 0.131 \\
F3 & 0.02 & 0.066 & 0.275 & 0.259 & 0.263 & 0.391 & 0.720 & 0.190 & 0.230 & 0.434 & 0.308 & 0.122 \\
F3 & 0.05 & 0.056 & 0.257 & 0.242 & 0.246 & 0.516 & 0.670 & 0.179 & 0.216 & 0.405 & 0.308 & 0.120 \\
F4 & 0.01 & 0.108 & 0.223 & 0.204 & 0.233 & 0.251 & 0.444 & 0.200 & 0.192 & 0.267 & 0.279 & 0.099 \\
F4 & 0.02 & 0.068 & 0.193 & 0.171 & 0.236 & 0.334 & 0.447 & 0.192 & 0.152 & 0.270 & 0.258 & 0.088 \\
F4 & 0.05 & 0.066 & 0.186 & 0.166 & 0.246 & 0.586 & 0.457 & 0.181 & 0.148 & 0.283 & 0.261 & 0.086 \\
F5 & 0.01 & 0.222 & 0.406 & 0.386 & 0.400 & 0.419 & 0.907 & 0.305 & 0.349 & 0.613 & 0.341 & 0.195 \\
F5 & 0.02 & 0.142 & 0.308 & 0.293 & 0.303 & 0.423 & 0.751 & 0.230 & 0.264 & 0.469 & 0.321 & 0.153 \\
F5 & 0.05 & 0.112 & 0.265 & 0.247 & 0.270 & 0.488 & 0.679 & 0.196 & 0.219 & 0.422 & 0.262 & 0.108 \\
\bottomrule
\end{tabular}
\end{adjustbox}
\end{table}

\begin{table}[tbp]
\centering
\scriptsize
\caption{E1 median width, $m=250$, $n=512$, $\alpha=0.05$.}
\label{tab:e1-m250-n512}
\begin{adjustbox}{max width=\textwidth}
\begin{tabular}{lrcccccccccc}
\toprule
$F$ & $\delta$ & VA & VA-iid & KL-Bonf & Hoeff-$k$ & Hoeff-cell & DKW-$4/\sqrt m$ & DKW-Massart & Berk--Jones & EB-DKW & $\chi^2$ \\
\midrule
F1 & 0.01 & 0.104 & 0.282 & 0.266 & 0.271 & 0.558 & 0.712 & 0.198 & 0.238 & 0.439 & 0.358 \\
F1 & 0.02 & 0.085 & 0.263 & 0.248 & 0.253 & 0.638 & 0.681 & 0.183 & 0.220 & 0.413 & 0.346 \\
F1 & 0.05 & 0.069 & 0.250 & 0.235 & 0.242 & 0.689 & 0.666 & 0.173 & 0.208 & 0.398 & 0.338 \\
F2 & 0.01 & 0.091 & 0.209 & 0.190 & 0.262 & 0.562 & 0.485 & 0.193 & 0.170 & 0.319 & 0.322 \\
F2 & 0.02 & 0.081 & 0.212 & 0.193 & 0.250 & 0.780 & 0.514 & 0.183 & 0.172 & 0.348 & 0.330 \\
F2 & 0.05 & 0.072 & 0.222 & 0.205 & 0.243 & 0.912 & 0.574 & 0.172 & 0.179 & 0.363 & 0.344 \\
F3 & 0.01 & 0.044 & 0.252 & 0.237 & 0.240 & 0.508 & 0.665 & 0.173 & 0.210 & 0.399 & 0.344 \\
F3 & 0.02 & 0.040 & 0.250 & 0.236 & 0.239 & 0.581 & 0.661 & 0.172 & 0.209 & 0.397 & 0.346 \\
F3 & 0.05 & 0.040 & 0.250 & 0.235 & 0.238 & 0.626 & 0.660 & 0.170 & 0.207 & 0.397 & 0.344 \\
F4 & 0.01 & 0.044 & 0.171 & 0.151 & 0.231 & 0.519 & 0.442 & 0.173 & 0.133 & 0.264 & 0.308 \\
F4 & 0.02 & 0.043 & 0.172 & 0.152 & 0.234 & 0.752 & 0.445 & 0.173 & 0.135 & 0.267 & 0.310 \\
F4 & 0.05 & 0.046 & 0.177 & 0.158 & 0.238 & 0.901 & 0.455 & 0.171 & 0.138 & 0.282 & 0.310 \\
F5 & 0.01 & 0.057 & 0.253 & 0.238 & 0.243 & 0.483 & 0.667 & 0.175 & 0.211 & 0.401 & 0.316 \\
F5 & 0.02 & 0.079 & 0.263 & 0.248 & 0.254 & 0.536 & 0.679 & 0.186 & 0.221 & 0.412 & 0.322 \\
F5 & 0.05 & 0.068 & 0.228 & 0.212 & 0.234 & 0.570 & 0.615 & 0.166 & 0.186 & 0.374 & 0.268 \\
\bottomrule
\end{tabular}
\end{adjustbox}
\end{table}

\begin{table}[tbp]
\centering
\scriptsize
\caption{E1 median width, $m=1000$, $n=32$, $\alpha=0.05$.}
\label{tab:e1-m1000-n32}
\begin{adjustbox}{max width=\textwidth}
\begin{tabular}{lrccccccccccccc}
\toprule
$F$ & $\delta$ & VA & VA$\cap$KL & VA-Bennett & VA-iid & KL-Bonf & Hoeff-$k$ & Hoeff-cell & DKW-$4/\sqrt m$ & DKW-Massart & Berk--Jones & EB-DKW & $\chi^2$ & AMARI$^\dagger$ \\
\midrule
F1 & 0.01 & 0.509 & 0.530 & 0.511 & 0.613 & 0.611 & 0.613 & 0.560 & 0.719 & 0.591 & 0.605 & 0.634 & 0.460 & 0.332 \\
F1 & 0.02 & 0.269 & 0.278 & 0.270 & 0.417 & 0.409 & 0.417 & 0.335 & 0.719 & 0.340 & 0.384 & 0.503 & 0.292 & 0.193 \\
F1 & 0.05 & 0.140 & 0.145 & 0.140 & 0.228 & 0.223 & 0.227 & 0.250 & 0.493 & 0.185 & 0.209 & 0.277 & 0.210 & 0.108 \\
F2 & 0.01 & 0.251 & 0.252 & 0.251 & 0.303 & 0.299 & 0.305 & 0.305 & 0.411 & 0.288 & 0.294 & 0.318 & 0.291 & 0.212 \\
F2 & 0.02 & 0.220 & 0.228 & 0.220 & 0.303 & 0.299 & 0.305 & 0.301 & 0.411 & 0.288 & 0.294 & 0.318 & 0.251 & 0.142 \\
F2 & 0.05 & 0.114 & 0.117 & 0.113 & 0.194 & 0.188 & 0.207 & 0.227 & 0.433 & 0.165 & 0.176 & 0.242 & 0.189 & 0.092 \\
F3 & 0.01 & 0.241 & 0.253 & 0.238 & 0.564 & 0.548 & 0.553 & 0.371 & 0.680 & 0.417 & 0.505 & 0.596 & 0.242 & 0.213 \\
F3 & 0.02 & 0.136 & 0.142 & 0.135 & 0.320 & 0.311 & 0.314 & 0.244 & 0.680 & 0.241 & 0.288 & 0.404 & 0.184 & 0.128 \\
F3 & 0.05 & 0.069 & 0.072 & 0.068 & 0.183 & 0.178 & 0.179 & 0.197 & 0.447 & 0.137 & 0.164 & 0.232 & 0.162 & 0.085 \\
F4 & 0.01 & 0.109 & 0.109 & 0.108 & 0.148 & 0.143 & 0.163 & 0.163 & 0.268 & 0.146 & 0.139 & 0.159 & 0.154 & 0.124 \\
F4 & 0.02 & 0.111 & 0.111 & 0.110 & 0.148 & 0.143 & 0.163 & 0.163 & 0.268 & 0.146 & 0.139 & 0.159 & 0.155 & 0.085 \\
F4 & 0.05 & 0.076 & 0.080 & 0.077 & 0.133 & 0.127 & 0.172 & 0.186 & 0.278 & 0.140 & 0.118 & 0.170 & 0.124 & 0.071 \\
F5 & 0.01 & 0.574 & 0.586 & 0.575 & 0.659 & 0.658 & 0.661 & 0.608 & 0.767 & 0.635 & 0.652 & 0.681 & 0.497 & 0.350 \\
F5 & 0.02 & 0.285 & 0.292 & 0.283 & 0.423 & 0.414 & 0.431 & 0.356 & 0.767 & 0.355 & 0.391 & 0.509 & 0.306 & 0.202 \\
F5 & 0.05 & 0.148 & 0.153 & 0.148 & 0.230 & 0.225 & 0.233 & 0.253 & 0.498 & 0.192 & 0.212 & 0.279 & 0.196 & 0.105 \\
\bottomrule
\end{tabular}
\end{adjustbox}
\end{table}

\begin{table}[tbp]
\centering
\scriptsize
\caption{E1 median width, $m=1000$, $n=128$, $\alpha=0.05$.}
\label{tab:e1-m1000-n128}
\begin{adjustbox}{max width=\textwidth}
\begin{tabular}{lrccccccccccccc}
\toprule
$F$ & $\delta$ & VA & VA$\cap$KL & VA-Bennett & VA-iid & KL-Bonf & Hoeff-$k$ & Hoeff-cell & DKW-$4/\sqrt m$ & DKW-Massart & Berk--Jones & EB-DKW & $\chi^2$ & AMARI$^\dagger$ \\
\midrule
F1 & 0.01 & 0.126 & 0.130 & 0.126 & 0.231 & 0.225 & 0.227 & 0.241 & 0.524 & 0.179 & 0.209 & 0.286 & 0.249 & 0.106 \\
F1 & 0.02 & 0.091 & 0.093 & 0.091 & 0.183 & 0.178 & 0.180 & 0.240 & 0.409 & 0.142 & 0.166 & 0.225 & 0.227 & 0.083 \\
F1 & 0.05 & 0.062 & 0.065 & 0.062 & 0.150 & 0.145 & 0.148 & 0.389 & 0.371 & 0.111 & 0.133 & 0.192 & 0.202 & 0.065 \\
F2 & 0.01 & 0.099 & 0.100 & 0.096 & 0.178 & 0.172 & 0.208 & 0.219 & 0.352 & 0.161 & 0.159 & 0.230 & 0.212 & 0.085 \\
F2 & 0.02 & 0.071 & 0.071 & 0.068 & 0.142 & 0.137 & 0.165 & 0.226 & 0.370 & 0.128 & 0.127 & 0.182 & 0.199 & 0.067 \\
F2 & 0.05 & 0.053 & 0.055 & 0.053 & 0.129 & 0.125 & 0.143 & 0.364 & 0.361 & 0.107 & 0.114 & 0.169 & 0.195 & 0.056 \\
F3 & 0.01 & 0.033 & 0.035 & 0.032 & 0.176 & 0.170 & 0.171 & 0.196 & 0.464 & 0.124 & 0.155 & 0.230 & 0.197 & 0.067 \\
F3 & 0.02 & 0.021 & 0.023 & 0.021 & 0.136 & 0.132 & 0.132 & 0.196 & 0.361 & 0.096 & 0.119 & 0.178 & 0.197 & 0.062 \\
F3 & 0.05 & 0.023 & 0.024 & 0.022 & 0.129 & 0.125 & 0.126 & 0.302 & 0.338 & 0.092 & 0.114 & 0.168 & 0.196 & 0.061 \\
F4 & 0.01 & 0.047 & 0.050 & 0.046 & 0.112 & 0.104 & 0.143 & 0.153 & 0.249 & 0.126 & 0.095 & 0.136 & 0.158 & 0.054 \\
F4 & 0.02 & 0.031 & 0.032 & 0.030 & 0.086 & 0.081 & 0.138 & 0.196 & 0.252 & 0.102 & 0.074 & 0.124 & 0.148 & 0.044 \\
F4 & 0.05 & 0.032 & 0.034 & 0.032 & 0.086 & 0.081 & 0.131 & 0.313 & 0.265 & 0.097 & 0.074 & 0.122 & 0.153 & 0.040 \\
F5 & 0.01 & 0.136 & 0.144 & 0.138 & 0.242 & 0.236 & 0.238 & 0.239 & 0.542 & 0.188 & 0.220 & 0.300 & 0.223 & 0.120 \\
F5 & 0.02 & 0.095 & 0.098 & 0.095 & 0.185 & 0.181 & 0.183 & 0.245 & 0.412 & 0.145 & 0.169 & 0.228 & 0.223 & 0.094 \\
F5 & 0.05 & 0.061 & 0.063 & 0.059 & 0.143 & 0.138 & 0.145 & 0.400 & 0.378 & 0.107 & 0.126 & 0.186 & 0.173 & 0.059 \\
\bottomrule
\end{tabular}
\end{adjustbox}
\end{table}

\begin{table}[tbp]
\centering
\scriptsize
\caption{E1 median width, $m=1000$, $n=512$, $\alpha=0.05$.}
\label{tab:e1-m1000-n512}
\begin{adjustbox}{max width=\textwidth}
\begin{tabular}{lrccccccccccc}
\toprule
$F$ & $\delta$ & VA & VA$\cap$KL & VA-iid & KL-Bonf & Hoeff-$k$ & Hoeff-cell & DKW-$4/\sqrt m$ & DKW-Massart & Berk--Jones & EB-DKW & $\chi^2$ \\
\midrule
F1 & 0.01 & 0.053 & 0.055 & 0.148 & 0.143 & 0.145 & 0.326 & 0.364 & 0.109 & 0.131 & 0.189 & 0.244 \\
F1 & 0.02 & 0.045 & 0.047 & 0.138 & 0.134 & 0.135 & 0.484 & 0.352 & 0.099 & 0.122 & 0.178 & 0.236 \\
F1 & 0.05 & 0.037 & 0.038 & 0.129 & 0.124 & 0.127 & 0.595 & 0.338 & 0.090 & 0.111 & 0.168 & 0.225 \\
F2 & 0.01 & 0.043 & 0.045 & 0.112 & 0.107 & 0.139 & 0.311 & 0.336 & 0.104 & 0.098 & 0.149 & 0.212 \\
F2 & 0.02 & 0.035 & 0.037 & 0.109 & 0.104 & 0.130 & 0.509 & 0.341 & 0.095 & 0.094 & 0.146 & 0.215 \\
F2 & 0.05 & 0.032 & 0.034 & 0.112 & 0.108 & 0.126 & 0.764 & 0.336 & 0.090 & 0.096 & 0.150 & 0.225 \\
F3 & 0.01 & 0.012 & 0.013 & 0.124 & 0.120 & 0.120 & 0.278 & 0.332 & 0.086 & 0.108 & 0.163 & 0.212 \\
F3 & 0.02 & 0.011 & 0.012 & 0.123 & 0.119 & 0.119 & 0.410 & 0.331 & 0.086 & 0.108 & 0.162 & 0.212 \\
F3 & 0.05 & 0.015 & 0.016 & 0.125 & 0.121 & 0.121 & 0.504 & 0.333 & 0.087 & 0.109 & 0.164 & 0.210 \\
F4 & 0.01 & 0.018 & 0.020 & 0.073 & 0.068 & 0.124 & 0.281 & 0.244 & 0.090 & 0.061 & 0.107 & 0.145 \\
F4 & 0.02 & 0.018 & 0.019 & 0.075 & 0.069 & 0.124 & 0.461 & 0.249 & 0.090 & 0.063 & 0.110 & 0.147 \\
F4 & 0.05 & 0.020 & 0.021 & 0.078 & 0.073 & 0.122 & 0.773 & 0.263 & 0.088 & 0.066 & 0.114 & 0.152 \\
F5 & 0.01 & 0.019 & 0.020 & 0.126 & 0.122 & 0.123 & 0.280 & 0.335 & 0.089 & 0.111 & 0.165 & 0.222 \\
F5 & 0.02 & 0.045 & 0.044 & 0.141 & 0.137 & 0.137 & 0.427 & 0.350 & 0.104 & 0.125 & 0.180 & 0.232 \\
F5 & 0.05 & 0.032 & 0.034 & 0.118 & 0.113 & 0.120 & 0.536 & 0.332 & 0.085 & 0.101 & 0.156 & 0.192 \\
\bottomrule
\end{tabular}
\end{adjustbox}
\end{table}

\begin{table}[tbp]
\centering
\scriptsize
\caption{E1 median width, $m=4000$, $n=32$, $\alpha=0.05$.}
\label{tab:e1-m4000-n32}
\begin{adjustbox}{max width=\textwidth}
\begin{tabular}{lrccccccccccc}
\toprule
$F$ & $\delta$ & VA & VA-iid & KL-Bonf & Hoeff-$k$ & Hoeff-cell & DKW-$4/\sqrt m$ & DKW-Massart & Berk--Jones & EB-DKW & $\chi^2$ & AMARI$^\dagger$ \\
\midrule
F1 & 0.01 & 0.294 & 0.459 & 0.454 & 0.459 & 0.360 & 0.630 & 0.384 & 0.432 & 0.507 & 0.271 & 0.203 \\
F1 & 0.02 & 0.157 & 0.256 & 0.253 & 0.256 & 0.203 & 0.499 & 0.210 & 0.241 & 0.286 & 0.173 & 0.124 \\
F1 & 0.05 & 0.090 & 0.142 & 0.141 & 0.143 & 0.152 & 0.277 & 0.120 & 0.134 & 0.157 & 0.129 & 0.066 \\
F2 & 0.01 & 0.237 & 0.270 & 0.269 & 0.273 & 0.266 & 0.328 & 0.262 & 0.266 & 0.275 & 0.236 & 0.157 \\
F2 & 0.02 & 0.143 & 0.228 & 0.225 & 0.244 & 0.193 & 0.328 & 0.199 & 0.213 & 0.256 & 0.148 & 0.100 \\
F2 & 0.05 & 0.081 & 0.124 & 0.123 & 0.132 & 0.141 & 0.266 & 0.110 & 0.117 & 0.138 & 0.110 & 0.058 \\
F3 & 0.01 & 0.113 & 0.305 & 0.300 & 0.301 & 0.211 & 0.596 & 0.231 & 0.280 & 0.350 & 0.127 & 0.124 \\
F3 & 0.02 & 0.065 & 0.181 & 0.178 & 0.179 & 0.135 & 0.410 & 0.137 & 0.166 & 0.206 & 0.098 & 0.076 \\
F3 & 0.05 & 0.039 & 0.102 & 0.100 & 0.101 & 0.107 & 0.236 & 0.078 & 0.094 & 0.116 & 0.089 & 0.047 \\
F4 & 0.01 & 0.127 & 0.148 & 0.147 & 0.156 & 0.156 & 0.209 & 0.147 & 0.145 & 0.152 & 0.138 & 0.095 \\
F4 & 0.02 & 0.081 & 0.145 & 0.143 & 0.156 & 0.147 & 0.209 & 0.146 & 0.134 & 0.152 & 0.091 & 0.062 \\
F4 & 0.05 & 0.046 & 0.082 & 0.080 & 0.105 & 0.111 & 0.218 & 0.083 & 0.076 & 0.095 & 0.075 & 0.043 \\
F5 & 0.01 & 0.363 & 0.508 & 0.503 & 0.522 & 0.414 & 0.714 & 0.443 & 0.481 & 0.558 & 0.312 & 0.227 \\
F5 & 0.02 & 0.184 & 0.269 & 0.266 & 0.278 & 0.223 & 0.517 & 0.230 & 0.254 & 0.298 & 0.192 & 0.138 \\
F5 & 0.05 & 0.091 & 0.146 & 0.145 & 0.152 & 0.156 & 0.284 & 0.127 & 0.138 & 0.162 & 0.114 & 0.064 \\
\bottomrule
\end{tabular}
\end{adjustbox}
\end{table}

\begin{table}[tbp]
\centering
\scriptsize
\caption{E1 median width, $m=4000$, $n=128$, $\alpha=0.05$.}
\label{tab:e1-m4000-n128}
\begin{adjustbox}{max width=\textwidth}
\begin{tabular}{lrccccccccccc}
\toprule
$F$ & $\delta$ & VA & VA-iid & KL-Bonf & Hoeff-$k$ & Hoeff-cell & DKW-$4/\sqrt m$ & DKW-Massart & Berk--Jones & EB-DKW & $\chi^2$ & AMARI$^\dagger$ \\
\midrule
F1 & 0.01 & 0.078 & 0.137 & 0.135 & 0.136 & 0.144 & 0.284 & 0.111 & 0.128 & 0.152 & 0.157 & 0.062 \\
F1 & 0.02 & 0.053 & 0.106 & 0.105 & 0.105 & 0.139 & 0.225 & 0.083 & 0.098 & 0.120 & 0.139 & 0.047 \\
F1 & 0.05 & 0.034 & 0.081 & 0.080 & 0.081 & 0.221 & 0.194 & 0.062 & 0.074 & 0.093 & 0.123 & 0.035 \\
F2 & 0.01 & 0.068 & 0.109 & 0.107 & 0.127 & 0.130 & 0.263 & 0.102 & 0.101 & 0.124 & 0.124 & 0.054 \\
F2 & 0.02 & 0.048 & 0.089 & 0.088 & 0.102 & 0.133 & 0.218 & 0.080 & 0.082 & 0.101 & 0.113 & 0.041 \\
F2 & 0.05 & 0.031 & 0.071 & 0.070 & 0.079 & 0.217 & 0.192 & 0.060 & 0.065 & 0.082 & 0.106 & 0.031 \\
F3 & 0.01 & 0.012 & 0.089 & 0.087 & 0.087 & 0.100 & 0.234 & 0.064 & 0.080 & 0.104 & 0.106 & 0.034 \\
F3 & 0.02 & 0.008 & 0.068 & 0.067 & 0.067 & 0.099 & 0.182 & 0.049 & 0.062 & 0.080 & 0.106 & 0.031 \\
F3 & 0.05 & 0.011 & 0.066 & 0.065 & 0.065 & 0.153 & 0.171 & 0.048 & 0.060 & 0.077 & 0.108 & 0.031 \\
F4 & 0.01 & 0.023 & 0.064 & 0.062 & 0.094 & 0.101 & 0.190 & 0.071 & 0.058 & 0.078 & 0.082 & 0.031 \\
F4 & 0.02 & 0.016 & 0.049 & 0.048 & 0.073 & 0.104 & 0.187 & 0.054 & 0.045 & 0.060 & 0.082 & 0.026 \\
F4 & 0.05 & 0.017 & 0.050 & 0.049 & 0.070 & 0.165 & 0.179 & 0.053 & 0.045 & 0.060 & 0.084 & 0.023 \\
F5 & 0.01 & 0.078 & 0.145 & 0.144 & 0.145 & 0.140 & 0.303 & 0.117 & 0.136 & 0.163 & 0.132 & 0.071 \\
F5 & 0.02 & 0.063 & 0.112 & 0.111 & 0.112 & 0.145 & 0.229 & 0.092 & 0.105 & 0.125 & 0.141 & 0.057 \\
F5 & 0.05 & 0.031 & 0.074 & 0.072 & 0.077 & 0.237 & 0.195 & 0.058 & 0.067 & 0.086 & 0.100 & 0.031 \\
\bottomrule
\end{tabular}
\end{adjustbox}
\end{table}

\begin{table}[tbp]
\centering
\scriptsize
\caption{E1 median width, $m=4000$, $n=512$, $\alpha=0.05$.}
\label{tab:e1-m4000-n512}
\begin{adjustbox}{max width=\textwidth}
\begin{tabular}{lrcccccccccc}
\toprule
$F$ & $\delta$ & VA & VA-iid & KL-Bonf & Hoeff-$k$ & Hoeff-cell & DKW-$4/\sqrt m$ & DKW-Massart & Berk--Jones & EB-DKW & $\chi^2$ \\
\midrule
F1 & 0.01 & 0.030 & 0.081 & 0.079 & 0.080 & 0.187 & 0.195 & 0.061 & 0.074 & 0.093 & 0.149 \\
F1 & 0.02 & 0.023 & 0.072 & 0.071 & 0.071 & 0.306 & 0.182 & 0.053 & 0.066 & 0.084 & 0.144 \\
F1 & 0.05 & 0.017 & 0.066 & 0.065 & 0.066 & 0.493 & 0.172 & 0.047 & 0.058 & 0.077 & 0.139 \\
F2 & 0.01 & 0.027 & 0.062 & 0.060 & 0.077 & 0.179 & 0.189 & 0.059 & 0.056 & 0.073 & 0.129 \\
F2 & 0.02 & 0.021 & 0.058 & 0.056 & 0.070 & 0.292 & 0.180 & 0.052 & 0.052 & 0.068 & 0.131 \\
F2 & 0.05 & 0.016 & 0.057 & 0.056 & 0.065 & 0.580 & 0.172 & 0.047 & 0.050 & 0.068 & 0.137 \\
F3 & 0.01 & 0.004 & 0.061 & 0.060 & 0.060 & 0.139 & 0.166 & 0.043 & 0.055 & 0.072 & 0.142 \\
F3 & 0.02 & 0.004 & 0.062 & 0.060 & 0.060 & 0.227 & 0.166 & 0.044 & 0.055 & 0.072 & 0.142 \\
F3 & 0.05 & 0.006 & 0.062 & 0.061 & 0.061 & 0.402 & 0.167 & 0.044 & 0.056 & 0.073 & 0.142 \\
F4 & 0.01 & 0.008 & 0.041 & 0.039 & 0.063 & 0.142 & 0.169 & 0.046 & 0.036 & 0.050 & 0.118 \\
F4 & 0.02 & 0.008 & 0.041 & 0.040 & 0.063 & 0.233 & 0.170 & 0.046 & 0.037 & 0.051 & 0.119 \\
F4 & 0.05 & 0.010 & 0.043 & 0.042 & 0.063 & 0.512 & 0.169 & 0.045 & 0.038 & 0.053 & 0.122 \\
F5 & 0.01 & 0.011 & 0.067 & 0.066 & 0.066 & 0.145 & 0.172 & 0.049 & 0.060 & 0.078 & 0.140 \\
F5 & 0.02 & 0.029 & 0.079 & 0.078 & 0.078 & 0.251 & 0.185 & 0.061 & 0.073 & 0.090 & 0.148 \\
F5 & 0.05 & 0.015 & 0.059 & 0.058 & 0.062 & 0.420 & 0.168 & 0.044 & 0.052 & 0.070 & 0.119 \\
\bottomrule
\end{tabular}
\end{adjustbox}
\end{table}

\begin{table}[tbp]
\centering
\scriptsize
\caption{E4: VA / best comparator width ratio, median (and maximum) over F1--F5.}
\label{tab:e4}
\begin{adjustbox}{max width=\textwidth}
\begin{tabular}{rrrcccc}
\toprule
$\delta$ & $m$ & $n$ & vs best F-loc. (median) & (max) & vs best incl.\ AMARI & vs best proven-valid \\
\midrule
0.01 & 250 & 32 & 0.903 & 1.004 & 1.363 & 0.841 \\
0.01 & 250 & 128 & 0.694 & 0.784 & 1.140 & 0.536 \\
0.01 & 250 & 512 & 0.333 & 0.536 & 0.333 & 0.295 \\
0.01 & 1000 & 32 & 0.998 & 1.157 & 1.186 & 0.838 \\
0.01 & 1000 & 128 & 0.621 & 0.723 & 1.132 & 0.558 \\
0.01 & 1000 & 512 & 0.299 & 0.485 & 0.299 & 0.272 \\
0.01 & 4000 & 32 & 1.002 & 1.164 & 1.445 & 0.865 \\
0.01 & 4000 & 128 & 0.668 & 0.700 & 1.106 & 0.561 \\
0.01 & 4000 & 512 & 0.231 & 0.493 & 0.231 & 0.196 \\
0.02 & 250 & 32 & 0.831 & 0.961 & 1.351 & 0.762 \\
0.02 & 250 & 128 & 0.589 & 0.623 & 0.926 & 0.458 \\
0.02 & 250 & 512 & 0.426 & 0.472 & 0.426 & 0.320 \\
0.02 & 1000 & 32 & 0.876 & 0.929 & 1.395 & 0.774 \\
0.02 & 1000 & 128 & 0.559 & 0.655 & 1.007 & 0.511 \\
0.02 & 1000 & 512 & 0.373 & 0.452 & 0.373 & 0.326 \\
0.02 & 4000 & 32 & 0.908 & 0.962 & 1.300 & 0.739 \\
0.02 & 4000 & 128 & 0.599 & 0.681 & 1.091 & 0.504 \\
0.02 & 4000 & 512 & 0.400 & 0.482 & 0.400 & 0.325 \\
0.05 & 250 & 32 & 0.700 & 0.707 & 1.199 & 0.597 \\
0.05 & 250 & 128 & 0.514 & 0.572 & 0.824 & 0.396 \\
0.05 & 250 & 512 & 0.396 & 0.417 & 0.396 & 0.291 \\
0.05 & 1000 & 32 & 0.689 & 0.771 & 1.234 & 0.604 \\
0.05 & 1000 & 128 & 0.497 & 0.566 & 0.949 & 0.427 \\
0.05 & 1000 & 512 & 0.362 & 0.406 & 0.362 & 0.284 \\
0.05 & 4000 & 32 & 0.731 & 0.798 & 1.370 & 0.630 \\
0.05 & 4000 & 128 & 0.516 & 0.553 & 0.970 & 0.430 \\
0.05 & 4000 & 512 & 0.350 & 0.372 & 0.350 & 0.266 \\
\bottomrule
\end{tabular}
\end{adjustbox}
\end{table}

\begin{table}[tbp]
\centering
\scriptsize
\caption{E5: band-event non-coverage by region, $R=10^5$ per cell.}
\label{tab:e5}
\begin{adjustbox}{max width=\textwidth}
\begin{tabular}{llrccccc}
\toprule
Region & laws & cells & max & mean & max CP upper & median rate$/\alpha$ & cells CP lower $>\alpha$ \\
\midrule
VA & F1--F5 & 45 & 0.00207 & 0.00048 & 0.00237 & 0.0062 & 0 \\
VA ($K_{\rm geo}$) & F1--F5 & 45 & 0.00207 & 0.00052 & 0.00237 & 0.0066 & 0 \\
VA ($K=[n]$) & F1--F5 & 45 & 0.00199 & 0.00031 & 0.00229 & 0.0018 & 0 \\
VA-Bennett & F1--F5 & 45 & 0.00216 & 0.00054 & 0.00247 & 0.0070 & 0 \\
VA$\cap$KL & F1--F5 & 45 & 0.00082 & 0.00020 & 0.00102 & 0.0022 & 0 \\
VA-iid & F1--F5 & 45 & 0.00003 & 0.00000 & 0.00009 & 0.0000 & 0 \\
KL-Bonf & F1--F5 & 45 & 0.00025 & 0.00002 & 0.00037 & 0.0000 & 0 \\
Hoeff-$k$ & F1--F5 & 45 & 0.00000 & 0.00000 & 0.00004 & 0.0000 & 0 \\
Hoeff-cell & F1--F5 & 45 & 0.00000 & 0.00000 & 0.00004 & 0.0000 & 0 \\
DKW-$4/\sqrt m$ & F1--F5 & 45 & 0.00000 & 0.00000 & 0.00004 & 0.0000 & 0 \\
DKW-Massart & F1--F5 & 45 & 0.00000 & 0.00000 & 0.00004 & 0.0000 & 0 \\
Berk--Jones & F1--F5 & 45 & 0.00202 & 0.00047 & 0.00232 & 0.0016 & 0 \\
EB-DKW & F1--F5 & 45 & 0.00000 & 0.00000 & 0.00004 & 0.0000 & 0 \\
$\chi^2$ & F1--F5 & 45 & 0.05917 & 0.01565 & 0.06065 & 0.2600 & 2 \\
\midrule
VA & A1--A2 & 18 & 0.00139 & 0.00036 & 0.00164 & 0.0057 & 0 \\
VA ($K_{\rm geo}$) & A1--A2 & 18 & 0.00139 & 0.00036 & 0.00164 & 0.0057 & 0 \\
VA ($K=[n]$) & A1--A2 & 18 & 0.00124 & 0.00040 & 0.00148 & 0.0069 & 0 \\
VA-Bennett & A1--A2 & 18 & 0.00200 & 0.00061 & 0.00230 & 0.0069 & 0 \\
VA$\cap$KL & A1--A2 & 18 & 0.00088 & 0.00028 & 0.00108 & 0.0039 & 0 \\
VA-iid & A1--A2 & 18 & 0.00053 & 0.00011 & 0.00069 & 0.0010 & 0 \\
KL-Bonf & A1--A2 & 18 & 0.00105 & 0.00031 & 0.00127 & 0.0027 & 0 \\
Hoeff-$k$ & A1--A2 & 18 & 0.00000 & 0.00000 & 0.00004 & 0.0000 & 0 \\
Hoeff-cell & A1--A2 & 18 & 0.00000 & 0.00000 & 0.00004 & 0.0000 & 0 \\
DKW-$4/\sqrt m$ & A1--A2 & 18 & 0.00000 & 0.00000 & 0.00004 & 0.0000 & 0 \\
DKW-Massart & A1--A2 & 18 & 0.00059 & 0.00023 & 0.00076 & 0.0031 & 0 \\
Berk--Jones & A1--A2 & 18 & 0.01010 & 0.00598 & 0.01074 & 0.1205 & 0 \\
EB-DKW & A1--A2 & 18 & 0.00000 & 0.00000 & 0.00004 & 0.0000 & 0 \\
$\chi^2$ & A1--A2 & 18 & 0.07930 & 0.04331 & 0.08099 & 0.7901 & 2 \\
\bottomrule
\end{tabular}
\end{adjustbox}
\end{table}

\begin{table}[tbp]
\centering
\small
\caption{E5: VA band-event non-coverage $1-\hat P(\mathcal E)$ by cell, $R=10^5$, rule-based $K$. A1: half the $p_i$ at 0, half at 0.5. A2: half at 0, half uniform on $(0.45,0.55)$.}
\label{tab:e5-va}
\begin{adjustbox}{max width=\textwidth}
\begin{tabular}{lrccc}
\toprule
Law & $m$ & $n=32$ & $n=128$ & $n=512$ \\
\midrule
A1 & 250 & 0.00049 & 0.00005 & 0.00005 \\
A1 & 1000 & 0.00061 & 0.00030 & 0.00003 \\
A1 & 4000 & 0.00139 & 0.00031 & 0.00008 \\
A2 & 250 & 0.00050 & 0.00011 & 0.00001 \\
A2 & 1000 & 0.00079 & 0.00027 & 0.00001 \\
A2 & 4000 & 0.00107 & 0.00038 & 0.00010 \\
F1 & 250 & 0.00008 & 0.00003 & 0.00000 \\
F1 & 1000 & 0.00046 & 0.00051 & 0.00012 \\
F1 & 4000 & 0.00130 & 0.00118 & 0.00061 \\
F2 & 250 & 0.00012 & 0.00006 & 0.00000 \\
F2 & 1000 & 0.00068 & 0.00026 & 0.00002 \\
F2 & 4000 & 0.00171 & 0.00129 & 0.00053 \\
F3 & 250 & 0.00010 & 0.00001 & 0.00000 \\
F3 & 1000 & 0.00053 & 0.00027 & 0.00002 \\
F3 & 4000 & 0.00151 & 0.00096 & 0.00022 \\
F4 & 250 & 0.00015 & 0.00002 & 0.00000 \\
F4 & 1000 & 0.00090 & 0.00033 & 0.00003 \\
F4 & 4000 & 0.00207 & 0.00130 & 0.00038 \\
F5 & 250 & 0.00007 & 0.00004 & 0.00006 \\
F5 & 1000 & 0.00031 & 0.00033 & 0.00031 \\
F5 & 4000 & 0.00101 & 0.00098 & 0.00058 \\
\bottomrule
\end{tabular}
\end{adjustbox}
\end{table}

\begin{table}[tbp]
\centering
\scriptsize
\caption{E6: VA width for each $K$ variant divided by the width with the rule-based $K$ (median over paired replicates; $m=1000$, $R=20$ at $n=128$ and $10$ at $n=512$; $K=[n]$: $R=10$ and $3$). Empty: no successful solve. \texttt{geo16} is the $K$ used in E1--E3.}
\label{tab:e6}
\begin{adjustbox}{max width=\textwidth}
\begin{tabular}{lrrcccccccc}
\toprule
$F$ & $n$ & $\delta$ & geo16 & rule & sqrt1 & sqrt2 & geo8 & dyadic & weighted & all \\
\midrule
F2 & 128 & 0.01 & 1.000 & -- & 1.000 & 0.987 & 0.985 & 0.930 & 0.986 & 1.070 \\
F2 & 128 & 0.02 & 1.000 & -- & 1.000 & 0.988 & 0.986 & 0.966 & 0.987 & 1.063 \\
F2 & 128 & 0.05 & 1.000 & -- & 1.000 & 0.984 & 0.981 & 1.045 & 0.983 & 1.087 \\
F2 & 512 & 0.01 & 0.952 & -- & 1.012 & 1.001 & 0.993 & 0.934 & 0.993 & 1.112 \\
F2 & 512 & 0.02 & 0.946 & -- & 1.014 & 1.001 & 0.992 & 1.061 & 0.992 & 1.150 \\
F2 & 512 & 0.05 & 0.993 & -- & 1.015 & 1.001 & 0.991 & 1.148 & 0.992 & -- \\
F3 & 128 & 0.01 & 1.000 & -- & 1.000 & 0.973 & 0.969 & 0.859 & 0.971 & 1.144 \\
F3 & 128 & 0.02 & 1.000 & -- & 1.000 & 0.974 & 0.969 & 0.911 & 0.972 & 1.142 \\
F3 & 128 & 0.05 & 1.000 & -- & 1.000 & 0.980 & 0.976 & 0.975 & 0.978 & 1.109 \\
F3 & 512 & 0.01 & 0.909 & -- & 1.024 & 1.002 & 0.986 & 0.811 & 0.987 & -- \\
F3 & 512 & 0.02 & 0.915 & -- & 1.023 & 1.002 & 0.987 & 0.867 & 0.987 & -- \\
F3 & 512 & 0.05 & 0.951 & -- & 1.018 & 1.001 & 0.989 & 1.059 & 0.990 & -- \\
F5 & 128 & 0.01 & 1.000 & -- & 1.000 & 0.986 & 0.983 & 0.985 & 0.985 & 1.077 \\
F5 & 128 & 0.02 & 1.000 & -- & 1.000 & 0.989 & 0.987 & 0.952 & 0.988 & 1.062 \\
F5 & 128 & 0.05 & 1.000 & -- & 1.000 & 0.982 & 0.979 & 1.047 & 0.980 & 1.099 \\
F5 & 512 & 0.01 & 0.925 & -- & 1.019 & 1.001 & 0.989 & 0.836 & 0.989 & 1.179 \\
F5 & 512 & 0.02 & 0.958 & -- & 1.011 & 1.001 & 0.994 & 0.934 & 0.994 & 1.099 \\
F5 & 512 & 0.05 & 0.980 & -- & 1.017 & 1.001 & 0.990 & 1.112 & 0.991 & -- \\
\midrule
\multicolumn{3}{l}{$|K|$ at $n=128$} & 31 & 31 & 31 & 24 & 23 & 8 & 31 & 128 \\
\multicolumn{3}{l}{$|K|$ at $n=512$} & 31 & 69 & 85 & 70 & 61 & 10 & 69 & 512 \\
\bottomrule
\end{tabular}
\end{adjustbox}
\end{table}

\begin{table}[tbp]
\centering
\small
\caption{Mean wall-clock seconds per replicate and region (three targets per solve), E1 under load (18 workers).}
\label{tab:runtime}
\begin{adjustbox}{max width=\textwidth}
\begin{tabular}{lccc}
\toprule
Region & $n=32$ & $n=128$ & $n=512$ \\
\midrule
VA & 12.48 & 32.40 & 32.20 \\
VA$\cap$KL & 12.66 & 34.34 & 42.15 \\
VA-Bennett & 14.67 & 38.91 & -- \\
VA-iid & 0.51 & 0.38 & 0.41 \\
KL-Bonf & 0.50 & 0.36 & 0.39 \\
Hoeff-$k$ & 0.41 & 0.33 & 0.38 \\
Hoeff-cell & 0.27 & 1.36 & 10.95 \\
DKW-$4/\sqrt m$ & 0.27 & 1.33 & 11.30 \\
DKW-Massart & 0.48 & 1.69 & 13.15 \\
Berk--Jones & 0.55 & 1.79 & 14.10 \\
EB-DKW & 0.39 & 0.32 & 0.36 \\
$\chi^2$ & 3.51 & 6.89 & 28.65 \\
AMARI$^\dagger$ & 3.30 & 11.87 & -- \\
\bottomrule
\end{tabular}
\end{adjustbox}
\end{table}

\clearpage
{\tiny
\setlength{\tabcolsep}{3pt}
\begin{longtable}{lrrllccccccc}
\caption{Width by grid on one realised data set per case. \texttt{atom} inserts the true $p_i$ as nodes (oracle); \texttt{env\_}$\eta$ is the data-independent envelope grid with tolerance $\eta$; \texttt{coarse} is a 2001+200-node grid with envelopes or as point values (no guarantee).}\label{tab:grid}\\
\toprule
$F$ & $m$ & $n$ & Region & Target & atom\_0.001 & env\_0.0005 & env\_0.001 & env\_0.003 & env\_0.01 & coarse\_env & coarse\_point \\
\midrule
\endfirsthead
\toprule
$F$ & $m$ & $n$ & Region & Target & atom\_0.001 & env\_0.0005 & env\_0.001 & env\_0.003 & env\_0.01 & coarse\_env & coarse\_point \\
\midrule
\endhead
\bottomrule
\endlastfoot
F1 & 1000 & 32 & DKW-Massart & $\theta_{0.01}$ & 0.5834 & 0.5837 & 0.5840 & 0.5853 & 0.5892 & 0.5961 & 0.5834 \\
F1 & 1000 & 32 & DKW-Massart & $\theta_{0.02}$ & 0.3311 & 0.3314 & 0.3316 & 0.3322 & 0.3345 & 0.3421 & 0.3311 \\
F1 & 1000 & 32 & DKW-Massart & $\theta_{0.05}$ & 0.1806 & 0.1807 & 0.1808 & 0.1814 & 0.1830 & 0.1844 & 0.1806 \\
F1 & 1000 & 32 & KL-Bonf & $\theta_{0.01}$ & 0.6145 & 0.6145 & 0.6146 & 0.6146 & 0.6146 & 0.6147 & 0.6145 \\
F1 & 1000 & 32 & KL-Bonf & $\theta_{0.02}$ & 0.4015 & 0.4017 & 0.4019 & 0.4025 & 0.4047 & 0.4121 & 0.4015 \\
F1 & 1000 & 32 & KL-Bonf & $\theta_{0.05}$ & 0.2185 & 0.2186 & 0.2187 & 0.2193 & 0.2210 & 0.2223 & 0.2185 \\
F1 & 1000 & 32 & VA & $\theta_{0.01}$ & 0.4750 & 0.4754 & 0.4758 & 0.4774 & 0.4825 & 0.4978 & 0.4750 \\
F1 & 1000 & 32 & VA & $\theta_{0.02}$ & 0.2615 & 0.2618 & 0.2622 & 0.2634 & 0.2670 & 0.2716 & 0.2615 \\
F1 & 1000 & 32 & VA & $\theta_{0.05}$ & 0.1337 & 0.1338 & 0.1340 & 0.1346 & 0.1365 & 0.1372 & 0.1337 \\
F1 & 1000 & 128 & DKW-Massart & $\theta_{0.01}$ & 0.1808 & 0.1811 & 0.1814 & 0.1827 & 0.1869 & 0.1862 & 0.1808 \\
F1 & 1000 & 128 & DKW-Massart & $\theta_{0.02}$ & 0.1419 & 0.1421 & 0.1423 & 0.1430 & 0.1444 & 0.1486 & 0.1419 \\
F1 & 1000 & 128 & DKW-Massart & $\theta_{0.05}$ & 0.1139 & 0.1140 & 0.1141 & 0.1147 & 0.1165 & 0.1172 & 0.1139 \\
F1 & 1000 & 128 & KL-Bonf & $\theta_{0.01}$ & 0.2259 & 0.2263 & 0.2266 & 0.2279 & 0.2322 & 0.2317 & 0.2259 \\
F1 & 1000 & 128 & KL-Bonf & $\theta_{0.02}$ & 0.1834 & 0.1836 & 0.1838 & 0.1843 & 0.1856 & 0.1905 & 0.1834 \\
F1 & 1000 & 128 & KL-Bonf & $\theta_{0.05}$ & 0.1474 & 0.1475 & 0.1476 & 0.1482 & 0.1500 & 0.1504 & 0.1474 \\
F1 & 1000 & 128 & VA & $\theta_{0.01}$ & 0.1298 & 0.1301 & 0.1304 & 0.1316 & 0.1349 & 0.1353 & 0.1298 \\
F1 & 1000 & 128 & VA & $\theta_{0.02}$ & 0.0913 & 0.0916 & 0.0918 & 0.0927 & 0.0949 & 0.0974 & 0.0913 \\
F1 & 1000 & 128 & VA & $\theta_{0.05}$ & 0.0660 & 0.0662 & 0.0663 & 0.0669 & 0.0687 & 0.0698 & 0.0660 \\
F1 & 1000 & 512 & DKW-Massart & $\theta_{0.01}$ & 0.1109 & 0.1110 & 0.1111 & 0.1115 & 0.1128 & 0.1150 & 0.1109 \\
F1 & 1000 & 512 & DKW-Massart & $\theta_{0.02}$ & 0.0989 & 0.0990 & 0.0991 & 0.0995 & 0.1005 & 0.1038 & 0.0989 \\
F1 & 1000 & 512 & DKW-Massart & $\theta_{0.05}$ & 0.0888 & 0.0888 & 0.0889 & 0.0892 & 0.0900 & 0.0929 & 0.0888 \\
F1 & 1000 & 512 & KL-Bonf & $\theta_{0.01}$ & 0.1463 & 0.1464 & 0.1466 & 0.1471 & 0.1483 & 0.1497 & 0.1463 \\
F1 & 1000 & 512 & KL-Bonf & $\theta_{0.02}$ & 0.1329 & 0.1330 & 0.1331 & 0.1335 & 0.1345 & 0.1380 & 0.1329 \\
F1 & 1000 & 512 & KL-Bonf & $\theta_{0.05}$ & 0.1251 & 0.1252 & 0.1253 & 0.1256 & 0.1266 & 0.1295 & 0.1251 \\
F1 & 1000 & 512 & VA & $\theta_{0.01}$ & 0.0542 & 0.0544 & 0.0545 & 0.0550 & 0.0567 & 0.0586 & 0.0542 \\
F1 & 1000 & 512 & VA & $\theta_{0.02}$ & 0.0452 & 0.0453 & 0.0455 & 0.0459 & 0.0473 & 0.0510 & 0.0452 \\
F1 & 1000 & 512 & VA & $\theta_{0.05}$ & 0.0365 & 0.0366 & 0.0367 & 0.0370 & 0.0382 & 0.0411 & 0.0365 \\
F2 & 1000 & 128 & DKW-Massart & $\theta_{0.01}$ & 0.1635 & 0.1637 & 0.1638 & 0.1646 & 0.1671 & 0.1678 & 0.1635 \\
F2 & 1000 & 128 & DKW-Massart & $\theta_{0.02}$ & 0.1356 & 0.1358 & 0.1359 & 0.1362 & 0.1372 & 0.1401 & 0.1356 \\
F2 & 1000 & 128 & DKW-Massart & $\theta_{0.05}$ & 0.1058 & 0.1058 & 0.1059 & 0.1063 & 0.1076 & 0.1082 & 0.1058 \\
F2 & 1000 & 128 & KL-Bonf & $\theta_{0.01}$ & 0.1712 & 0.1714 & 0.1715 & 0.1723 & 0.1749 & 0.1758 & 0.1712 \\
F2 & 1000 & 128 & KL-Bonf & $\theta_{0.02}$ & 0.1438 & 0.1438 & 0.1439 & 0.1441 & 0.1447 & 0.1473 & 0.1438 \\
F2 & 1000 & 128 & KL-Bonf & $\theta_{0.05}$ & 0.1215 & 0.1216 & 0.1217 & 0.1221 & 0.1234 & 0.1239 & 0.1215 \\
F2 & 1000 & 128 & VA & $\theta_{0.01}$ & 0.1054 & 0.1056 & 0.1057 & 0.1064 & 0.1086 & 0.1096 & 0.1054 \\
F2 & 1000 & 128 & VA & $\theta_{0.02}$ & 0.0789 & 0.0791 & 0.0792 & 0.0797 & 0.0811 & 0.0837 & 0.0789 \\
F2 & 1000 & 128 & VA & $\theta_{0.05}$ & 0.0544 & 0.0545 & 0.0546 & 0.0550 & 0.0564 & 0.0573 & 0.0544 \\
F3 & 4000 & 512 & DKW-Massart & $\theta_{0.01}$ & 0.0432 & 0.0433 & 0.0433 & 0.0433 & 0.0433 & 0.0438 & 0.0432 \\
F3 & 4000 & 512 & DKW-Massart & $\theta_{0.02}$ & 0.0437 & 0.0437 & 0.0437 & 0.0437 & 0.0438 & 0.0441 & 0.0437 \\
F3 & 4000 & 512 & DKW-Massart & $\theta_{0.05}$ & 0.0443 & 0.0443 & 0.0443 & 0.0443 & 0.0444 & 0.0446 & 0.0443 \\
F3 & 4000 & 512 & KL-Bonf & $\theta_{0.01}$ & 0.0600 & 0.0600 & 0.0600 & 0.0600 & 0.0601 & 0.0605 & 0.0600 \\
F3 & 4000 & 512 & KL-Bonf & $\theta_{0.02}$ & 0.0605 & 0.0605 & 0.0605 & 0.0605 & 0.0606 & 0.0609 & 0.0605 \\
F3 & 4000 & 512 & KL-Bonf & $\theta_{0.05}$ & 0.0614 & 0.0614 & 0.0614 & 0.0614 & 0.0615 & 0.0618 & 0.0614 \\
F3 & 4000 & 512 & VA & $\theta_{0.01}$ & 0.0035 & 0.0038 & 0.0038 & 0.0039 & 0.0039 & 0.0050 & 0.0035 \\
F3 & 4000 & 512 & VA & $\theta_{0.02}$ & 0.0038 & 0.0039 & 0.0039 & 0.0040 & 0.0040 & 0.0046 & 0.0038 \\
F3 & 4000 & 512 & VA & $\theta_{0.05}$ & 0.0060 & 0.0061 & 0.0061 & 0.0061 & 0.0062 & 0.0066 & 0.0060 \\
\end{longtable}
}

\begin{figure}[tbp]
\centering
\includegraphics[width=\textwidth]{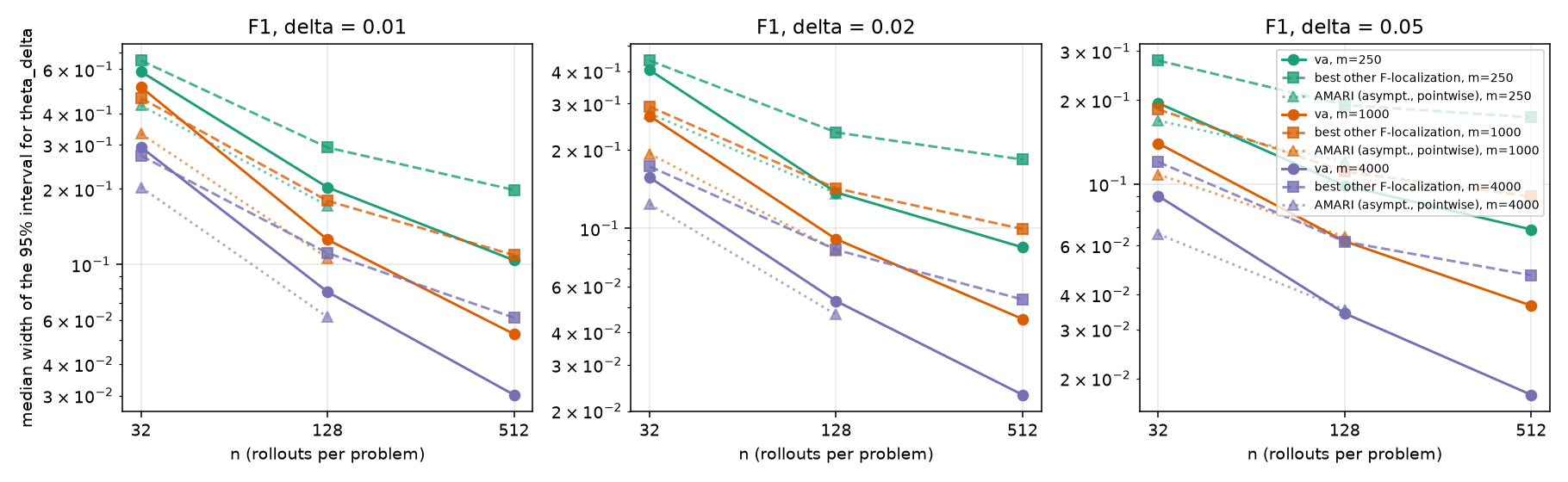}\\
\includegraphics[width=\textwidth]{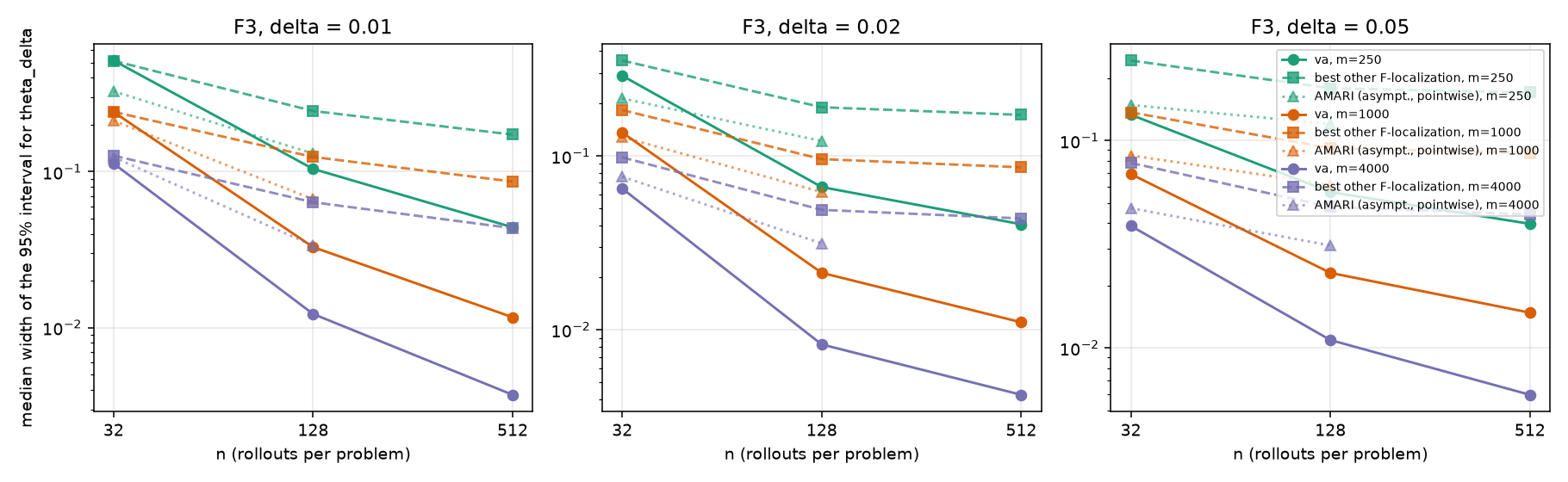}
\caption{Preliminary run. Median width of the 95\% interval for $\theta_\delta$ against $n$ for F1 (top) and F3 (bottom): VA (solid), best other localization among those of that run, which did not include the two-stage and exact-binomial bands (dashed), and AMARI (dotted, model-based variant, $n\le128$).}
\label{fig:width-F1F3}
\end{figure}
\begin{figure}[tbp]
\centering
\includegraphics[width=\textwidth]{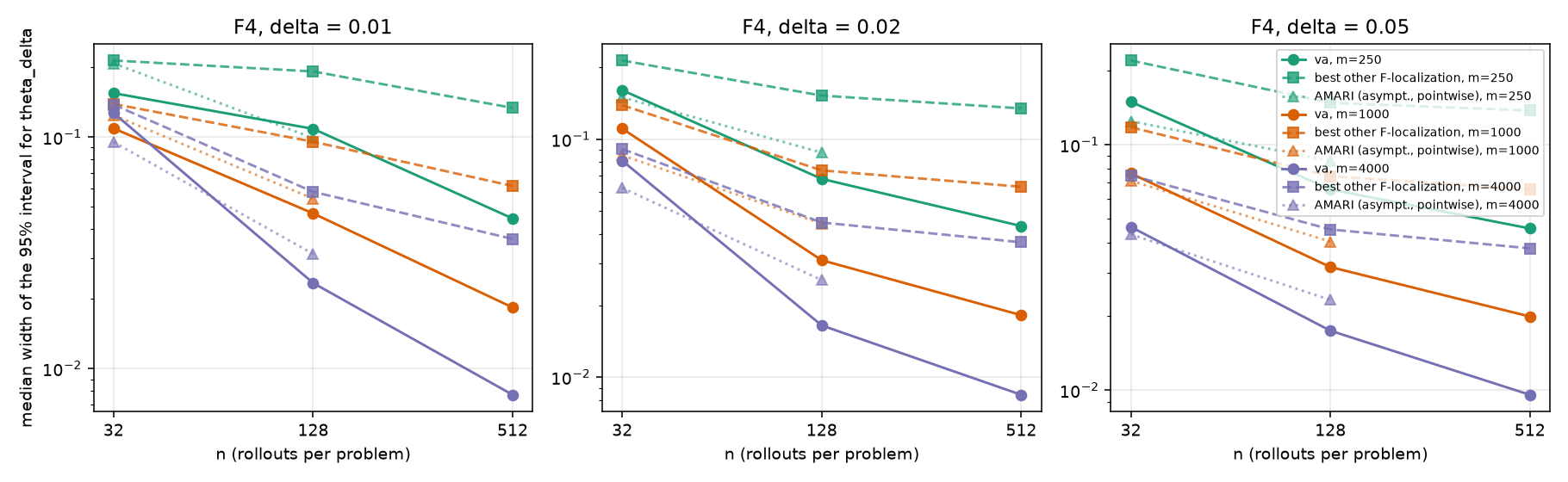}\\
\includegraphics[width=\textwidth]{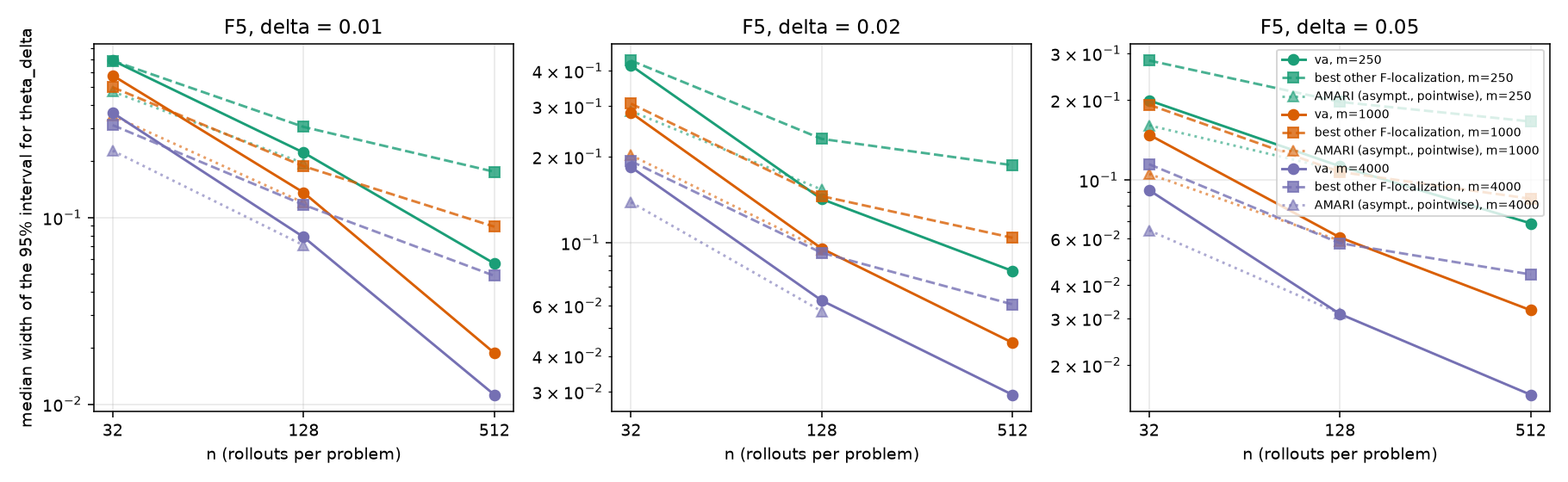}
\caption{As Figure~\ref{fig:width-F1F3}, for F4 (top) and F5 (bottom).}
\label{fig:width-F4F5}
\end{figure}
\begin{figure}[tbp]
\centering
\includegraphics[width=\textwidth]{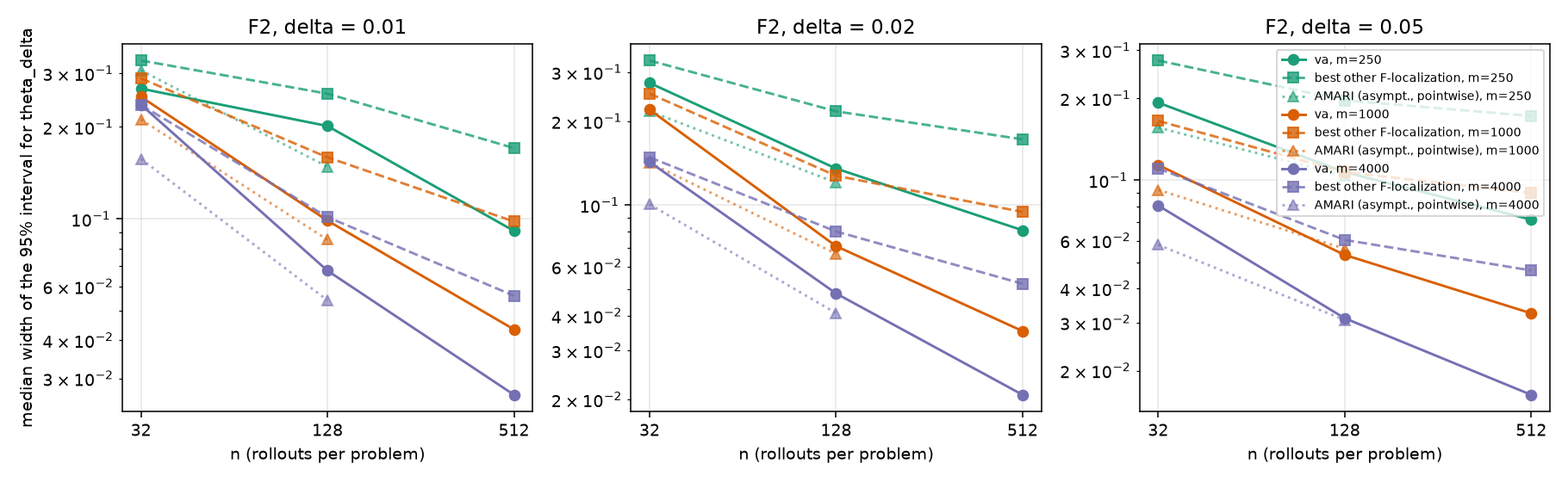}\\
\includegraphics[width=\textwidth]{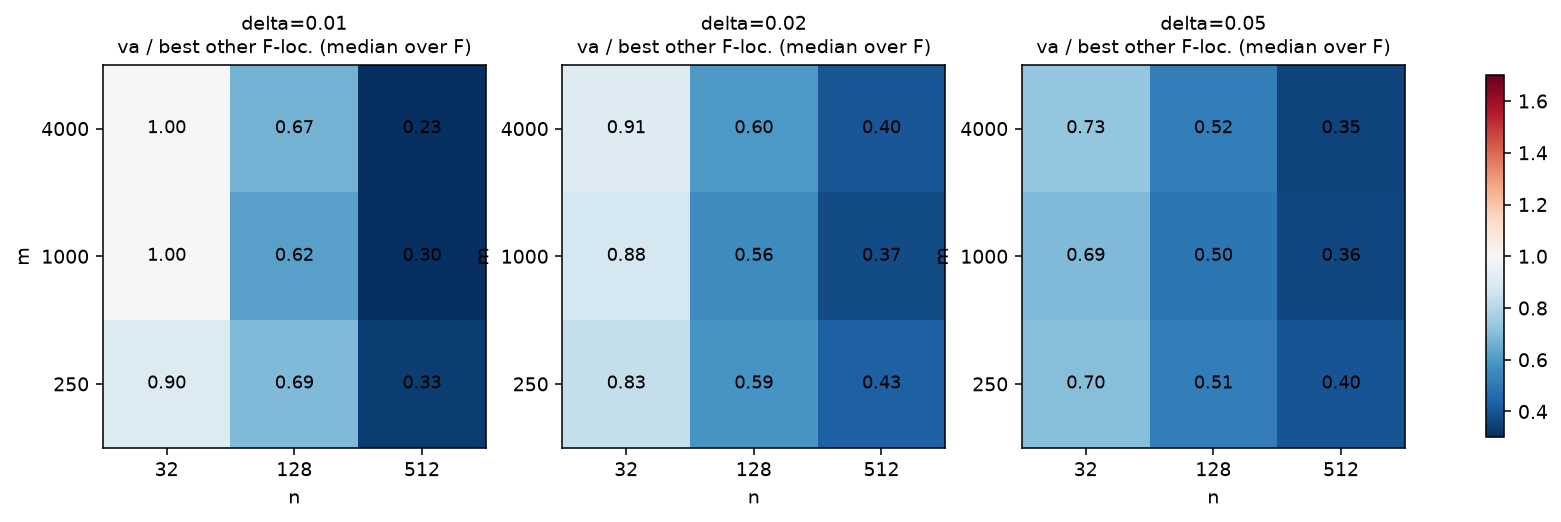}
\caption{Preliminary run. Top: as Figure~\ref{fig:width-F1F3}, for F2. Bottom: median over F1--F5 of the width ratio of VA to the best other localization of that run, on the $(m,n)$ grid, one panel per $\delta$.}
\label{fig:width-F2}
\end{figure}
\clearpage

\subsection{Monkey Business (E2)}
The pooled VA widths of the main run equal those of the preliminary with-replacement run.
\begin{table}[tbp]
\centering
\footnotesize
\caption{Monkey Business, pooled data ($m=1022$ problem--model units), median width of the 95\% interval for $\theta_\delta(\hat p)$; rollouts subsampled without replacement. Index set $K_{\rm geo}$ ($|K|=30$--$31$). $^\dagger$AMARI, model-based SE variant, shown for comparison. AMARI was run at $n\le 128$ only.}
\label{tab:mb-wor}
\begin{adjustbox}{max width=\textwidth}
\begin{tabular}{rrccccccccc}
\toprule
$n$ & $\delta$ & VA & VA-iid & KL-Bonf & Hoeff-$k$ & DKW-Massart & Berk--Jones & EB-DKW & $\chi^2$ & AMARI$^\dagger$ \\
\midrule
32 & 0.01 & 0.283 & 0.339 & 0.336 & 0.340 & 0.321 & 0.330 & 0.355 & 0.312 & 0.248 \\
32 & 0.02 & 0.276 & 0.339 & 0.336 & 0.340 & 0.321 & 0.330 & 0.355 & 0.300 & 0.165 \\
32 & 0.05 & 0.147 & 0.221 & 0.216 & 0.230 & 0.188 & 0.203 & 0.269 & 0.216 & 0.104 \\
128 & 0.01 & 0.138 & 0.200 & 0.193 & 0.230 & 0.184 & 0.180 & 0.245 & 0.230 & 0.103 \\
128 & 0.02 & 0.100 & 0.164 & 0.160 & 0.187 & 0.148 & 0.150 & 0.204 & 0.209 & 0.078 \\
128 & 0.05 & 0.067 & 0.139 & 0.134 & 0.148 & 0.112 & 0.123 & 0.179 & 0.200 & 0.059 \\
512 & 0.01 & 0.057 & 0.116 & 0.111 & 0.145 & 0.110 & 0.102 & 0.153 & 0.197 & -- \\
512 & 0.02 & 0.048 & 0.114 & 0.109 & 0.135 & 0.100 & 0.100 & 0.152 & 0.203 & -- \\
512 & 0.05 & 0.039 & 0.117 & 0.112 & 0.126 & 0.090 & 0.100 & 0.155 & 0.215 & -- \\
\bottomrule
\end{tabular}
\end{adjustbox}
\end{table}

\begin{table}[tbp]
\centering
\small
\caption{Monkey Business: median width with replacement / without replacement, per (configuration, $n$, $\delta$) cell, and minimum coverage under each scheme. Both runs use $K_{\rm geo}$. $^\dagger$AMARI, model-based SE variant.}
\label{tab:mb-wr-wor}
\begin{adjustbox}{max width=\textwidth}
\begin{tabular}{lcccccc}
\toprule
Region & cells & median & min & max & min cov. (with) & min cov. (without) \\
\midrule
VA & 81 & 1.001 & 0.937 & 1.122 & 1.000 & 1.000 \\
VA-iid & 81 & 1.001 & 0.961 & 1.077 & 1.000 & 1.000 \\
KL-Bonf & 81 & 1.001 & 0.901 & 1.203 & 1.000 & 1.000 \\
Hoeff-$k$ & 81 & 1.002 & 0.977 & 1.062 & 1.000 & 1.000 \\
Hoeff-cell & 81 & 1.000 & 0.970 & 1.042 & 1.000 & 1.000 \\
DKW-$4/\sqrt m$ & 81 & 1.001 & 0.987 & 1.028 & 1.000 & 1.000 \\
DKW-Massart & 81 & 1.002 & 0.970 & 1.076 & 1.000 & 1.000 \\
Berk--Jones & 81 & 1.000 & 0.887 & 1.259 & 1.000 & 1.000 \\
EB-DKW & 81 & 1.002 & 0.947 & 1.114 & 1.000 & 1.000 \\
$\chi^2$ & 81 & 0.993 & 0.946 & 1.055 & 1.000 & 1.000 \\
AMARI$^\dagger$ & 54 & 1.004 & 0.771 & 1.150 & 0.700 & 0.800 \\
\bottomrule
\end{tabular}
\end{adjustbox}
\end{table}

{\scriptsize
\setlength{\tabcolsep}{3pt}
\begin{longtable}{lrcccccc}
\caption{Monkey Business with replacement: median width with the rule $K$ ($|K|=69$ at $n=512$) / with $K_{\rm geo}$ ($|K|=31$), per (configuration, $n$, $\delta$) cell. At $n\le128$ the two index sets coincide. $^\dagger$AMARI, model-based SE variant.}\label{tab:mb-wr-K}\\
\toprule
Region & $n$ & cells & median & min & max & min cov. (rule $K$) & min cov. ($K_{\rm geo}$) \\
\midrule
\endfirsthead
\toprule
Region & $n$ & cells & median & min & max & min cov. (rule $K$) & min cov. ($K_{\rm geo}$) \\
\midrule
\endhead
\bottomrule
\endlastfoot
VA & 32 & 27 & 1.000 & 1.000 & 1.000 & 1.000 & 1.000 \\
VA & 128 & 27 & 1.000 & 1.000 & 1.000 & 1.000 & 1.000 \\
VA & 512 & 27 & 1.062 & 1.005 & 1.083 & 1.000 & 1.000 \\
VA-iid & 32 & 27 & 1.000 & 1.000 & 1.000 & 1.000 & 1.000 \\
VA-iid & 128 & 27 & 1.000 & 1.000 & 1.000 & 1.000 & 1.000 \\
VA-iid & 512 & 27 & 1.051 & 1.033 & 1.100 & 1.000 & 1.000 \\
KL-Bonf & 32 & 27 & 1.000 & 1.000 & 1.000 & 1.000 & 1.000 \\
KL-Bonf & 128 & 27 & 1.000 & 1.000 & 1.000 & 1.000 & 1.000 \\
KL-Bonf & 512 & 27 & 1.049 & 1.030 & 1.109 & 1.000 & 1.000 \\
Hoeff-$k$ & 32 & 27 & 1.000 & 1.000 & 1.000 & 1.000 & 1.000 \\
Hoeff-$k$ & 128 & 27 & 1.000 & 1.000 & 1.000 & 1.000 & 1.000 \\
Hoeff-$k$ & 512 & 27 & 1.048 & 1.026 & 1.055 & 1.000 & 1.000 \\
Hoeff-cell & 32 & 27 & 1.000 & 1.000 & 1.000 & 1.000 & 1.000 \\
Hoeff-cell & 128 & 27 & 1.000 & 1.000 & 1.000 & 1.000 & 1.000 \\
Hoeff-cell & 512 & 27 & 1.000 & 1.000 & 1.000 & 1.000 & 1.000 \\
DKW-$4/\sqrt m$ & 32 & 27 & 1.000 & 1.000 & 1.000 & 1.000 & 1.000 \\
DKW-$4/\sqrt m$ & 128 & 27 & 1.000 & 1.000 & 1.000 & 1.000 & 1.000 \\
DKW-$4/\sqrt m$ & 512 & 27 & 1.000 & 1.000 & 1.000 & 1.000 & 1.000 \\
DKW-Massart & 32 & 27 & 1.000 & 1.000 & 1.000 & 1.000 & 1.000 \\
DKW-Massart & 128 & 27 & 1.000 & 1.000 & 1.000 & 1.000 & 1.000 \\
DKW-Massart & 512 & 27 & 1.000 & 1.000 & 1.000 & 1.000 & 1.000 \\
Berk--Jones & 32 & 27 & 1.000 & 1.000 & 1.000 & 1.000 & 1.000 \\
Berk--Jones & 128 & 27 & 1.000 & 1.000 & 1.000 & 1.000 & 1.000 \\
Berk--Jones & 512 & 27 & 1.000 & 1.000 & 1.000 & 1.000 & 1.000 \\
EB-DKW & 32 & 27 & 1.000 & 1.000 & 1.000 & 1.000 & 1.000 \\
EB-DKW & 128 & 27 & 1.000 & 1.000 & 1.000 & 1.000 & 1.000 \\
EB-DKW & 512 & 27 & 1.058 & 1.037 & 1.102 & 1.000 & 1.000 \\
$\chi^2$ & 32 & 27 & 1.000 & 1.000 & 1.000 & 1.000 & 1.000 \\
$\chi^2$ & 128 & 27 & 1.000 & 1.000 & 1.000 & 1.000 & 1.000 \\
$\chi^2$ & 512 & 27 & 1.000 & 1.000 & 1.000 & 1.000 & 1.000 \\
AMARI$^\dagger$ & 32 & 27 & 1.000 & 1.000 & 1.000 & 0.900 & 0.900 \\
AMARI$^\dagger$ & 128 & 27 & 1.000 & 1.000 & 1.000 & 0.700 & 0.700 \\
\end{longtable}
}

\section{Identification constant, design rule and separated-law checks}\label{supp:spike}

\begin{table}[tbp]
\centering
\small
\caption{Certified constant $c_n(\delta)=2E_n(h_\delta)\,n\sqrt\delta/\sqrt{1-\delta}$: all 40 cases computed (check CHK-7, script \texttt{checks/chk7\_cn\_brackets.py}). Upper end: the maximum error of a Chebyshev-basis minimax-LP polynomial on a fine grid plus a rigorous curvature remainder and a floating-point allowance; lower end: de la Vall\'ee Poussin alternation bound. Brackets are rounded outward; $c_n/\beta$ is the lower end. $\beta=0.2801694990\ldots$ is Bernstein's constant.}
\label{tab:cn}
\begin{adjustbox}{max width=\textwidth}
\begin{tabular}{rrrcc}
\toprule
$n$ & $\delta$ & $n\sqrt\delta$ & $c_n$ bracket & $c_n/\beta$ \\
\midrule
16 & 0.002 & 0.72 & [0.44584, 0.44587] & 1.5913 \\
16 & 0.005 & 1.13 & [0.38613, 0.38615] & 1.3782 \\
16 & 0.01 & 1.60 & [0.27365, 0.27367] & 0.9767 \\
16 & 0.02 & 2.26 & [0.34194, 0.34196] & 1.2204 \\
16 & 0.05 & 3.58 & [0.31000, 0.31002] & 1.1065 \\
16 & 0.1 & 5.06 & [0.30121, 0.30123] & 1.0751 \\
16 & 0.25 & 8.00 & [0.29755, 0.29757] & 1.0620 \\
16 & 0.5 & 11.31 & [0.27948, 0.27950] & 0.9975 \\
32 & 0.002 & 1.43 & [0.30675, 0.30680] & 1.0948 \\
32 & 0.005 & 2.26 & [0.34117, 0.34123] & 1.2177 \\
32 & 0.01 & 3.20 & [0.27882, 0.27887] & 0.9952 \\
32 & 0.02 & 4.53 & [0.29069, 0.29073] & 1.0375 \\
32 & 0.05 & 7.16 & [0.30001, 0.30006] & 1.0708 \\
32 & 0.1 & 10.12 & [0.29486, 0.29490] & 1.0524 \\
32 & 0.25 & 16.00 & [0.28901, 0.28905] & 1.0315 \\
32 & 0.5 & 22.63 & [0.27999, 0.28004] & 0.9993 \\
64 & 0.002 & 2.86 & [0.30718, 0.30736] & 1.0964 \\
64 & 0.005 & 4.53 & [0.29139, 0.29156] & 1.0400 \\
64 & 0.01 & 6.40 & [0.28393, 0.28409] & 1.0134 \\
64 & 0.02 & 9.05 & [0.29027, 0.29044] & 1.0360 \\
64 & 0.05 & 14.31 & [0.28550, 0.28567] & 1.0190 \\
64 & 0.1 & 20.24 & [0.28244, 0.28259] & 1.0081 \\
64 & 0.25 & 32.00 & [0.28453, 0.28469] & 1.0155 \\
64 & 0.5 & 45.25 & [0.28012, 0.28028] & 0.9998 \\
128 & 0.002 & 5.72 & [0.30309, 0.30378] & 1.0818 \\
128 & 0.005 & 9.05 & [0.29075, 0.29141] & 1.0377 \\
128 & 0.01 & 12.80 & [0.28516, 0.28581] & 1.0178 \\
128 & 0.02 & 18.10 & [0.28791, 0.28856] & 1.0276 \\
128 & 0.05 & 28.62 & [0.28480, 0.28543] & 1.0165 \\
128 & 0.1 & 40.48 & [0.28242, 0.28305] & 1.0080 \\
128 & 0.25 & 64.00 & [0.28233, 0.28294] & 1.0077 \\
128 & 0.5 & 90.51 & [0.28014, 0.28076] & 0.9999 \\
256 & 0.002 & 11.45 & [0.28966, 0.29057] & 1.0339 \\
256 & 0.005 & 18.10 & [0.28796, 0.28886] & 1.0278 \\
256 & 0.01 & 25.60 & [0.28479, 0.28566] & 1.0165 \\
256 & 0.02 & 36.20 & [0.28159, 0.28249] & 1.0050 \\
256 & 0.05 & 57.24 & [0.28189, 0.28277] & 1.0061 \\
256 & 0.1 & 80.95 & [0.28192, 0.28279] & 1.0062 \\
256 & 0.25 & 128.00 & [0.28123, 0.28210] & 1.0038 \\
256 & 0.5 & 181.02 & [0.28014, 0.28101] & 0.9999 \\
\bottomrule
\end{tabular}
\end{adjustbox}
\end{table}

Table~\ref{tab:cn} reports $c_n(\delta)$ from rigorous brackets: the upper end is the error of a Chebyshev-basis LP polynomial on a fine grid plus a curvature remainder and a floating-point allowance, and the lower end is the de la Vall\'ee Poussin alternation bound (CHK-7). The values at $n=128$, $\delta=0.01$ give $2E_n\in[0.022167,0.022217]$. The convergence towards $\beta$ is slow and not monotone in $n\sqrt\delta$. We know of no verified source for the limit $c_n\to\beta$ for the shifted kink $\min(p/\delta,1)$ (Remark~\ref{M-rem:cn}).

\subsection{A design rule}\label{supp:design}
To certify a paired difference $\epsilon$ in $\theta_\delta$ from two marginal regions, the condition $2\cdot2E_n(h_\delta)\le\epsilon$ is sufficient for the identification part (Theorem~\ref{M-thm:id}(i), $K=\{1,\dots,n\}$). It is also a worst-case benchmark: by Theorem~\ref{M-thm:id}(ii), as $m\to\infty$, every procedure for one marginal has width close to $2E_n(h_\delta)$ or more at some $F$. It is not a necessary condition at the user's two laws, nor for joint procedures.

\begin{corollary}[Identification part of the design rule]\label{cor:design}
With $K=\{1,\dots,n\}$, the worst-case identification width $2E_n(h_\delta)$ of each marginal is at most $\epsilon/2$ if and only if $n\ge2c_n(\delta)\sqrt{1-\delta}/(\epsilon\sqrt\delta)$, with $c_n$ as in Remark~\ref{M-rem:cn}.
\end{corollary}

Since $c_n$ depends on $n$, the rule is solved with LP values of $E_n$, not with $\beta$: in Table~\ref{tab:cn}, $c_n/\beta$ lies in $[0.976,1.221]$ (rounded outward) for $n\sqrt\delta\in[1.6,6]$ and reaches 1.591 at $n\sqrt\delta=0.72$. The estimation part has only the upper bound of Corollary~\ref{M-cor:oracle}, which can be evaluated cheaply by an oracle-centred VA LP at a pilot $F$, that is, the VA LP with $T_k$ replaced by $\mu_k(F)$.

\emph{Worked example.} Take the two models of Section~\ref{M-sec:pair}, with the plug-in laws $c_i/128$ as pilots, and ask which budget would certify a difference of $0.025$ in $\theta_{0.01}$ by the Bonferroni difference of two VA marginals at level $\alpha/2$ each. This needs $e_{\rm lo}+e_{\rm hi}<0.025$, where $e_{\rm lo}$ is the distance from the pilot value to the lower endpoint for one model and $e_{\rm hi}$ the distance to the upper endpoint for the other. At the observed design, $m=1060$ and $n=128$, the sum is $0.0587$, so the plug-in difference of $0.0186$ is not certified. On a grid of $m\in\{1060,\dots,16960\}$ and $n\in\{128,\dots,1024\}$ (Table~\ref{tab:s4-design}), the smallest budgets that certify $0.025$ use $mn=542{,}720$ rollouts, four times the observed budget: $(m,n)=(1060,512)$ with sum $0.0237$, or $(2120,256)$ with sum $0.0223$. Keeping $n=128$ needs $m=16{,}960$, sixteen times as many problems. The oracle-centred width is a worst-case-benchmark proxy, not a guarantee for the data that will be observed.

\begin{table}[tbp]
\centering\small
\caption{Approximation error $2E_K(B)$, the best $2\|h_\delta-c-\lambda\cdot P\|_\infty$ subject to $\|\lambda\|_1\le B$, at $n=128$, $\delta=0.01$ and the 31-element $K$. The $B=\infty$ entry is the true $2E_n$ from a Chebyshev-basis LP; the LP in the $P_k$ basis could not reach it.}
\label{tab:tradeoff}
\begin{tabular}{lcccccccc}
\toprule
$B$ & 1 & 2 & 5 & 10 & 30 & 100 & 1000 & $\infty$ \\
\midrule
$2E_K(B)$ & 0.302 & 0.170 & 0.131 & 0.107 & 0.086 & 0.081 & 0.067 & 0.022 \\
\bottomrule
\end{tabular}

\end{table}

\subsection{Separated laws: numerical illustration}\label{supp:sepnum}

\begin{figure}[tbp]
\centering
\includegraphics[width=\textwidth]{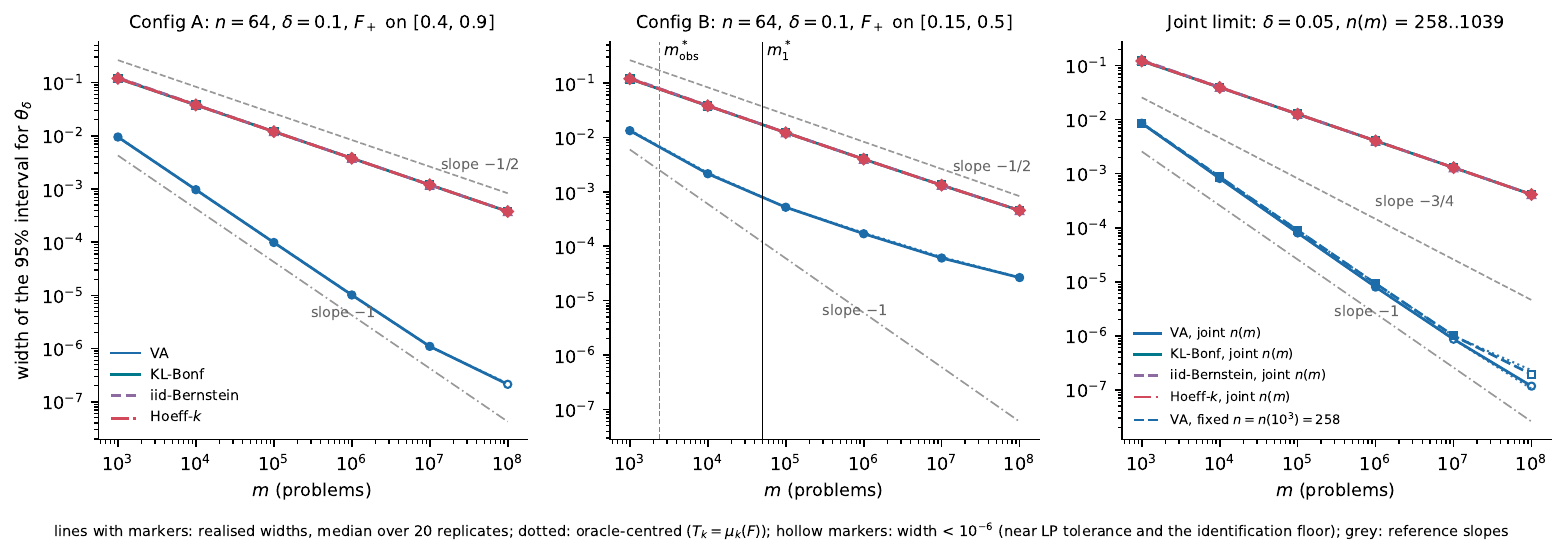}
\caption{Separated laws ($\pi_0=0.5$). Width of the 95\% interval for $\theta_\delta$ against $m$: VA, KL-Bonf, iid-Bernstein and Hoeff-$k$ (realised widths, median over 20 replicates; dotted: widths with $T_k$ replaced by $\mu_k(F)$). Left: config A, pre-crossover regime. Middle: config B, with the crossover $m^*_1$ of Corollary~\ref{M-cor:sepA} and the heuristic $m^*_{\rm obs}$; constant-factor regime beyond $m^*_1$. Right: joint path $n(m)$ and, for comparison, fixed $n=258$; both are pre-crossover in this range. Index set $K_{\rm rule}\cup\{1,2,k_*,2k_*\}$. Hollow markers: widths below $10^{-6}$.}
\label{fig:separated}
\end{figure}
Figure~\ref{fig:separated} shows realised widths, the median over 20 replicates, for three separated laws with $\pi_0=0.5$ (E8). Every region covered $\theta_\delta(F)$ in every replicate. The index set is $K_{\rm rule}\cup\{1,2,k_*,2k_*\}$, because Corollaries~\ref{M-cor:sepA} and~\ref{M-cor:sepB} need $1$, $2$, $k_*$ and its partner in $K$; $K_{\rm geo}$ lacks $2k_*$ at every $n$ on the joint path. This gives $|K|=31$ at $n=64$ (equal to $K_{\rm geo}$) and $|K|$ from 43 to 121 on the joint path. The LPs use an atomic support grid that contains the atoms of $F$, so the figure illustrates rates; it does not use the envelope grid of the other experiments, and we have not rerun these laws on the envelope grid. Realised and oracle-centred widths agree: the ratio VA/Hoeff-$k$ differs between the two by at most 0.003 at every point. Widths below $10^{-6}$ are within about ten times the LP tolerance and the identification floor, and are certified upper bounds rather than precise values. In the bound of Corollary~\ref{M-cor:sepA}, the third term is $O((L/m)^{1/4})$ relative to the leading term and dominates below the crossover $m^*_1=2L\kappa'^4/A_{\VA}^4$ defined in the proof of Corollary~\ref{M-cor:sepA}, where the VA bound behaves like $\kappa'(2L/m)^{3/4}$.

\emph{Config A ($n=64$, $\delta=0.1$, $F_+$ on $[0.4,0.9]$): pre-crossover regime.} The law is strongly separated, $m^*_1=6.3\times10^{11}$ lies far outside the plotted range, and VA's width falls with slope $-0.99$ up to $m=10^6$, against $-0.50$ for KL-Bonf, iid-Bernstein and Hoeff-$k$.

\emph{Config B ($n=64$, $\delta=0.1$, $F_+$ on $[0.15,0.5]$): beyond the crossover.} Here $A_{\VA}=0.475$ and $m^*_1=5.0\times10^4$ (Table~\ref{tab:e8-const}). Beyond the crossover VA's width falls at the competitors' rate, and VA/Hoeff-$k$ stays between 0.043 and 0.058 for $m=10^5,\dots,10^8$. Corollary~\ref{M-cor:sepA} does not explain this gap. Its ratio bound divides by the computed $\epsilon_{\rm real}/a$, which lies in $[0.4797,0.4975]$ for Hoeff-$k$ in config B (computed as in Table~\ref{tab:realised}). With $A_{\VA}=0.475$, the bound for Hoeff-$k$ is at least $0.955$--$0.990$ even before the window is imposed, so in config B the corollary guarantees no gain. The window makes it weaker still: $m_{\rm ID}(\gamma)=3008\,\gamma^2$ in config B, so $m\le m_{\rm ID}(\gamma)$ forces $\gamma\ge0.58$, $5.8$ and $182$ at $m=10^3$, $10^5$ and $10^8$, and the Hoeff-$k$ ratio bound is then at least 4.83, 13.4 and 368. In config A, $m_{\rm ID}(\gamma)=4.09\times10^7\gamma^2$; the smallest admissible $\gamma$ is 0.005, 0.05 and 1.56 at the same $m$, and the Hoeff-$k$ ratio bound is 1.62, 0.60 and 3.24, below $1$ only in the middle of the range. With the oracle-centred constant $c_{\rm comp}=\frac12$ in place of $\epsilon_{\rm real}/a$, which the hypothesis of the corollary would not deliver, these bounds would be 0.95, 4.64, 13.3 and 366 (config B) and 1.56, 0.60 and 3.22 (config A); no conclusion changes. The observed constant factor in config B is a property of the computed widths, not of the bound. What the theory does explain there is the slope: the competitors' realised widths fall like $m^{-1/2}$ with constant at least $0.48\,a$ (Table~\ref{tab:realised}), and VA's width is $O(\sqrt{L/m})$ for fixed $(n,F)$ (Theorem~\ref{M-thm:width}), so a constant-factor gap with no rate separation is what one should see beyond $m^*_1$. The bound $A_{\VA}$ is loose there because the crude certificate of Lemma~\ref{lem:certificate} charges $\sqrt{g_{k_*}}$, and $g_{k_*}(F)=0.090$ in config B through mass of $F_+$ near $k_*/n$.

\emph{Joint path ($\delta=0.05$, $F_+$ on $[0.15,0.5]$).} Along the smallest $n$ satisfying (J1)--(J2), $n(m)=258,417,560,720,879,1039$ for $m=10^3,\dots,10^8$, VA's width falls with slope $-1.01$ up to $m=10^6$, against $-0.49$ for the competitors. At fixed $n=258$, however, VA's width also falls with slope $-0.99$ over the same range. For that law $m^*_1=7.6\times10^9$, far beyond $10^8$, as the formula for $m^*_1$ predicts. So in the plotted range the figure does not separate the joint-limit rate of Corollary~\ref{M-cor:sepB} from the fixed-$n$ pre-crossover regime of Corollary~\ref{M-cor:sepA}. What it shows is the pre-crossover $1/m$ regime (config A and the joint path) and a constant-factor gap once the crossover has passed (config B).

\paragraph{Constants and widths.} Tables~\ref{tab:e8-const} and~\ref{tab:e8} give the constants and widths behind Figure~\ref{fig:separated}. The VA tangent set is $\mathrm{geomspace}(10^{-7},0.5,28)$ ($\kappa=1.041$) plus refinement, and $K$ is the default set together with $\{1,2,k_*,2k_*\}$. An independent implementation of the oracle-centred widths, which does not use our LP code, agrees with it on configs A and B: the ratio of the two is 0.999993--1.000000 for the competitors and 0.9904--1.0036 for VA (CHK-6, \texttt{checks/chk6\_e8\_crosscheck.py}). A dash marks Berk--Jones, which was run only for $m\le10^5$.

\begin{table}[tbp]
\centering
\small
\caption{Constants of Corollary~\ref{M-cor:sepA} for the E8 laws ($t_{\min}=10^{-7}$). $m^*_{\rm obs}=2L/V_{k_*}(F)$ is a heuristic crossover. Row J is the joint-path law at $n=258$.}
\label{tab:e8-const}
\begin{adjustbox}{max width=\textwidth}
\begin{tabular}{lrrrrrrrrrr}
\toprule
Config & $n$ & $\delta$ & $|K|$ & $L$ & $\kappa$ & $A_{\rm VA}$ & $\kappa'$ & $m^*_1$ & $C_{\rm ID}$ & $m^*_{\rm obs}$ \\
\midrule
A & 64 & 0.10 & 31 & 7.123 & 1.041 & 0.0049 & 2.229 & 6.3e+11 & 0.00059 & 1e+10 \\
B & 64 & 0.10 & 31 & 7.123 & 1.041 & 0.4751 & 3.664 & 5e+04 & 0.069 & 2.4e+03 \\
J & 258 & 0.05 & 43 & 7.450 & 1.041 & 0.0156 & 2.343 & 7.6e+09 & 9.5e-07 & 5.4e+09 \\
\bottomrule
\end{tabular}
\end{adjustbox}
\end{table}

\begin{table}[tbp]
\centering
\scriptsize
\caption{E8 (Figure~\ref{fig:separated}): separated laws, $\pi_0=0.5$. Configs A and B: $n=64$, $\delta=0.1$, $F_+$ on $[0.4,0.9]$ and $[0.15,0.5]$. J: $\delta=0.05$, $F_+$ on $[0.15,0.5]$, along the joint path $n(m)$ and at fixed $n=258$. Median realised width over 20 replicates; point-value LPs on an atomic support grid containing the atoms of $F$. $^\circ$Width below $10^{-6}$: within about ten times the LP tolerance and the identification floor, a certified upper bound rather than a precise value.}
\label{tab:e8}
\begin{adjustbox}{max width=\textwidth}
\begin{tabular}{llrrcccccc}
\toprule
Panel & Config & $n$ & $m$ & VA & KL-Bonf & VA-iid & Hoeff-$k$ & VA/Hoeff-$k$ (realised) & (oracle) \\
\midrule
joint $n(m)$ & J & 258 & $10^{3}$ & 0.00858 & 0.122 & 0.126 & 0.122 & 0.0703 & 0.0704 \\
joint $n(m)$ & J & 417 & $10^{4}$ & 0.000815 & 0.0394 & 0.0399 & 0.0394 & 0.0207 & 0.0208 \\
joint $n(m)$ & J & 560 & $10^{5}$ & 8e-05 & 0.0126 & 0.0127 & 0.0126 & 0.0063 & 0.0063 \\
joint $n(m)$ & J & 720 & $10^{6}$ & 7.95e-06 & 0.00404 & 0.00405 & 0.00404 & 0.0020 & 0.0020 \\
joint $n(m)$ & J & 879 & $10^{7}$ & 8.63e-07$^\circ$ & 0.00129 & 0.00129 & 0.00129 & 0.0007 & 0.0007 \\
joint $n(m)$ & J & 1039 & $10^{8}$ & 1.17e-07$^\circ$ & 0.000412 & 0.000412 & 0.000412 & 0.0003 & 0.0003 \\
\midrule
fixed $n=258$ & J & 258 & $10^{4}$ & 0.000872 & 0.0386 & 0.0391 & 0.0386 & 0.0226 & 0.0228 \\
fixed $n=258$ & J & 258 & $10^{5}$ & 8.94e-05 & 0.0122 & 0.0123 & 0.0122 & 0.0073 & 0.0074 \\
fixed $n=258$ & J & 258 & $10^{6}$ & 9.23e-06 & 0.00386 & 0.00387 & 0.00386 & 0.0024 & 0.0024 \\
fixed $n=258$ & J & 258 & $10^{7}$ & 1e-06 & 0.00122 & 0.00122 & 0.00122 & 0.0008 & 0.0008 \\
fixed $n=258$ & J & 258 & $10^{8}$ & 1.94e-07$^\circ$ & 0.000386 & 0.000386 & 0.000386 & 0.0005 & 0.0006 \\
\midrule
fixed $n$ & A & 64 & $10^{3}$ & 0.00946 & 0.119 & 0.123 & 0.119 & 0.0792 & 0.0796 \\
fixed $n$ & A & 64 & $10^{4}$ & 0.000968 & 0.0378 & 0.0382 & 0.0378 & 0.0256 & 0.0257 \\
fixed $n$ & A & 64 & $10^{5}$ & 9.87e-05 & 0.0119 & 0.012 & 0.0119 & 0.0083 & 0.0083 \\
fixed $n$ & A & 64 & $10^{6}$ & 1.02e-05 & 0.00378 & 0.00378 & 0.00378 & 0.0027 & 0.0027 \\
fixed $n$ & A & 64 & $10^{7}$ & 1.1e-06 & 0.00119 & 0.0012 & 0.00119 & 0.0009 & 0.0009 \\
fixed $n$ & A & 64 & $10^{8}$ & 2.14e-07$^\circ$ & 0.000378 & 0.000378 & 0.000378 & 0.0006 & 0.0006 \\
\midrule
fixed $n$ & B & 64 & $10^{3}$ & 0.0133 & 0.119 & 0.123 & 0.12 & 0.1113 & 0.1139 \\
fixed $n$ & B & 64 & $10^{4}$ & 0.00215 & 0.038 & 0.0385 & 0.0381 & 0.0565 & 0.0587 \\
fixed $n$ & B & 64 & $10^{5}$ & 0.000521 & 0.0122 & 0.0123 & 0.0122 & 0.0426 & 0.0426 \\
fixed $n$ & B & 64 & $10^{6}$ & 0.00017 & 0.00398 & 0.00399 & 0.00398 & 0.0426 & 0.0445 \\
fixed $n$ & B & 64 & $10^{7}$ & 6.07e-05 & 0.00132 & 0.00133 & 0.00133 & 0.0458 & 0.0479 \\
fixed $n$ & B & 64 & $10^{8}$ & 2.66e-05 & 0.000455 & 0.000455 & 0.000455 & 0.0583 & 0.0590 \\
\bottomrule
\end{tabular}
\end{adjustbox}
\end{table}

\begin{table}[tbp]
\centering\small
\caption{Numerical check of Proposition~\ref{prop:compexcess} (CHK-2, \texttt{checks/chk2\_prop56.py}): ratio of (oracle-centred width $-\ID_K(F)$) to the lower bound over three separated laws and $m=10^3,\dots,10^7$ (45 cases, no violation).}
\label{tab:propcheck}
\begin{tabular}{lccc}
\toprule
Region & cases & ratio range & violations (ratio $<1$) \\
\midrule
KL & 15 & 4.00--8.22 & 0 \\
iid-Bernstein & 15 & 2.89--5.81 & 0 \\
Hoeffding & 15 & 2.01--4.04 & 0 \\
\bottomrule
\end{tabular}

\end{table}

\paragraph{Spike plus Beta.}\label{supp:spikebeta}
Let $F=\pi_0\delta_0+(1-\pi_0)\mathrm{Beta}(a_0,b_0)$ and $\mathrm B$ the Beta function. Since $u(P_1)=(1-p)^n-(1-p)^{2n}$,
\[
V_1=(1-\pi_0)\frac{\mathrm B(a_0,b_0+n)-\mathrm B(a_0,b_0+2n)}{\mathrm B(a_0,b_0)},\qquad q_1=(1-\pi_0)\Big\{1-\frac{\mathrm B(a_0,b_0+n)}{\mathrm B(a_0,b_0)}\Big\}.
\]
As $n\to\infty$, $\mathrm B(a_0,b_0+n)/\mathrm B(a_0,b_0)\sim\Gamma(a_0+b_0)\Gamma(b_0)^{-1}n^{-a_0}$, so
\[
\sqrt{\frac{V_1}{q_1(1-q_1)}}\sim\Big\{\frac{(1-2^{-a_0})\Gamma(a_0+b_0)}{\pi_0\Gamma(b_0)}\Big\}^{1/2}n^{-a_0/2}.
\]
The $k=1$ gain grows like $n^{-a_0/2}$, where $a_0$ is the exponent of the density at $0^+$. For F1 at $n=128$ the formula gives $V_1=0.0319$, matching the numerical value $0.03194$ from quadrature (CHK-8, \texttt{checks/chk8\_spikebeta\_V1.py}).

\section{Data provenance and reproducibility}\label{supp:data}

\paragraph{Monkey Business.} Hugging Face dataset \texttt{ScalingIntelligence/monkey\_business}, parquet branch \texttt{refs/convert/parquet}, revision recorded in \texttt{results/data/monkey\_business/manifest.json}. Only the \texttt{is\_corrects} field is read. The dataset is MIT-licensed; the terms for the underlying problems and model outputs may differ.

\paragraph{Counts for the distilled and RL models.} \texttt{github.com/cafferychen777/passk-inference} (MIT), commit \texttt{ac7c3fc58d5e325447607e778bdcbc897182cf7e} (2026-09-23), files \texttt{examples/deepscaler32k/\{base,rl\}.jsonl}; $m=1060$, $n=128$, identical problem identifiers in both files.

\paragraph{Seeds.} $F_m$ draws use seed $[20261112,F,m]$; counts $[20261112,7,F,m,n,\mathrm{rep}]$; Monkey Business subsamples $[20261113,\mathrm{config},n,\mathrm{rep}]$; Berk--Jones thresholds $[20261112,m]$; E5 $[20261114,\dots]$; E7 replicate offset 5000.

\paragraph{Commands} (from \texttt{code/}, with the pinned environment):
\begin{verbatim}
.venv/bin/python -m pytest -q tests/
.venv/bin/python experiments/e1_simulation.py run --workers 18
.venv/bin/python experiments/e2_monkey_business.py run --workers 6
.venv/bin/python experiments/e3_chen_yang.py run --workers 5
.venv/bin/python experiments/e5_band_mc.py run --workers 24
.venv/bin/python experiments/e6_K_sensitivity.py run --workers 24
.venv/bin/python experiments/e7_amari_hiR.py run --workers 24
.venv/bin/python experiments/e8_separated.py run
.venv/bin/python experiments/make_paper_tables.py   # paper/tables/*.tex
.venv/bin/python experiments/make_paper_figs.py     # results/figs/fig_*.pdf
\end{verbatim}
These commands produce the preliminary runs; the preliminary with-replacement Monkey Business run is \texttt{experiments/e2\_monkey\_business.py run --wr --K geo16}. Numerical checks: \texttt{checks/chk1\_dkw\_EZ.py} to \texttt{checks/chk8\_spikebeta\_V1.py}, and CHK-9 as described in Section~\ref{supp:comparators}. Main runs (each resumable from its cache):
\begin{verbatim}
.venv/bin/python experiments/e1s4_sweep.py run --workers 22
.venv/bin/python experiments/e1s4_sweep.py summarize
.venv/bin/python experiments/e9_amari_s4.py run
.venv/bin/python experiments/e5_band_mc_s4.py run --workers 2
.venv/bin/python experiments/e2_monkey_business.py run --wr --K geo16 \
        --s4 --workers 22
.venv/bin/python experiments/e3_chen_yang.py run --s4
.venv/bin/python experiments/glevel_leb_s4.py run --workers 2
.venv/bin/python experiments/glevel_s4.py
.venv/bin/python experiments/design_rule_s4.py run
.venv/bin/python experiments/oracle_calib_s4.py run --workers 2 --reps 10
.venv/bin/python experiments/mc_errors.py --raw ../results/e1s4_raw.csv \
        --keys F,m,n,target --out ../results/e1s4_mc
.venv/bin/python experiments/verdict_s4.py
.venv/bin/python experiments/make_paper_tables_s4.py   # paper/tables/s4_*.tex
.venv/bin/python experiments/make_paper_figs.py --s4
.venv/bin/python experiments/e3_bracket_source.py --workers 8
.venv/bin/python experiments/make_paper_tables_e3extra.py
\end{verbatim}
The second-to-last group writes the tables of the main text and the Supplement; \texttt{make\_paper\_figs.py --s4} draws Figure~\ref{M-fig:forest} of the main text. The exact options of each script are in the release README. The release is at \url{https://github.com/YYKKCC-kevin/valoc} (MIT licence).

\bibliographystyle{plainnat-nonote}
\bibliography{refs}